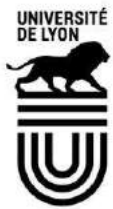 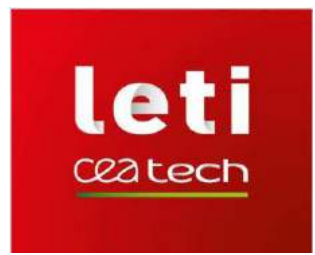

N°d'ordre NNT : 2017LYSEI019

## PhD THESIS of the UNIVERSITY of LYON
performed at
**The Commission for Atomic Energy and Alternative Energies**

**Doctoral School** N° 160
**Automation, Electrical Engineering, Signal and Image Processing, Electronics, Micro and Nanoelectronics, Optics and Laser, Engineering for Living**

**Specialty / Ph.D. discipline:**
Electronics, micro and nano-electronics, optics and laser

Defended publicly on 10 / 03 / 2017, by:
**Ainur Koshkinbayeva**

# New photonic architectures for mid-infrared gas sensors integrated on silicon

In front of the jury composed of:

| | |
|---|---|
| Pr. Ségolène Callard, Professeur des Universités, ECL de Lyon, Institut des Nanotechnologies de Lyon | President |
| Dr. Laurent Vivien, Directeur de Recherches CNRS, Institut d'Electronique Fondamentale | Rapporteur |
| Pr. Jean-Emmanuel Broquin, Professeur des Universités INPG, Institut de Microélectronique Electromagnétisme et Photonique | Rapporteur |
| Pr. Eric Tournié, Professeur des Universités de Montpellier, Institut d'Electronique et des Systèmes | Examiner |
| Dr. Régis Orobtchouk, Maître de conférences de l'INSA de Lyon, Institut des Nanotechnologies de Lyon | Director of thesis |
| Dr. Labeye Pierre, Chef de projet, Commissariat à l'énergie atomique et aux énergies alternatives | Coordinator of thesis |

# Publications

1. A. Koshkinbayeva, R. Orobtchouk, M. Brun, M. Carras, and P. Labeye, "Broad-band Source for Optical Gas Sensing at 5.6 — 5.9 µm," in 2015 European Conference on Lasers and Electro-Optics - European Quantum Electronics Conference (2015), paper CH-3.5, 2015, p. CH-3.5.

2. A. Koshkinbayeva, P. Barritault, S. Ortiz, S. Boutami, j m hartmann, P. Brianceau, O. Lartigue, R. Orobtchouk, M. Brun, F. Boulila, and P. Labeye, "Impact of non-central input in NxM mid-IR arrayed waveguide gratings integrated on Si ," IEEE Photonics Technology Letters, vol. PP, no. 99, pp. 1–1, 2016.

3. A. Koshkinbayeva, R. Orobtchouk, P. Labeye, "Phase Errors in Arrayed Waveguide Gratings based on SiGe/Si," in process of preparation.

New photonic architectures for mid-infrared gas sensors integrated on silicon

# Acknowledgements

Firstly, I would like to express my sincere gratitude to my advisors Dr. Pierre Labeye and Prof. Régis Orobtchouk for granting me their support and guidance in pursue of my doctoral degree. I thank Dr. Pierre Labeye for setting high standards, patience while I was discovering new ways to look at the problem, and pushing me to grow both as a professional and as a personality. I thank Prof. Régis Orobtchouk for continuous support of my Ph.D. study and related research, for his patience, motivation, and immense knowledge. I highly appreciate the time and effort that were granted to me at every stage, from solving numerous administrative matters to developing analytical model.

I would like to thank Dr. Laurent Vivien, Prof. Jean-Emmanuel Brouquin, Prof. Segolene Callard and Prof. Eric Tournié for kindly accepting to be members of the dissertation committee. I would like especially express my appreciation of the advices and suggestions given in the reviews by Dr. Laurent Vivien and Prof. Jean-Emmanuel Brouquin.

My stay in Grenoble would not have been that much fulfilling without Ph.D. fellows of the DOPT department, Justin Rouxel, Cédric Durantin, Hélène Duprez, Anthony Lefebvre, Julien Favreau and Shayma Bouanani, who have created a friendly atmosphere helping me to integrate into a new environment and improve my French.

This work would have lacked an important part without devices fabricated by Silicon Technology Department team from CEA-Leti. I would like to especially thank Sophie Ortiz, for fabricating high-quality devices and thus enabling the verification of the analytical model, and Stéphanie Garcia for arranging and taking high resolution SEM photos of the samples. I thank Emerick Lorent for thorough guidance in understanding the nuances of fabrication process which was of great help in analysis of characterization results.

I am deeply grateful to Olivier Lartigue and Pierre Barritault, for making the experimental verification of the device function possible. I appreciate all the time and effort Olivier Lartigue invested into preparation of the test bench and his guidance. I thank Pierre Barritault for providing the home-made tool for analyzing the measured spectra that allowed to implement rapidly the complete analysis of the experimental data including the observation of second-order spectral shift of AWGs which was later theoretically explained and presented in Photonics Technology Letters. I would particularly like to commend Fahem Boulila for assistance in conduction measurements of AWGs' spectra.

My sincere gratitude goes to all colleagues from LCNA. In particular, I would like to thank Jean-Marc Fedeli for a fascinating clean room tour and, especially, for constant support during the three years, and Serge Gidon for fruitful discussions. I am grateful to head of the laboratory, Laurent Duraffourg, for supporting my trainings on the work with lasers and courses of French.

I am sincerely grateful to my friends, whom I met in Grenoble, for enormous support, Ekaterina Pavlenko, Vasily Tarnopolskiy, Irina Profatilova, Olena Kraieva, Elena Zvereva, Andriy Yelisyeyev, Dilyara Timerkaeva and Anna Mukhtarova.

My journey would not have been possible without the support of National Laboratory Astana in Kazakhstan, whom I thank for granting me a scholarship to pursue my doctoral studies.

I would like to express my deepest appreciation to my parents, siblings for infinite source of love and support that inspired me during these three years. Finally, I dedicate the words of highest gratitude to my beloved, friend and husband, Zhanat Kappassov, who has embarked on this ambitious and challenging path with me, sharing all the sweet and bitter moments in this journey.

# Résumé

**Table des matières**

## Introduction

Le besoin en détecteurs de gaz à l'heure actuelle couvre un grand nombre de domaines. Les dangers potentiels inhérents aux gaz, comme la combustion, l'inflammabilité, la toxicité ou la diminution en oxygène sont à l'origine d'une grande activité de contrôle allant de la détection des émissions industrielles jusqu'aux analyses respiratoires dans le domaine de la santé.

Les détecteurs de gaz optiques ont de nombreux avantages comparés aux autres types de capteurs, comme leurs grandes sensibilités et sélectivité, leur rapidité qui permet de réaliser des analyses en temps réels, leur immunité aux perturbations extérieures, ainsi que leur possibilité de détecter plusieurs espèces dans les analyses chromatographiques. Leurs principaux inconvénients sont les grandes dimensions des appareillages et leurs coûts. Dans ces conditions, il est important de développer de nouvelles solutions permettant de miniaturiser les systèmes de détection optique tout en conservant leurs principales spécificités telles que la détection sélective, la faible puissance consommée et la fiabilité. Les études récentes de ce dernier quart de siècle ont porté sur l'intégration et la miniaturisation de détecteurs de gaz sur une puce. La brique de base clef d'un système de détection optique est la source optique large bande dans le domaine du moyen infra-rouge. Son intégration peut être réalisée grâce au récent développement de sources lasers à puits quantiques cascadés miniaturisés qui sont hybridés sur des multiplexeurs optiques passifs. Les travaux de recherches de cette thèse sont consacrés aux développement de nouvelles architectures de multiplexeurs utilisant une plateforme d'intégration sur substrats de silicium dans le domaine de la détection de gaz par voie optique à des longueurs d'onde du moyen infra-rouge.

# 1 Etat de l'art

## 1.1 Principe de fonctionnement des détecteurs de gaz optiques

Le principe de fonctionnement d'un détecteur de gaz utilise l'absorption spécifique des molécules de gaz liée aux transitions énergétiques des niveaux électriques et des modes de vibration les liaisons atomiques. La présence ou la quantité de gaz peut être détectée ou quantifié en éclairant l'échantillon et mesurant la quantité de lumière transmise en fonction de la longueur d'onde. Une absorption spécifique se traduira par un creux dans le signal de transmission. La position en longueur d'onde de ce creux et son amplitude fournissent des informations respectivement sur la nature du gaz et sa concentration. Les niveaux des transitions énergétiques étant très différents suivant la nature des liaisons et niveaux électroniques des molécules, le domaine spectrale des sources utilisée doit être adapté à ce que l'on cherche a étudier [2]. Le moyen infra-rouge, qui couvre les longueurs d'onde allant de 2,5 à 14 µm, permet de détecter la majorité des modes de vibrations et de rotations des molécules [3] de façon précise et est à l'origine de la technique de spectroscopie infra-rouge pour la détermination de la composition chimique d'un échantillon inconnu.

Dans la spectroscopie infra-rouge, le gaz et sa concentration sont reliés à l'absorption par la loi de Lambert-Beer :

$$P_1 = P_0 \exp(-\alpha L), \qquad (1.1)$$

où $P_0$ est la puissance du faisceau incident, $P_t$ est la puissance transmise, α est le coefficient d'absorption et L l'épaisseur de l'échantillon.

Le coefficient $\alpha$ est définit comme :

$$\alpha = c \cdot \varepsilon, \quad (1.2)$$

où c est la concentration en gaz (atm) et ε est l'absorption spécifique du gaz (atm.cm$^{-1}$).

L'utilisation de la loi de Lambert-Beer dans les mesures de spectre d'absorption et la comparaison avec la base de données HITRAN [3] permet à la fois de déterminer la présence de différentes espèces de gaz et leurs concentrations.

## 1.2  Détecteurs de gaz sur puce

Les différentes briques de base d'une cellule d'analyse de gaz et leur agencement sur une puce est représenté sur la figure 1.1. Une série de sources lasers sont couplées à des guides d'onde individuels. Ces guides d'ondes sont raccordés à un seul guide par l'intermédiaire d'un multiplexeur pour constituer un signal large bande. Ce signal constitue le faisceau incident de la cellule d'analyse de gaz. Le photo-détecteur va permettre de mesurer le spectre d'absorption.

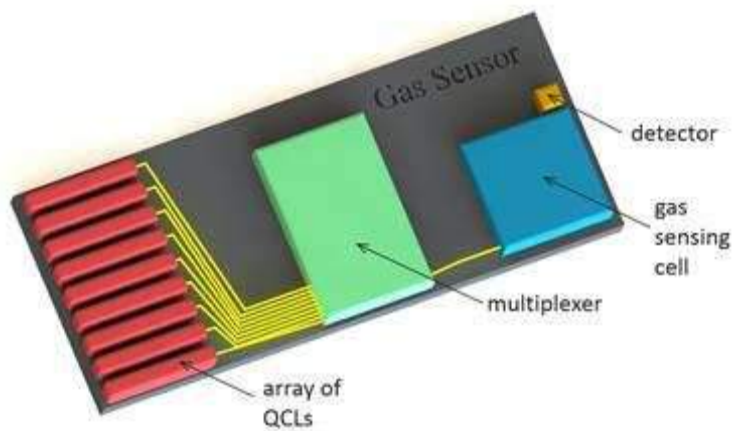

Fig 1.1 Schéma d'une cellule d'analyse de gaz intégrée sur une puce.

L'intégration de l'ensemble des fonctions optiques sur une puce a fait l'objet d'un grand nombre de travaux de recherches ces vingt dernières années [10]-[17]. Les travaux les plus récents dans le domaine de la détection de gaz ont été réalisés par Stanton et al. [17]. Leur plateform intègre une source large bande qui couvre une large gamme spectrale allant de l'ultra-violet au moyen infra-rouge déclinée en deux configurations, 350 - 1500 nm et 1500- 6500 nm. La puce est une hybridation de guides en nitrure de silicium et en silicium sur isolant. Elle contient 4 AWGs pour adresser une large gamme spectrale. Les résultats expérimentaux de cette technologie ont permis d'obtenir une transmission de 0.9 sur une gamme spectrale comprise entre 780 et 3600 nm. Ces premiers résultats montrent que l'intégration de sources larges bandes efficaces adressant l'ensemble du spectre moyen infra-rouge pour la réalisation de systèmes d'analyse de gaz sur puce est envisageable.

## 1.3  Guides d'ondes SiGe

Avec l'augmentation des fonctionnalités et la réduction des composants d'optique intégrée, un grand nombre de technologies ont été développées ces trente dernières années dans le but de diminuer le prix des éventuels produits commerciaux. La possibilité d'utilisé du matériau SiGe pour la réalisation de circuits en micro-électronique a été envisagée dans les années 1950 [27]. La fabrication et la caractérisation de guides SiGe fonctionnant à différentes longueurs d'onde dans le moyen infra-rouge c'est quant à elle fortement développée ces cinq dernières années. Les performances des principales réalisations sont résumées dans le tableau 1.1. Les guides ont été réalisés par un dépôt chimique en phase vapeur du matériau sur un substrat de silicium (CVD pour Chemical Vapor Deposition) et une gravure ionique réactive (RIE pour Reactive Ionique Etching) pour la définition des guides d'onde optique. L'état de l'art en termes de niveau de pertes autour de la longueur d'onde de 3,7 µm est respectivement de 7 – 8 dB/cm et de 0,6 dB/cm respectivement pour des configurations de guides partiellement gravés monomodes et multimodes. Les guides partiellement gravés sont communément appelés guides rib dans la litérature. Dans le cas d'une longueur d'onde plus élevée de 5 µm, les pertes de propagations chutent à 2 – 3 dB/cm quel que soit la nature du guide.

| Waveguide | Reported by / in | Propagation loss | Operational λ | Modal property |
|---|---|---|---|---|
| Air/Ge/Si | Chang et al. 2012 | 2.5 dB/cm (TM) | 5.8 µm | single mode |
| Si/Si$_{0.72}$Ge$_{0.28}$/Si | Kim et al. 2013 | 5.4 dB/cm (TE) | 1.55 µm | single mode |
| Si/Ge/Si | Malik et al. 2013 | 7 dB/cm / 3dB/cm (TE) | 3.7 – 3.8 µm / 5.2 – 5.4 µm | single mode |
| Air/Ge/Si | Nedeljkovic et al. 2015 | 0.6 dB/cm (TE) | 3.8 µm | multi mode |
| InGaAs/InP | Gilles et al. 2015 | 2.9 dB/cm (quasi TM) | 7.4 µm | single mode |
| Ge/SOI | Younis et al. 2016 | 8 dB/cm (TE) | 3.68 µm | single mode |
| Si/Si$_x$Ge$_{0.4-x}$/Si | Brun et al. 2014 | 1 dB/cm / 2 dB/cm (TM) | 4.5 µm / 7.4 µm | single mode |

Tab. 1.1. Mid-infrared waveguides.

## 1.4  Systèmes de multiplexage envisageables

Le besoin d'intégration à la fois de fonctions actives et passives sur une même puce est motivé par l'accroissement constant de la demande en moyen de communication qui requiert une augmentation constante des capacités des systèmes de transmission par voies optiques. L'énorme avantage des systèmes optiques face aux supports électriques est le multiplexage en longueur d'onde qui fait qu'avec la complexité des circuits, l'optique occupe une place de plus en plus importante dans les systèmes de communication et a supplanté les autres solutions jusqu'aux connections puce à puce de nos jours. Les différentes architectures utilisables sont les réseaux de diffraction, les faisceaux de guides couplés (AWG pour Arrayed Waveguide Grating) (Fig. 1.2), les réseaux de diffraction planaires à échelle concave (Fig. 1.3), les matrices de résonateurs en anneaux (Fig. 1.4) et les imageurs multiples utilisant un guide multimode avec des sorties déportées (Fig. 1.5). Quel que soit l'architecture envisagée, la fonction optique de base à satisfaire reste la même, adresser différentes longueurs d'onde dans différents canaux de transmission. Le principe de fonctionnement des deux solutions étudié dans ce travail de thèse utilisant des AWG et des réseaux de diffraction planaires concaves pour disperser la lumière seront décrites en détail dans les chapitres suivants.

*AWG.* Le multiplexeur AWG a été inventé en 1988 par Smit [31]. Son concept a été démontré expérimentalement par Vellekoop et Smit [32]. Takahashi et coll. ont obtenu un composent avec un espacement entre canaux de 1 nm [33] compatible avec les normes des télécommunications optiques. Une étape a ensuite été franchie par Dragon [34] qui a étendu le concept à des dispositifs de multiplexage de N vers N. Les performances des AWG ont été améliorées par Adar et coll. [35] qui ont introduit une asymétrie dans le routage permettant d'élargir la bande spectrale et de diminuer le couplage entre les voies de sortie. Amersfoot a ensuite réalisé un AWG sur substrat InP avec une réponse apodisée se rapprochant d'un créneau en ajoutant des coupleurs MMIs sur les guides d'entrée en dépit d'une pertes de transmission de 3,5 dB (augmentation de 3 dB) et en intégrant également le dispositif avec des photo-détecteurs [36]. Le niveau de seuil du au couplage entre canaux est de 12 dB. Les premiers AWGs étaient fabriqués dans des technologies utilisant des substrats de silice ou d'InP principalement parce qu'ils étaient destinés à des applications dans le domaine des télécommunications optiques [32 – 36]. C'est plus récemment que le domaine spectral s'est déplacé vers le moyen infra-rouge. Des AWG utilisant une technologie de guide SiGe/Si à gradient d'indice de réfraction travaillant autour d'une longueur d'onde de 4,5 μm ont été démontré par P. Barritault et Coll. au CEA-LETI [40]. L'IMEC a également démontré un dispositif à 6 canaux travaillant à 3,8 μm sur substrat silicium sur isolant [38] et un AWG à 5 canaux utilisant des guides germanium sur silicium pour atteindre une longueur d'onde de fonctionnement de 5,3 μm [39].

*PCG.* Les premier PCG (Planar Concave Grating) ont été fabriqués pour des applications aux télécommunications optiques. Ils étaient constitués de composants massifs et de fibres optiques [30]. De tels montages optiques étaient difficiles à obtenir du fait des faibles tolérances d'alignement de ces systèmes. C'est en 1970 que Tomlinson [48] a obtenu le premier dispositif monolithique pour lequel le réseau de diffraction miniaturisé était utilisé conjointement avec un prisme pour obtenir une variation de l'angle d'incidence. Le couplage avec les fibres optiques d'entrée et de sortie était réalisé par l'intermédiaire de lentilles à gradient d'indice pour assurer la collimation. Tangonan et Coll. ont été les premier a réaliser un composant en optique intégrée en configuration de Rowland [49]. Avec les premières réalisations de guides dans le moyen infra-rouge, il est possible maintenant de concevoir des réseaux de diffraction dédiés aux applications de détection de gaz. L'IMEC a réalisé des premiers dispositifs fonctionnant à 3,8 μm en silicium sur isolant et à 5,25 μm en guide germanium sur silicium respectivement par Muneb et Coll. [38] et Malik et Coll. [56].

*Configurations alternatives.* Les configurations alternatives aux AWGs et PCGs sont les matrices de résonateurs en anneaux et les MMIs à sorties déportées. Des multiplexeurs utilisant des résonateurs en anneaux pour les télécommunications optiques ont été réalisés au CEA [65] et dans d'autres laboratoires [66]. Dans le moyen infra-rouge, une première réalisation utilisant un effet de Vernier avec des résonateurs de type stade a été obtenue par Tria et Coll. [67] pour des longueurs d'ondes comprises entre 3,7 et 3,8 µm. Cette configuration peut être adaptée pour du multiplexage multi-canaux avec un encombrement avec un encombrement relativement réduit à comparer aux AWGs et PCGs lorsqu'on utilise des guides à forts contrastes d'indice de réfraction.

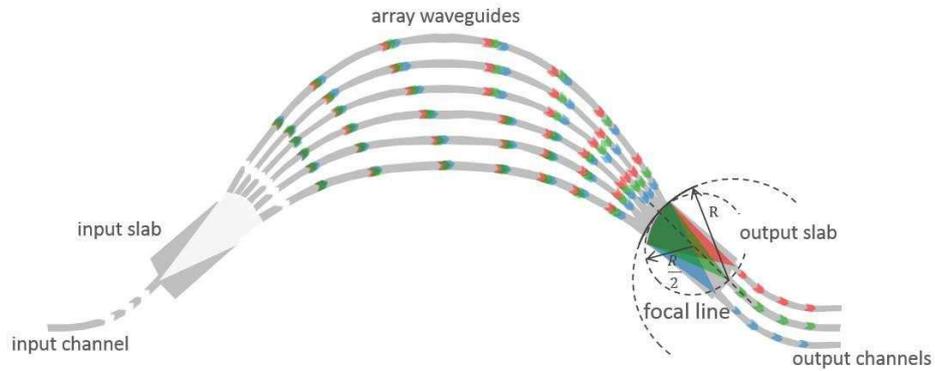

Fig. 1.2. Schéma d'un AWG.

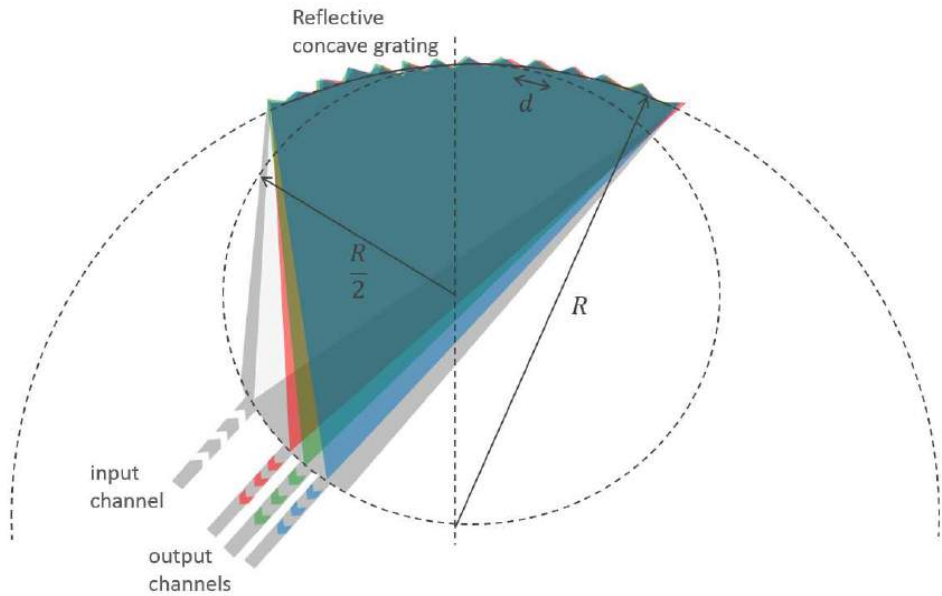

Fig. 1.3. Schéma d'un PCG.

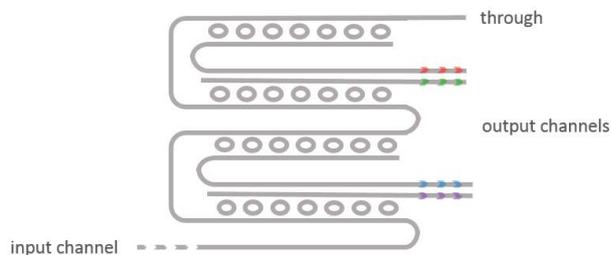    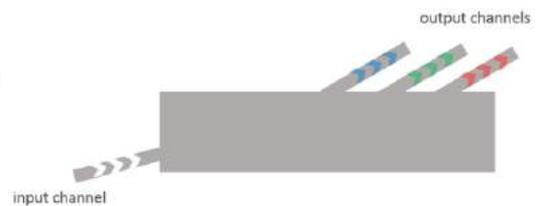

Fig. 1.4. Schéma d'un multiplexeur RR.    Fig. 1.5. Schéma d'un multiplexeur MMI asymétrique.

| Lab | year | Type | Technology | Δλ | $N_{ch}$ |
|---|---|---|---|---|---|
| IMEC | 2013 | AWG | Si/SiO$_2$ | 3.75 - 3.80 µm | 6 |
| IMEC | 2013 | PCG | Si/SiO$_2$ | 3.77 - 3.80 µm | 8 |
| IMEC | 2013 | AWG | Ge/Si | 5.15 – 5.4 µm | 5 |
| IMEC | 2013 | PCG | Ge/Si | 5.14 – 5.32 µm | 6 |
| ORC | 2014 | MMI | Ge/Si | 3.73 - 3.81 µm | 6 |
| US | 2014 | mRR | Si/SiO$_2$ | 3.72-3.80 µm | 2 |
| CEA | 2015 | AWG | SiGe/Si | 4.37-4.58 µm | 35 |

Tab. 1.2. Multiplexers operating in mid-IR.

Le principe de fonctionnement des multiplexeurs de type MMI à sortie déportée utilise l'effet de dispersion des images multiples dans une section de guide d'onde multimode. Cette approche est relativement simple à modéliser et à réaliser. Quoi qu'il en soit, la qualité des images se détériore avec les erreurs de phase lorsque des canaux augmentent. Une première démonstration de ce nouveau concept dans l'infra-rouge a été faite par Hu et Coll. [68] qui présente un MMI à 6 canaux aux longueurs d'onde de fonctionnement comprises entre 3,73 et 3,81 µm. La dernière configuration est très récente et nécessite une étude préliminaire de faisabilité pour adapter ce nouveau concept dans le domaine du moyen infra-rouge et pour un fonctionnement large bande. Les performances des différents multiplexeurs travaillant dans le moyen infra-rouge sont répertoriées dans le tableau 1.2.

## 2   Théorie

Les notions nécessaires à la compréhension du fonctionnement des multiplexeurs AWG et PCG, ainsi que leur modélisation seront exposées dans ce chapitre. Pour chaque configuration, les paramètres géométriques seront déterminés et leurs performances spectrales seront discutées. La méthode de calcul du champ électromagnétique utilisé lors de ces travaux utilise une approximation gaussienne pour la représentation des modes qui se propagent dans les guides et dans les parties planaires des dispositifs et une transformée de Fourier optique.

L'optimisation des réponses spectrales des AWGs utilisant des coupleurs MMI sur les guides de sortie pour leur apodisation et des élargissements aux deux extrémités des réseaux de guides pour augmenter le couplage afin d'accroître la transmission du dispositif. Une analyse de l'erreur de phase sera également discutée. Cette étude a pu être réalisée grâce à l'introduction dans le modèle du concept de dérive standard de l'erreur de phase. Les répercussions des approximations faites sur la description de la puissance guidée ont été quantifiées pour les AWGs. Les variations des réponses spectrales en fonction de la température seront également évaluées théoriquement. Cette dernière étude sur les AWGs utilise une approximation empirique développé par Li [84].

Les géométries des PCGs fonctionnant en configuration de Rowland seront exposées en détail. L'optimisation des réponses spectrales de ces dispositifs repose sur l'ajustement des positions des facettes des réseaux de diffraction concaves. Les erreurs de phase induites par la résolution du masque de lithographie seront également discutées.

## 2.1 Multiplexeurs de type AWG

Un AWG (Arrayed Waveguide Grating) est composé de canaux d'entrée/sortie à ces extrémités. Le signal de chaque canal est couplé à un réseau de guides par l'intermédiaire d'un coupleur planaire. Les guides du réseau ont des longueurs différentes, ce qui crée un déphasage entre les signaux lors de la propagation. Ce déphasage est exploité en sortie du dispositif dans le deuxième coupleur planaire. Les interférences ainsi créées vont produire un décalage angulaire du signal de sortie qui est variable en fonction de la longueur d'onde et qui va permettre de séparer les longueurs d'ondes dans les guides de sortie. Le composant AWG représenté sur la figure 2.1 exploite ce principe de façon réciproque. Il est constitué de plusieurs guides d'entrée pour pouvoir multiplexer les n sources en une seule sortie pour la réalisation de la source large bande.

Les conditions pour obtenir des interférences constructives sont données par la relation :

$$d\beta_s \sin\theta_{in} + \beta_e \Delta L + d\beta_s \sin\theta_{out} = 2\pi m \tag{2.1}$$

où $d$ est l'espacement entre les guides au niveau du coupleur planaire, $\beta_e = 2\pi n_e/\lambda$ ($\beta_s = 2\pi n_s/\lambda$) sont respectivement les constantes de propagation dans la partie réseau de guides et coupleur planaire, $n_e$ ($n_s$) sont respectivement les indices effectifs dans ces deux parties du dispositif, $\lambda$ est la longueur d'onde dans le vide, $\Delta L$ est la variation de longueur entre les guides du réseau, $\theta_{in}$ ($\theta_{out}$) sont les positions angulaire des faisceaux d'entrée (de sortie), et $m$ est l'ordre de diffraction (Fig. 2.2).

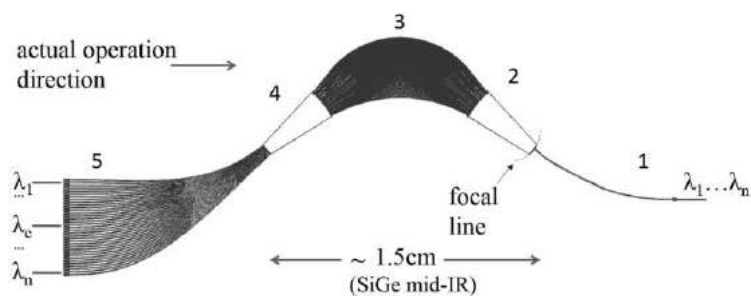

Fig. 2.1. Schéma d'un multiplexeur AWG. Le calcul analytique est effectué réciproquement en considérant: 1- un guide d'entrée, 2- coupleur planaire, "- faisceau de guides, 4-coupleur planaire de sortie et 5- guides de sortie. Les encarts représentent les coupleurs planaires d'entrée et de sortiede sortie.

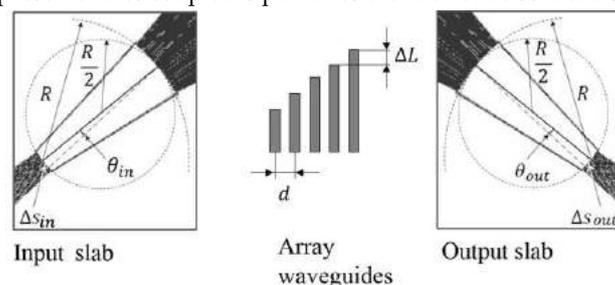

Fig. 2.2. Schémas des coupleurs planaires d'entrée et de sortie d'un AWG.

9### 2.1.1 Grandeurs caractéristiques d'un AWG

*La variation de longueur* ΔL entre les guides d'onde du réseau est constante. Elle est égale au produit entre l'ordre de diffraction et l'ordre de diffraction divisé par la longueur d'onde centrale de fonctionnement $\lambda_c$ pour un guide d'onde d'indice effectif $n_{ec}$:

$$\Delta L = m \frac{\lambda_c}{n_{ec}}, \qquad (2.2)$$

*La dispersion angulaire* donne la dépendance angulaire du faisceau dans le coupleur planaire de sortie comme (l'approximation est valable pour les petits angles) :

$$\theta_{out} \approx \frac{\left(\frac{\lambda}{\lambda_c} \cdot n_{ec} - n_e\right) \cdot \Delta L}{d \cdot n_s} - \theta_{in}. \qquad (2.3)$$

La dispersion peut être également représentée par deux paramètres linéairement dépendent – angulaire $\frac{d\theta_{out}}{d\lambda}$ and latérale $\frac{dx}{d\lambda}$. La variation de la dispersion angulaire est donnée par la relation :

$$\frac{d\theta_{out}}{d\lambda} = \frac{m n_g}{n_{ec} n_s d} \quad \text{or} \quad \frac{\Delta L n_g}{\lambda_c n_s d} \qquad (2.4)$$

où $n_g = n_{ec} - \lambda_c \frac{dn_e}{d\lambda}$ est l'indice de groupe. La variation de la dispersion latérale est $\frac{dx}{d\lambda} = R \frac{d\theta_{out}}{d\lambda}$, R est la longueur du coupleur planaire.

*L'intervalle spectral libre Free spectral range (FSR)* représente la variation en longueur d'onde (ou fréquence) entre deux ordres de diffraction contiguës. Il s'exprime comme:

$$\Delta \lambda_{FSR} = -\frac{\lambda_c^2}{n_g \Delta L} \qquad (2.5)$$

### 2.1.2 Propriétés Spectrales

Les pertes d'insertion $L_c$ correspondent aux pertes du guide central. Les principales sources de pertes sont liées aux pertes de diffraction lors du couplage de la lumière entre les coupleurs planaires d'entrée et de sortie et les guides du faisceau, aux pertes de radiation lors de la propagation dans les parties courbes des guides du faisceau et les pertes de propagation dues à l'absorption du matériau et à la diffraction sur les rugosités des facettes latérales des guides. Durant la phase de conception et de modélisation du multiplexeur AWG, un soin tout particulier a été promulgué pour réduire les pertes de couplage aux interfaces entre les coupleurs planaires et le faisceau de guide par l'ajout d'élargissement. En choisissant des rayons de courbure suffisamment grands, les pertes de radiations peuvent être réduites à un niveau négligeable.

La *non-uniformité* $L_u$ représente le rapport de puissance entre le guide à l'extrémité du faisceau de guide et le guide central. Dans des AWG à 35 canaux, les valeurs sont comprises entre 1 et 3 dB. La non-uniformité peut être réduite lorsque la fréquence du FSR $\Delta f_{FSR}$ est augmentée, ce qui est obtenu en diminuant la différence de longueur dans le faisceau de guides ΔL [37].

Pour des applications large bande, l'apodisation des réponses spectrales des guides de sortie est très importante, ainsi que les interférences entre canaux $X_{ch}$. L'introduction de coupleurs MMIs à la sortie du faisceau de guides est utilisée pour le réaliser. On peut noter que l'optimisation d'un tel composant est un compromise entre l'apodisation et la diminution des pertes d'insertion. L'introduction de coupleurs MMI entraine généralement une augmentation des pertes d'insertion allant de 3 à 5 dB.

Il y a plusieurs causes qui peuvent provoquer une augmentation de la diaphonie entre canaux $L_x$. Les effets couplage de lumière entre guide adjacent du faisceau de guides, d'un mauvais aiguillage dans les coupleurs planaires et d'étalement des faisceaux dans les coupleurs planaires doivent être maîtrisés pour un dimensionnement correct des AWGs. La principale cause d'erreur de phase est quant à elle essentiellement due aux variation de la largeur des guides induit par la rugosité des flancs, de la variation de concentration en germanium dans le multi-couche qui constitue le guide d'onde, ce qui induit une variation de l'indice effectif du mode guidé. L'analyse détaillée de l'erreur de phase sera présentée dans le chapitre "2.1.6 erreurs de phase".

### 2.1.3 Calcul analytique du champ électromagnétique

Le calcul de la réponse spectrale du dispositif utilise une approximation gaussienne des modes qui se propagent dans les différentes parties de l'AWG et la théorie scalaire de la diffraction puisque l'on considère uniquement la propagation de la puissance lumineuse dans le dispositif. Les effets de conversion de polarisation sont négligés. Le transfert de lumière à la transition entre deux guides d'onde différents est décrit par le coefficient donnant le recouvrement entre les deux modes se propageant dans les deux guides (tous deux normalisés) [70]:

$$\eta(x,y) = \left| \iint_{-\infty}^{+\infty} dxdy \Psi_1 \Psi_2^* \right|^2 \tag{2.6}$$

Le profil du champ $\Psi(x,y)$ se décompose comme le produit d'une composante verticale $\Psi(y)$ et une composante horizontale $\Psi(x)$ indépendantes l'une de l'autre. L'épaisseur des guides étant la même, on peut considérer en première approximation que la composante vertical est identique dans tout le dispositif. Elle sera approximée par le profil du mode TM du guide plan infini défini par l'empilement de couches du substrat utilisé. Dans cette étude, on ne considérera que la variation de la composante horizontale. Les différentes étapes du calcul sont décrites dans les paragraphes suivant.

Le champ du guide d'entrée (1) de la Fig. 2.1 est représenté par un faisceau gaussien de puissance normalisée $\Psi_1$ avec un waist ou largeur à 1/e² $w_{ex}$. Ce faisceau lumineux diverge dans le coupleur planaire (2) du fait du manque de confinement latéral. Son champ $\Psi_2$ correspond à un faisceau gaussien qui a divergé avec une largeur de waist qui dépend de la distance de propagation. A la distance $z = R$ par rapport à l'extrémité du guide d'entrée, la largeur du waist correspond à $w_{sx} = w_{ex}\sqrt{1+\left(\frac{\lambda R}{n_e \pi w_{ex}^2}\right)^2}$. On considère que la distance de propagation est inférieure à la zone de Rayleigh et que la distorsion de phase est négligeable. Ce faisceau va ensuite se coupler au faisceau de guides (3). Le champ dans le faisceau de guide $\Psi_3$ est la somme des champs se propageant dans chaque guide individuel du faisceau de guide qui sont également représenté par des gaussiennes. Le chemin optique dans les différents guides change d'un multiple de la longueur d'onde centrale pour que les différents faisceaux qui arrivent sur le coupleur planaire de sortie (4) est une différence de phase de $2\pi$ phase dans le plan de focalisation correspondant à l'entrée des guides de sortie. Le champ diffracté $\Psi_4$ est calculé à l'aide d'une transformée de Fourier des champs provenant du coupleur planaire de sortie.

$$\Psi_4(x) = a \sum_{p=-M}^{M} A_p \exp(-jp\Delta\varphi) \qquad (2.7)$$

où $a = \exp\left(j2\pi R \frac{n_e}{\lambda}\right)/(j\frac{\lambda}{n_e}R)$ est un facteur commun resultant de transformée de Fourier, $M = \frac{N-1}{2}$, $N$ est le nombre total de guide du faisceau de guides, $A_p = \sqrt[4]{2\pi w_{tx}^2}\exp\left(-\frac{(\pi w_{tx}x)^2}{\alpha^2}\right)\sqrt{\eta_p(x)}$ est l'amplitude, $w_{tx}$ est le waist du faisceau au niveau de l'élargissement à l'entrée du faisceau de guides, $\alpha = \frac{\lambda}{n_s}R$, $\eta_p(x)$ est le facteur de couplage, $\Delta\varphi = (\beta_e \Delta L + 2\pi\chi\frac{x}{\alpha})$ est la phase, $\chi \approx d$ est la longueur de l'arc entre deux guides voisins du faisceau de guides.

Finalement, le champ injecté dans un des guides de sortie $\Psi_5$ est obtenu. La réponse spectrale de l'AWG est obtenue en utilisant l'équation (2.6). On peut noter que le calcul est plus simple à réaliser en ne considérant qu'un seul guide d'entrée. Le principe du retour inverse de la lumière permet d'obtenir la réponse spectrale des multiplexeurs étudiés.

### 2.1.4 Simulations

Un outil semi-analytique a été développé sous matlab pour déterminer numériquement la réponse spectrale d'un AWG en utilisant le formalisme du calcul du champ détaillé dans le chapitre précédent. Le résultat de simulation d'un AWG à 33 canaux centré sur une longueur d'onde de 5,7 µm est représenté sur la Fig. 2.3. Les paramètres géométriques sont donnés dans l'annexe I. La Fig. 2.3 montre les positions et les amplitudes normalisées $\lambda_{min}$, $\lambda_c$ et $\lambda_{max}$ des faisceaux dans le plan de focalisation selon l'équation (2.7).

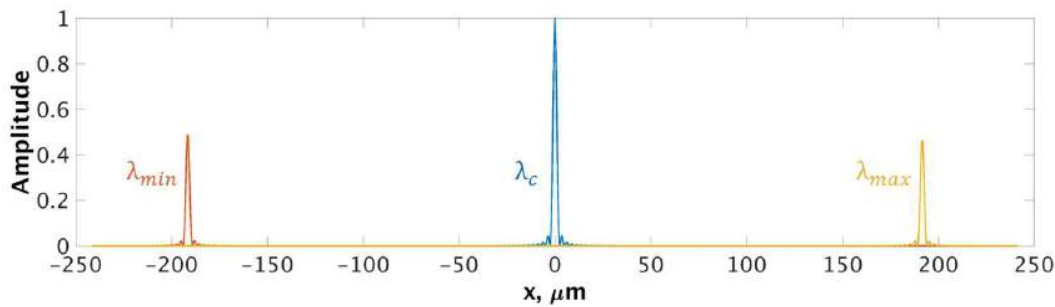

Fig. 2.3. Amplidude normalize et positions def $\lambda_{min}$, $\lambda_c$ and $\lambda_{max}$ dans le plan de focalisation.

Le spectre de transmission de l'AWG est reporté sur la Fig. 2.4. Les pertes d'insertion du canal central sont de -2.2 dB, la diaphonie est inférieure à -75 dB et le déséquilibre est de 3.6 dB.

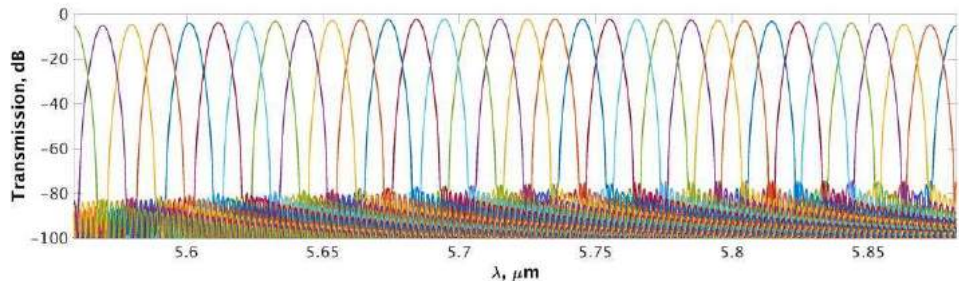

Fig. 2.4. Réponse spectrale d'un AWG à 33 canaux centrés sur une longueur d'onde de 5,7 μm.

## 2.1.5 Optimisation

Le but est d'hybrider le multiplexeur avec une batterie de source QCLs sur une large bande spectrale tout en ayant une bonne uniformité des réponses individuelles des canaux.

Pour satisfaire ce critère, un élargissement de 6 μm à la sortie des guides du faisceau a été ajouté. Le dimensionnement de cet élargissement et la réponse spectrale de l'AWG sont reportés sur la Fig. 2.5. Si on compare la réponse obtenue à celle de la fig. 2.4 d'un AWG standard, l'uniformité a été améliorée de 2,5 dB et la diaphonie réduite de 2 dB. En contrepartie, les pertes d'insertion ont augmenté faiblement de 0,4 dB.

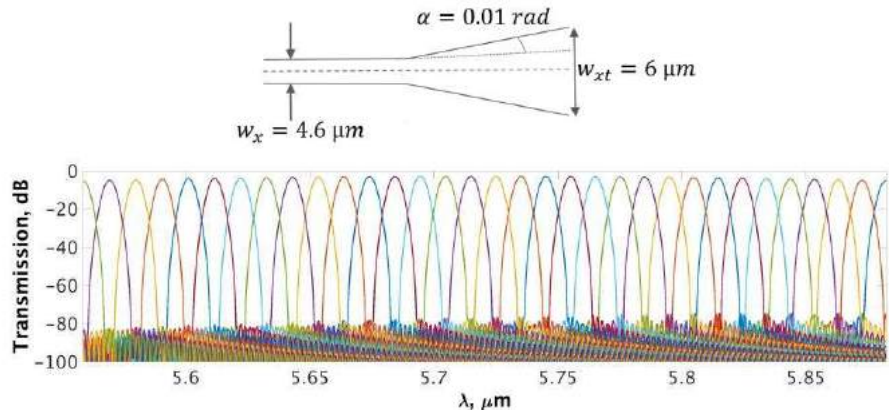

Fig. 2.5. Réponse spectrale d'un AWG à 33 canaux centré sur la longueur d'onde de 5,7 μm avec des élargissements de 6 μm (encart au-dessus).

Nous avons également étudié l'impact de coupleurs MMI de 1 vers 2 placés à l'entrée des canaux de sortie. La longueur des coupleurs MMI a été obtenu grâce à des outils de simulation développés sous Matlab par R. Orobtchouk et le module Beamprop de R-soft (voir Fig. 2.6). L'image double du guide d'entrée est obtenue à la moitié de la position de l'image simple du MMI.

Figs. 2.7 and 2.8 present the spectral responses of AWGs with 9 μm and 11 μm MMIs, and their impact on flattening of the top of the response is compared in Fig. 2.9. The insertion losses of AWGs with 9 μm and 11 μm MMIs are -4.8 dB and -6.3 dB, respectively; inter-channel crossing reduced from around 20 dB to 7dB and 8 dB, respec-tively.

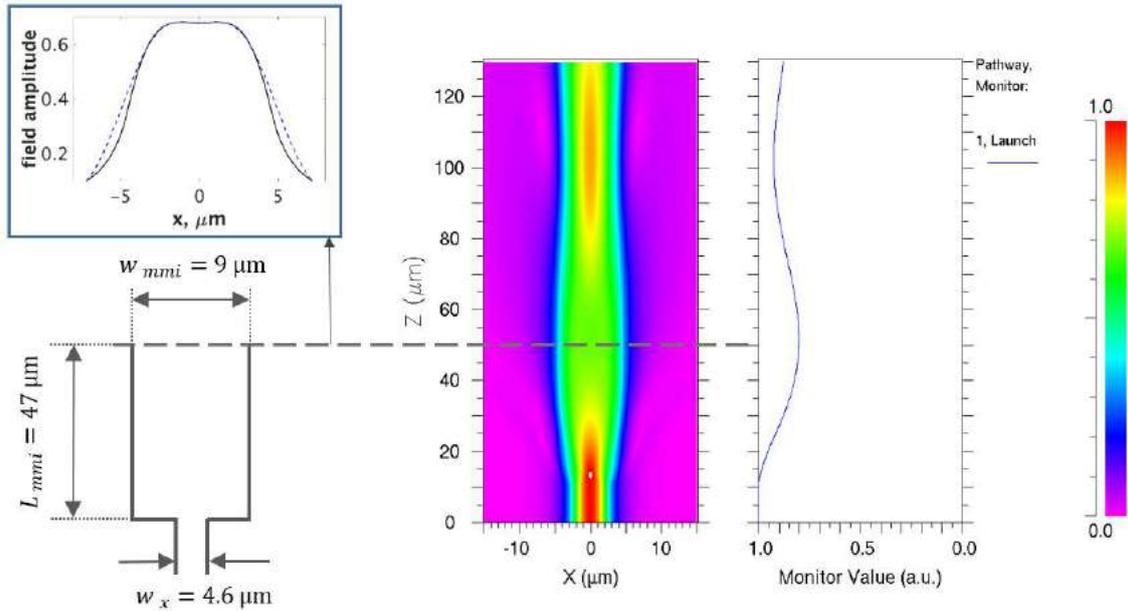

Fig. 2.6. Coupleur MMI de 9 μm de largeur et représentation de la lumière qui se propage à l'intérieur. L'encart en haut à gauche représente le profil du mode avec une largeur du waist élargie par rapport à celle du guide monomode. La ligne en pointillée représente l'approximation du mode par deux gaussiennes.

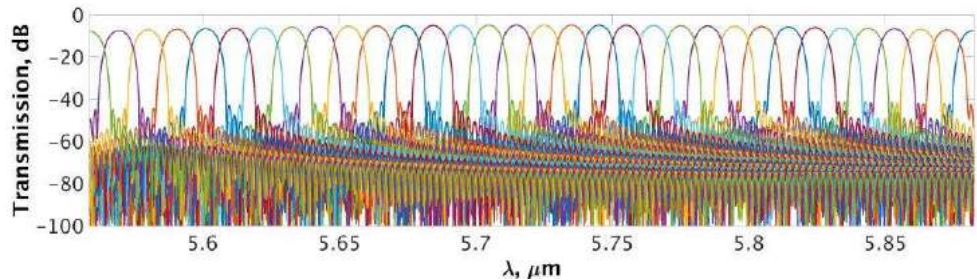

Fig. 2.7. Réponse spectrale de l'AWG à 33 canaux centré sur $\lambda$ = 5.7 μm avec des MMI de 9 μm.

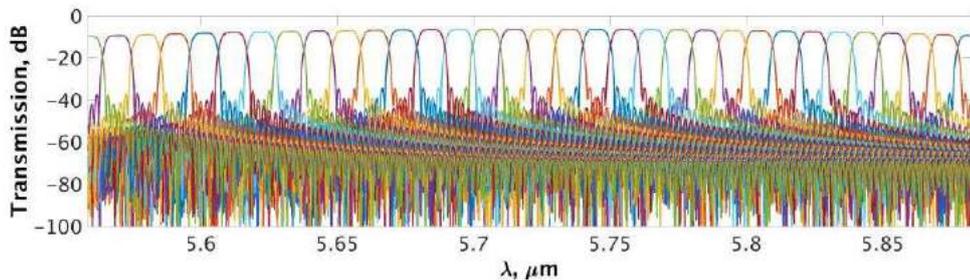

Fig. 2.8. Réponse spectrale de l'AWG à 33 canaux centré sur $\lambda$ = 5.7 μm avec des MMI de 11 μm.

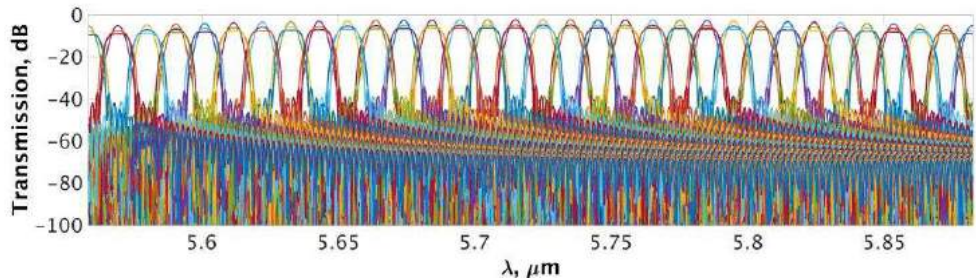

Fig. 2.9. Comparaison pour évaluer l'impact de l'ajout des MMIs.

### 2.1.6 Analyse de l'erreur de phase

La lumière incidente aux différentes longueur d'onde ($\lambda_1,…, \lambda_n$) accumule une différence de phase $\Delta\varphi(\lambda)$ dans le faisceau de guides qui induit une variation du plan de phase qui permet de définir l'angle du faisceau diffracté au niveau du plan de focalisation incurvé comme :

$$\Delta\varphi(\lambda) = \beta_e(\lambda) \cdot \Delta L = 2\pi m - d \cdot \beta_s(\lambda) \cdot (\sin\theta_{in} + \sin\theta_{out}) \quad (2.8)$$

Nous avons introduit la deviation standard d'erreur de phase $\varphi = \sum_{i=1}^{N}\exp(-j(\Delta\varphi(\lambda)_i + \delta_i))$ à notre modèle dans le but d'étudier l'impact de l'erreur de phase sur la diaphonie. La dépendance de la diaphonie en fonction de l'écart de la variation de l'indice effectif est représentée sur la Fig. 2.10 pour différentes valeurs de la puissance répartie qui est une conséquence directe de la réduction du nombre de guide dans le faisceau de guides. Pour un niveau de puissance répartie de 1 %, on peut estimer que la variation correspondante de l'indice effectif est de 3.6×10$^{-5}$ pour les AWGs conçus à 4.5 µm and 6.1×10$^{-5}$ pour ceux conçus à 7.6 µm. L'écart de l'erreur de phase peut aussi être représentée comme la longueur équivalent du chemin optique qui est respectivement de 0.10 µm et 0.20 µm pour les AWGs conçus à 4,5 et 7,6 µm ( ≈λ/40), ce qui est considérablement faible comparé à la longueur de 10 mm du composant. La variation de la diaphonie induit par les erreurs de phase et la puissance distribuée pour AWG conçu à 5,7 µm est décrite sur la Fig. 2.11.

Nous avons également étudié l'influence des MMIs sur la sensibilité à la diaphonie des AWGs. La comparaison entre les AWGs (λ = 5.7 µm) avec et sans MMIs montre que malgré la dégradation initiale de la diaphonie induite par l'introduction des MMIs, elle reste à des niveaux de -22 dB and -38 dB respectivement lorsque des erreurs de phase, i.e. $\sigma_\varphi$ = 0.0002 rad et $\sigma_\varphi$ = 0.0009 rad, sont introduites.

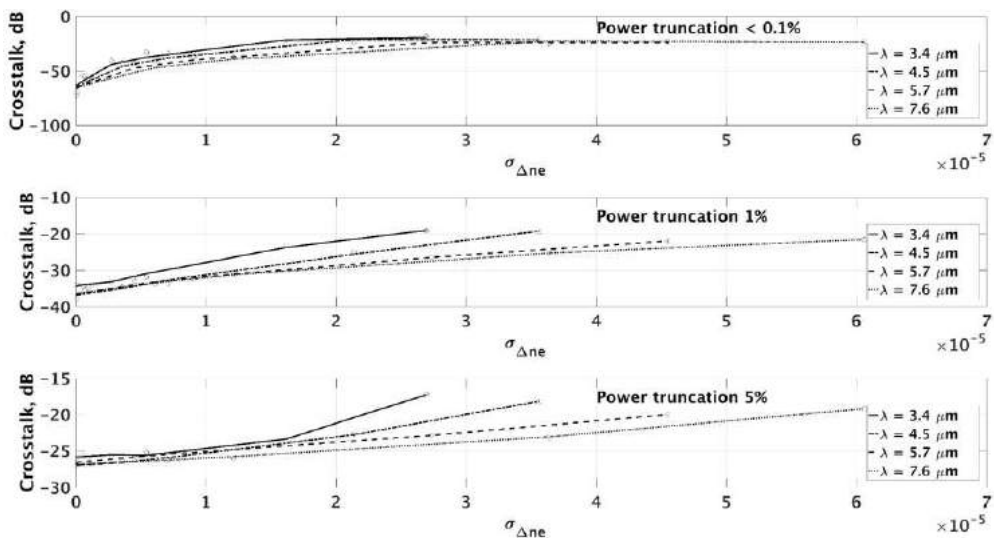

Fig. 2.10. Influence de la deviation standard de l'indice effectif sur la diaphonie pour différentes valeurs de la puissance répartie (guide de 10 mm de long).

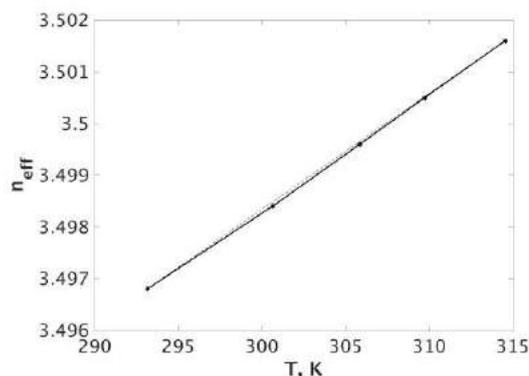

Fig. 2.12. Variation de l'indice effectif du guide SiGe (λ = 5.7 μm) en fonction de la température.

### 2.1.7 Dépendance en température

La dépendance de la réponse spectrale à la température a été modélisée en utilisant l'approche développée par Li [84]. Les calculs ont été réalisés avec le logiciel Rsoft. Les indices de réfraction en fonction du pourcentage de germanium des 9 couches du guide SiGe à gradient d'indice sont obtenus par une approximation linéaire comme :

$$n_{Si_{0.6+x}Ge_{0.4-x}} = (1-x)n_{Si} + xn_{Ge}, \qquad (2.9)$$

où $x$ est la concentration de Ge.

Cette relation reste valable jusqu'à une concentration de 40 %. La détermination de la variation de l'indice de réfraction en fonction de la température est calculée à l'aide de la formule empirique de Li. La variation de l'indice effectif du mode guidé en fonction de la température est illustrée sur la Fig. 2.12.

La variation de l'indice effectif permet d'obtenir le décalage en longueur d'onde qui correspond à $\Delta\sigma = 0.08$ cm$^{-1}$ par °C. La comparaison avec les données expérimentales est détaillée dans le chapitre "Caractérisation".

## 2.2 Multiplexeur de type PCG

Contrairement aux AWGs, dans les PCGs, la variation de phase et la focalisation sont réalisées par le réseau concave réfléchissant. Comme la variation de phase ne dépend pas de la géométrie des guides, les multiplexeurs PCG doivent être plus robuste et également plus compacts. Un schéma de la géométrie d'un multiplexeur PCG est donné sur la Fig. 2.13. Il est constitué d'un guide d'entrée, d'une série de guides de sortie et d'un réseau de diffraction en réflexion concave. Tous les éléments sont localisés sur le cercle de Rowland. Le faisceau lumineux comprenant l'ensemble des longueurs d'onde va diverger jusque sur le réseau de diffraction concave. La condition d'interférence constructive des faisceaux réfléchis est donnée par la relation de Bragg :

$$d(sin\theta_i + sin\theta_d) = m\frac{\lambda}{n_s}, \qquad (2.10)$$

où $\theta_i$ et $\theta_d$ sont les angles d'incidence et de diffraction, respectivement, $d$ est la période du réseau, $m$ est l'ordre de diffraction, $\lambda$ est la longueur d'onde et $n_s$ est l'indice effective du mode du guide plan infini.

Les faisceaux réfléchis n'auront pas le même angle de diffraction pour les différentes longueurs d'onde et seront séparés et refocalisés dans un guide de sortie par longueur d'onde.

### 2.2.1 Grandeurs caractéristiques du PCG

La *Dispersion* décrit l'inclinaison des faisceaux pour les différentes longueurs d'onde dans le plan de focalisation. Pour un angle d'incidence donné, la dispersion angulaire s'obtient à partir de l'équation (2.10) en prenant la dérivée de l'angle de diffraction par rapport à la longueur d'onde.

$$\frac{d\theta_d}{d\lambda} = \frac{mn_g}{n_s^2 d\cos\theta_d}, \qquad (2.11)$$

où $n_g = n_{ec} - \lambda_c \frac{dn_e}{d\lambda}$ est l'indice effectif de groupe du guide plan infini.

A partir de la Fig. 2.14, on peut montrer que la dispersion linéaire sur le cercle de Rowland est $r_R \frac{\varphi_{r_R}}{\Delta\lambda_{P_1 P_c}}$, où $\Delta\lambda_{P_1 P_c} = \lambda_{P_c} - \lambda_{P_1}$ est l'intervalle spectrale entre les canaux $P_1$ et $P_c$, et $\varphi_{r_R} = 2\varphi_R = 2(\theta_{dP_c} - \theta_{dP_1})$, où, $r_R$ est le rayon du cercle de Rowland, $R$ est le rayon du réseau concave. Les 2 angles sont liés par la relation $\varphi_{r_R} r_R = \varphi_R R$. On obtient à partir de ces relations, l'expression de la dispersion :

$$D = R\frac{\varphi_R}{\Delta\lambda_{P_1 P_c}} = \frac{Rmn_g}{n_s^2 d\cos\theta_d} \qquad (2.12)$$

*L'Intervalle spectrale libre (FSR pour Free Spectral Range)* est l'intervalle en longueur d'onde ou en fréquence séparant deux ordres de diffractions adjacents. Il est défini par la relation :

$$\Delta\lambda_{FSR} = \frac{\lambda}{m}\left[1 - \frac{(m+1)}{m}\left(1 - \frac{n_g}{n_s}\right)\right]^{-1} \qquad (2.13)$$

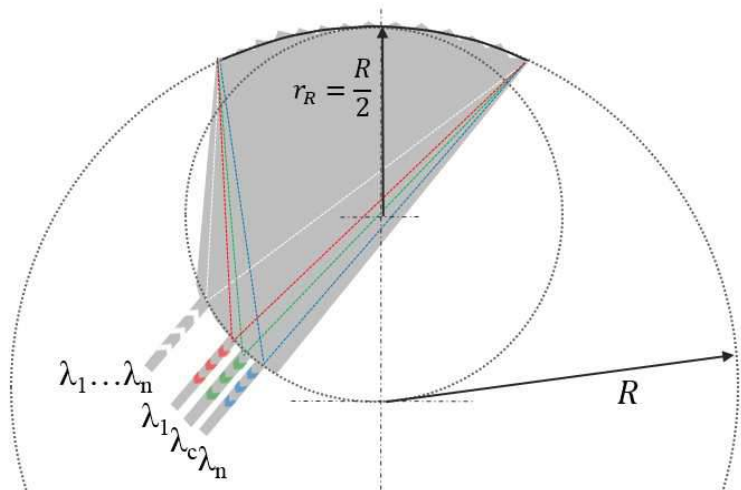

Fig. 2.13. Schéma et illustration du principe de fonctionnement d'un PCG.

*L'ordre de* diffraction apparait dans l'équation (2.13). Il est inversement proportionnel à l'intervalle spectral libre.

$$m = \left(\frac{\lambda_c}{\Delta\lambda_{FSR}} + 1\right)\frac{n_s}{n_g} - 1 \tag{2.14}$$

Si on cherche à réduire la taille du dispositif avec une grande dispersion, il faut utiliser un ordre de diffraction important.

### 2.2.2 Propriétés Spectrales

Les *pertes d'insertion* $L_c$ d'un PCG dépendent des paramètres suivant : les pertes de couplage entre le guide monomode d'entrée et la partie planaire du composant, les pertes liées au couplage vers les autres ordres de diffraction, les pertes de réflexion, les pertes liées aux imperfections du réseau concave comme la rugosité ou l'arrondi des facettes du réseau et les pertes de diffraction qui dépendent de la conception du PCG [64]. Les outils de simulations permettent d'adapter la géométrie des guides de manière à réduire les pertes de couplage. L'inclinaison des facettes du réseau concave permet de diminuer la quantité de lumière transférée sur les autres ordres de diffraction par un effet de Blaze.

Les *pertes de non-uniformité* $L_u$ sont liées à la période du réseau concave $d$. Plus la période est petite plus les pertes $L_u$ sont réduites. Ce paramètre intéinsèquement dépend du choix des angles d'entrée $\theta_i$ / de sortie $\theta_d$ et de l'ordre de diffraction. Plus l'ordre de diffraction est élevé, plus la différence des pertes d'insertion entre les longueurs d'onde centrale et périphériques sera réduite.

La réduction de la *diaphonie entre canaux* $X_{ch}$ est privilégiée. Celle-ci s'accompagne de l'apodisation de la réponse spectrale. Comme pour l'AWG, on peut utiliser des coupleurs MMI. Une des conséquences de l'apodisation réside également dans la réduction des pertes d'insertion.

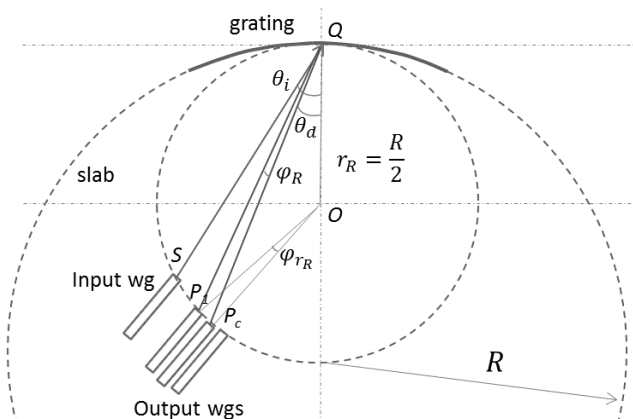

Fig. 2.14. Schéma du multiplexeur PCG.

La diaphonie $L_x$ représente la quantité de signal couplée sur le canal. Il est en partie dû à l'arrondi des facettes du réseau concave qui provoque une diffraction de la lumière sur une variété d'angles qui vont se coupler dans des guides de sorties non désirés. L'autre grandeur qui va induire de la diaphonie est la distance entre les guides de sortie $d_r$.

### 2.2.3 Calcul analytique du champ dans le PCG

Comme pour l'AWG, le calcul du champ du PCG utilise l'approximation Gaussienne et la théorie scalaire de la diffraction. Le profil de l'énergie du mode guidé fondamental est approximé par une fonction gaussienne normalisée. On considère que le faisceau incident se propage suivant la direction z et que le profil du mode est situé dans le plan (x, y). Comme pour le calcul de l'AWG, le profil du mode dans la direction y est considéré comme constant, ce qui permet de réduire le calcul à un problème à deux dimensions. Les coefficients de transmission sont obtenus en utilisant l'équation (2.6).

Le profil de l'énergie du mode $\Psi(x, y)$ est représenté comme le produit de deux composantes verticale $\Psi(y)$ et horizontale $\Psi(x)$ indépendante l'une de l'autre. Comme la variation de la composante verticale est relativement faible entre les guides d'entrée et de sortie et le guide plan infini, elle sera supposée comme constante et négligée en première approximation dans le calcul. On ne considérera que la divergence de la composante horizontale dans la suite de ce chapitre.

Le faisceau incident de largeur $w_{ex}$ du guide d'entrée n'est plus confine dans la partie guide plan infini du dispositif et diverge jusqu'au réseau concave, comme le montre la ligne en tiret de la Fig. 2.13. Son profil $\Psi_2$ est décrit par une fonction gaussienne dont la largeur dépend de la distance de propagation par rapport à l'interface du guide d'entrée. A la distance $z = R$, sa largeur est donnée par la relation $w_{sx} = w_{ex}\sqrt{1 + \left(\frac{\lambda R}{n_e \pi w_{ex}^2}\right)^2}$, où $R$ est le rayon de courbure du réseau concave. Chaque facette du réseau concave agit comme un miroir, diffractant le faisceau incident suivant une multitude d'angles. La superposition des réponses de chaque facette va produire des interférences constructives qui du fait de la périodicité de leurs positions lorsque la différence de chemin optique est constante. La différence de longueur du chemin optique $\Delta L$ est choisie de façon à être égal à un multiple de la longueur d'onde centrale de fonctionnement du dispositif $\Delta L = m\frac{\lambda_c}{n_s}$. Dans ce cas, chaque composante du faisceau à la longueur d'onde centrale possède une différence de phase de $2\pi m$ de son voisin immédiat ce qui assure sa condition d'interférence constructive et sa focalisation sur la partie centrale du plan de focalisation. Alors que les autres longueurs d'onde auront un décalage de la différence de phase et seront donc décalées du centre du dispositif. Aussi, le faisceau réfléchi sera focalisé dans un autre guide de sortie adjacent de celui du centre du faisceau de guides de sortie. Le profil diffracté est calculé à l'aide de la transformé de Fourier. Il est exprimé par la relation suivante :

$$\Psi_{os}(x) = \frac{\eta_r \eta(x)}{2}\frac{d_M}{\sqrt{2\pi}} \sum_{M=-\frac{N-1}{2}}^{\frac{N-1}{2}} (\cos\theta_{iM} + \cos\theta_{dM})\exp\left(-j\left(M\beta_s \Delta L - 2\pi\frac{x_M x_4}{\alpha}\right)\right) sinc\left(\frac{d_M}{2}\frac{x_4}{\alpha}\right) \quad (2.15)$$

où $\eta_r$ est le coefficient de réflexion du réseau concave, $\eta(x)$ est le coefficient de couplage, $N$ est le nombre de facettes du réseau concave, $d_M$ est la largeur de la M[ième] facette, , $\theta_{iM}$ et $\theta_{dM}$ sont les angles d'incidence et réfléchis par rapport à la normale de la $M$[ième] facette, $\beta_s$ est la constante de propagation du guide plan infini the propagation constant. Le résultat de la diffraction sur une facette est que le champ dans le plan focal est une somme de fonction sinus cardinal. Finallement, l'efficacité de sortie $\eta_{out}(x)$ du PCG est l'intégralle de recouvrement entre le champ diffracté avec le mode d'un guide de sortie. Les paramètres géométriques résultant du travail de conception sont donnés dans l'annexe II, ainsi que le détail des calculs.

### 2.2.4 Simulations

Dans cette partie, nous présentons les résultats de simulation obtenus pour les multiplexeurs PCG fonctionnant à trois différentes longueurs d'onde. Les grandeurs caractéristiques sont répertoriées dans le tableau 2.1. La méthode de conception des PCGs est exposée dans l'annexe II.

Les spectres de transmission en fonction de la longueur d'onde des trois PCGs sont reportés sur la figure 2.15. A partir des simulations, nous avons obtenus les résultats suivants. Le PCG fonctionnant à 3.38 µm a été conçu pour que la longueur d'onde centrale soit exactement de 3.38 µm, aussi l'espacement en longueur d'onde des guides de sortie n'est pas constant.

Les pertes d'insertion, la diaphonie et le déséquilibre sont respectivement de -4 dB, -50 dB et 2 dB pour le PCG à 3.38 µm, de -2.3 dB, -39 dB et 0.3 dB pour le PCG à 6.85 µm et de -2.3 dB, -39 dB et de 0.3dB pour le PCG à 8.76 µm.

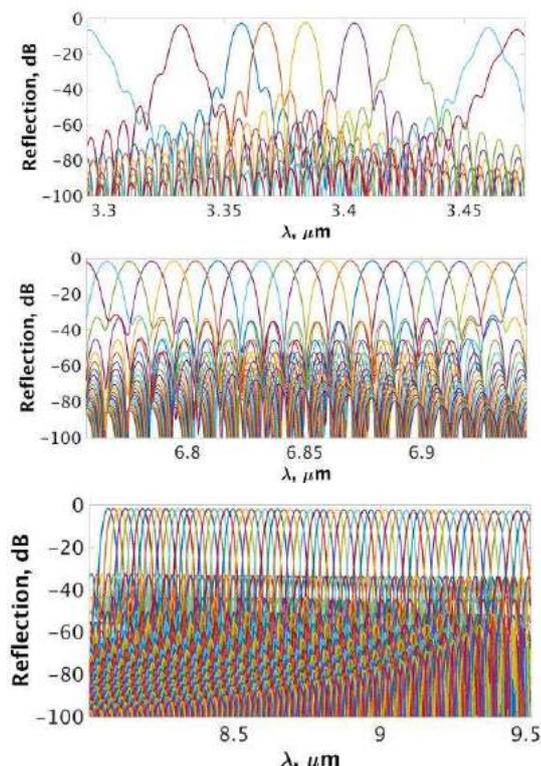

Fig. 2.15. Réponse spectrale des PCGs à λ = 3.38 µm, 6.85 µm and 8.76 µm.

| notation | 3.38 μm | 6.85 μm | 8.76 μm |
|---|---|---|---|
| $N_{ch}$ | 9 (espacement irrégulier) | 21 | 67 |
| $\sigma_{min}$ - $\sigma_{max}$ | 2878 – 3037 cm$^{-1}$ | 1440 – 1480 cm$^{-1}$ | 1051 – 1249 cm$^{-1}$ |
| technology | Si$_{0.6+x}$Ge$_{0.4-x}$/Si | Si$_{0.6+x}$Ge$_{0.4-x}$/Si | Ge/Si$_{0.6}$Ge$_{0.4}$ |
| $w_x$, $w_y$ | 2.4 μm | 6.8 μm | 2.8 μm |
| $n_s$ | 3.5146 | 3.4726 | 4.0493 |
| $n_g$ | 3.5947 | 3.4835 | 4.0522 |
| $w_e$ | 1.52 μm | 4.11 μm | 1.88 μm |
| $\Delta\lambda_{FSR}$ | 0.3836 μm | 1.1824 μm | 4.5936 μm |
| m | 8 | 5 | 1 |
| $dr$ | 5.72 μm | 15.45 μm | 7.07 μm |
| D | 566.2184 | 1646.7 | 309.3022 |
| $\Delta\theta$ | 4° | 2° | 3° |
| $SP_c$ | 85.73 μm | 247.25 μm | 353.43 μm |
| $\theta_i$ / $\theta_d$ | 40° / 36° | 39.8° / 37.8° | 12.8° / 9.8° |
| $r_R$ | 613.9722 μm | 3.5416e+03 μm | 3.3751e+03 μm |
| $r_i$ | 939.4697 μm | 5.4398e+03 μm | 6.5819e+03 μm |
| $r_d$ | 992.3393 μm | 5.5948e+03 μm | 6.6513e+03 μm |
| d | 6.2476 μm | 7.8355 μm | 5.4735 μm |
| N | 62 | 141 | 200 |

Tab. 2.1. Grandeurs caractéristiques des PCGs optimisées par longueur d'onde centrale de fonctionnement 3.4 μm, 6.9 μm and 8.7 μm. Les labels désignent: $N_{ch}$ est le nombre de canaux de sortie, $\sigma_{min}$ - $\sigma_{max}$ est la bande spectrale in nombre d'onde, $w_x$, $w_y$ sont les largeurs des guides d'onde, $w_e$ est la largeur du faisceau, $dr$ est la distance entre les guides de sortie, $\Delta\theta = \theta_i - \theta_d$, $r_R$ est le rayon du cercle de Rowland, $r_i$ ($r_d$) – est la distance entre l'interface du guide central de sortie et le milieu du réseau concave, $N$ est le nombre de facette.

### 2.2.5 Optimisation

Initialement, les positions des facettes ont été considérées comme équidistantes, ce qui fournit une première valeur approché du chemin optique dans le dispositif. Nous avons modifiés la positions des facettes en prenant un développement limité au cinquième ordre du chemin optique donné par la relation :

$$\Delta L = -d_M(\sin\theta_i + \sin\theta_d) + \frac{d_M^4}{8R^3}\left[\frac{sin^2\theta_i}{cos\theta_i} + \frac{sin^2\theta_d}{cos\theta_d}\right]$$
$$+ \frac{d_M^5}{8R^4}\left[\frac{sin^3\theta_i}{cos^2\theta_i} + \frac{sin^3\theta_d}{cos^2\theta_d}\right] \quad (2.17)$$

En égalisant le terme de droite de l'équation (2.17) à $mM\frac{\lambda_c}{n_{sc}}$, on obtient la modification des positions des facettes $d_M$, ce qui permet de corriger les abbérations jusqu'à

21l'ordre 5. Pour les composants étudiés, le plus grand écart par rapport au cahier des charges est de 60 nm (PCG à 3.38 µm avec $d$ = 6.25 µm), inférieure à 30 nm (PCG à 6.85 µm avec $d$ = 7.84 µm) et inférieure à 10 nm (PCG à 8.76 µm avec $d$ = 5.47 µm), ce qui très faible comparé à la période du réseau concave. Les réponses spectrales des dispositifs avec les facettes modifiées ne présentent pas de différences par rapport aux dispositifs initiaux.

Nous avons également étudié l'influence d'un déplacement longitudinal de la facette $\Delta h$ pour évaluer les différences entre une facette concave ou plan. Ce paramètre est introduit dans le programme de calcul en considérant la condition suivante :

$$|(SQ_{Ml} + PQ_{Ml}) - (SQ_M + PQ_M)| \qquad (2.18)$$
$$= |(SQ_{Mr} + PQ_{Mr}) - (SQ_M + PQ_M)|$$

où $Q_M$ est le centre de la $M^{\text{ième}}$ facette, $Q_{Ml}$ ($Q_{Mr}$) est la limite gauche (droite) de la $M^{\text{ième}}$ facette. Pour les trois PCGs, $\Delta h$ ne doit pas excéder 3 nm pour respecter la limite de résolution de 5 nm de la fabrication des masques. La différence entre la configuration droite et concave est négligeable.

### 2.2.6 Analyse de l'erreur de phase

Les erreurs de phase dégradent les performances des multiplexeurs PCG, principalement en ce qui concerne la diaphonie. Les principales causes de l'erreur de phase est induite par la résolution du masque, l'inclinaison et l'arrondi des coins des facettes et la variation d'épaisseur du guide plan [64]. Nous avons étudié l'effet de la résolution du masque en modifiant la valeur de la grille élémentaire. Comme les positions des facettes sont définis par les coordonnées de leurs deux extrémités si un des points ne coïncide pas avec la grille, il sera automatiquement repositionné, ce qui conduit à un décalage en translation et angulaire de la facette par rapport à sa position optimale. Nous avons étudié cet effet lorsque la résolution de la grille est fixée à 1 nm, 5 nm et 10 nm lors de la réalisation du masque et son influence sur la diaphonie pour les trois différents PCGs fonctionnant aux longueurs d'onde centrales de 3.38 µm, 6.85 µm, et 8.76 µm.

Le décalage lateral au centre des facettes contribue à une erreur de phase donnée par la relation :

$$\sigma_\varphi = 2k\sigma_x \qquad (2.127)$$

où $\sigma_\varphi$ est l'écart type de la distribution normal de l'erreur de phase, $k = \frac{2\pi}{\lambda} n_s$ est le nombre d'onde du guide plan, $\sigma_x$ est l'écart type de la coordonnée horizontale de la facette.

L'erreur de phase maximale introduite par la resolution intermédiaire de 5 nm est de 5.7e-4 $rad$ à 3.38 µm, 2.8e-4 $rad$ à 6.85 µm et 2.5e-4 $rad$ à 8.76 µm. La figure 2.16 donne les réponses spectrales du PCG à 6,85 µm pour les trois canaux centraux et pour différents niveaux de l'erreur de phase.

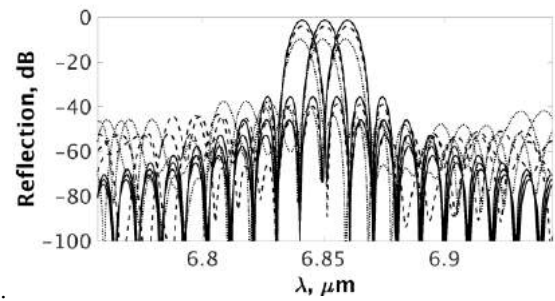

Fig. 2.16. Impact of phase errors in 6.85 μm PCG: solid line $\sigma_\varphi = 0$ rad, dashed line $\sigma_\varphi = $ 9e-3 rad, dotted-line $\sigma_\varphi = $ 17e-3 rad.

# 3 Fabrication

Les composants sont fabriqués sur une plateforme de guides à gradient d'indice de SiGe déposés sur un substrat de silicium et encapsulés dans une couche de silicium sur la ligne pilote 200 mm du CEA-LETI [26]. Les masques sont déssinés au format "gds" à partir des outils développés C++ par Dr. P. Labeye pour les AWGs, and des outils développés en Matlab durant ce travail de thèse pour les PCGs.

*Croissance épitaxiale des guides à gradient d'indice*. La croissance des guides SiGe est réalisée dans un bâti RP-CVD de la compagnie Applied Materials. Une pression de 20 Torr permet d'obtenir des vitesses de croissance élevées à la temperature de 850°C, tout en préservant une qualité cristalline inhérente à l'obtention de guides à faibles pertes de propagation (voir Fig. 3.1 (a)-(b)). A la surface de la couche de SiGe de 3 μm d'épaisseur, une couche additionnelle épaisse de silicium est ajoutée et planarisée de manière à protéger les guides de SiGe.

*Planarisation par polissage mécano-chimique (CMP pour Chemical-mechanical polishing)*. Le proceed de planarization CMP est une technique qui permet de réduire la rugosité de surface des guides et donc d'en limiter les pertes de propagation.

*Lithographie des guides d'onde*. Les composants d'optique intégrée réalisés durant ce travail de thèse sont définis par un procédé de lithographie optique. Le plus petit motif figurant sur le masque à une dimension de 1 μm ce qui requiert l'utilisation d'un masqueur Iline 365 nm ou DUV 248 nm. Compte tenu de la résolution de ces équipements, une grille de 5 nm pour la résolution du masque est suffisante pour obtenir l'ensemble des motifs, y compris les virages. Cette résolution permet de diminuer fortement les coûts de fabrication des masques.

Les substrats sont nettoyés après l'étape de pollissage CMP et la résine photosensible est déposée par centrifugation ou spin coating (voir Fig. 3.1 (c)). La vitesse de rotation de la tournette varie entre 1000 et 5000 tr/mn et le temps de dépôt dépend de l'épaisseur de résine souhaité. Pour obtenir un procédé reproductible, le temps est en général supérieur ou de l'ordre de la minute. Pour nos échantillons, l'épaisseur de résine visée est de 820 nm. Les motifs des composants sont insolés dans la résine par le masqueur DUV à 248 nm. La résine est ensuite développée pour obtenir les motifs représentés sur la figure 3.1 (d).

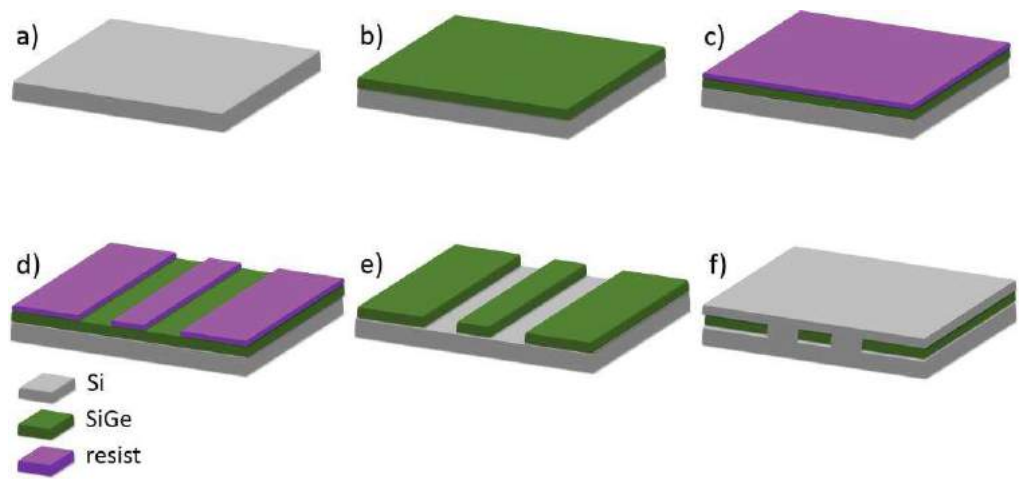

Fig. 3.1. Schéma des différentes étapes de fabrication.

*Gravure profonde ionique reactive (DRIE pour Deep Reactive Ion Etching)*. Compte tenu de la forte épaisseur des guides SiGe de l'ordre de 3 µm, il est nécessaire d'utiliser un bâti de gravure DRIE pour obtenir des flancs de guides verticaux avec une faible rugosité latérale. L'épaisseur de résine a été choisie pour que compte des sélectivités de gravure des différents matériaux, le procédé soit réalisable. Comme les sélectivités de gravure entre le SiGe et la résine sont d'un facteur 4, une épaisseur de résine supérieure à 800 nm est nécessaire auquel il faut rajouter une surépaisseur pour éviter d'avoir des profils arrondis. Les motifs après gravure sont représentés sur les figures 3.1 (e).

L'étape finale du procédé consiste à déposer une couche épaisse de silicium (voir Fig. 3.1 (f)). Une autre étape de gravure DRIE est utilisée pour définir les facettes d'entrée et de sortie des dispositifs de test, ainsi qu'un dépôt de Nitrure pour la réalisation d'un anti-reflet. Cette dernière étape achève le cycle de fabrication en salle blanche. Les puces sont ensuite découpées individuellement à la scie diamant pour être caractérisée expérimentalement.

*Métallisation des facettes du réseau concave.* La réalisation des multiplexeurs PCGs nécessite une étape supplémentaire de métallisation des facettes du réseau de diffraction concave pour augmenter le coefficient de réflexion et ainsi diminuer drastiquement les pertes d'insertions. Le dépôt métallique (TiN/W ou Ti/Au) est réalisé par PVD (Plasma Vapour Deposition). Le Titane est nécessaire pour obtenir une bonne adhérence de la couche sur le silicium. En contrepartie, il faut utiliser une couche de Titane la plus fine possible de l'ordre du nm pour ne pas augmenter l'absorption et ainsi dégrader le coefficient de réflexion des dispositifs.

# 4 Caractérisations optiques dans le moyen infra-rouge

## 4.1 Mesures expérimentales

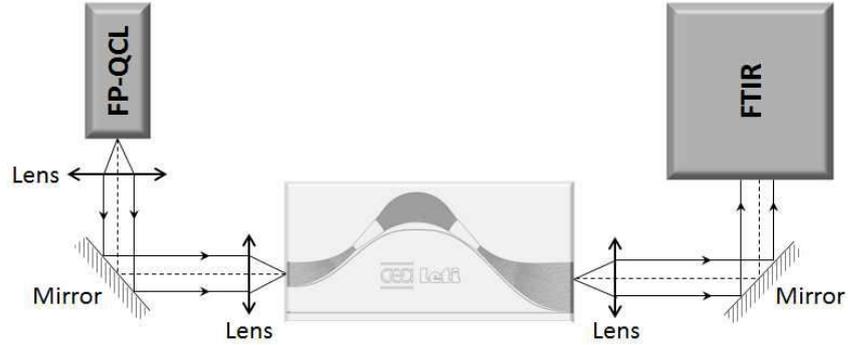

Fig. 4.1. Schéma du banc expérimental.

|  | $w_x$ | $w_{MMI}$ | $\Delta\sigma$ | $N_{ch}$ | $\sigma_{max} - \sigma_{min}$ | **puce** |
|---|---|---|---|---|---|---|
| AWG 1 | 4.6 µm | 9 µm | | 17/35 | | |
| AWG 2 | 4.6 µm | 11 µm | 3 cm$^{-1}$ | 17/35 | (1700-1799) cm$^{-1}$ | 36 / 32 |
| AWG 3 | 4.6 µm | 11 µm | | 10/35 | | |
| AWG 4 | 4.8 µm | 9 µm | | 17/35 | | |

Tab. 4.1. Caractéristiques des AWGs: largeurs des guides $w_x$ et MMIs $w_{MMI}$, espacement entre canaux $\Delta\sigma$, nombre de canaux $N_{ch}$, bande spectral, puce.

Un schéma du banc expérimental est représenté sur la figure 4.1. La source QCL dans le moyen infra-rouge est fabriquée par le III-V Lab [4]. Il délivre une puissance moyenne de 50 mW quand elle piloté par un générateur de pulse HP 8114 de Hewlett Packard avec une largeur d'impulsion de 1 µs et une fréquence de répétition de 100 kHz. Les composants à caractériser (DUT pour Device Under Test) sont détaillés dans le tableau 4.1. Les mesures des réponses spectrales sont représent ées sur la figure 4.2. Les performances spectrales des dispositifs sont résumées dans le tableau 4.2.

|  | $L_c$, dB | $L_u$, dB | $X_{ch}$, dB | $L_x$, dB | $\Delta f_{FSR}$, cm$^{-1}$ |
|---|---|---|---|---|---|
| | simulation | | | | |
| AWG1 | -4.8 | 3.5 | 7.8 | 37 | 157 |
| AWG2 | -6.3 | 3.6 | 6.7 | 28 | 157 |
| AWG4 | -5.2 | 3.4 | 7.6 | 38 | 157 |
| | characterization, chip 36 | | | | |
| AWG1 | -3.3 | 2.9 | 4.6 | 20 | 134 |
| AWG2 | -3.4 | 2.7 | 4 | 18 | 132 |
| AWG4 | -6 | 0.5 | 4.4 | 18 | 132 |
| | characterization, chip 32 | | | | |
| AWG1 | -4.8 | 3.1 | 4 | 15 | - |
| AWG2 | -5.7 | 2.2 | 4 | 19 | - |
| AWG4 | -4 | 3.8 | 4.7 | 20 | - |

Tab. 4.2. Spectral parameters at central insertion case of AWG 1, AW2, and AWG4 (chip36) experimental and theoretical.

En premier, nous comparons les dispositifs AWG1 et AWG2 qui sont similaires hormis la largeur des coupleurs MMI des guides de sortie, qui sont de 9 μm et 11 μm, respectivement. Si on se réfère aux résultats des simulations, le MMI le plus large doit être choisi pour satisfaire au gabarit, i.e. une réponse plate apodisée des sorties. La théorie prévoit une diaphonie $X_{ch}$ améliorée de 7.8dB à 6.7dB et en contrepartie une légère augmentation des pertes d'insertion $L_c$ de -4.8dB à -6.3dB avec l'augmentation de la largeur des MMIs. Expérimentalement, on observe cette tendance sur une puce seulement. La diaphonie décroit de 4.6 dB à 4 dB sur la puce 36, mais sont égaux à 4 dB sur la puce 32.

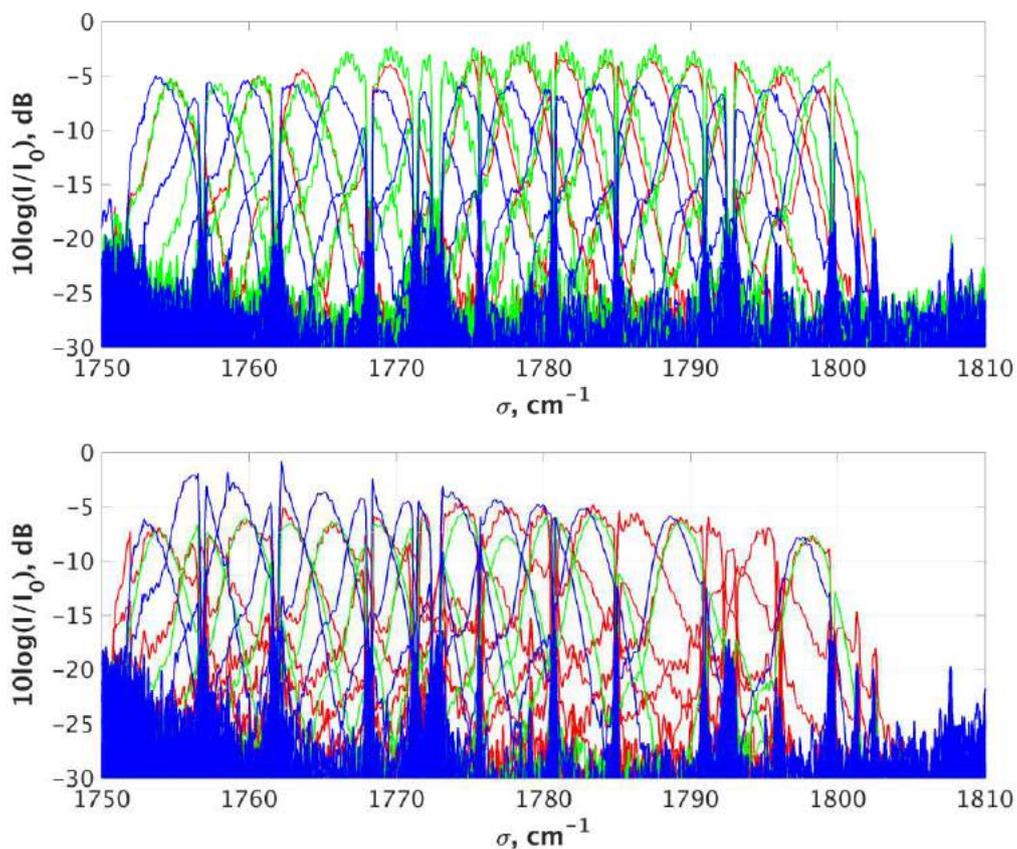

Fig. 4.2. Réponse spectrale duesDUTs: puce 36 (haut) et puce 32 (bas).

La détérioration des pertes d'insertion $L_c$ sont supérieures de 0.1dB pour la puce 36, et équivalentes aux simulations soit une valeur de 0.9dB pour la puce 32. Tous les DUTs de la puce 36 ont des pertes d'insertion au-dessus des prédictions théoriques, ce qui est due à un défaut de fabrication des guides de références et des coupleurs MMI. Pour les DUTs de la puce 32, la corrélation entre la largeur du MMI et les pertes d'insertion est évidente et proche des valeurs attendues.

La variation des pertes d'insertion est dépendante du choix du FSR, qui est lui-même lié aux paramètres géométriques comme la longueur du guide plan $R$ et la différence de longueur du faisceau de guides $\Delta L$, et est pratiquement indépendant des largeurs des guides d'onde et MMI. Aussi pour les trois DUTs, la variation des pertes d'insertion est attendue à un niveau de 3.5dB ce qui est vérifié expérimentalement sauf le composant AWG4 de la puce 36. Les plus faibles pertes d'insertion obtenues sont de -6dB.

le FSR calculé est de 157 cm$^{-1}$. Le fait d'avoir un nombre suffisent de guides d'entrée et la grande largeur spectrale de la source QCL utilisée pour la caractérisation, nous a permis de mesurer les ordres de diffraction consécutifs des AWGs de la puce 36. En moyenne, the FSR expérimental est de 133 cm$^{-1}$.

Il est clairement visible que la réponse spectrale des composants AWG4 se décale vers les grandes longueurs d'onde (faibles nombre d'onde) sur toutes les puces, ce qui vérifie que le décalage est lié directement à la variation de l'indice effectif des modes qui se propage dans le faisceau de guide. Les variations moyennes des nombres d'ondes du guide central par rapport à aux valeurs visées sont de -2.5 cm$^{-1}$, -2.4 cm$^{-1}$ and -3.2 cm$^{-1}$ pour les AWG1, AWG2 et AWG4, respectivement. Pour, l'AWG4, les pics se décalent de -0.7 cm$^{-1}$ – -0.8 cm$^{-1}$, ce qui correspond à une variation de l'indice effectif du mode guide de 0.0014 – 0.0016. On rappelle que les calculs préliminaires réalisée avec le logiciel RSoft prédisaient une variation de 0.001 sur l'indice effectif induite par les tolérances de fabrication et notamment pour une variation de 10 nm de la largeur du guide.

### 4.2 Réponses des guides périphériques

Dans le faisceau de guides, les modes des guides avec des nombres d'onde autres que la valeur centrale possèdent une avance ou un retard de phase, ce qui entraine un décalage du front d'onde à sa sortie. En conséquence, un tel faisceau va se focaliser sur un point qui sera incliné par rapport à au centre du cercle de Rowland d'un angle dans l'approximation des petits angle correspondant à :

$$\theta_{out}(\sigma) \approx \frac{\left(\frac{\sigma_c}{\sigma} \cdot n_{effc} - n_{eff}(\sigma)\right) \cdot \Delta L}{d \cdot n_s(\sigma)} - \theta_{in} \tag{4.1}$$

où $\sigma\sigma$ est le nombre d'onde, $d$ $d$ est l'écartement entre guides du faisceau de guides à son extrémité avec le guide plan, $n_{eff}(\sigma)$ et $n_s(\sigma)$ sont les indices effectifs dans le faisceau de guides et dans le guide plan respectivement, $\theta_{in}$ et $\theta_{out}$ sont les positions angulaires des canaux d'entrée et de sortie respectivement, $\Delta L$ est la différence de longueur du chemin optique dans le faisceau de guides, l'indice "c" fait référence au nombre d'onde central et le terme $\theta_{out}$ représente la dispersion angulaire. La position du faisceau diffracté dans le plan de focalisation est obtenue par la formule $\Delta s = R \cdot \theta_{out}$.

La Figure 4.3 représente le décalage en nombre d'onde par rapport à l'espacement entre canal théorique de 3 cm$^{-1}$ des différents guides de sortie de l'AWG à 5,7 µm par rapport au nombre d'onde du pic de résonance de la sortie correspondante. Les points correspondent aux décalages spectraux des sorties tandis que les lignes représentent la position de l'entrée avec comme position centrale le canal 9. Les pentes négatives indiquent que l'écart spectrale entre canaux est plus faible que celui prévu lors de la conception $\Delta\sigma_{ch}$, et les pentes positives le contraire.

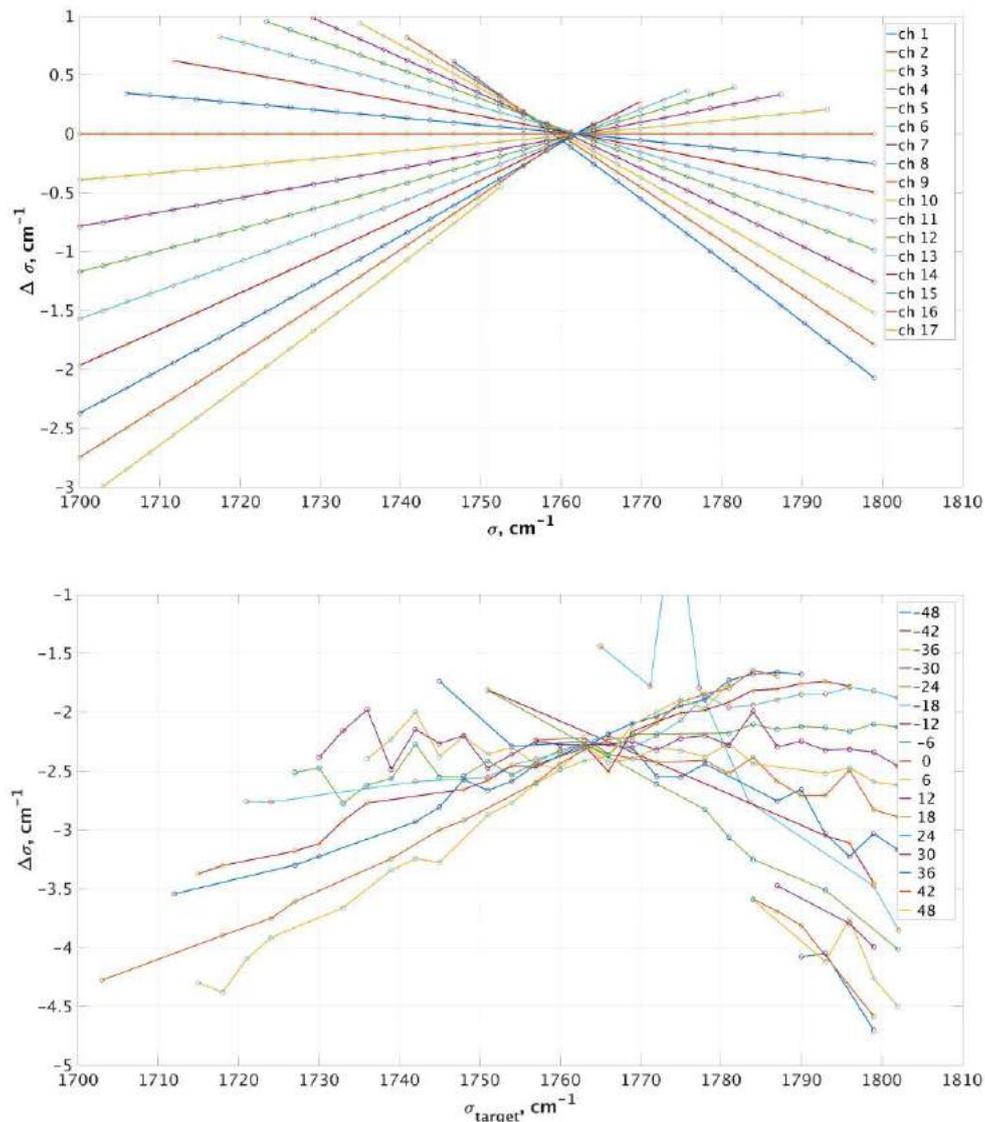

Fig. 4.3. Décalage spectrale de l'espacement entre canaux pour les entrées périphériques de l'AWG à 5.7 µm: Calculés (haut) et expérimentaux (bas).

La valeur de la pente est proportionnelle au décalage de l'espacement entre canaux $\Delta\sigma$. L'écart maximal est obtenu pour le canal 17. Il est égal à $\Delta\sigma_{Input17}$ = 0.16 cm$^{-1}$, ce qui représente une dérive de 5% avec les calculs de l'optimisation. Quoiqu'il en soit, la dispersion de la variation de l'espacement entre canaux est égale à 4 cm$^{-1}$. On constate que les lignes ne se croisent pas au milieu de la courbe. On l'attribue au fait que la loi de variation de l'indice de réfraction du SiGe linéaire utilisée lors de la conception n'est pas suffisamment précise. Des mesures plus récentes ont montré que la relation est quadratique. De plus, la position des guides de sortie qui permettent de conserver un écart spectral constant de 6 cm$^{-1}$ correspond à des écartements entre guides qui n'est pas parfaitement équidistant.

### 4.3 Erreur de phase

Nous avons déterminé le niveau de diaphonie induit par une valeur de l'écart type des erreurs de phase mesurée avec l'outil semi-analytique développé sous Matlab lor de ce travail de thèse. Les mêmes valeurs de diaphonies sont observées expérimentalement. Les mesures du spectre du canal 27 de AWG2 (trait bleu) est reporsur la figure 4.4. La courbe simulée est représenté en tiret avec un écart type de l'erreur de 0.0068 rad.

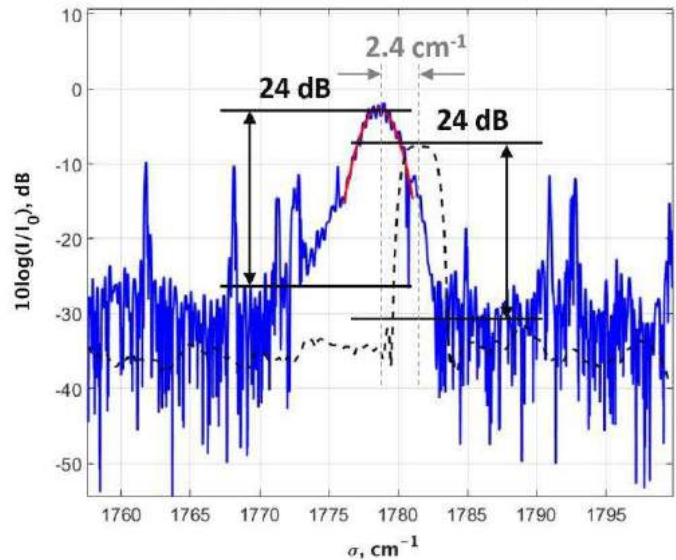

Fig. 4.4. haut : histogramme de la distribution de l'erreur de phase; bas : réponse spectrale du canal 27 de l'AWG2 (puce 36).

La courbe rouge représente l'ajustement avec une gaussienne. La variation équivalente de l'indice effectif peut être évalué par la relation : $\Delta n_{eff} = \Delta\varphi\lambda/(2\pi l_x)$, où $l_x$ est la longueur du guide d'onde. Pour les dispositifs AWGs, cette variation est de $3.5 \cdot 10^{-5}$ pour un guide d'onde de 1 cm de long.

## 4.4 Mesures des variations des performances des AWGs avec la température

Les réponses spectrales des AWG2 et AWG4 ont été mesurées à cinq températures différentes 20°, 27.5°, 32.7°, 36.6° et 41.4°. Le choix des DUTs est lié au fait que ces deux dispositifs ont des faisceaux de guides de largeurs de guides différentes. Les résultats sont résumés dans le tableau 4.3. Les mesures sont en bon accord avec les simulations et montrent une dépendance quasi-linéaire avec la température (voir Fig. 4.5).

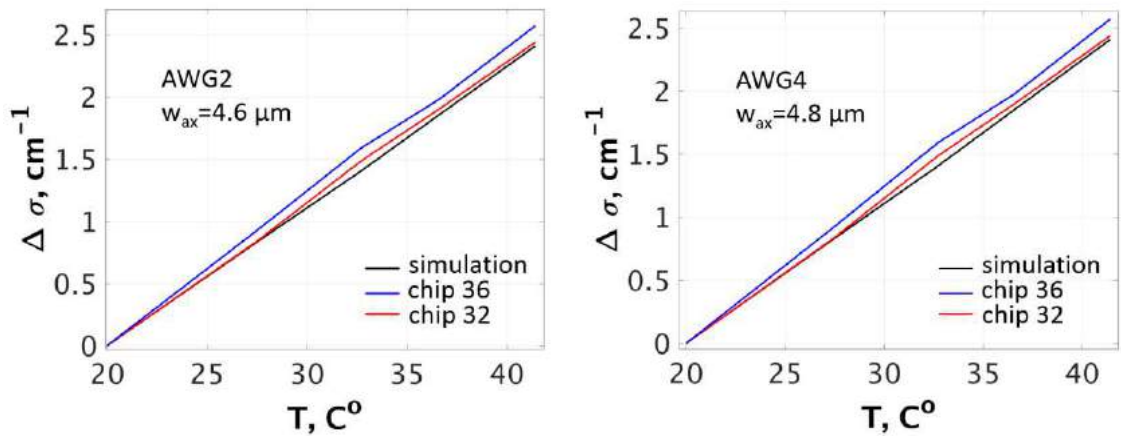

Fig. 4.5. Variation du décalage spectral en fonction de la température des AWG2 et AWG4 ayant des largeurs de guides du faisceau de 4.6 et 4.8 µm, respectivement.

|  |  | $<\Delta\sigma>$, 27.5°C | $<\Delta\sigma>$, 32.7°C | $<\Delta\sigma>$, 36.6°C | $<\Delta\sigma>$, 41.4°C |
|---|---|---|---|---|---|
| AWG2 | simulation | 0.834 | 1.405 | 1.856 | 2.408 |
|  | chip 32 | 0.840 | 1.485 | 1.902 | 2.438 |
|  | chip 36 | 0.927 | 1.589 | 1.980 | 2.570 |
| AWG4 | simulation | 0.836 | 1.407 | 1.859 | 2.411 |
|  | chip 32 | 0.897 | 1.723 | 2.037 | 2.485 |
|  | chip 36 | 0.912 | 1.533 | 2.073 | 2.527 |

Tab. 4.3. Valeur moyenne des décalages spectraux induit par la variation de température en cm$^{-1}$.

## 5  Améliorations potentielles

Si on veut diminuer la diaphonie des AWGs, une des voies pour y parvenir est de réduire le nombre de guides du faisceau. Nous proposons d'utiliser un AWG en forme de U avec des jonctions Y à l'entrée du faisceau de guides (voir Fig. 5.1), ce qui permet de diminuer d'un facteur 2 le nombre de guides et de les éloigner les uns des autres pour réduire l'effet de couplage guide à guide. Les améliorations de ce nouveau dispositif sont : 1) Les entrées et les sorties des coupleurs planaires ont leurs axes de symétrie parallèle; 2) Le plus petit guide du faisceau est en arc de cercle avec un rayon de courbure $R_{min}$ tel que les pertes de radiation soient inférieures à 0,1 dB dans la totalité de la bande spectrale de l'AWG; 3) Les autres guides sont constitués de partie droites et courbes de tel manière qu'en leur centre, les distances de séparation soient pratiquement équidistantes; 4) Pour améliorer le couplage dans le coupleur planaire d'entrée et réduire le couplage guide à guide, on utilise des jonctions Y. La distance de séparation est la même des deux côtés du faisceau de guides et est égale à la période du réseau de diffraction $d$.

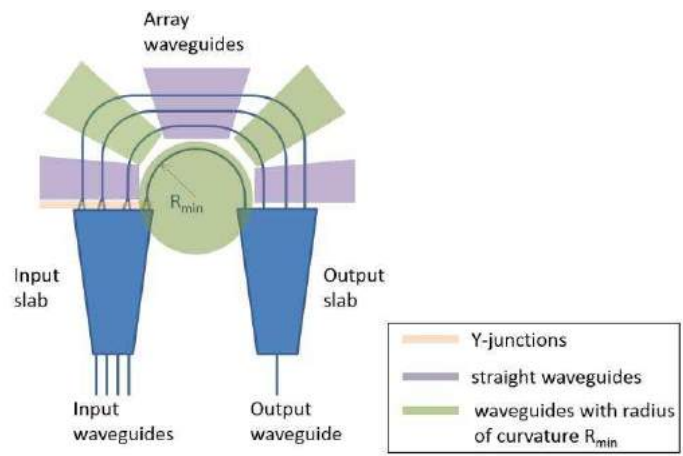

Fig. 5.1. Description d'un AWG en forme de U comprenant des jonctions Y.

Les pertes des jonctions Y et des courbures sont de 0.1 dB et 0.4 dB, respectivement.

Les calculs sur l'AWG en forme de U ont été réalisés pour tester sa stabilité vis-à-vis de l'erreur de phase. Pour cela, nous avons comparé les réponses spectrales théoriques de deux dispositifs fonctionnant à 5.7 µm en configuration classique et en forme de U avec un nombre de canaux optimal $N$. Les calculs ont été effectués avec l'outil analytique développé sous Matlab. Les résultats sont résumés dans le tableau 5.1.

|  | AWG classic, N=243 | AWG classic, N=130 | AWG w/Y-junction, N=260/130 |
|---|---|---|---|
| $L_c$ | -2.27 | -2.69 | -2.67 |
| Crosstalk $L_x$ at $<\varphi> = 0$ rad | 73.76 | 29.88 | 30.53 |
| Crosstalk $L_x$ at $<\varphi> = 0.0068$ rad | 23.76 | 23.28 | 27.81 |
| $\Delta L_x$ | 50.00 | 6.60 | 2.72 |

Tab. 5.1. Comparaison des performances des 2 configurations d'AWG (standard et en forme de U).

## Conclusion

L'objectif de ces travaux de thèse a été de développer des composants d'optique intégrée de multiplexage en longueur d'onde pour la détection de gaz dans le moyen infra-rouge sur des substrats de silicium.

La première partie de cette étude a été consacrée à la conception de multiplexeurs AWGs. Les outils semi-analytiques développés préalablement par le Dr. P. Labeye en C++ ont été adaptés sous Matlab. L'analyse de la propagation de la lumière dans le dispositif utilise le formalisme de transformée de Fourier optique qui conduit à des expressions analytiques dans le cas de faisceaux gaussiens. Une routine d'ajustement a été également ajoutée pour approximer le profil de la puissance des modes guides dans les différentes parties du dispositif à des faisceaux gaussiens. Les améliorations de l'outil ont également permis d'étudier les dégradations des performances induites par l'erreur de phase. Un AWG à 35 canaux fonctionnant à la longueur d'onde de 5.7 µm pour la détection optique de gaz comme le nitrate de chlore, l'acide nitrique, l'acide formique et l'acide formaldéhyde.

Les configurations optimales des AWGs, i.e. avec des coupleurs MMI intégrés au faisceau de guides ont été fabriqués et caractérisés optiquement dans le moyen infra-rouge. La fabrication a été réalisée dans le département Technologie du silicium du CEA-Leti MINATEC. Les mesures optiques ont été faites sur deux puces provenant de périphérie et du centre du substrat pour évaluer la robustesse de la technologie de fabrication. Nous avons étudié l'impact de la largeur des coupleurs MMI sur l'apodisation des réponses spectrales des dispositifs. La dépendance à la température des réponses spectrales des AWGs a également été étudiée. Les mesures sur cinq valeurs de température ont montré une réponse quasi-linéaire en accord avec le modèle théorique développé au préalable dans ce travail de thèse.

Une configuration alternative aux AWGs de multiplexeurs PCG a été envisagée. Un outil de simulation analytique utilisant le même formalisme que celui utilisé pour les AWGs a été conçu sous Matlab. Il permet d'obtenir les réponses spectrales des dispositifs en fonctions des paramètres géométriques. Afin de pouvoir fabriquer ces multiplexeurs, un outil de génération automatique de masque de PCGs en configuration de Rowland a été développé. Les dessins de masque ainsi générés sont directement importables sous CleWin, le logiciel utilise pour la préparation des masques de fabrication. Ces nouveaux outils numériques, nous ont permis de concevoir des multiplexeurs PCG fonctionnant aux longueurs d'onde de 3.4 µm, 6.9 µm et 8.7 µm, où les deux premiers utilisent des guides à gradient d'indice SiGe/Si et le dernier un guide à saut d'indice de Ge/SiGe. Les composants sont en cours de fabrication et n'ont pu être testés à ce jour.

Dans l'expectative de l'amélioration des performances des multiplexeurs étudiés, nous avons proposé une nouvelle architecture d'AWG utilisant des jonctions Y dont le but est de réduire le nombre de guides d'onde du faisceau de guides afin d'augmenter la robustesse du dispositif à l'erreur de phase induite par la rugosité des flancs. L'intégration de sources QCL miniaturisées aux guides d'entrées du multiplexeur permettra à terme de réaliser une source très large bande qui constitue à lors actuelle un des verrous à la réalisation d'une plateforme intégrée d'un système de détection de gaz dans le moyen infra-rouge.

# Table of contents

# Introduction

A need in gas detection systems nowadays is traced in a wide range of fields. The potential hazards caused by gases such as combustibility, flammability, toxicity or depletion of oxygen are important to be forewarned in vast variety of human activity from monitoring industrial emissions to respiratory examination in health-care including greenhouse studies.

Historically, the first approaches to detect the presence of methane and depletion of oxygen were taken in the coal mining industry. Long before any gas sensing technologies were available, the ambient air in the mine was examined by a canary in the cage, since the bird has a higher susceptibility than a regular human. At the beginning of the 19$^{th}$ century, the more benign technique, a flame safety lamp, became popular. By high flame, the lamp indicated the presence of methane and by low flame – the depletion of oxygen. The method remained unsafe, since a small crack of lamp's fragile glass could cause heavy losses. In 20s of 20$^{th}$ century, with high growth of industrialization, the gas detecting systems started actively developing. During the following years, a wide range of gas sensing technologies have been established.

Distinguished by their operation principle, the following types of gas detectors are known [1]: catalytic sensors for detecting combustible gases; thermal conductivity sensor for detecting gases that have greater thermal conductivity than the air; electrochemical gas sensors for detecting low concentrations of combustible and reducing concentrations of gases; optical gas sensors for gases possessing a characteristic absorption spectra in the infrared; acoustic gas sensors for detecting leaks from pressurized gases; and semiconductor sensors for detecting hydrogen, oxygen, alcohol and carbon monoxide.

Optical gas sensors offer a range of advantages compared to other types of gas detecting devices, such as higher sensitivity and selectivity, short time responsivity allowing real time gas detection, low sensitivity to environmental changes, as well as potential detection of multiple gases. The principle of operation is based on absorption specificity of gas molecules in mid-infrared spectral range that allows the concentration of particular chemical compounds to be precisely determined. The optical sensing devices are preferable where indirect interaction with gases is needed. They are often used for monitoring toxic and combustible gases. The major drawbacks of traditional optical gas sensors are the large packaging dimensions and high cost. So the practical task is the miniaturization without sacrificing the functionality such as gas specificity, low power consumption and reliability. The recent trend arising in the course of the last quarter of the century is the development of a miniature gas sensor assembled on a chip. The gas sensing based on integrated optics greatly benefits from two factors. In respect of theory and numerical tools, it was triggered by the flourishing optical communications, and in fabrication, it uses the basis of technologies developed earlier for semiconductor electronics.

The important part of the optical gas sensor is a powerful mid-infrared broad-band source. It can be achieved by integration of recently developed miniature powerful quantum cascade lasers with an optical multiplexer. The present work is devoted to a study of new multiplexer architectures for mid-infrared gas sensors integrated on silicon.

The remainder of this document is organized into five chapters. In the first chapter, the optical gas sensing principle is addressed, followed by description of the concept of gas sensor integrated on a chip. The current progress on mid-infrared waveguides and multiplexing devices is presented. In the second chapter, we cover the theoretical basics supported by numerical calculations. The waveguides, basic building blocks such as tapers, Y-junctions and MMI couplers, as well as two multiplexer configurations are discussed. For the latter, in particular, simulation results based on the semi-analytical tool developed in MATLAB™ are presented. The numerical tool is based on analytical calculation of the field using scalar diffraction theory, which allows to demonstrate the correlation of crosstalk at varying levels of phase errors. In the third chapter, we discuss the fabrication process and technology used for waveguide based devices, followed by the characterization results of the devices in the fourth chapter. The fifth chapter describes potential design alternatives that could improve the spectral responses of two multiplexer configurations.

# 1 State of the art

In this chapter, we discuss the principle of mid-infrared gas sensing, mid-IR waveguides presented earlier and waveguide based optical multiplexers. Different configurations of multiplexers, the history of their evolution and the current progress are reviewed.

**Contents**

## 1.1 Operational principle of optical gas sensors

The operational principle of optical gas sensors is based on absorption specificity of gas molecules. The presence of particular amount of gas is defined by illuminating the sample and measuring the residual light power as a function of wavelength. Origins of the absorption spectra differ depending on the spectral range of the source [2]. In 200-400 nm range, the gases illuminated by UV laser absorb due to electronic transitions. In near infrared (IR), 700 nm-2.5 µm, the first harmonic of molecular vibration and rotation play the key role of indicator. The mid-IR spectral range, 2.5-14 µm, covers the majority of fundamental modes of molecular vibrations and rotations. Mainly, optical gas sensors are based on mid-IR spectroscopy. Since majority of target gases possess inherit property of strong distinct absorption in infrared (Tab.1.1), IR sensing devices are highly selective allowing precise determination of the chemical composition.

The principle is explained as follows. The atoms constituting a molecule are consistently vibrating about their equilibrium positions. The frequency of the vibrations depend on the molecule, in particular, its mass and the length and strength of inter-atomic bonds. Specific molecule absorbs radiation of the frequency that corresponds to the natural frequency of vibration of the atomic bonds (in the range of 12 – 120 THz) which fall into infrared region.

In general, it is assumed that vibrations of certain type occur independently of the others around it where the atoms are considered as simple harmonic oscillators. Major vibration types are shown in Fig. 1.1. The mass center of the molecules is taken constant during these vibrations. It should be noted that only vibrations resulting in the change of a dipole moment possess resonant frequencies in the infrared region of the spectrum. In this light, homonuclear gas molecules, such as $H_2$ and $O_2$, do not exhibit distinguishable absorption peaks in infrared.

| List of common target gases | mid-IR absorption peaks |
|---|---|
| Alkanes / saturated hydrocarbons (methane, ethane, propane, butane, pentane, hexane, heptane, etc.) | 3.38 – 3.5 µm / 6.8 – 6.9 µm |
| Cycloalkanes (cyclopropane, cyclohexane, methyl cyclohexane, etc.) | 6.9 µm / 7.4 µm |
| Alkenes / unsaturated hydrocarbons (ethylene, propylene, butene, pentene, hexene, octene, etc.) | 5.95 – 6.10 µm |
| Cycloalkenes (cyclohexene and pinene) | 3.30 µm / 3.75 µm / 6.01 µm / 6.96 µm |
| Aromatics (benzene, toluene, and xylene) | ≈ 3.30 µm / 5.88 – 6.67 µm |
| Alcohols (methanol, ethanol, propanol, and allyl alcohol) | 2.82 – 3.13 µm |
| Amines (dimethyl amine, trimethyl amine, butanamine, cyclopropanamine, and pyridines) | 2.86 – 3.03 µm |
| Ethers (dimethyl ether, ethyl ether, n-propyl ether, methylvinyl ether, vinyl ether, ethylene oxide, tetrahydrofuran, furan, and 1,4-dioxane) | 5.71 – 5.76 µm |
| Ketones (acetone, methyl ethyl ketone, pentanone, methyl isobutyl ketone and heptanone) | 5.71 – 5.95 µm |
| Aldehydes (formaldehyde, ethanal, propanal, butanal, pentanal, etc. ) | 3.4 – 3.55 µm |
| Carbon dioxide, carbon monoxide | 4.3 – 4.6 µm |
| Ozone | 4.6 – 5.0 µm |
| Nitrogen dioxide, nitrous oxide, nitric oxide | 3.42 µm, 4.47 µm, 5.26 µm |

Tab. 1.1. Types of optical gas sensors. Adopted from http://webspectra.chem.ucla.edu//irtable.html and [3].

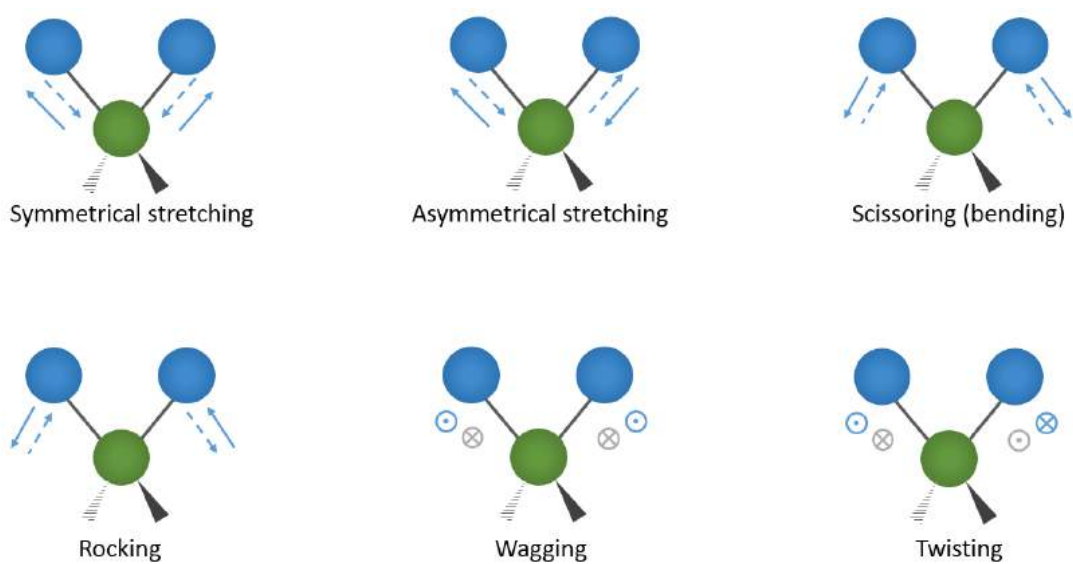

Fig. 1.1. Sketch of various vibration modes in molecules: blue and green spheres correspond to atoms of molecules, grey lines – indicate bonds between atoms, arrows show vibration directions, ⊙ - refers to vibration direction pointing towards the reader, ⊗ - refers to direction pointing in the opposite direction to the latter.

In infrared spectroscopy, the gas and its concentration are derived using Beer-Lambert law:

$$P_1 = P_0 \exp(-\alpha L), \tag{1.1}$$

where $P_0$ is the power of the illumination received by the sample, $P_1$ is the power of illumination after the absorption by the sample, $\alpha$ is the linear absorption coefficient per unite length, and $L$ is the length of the measured piece.

In some cases, it is convenient to express the absorption coefficient in dB $\alpha_{dB}$:

$$\alpha = \frac{\ln(10)}{10} \alpha_{dB}. \tag{1.2}$$

The absorption coefficient $\alpha$ is defined as follows:

$$\alpha = c \cdot \varepsilon, \tag{1.3}$$

where $c$ is the gas concentration and $\varepsilon$ is the specific absorptivity of gas. The concentration can be expressed in atm, the partial pressure in atmosphere, then the specific absorptivity is taken in cm$^{-1}\cdot$atm$^{-1}$ and length in cm, respectively.

The analysis of measured absorption spectra by applying Beer-Lambert law and using HITRAN data base [3], allows to determine the presence of specific gases as well as their concentrations.

In our case, the optical gas sensor is based on recently developed concept of photonic devices embedded on a single chip. Among advantages of such gas detector are the high compactness and capability of multiple gas sensing. Its concept is described in the following chapter.

## 1.2 Gas sensor integrated on chip

The scheme of potential optical gas sensor integrated on a chip is shown in Fig 1.2. It has a compact design consisting of a broad-band source and a sensing part. It is important to have a powerful mid-infrared multi-wavelength source with controllable wavelength. Such source can be built by integration of an array of quantum cascade lasers (QCL) [4] with each laser having a distinct wavelength, together with a multiplexer, that allows scanning a broad spectral region.

Traditional mid-IR laser sources have some major drawbacks: lack of continuous wavelength tunability, low output power and cooling requirement of lead salt diode lasers. Recently developed quantum cascade lasers (QCLs) [4] provide the tunability in the region important for spectroscopy, i.e. from 3 to 20 µm, with excellent properties in terms of narrow line width, the average power as low as tens of milliwatts, and room temperature operation.

The mid-IR broadband source made of QCLs with the multiplexer can be used for on-chip gas sensors. The sensing and detecting can be based on a number of existing and developing techniques, such as photo-acoustic cell [5], slot waveguides used for direct evaluation of absorbed portion of illumination allowing the enhancement of the field amplitude in the gap up to 50 times and higher [7], or cells utilizing Vernier effect for ultra-high performance sensors [8]. The intensity can be registered by miniature mid-IR photodetectors made of group III–V compound semiconductors such as AlGaAs-GaAs [9].

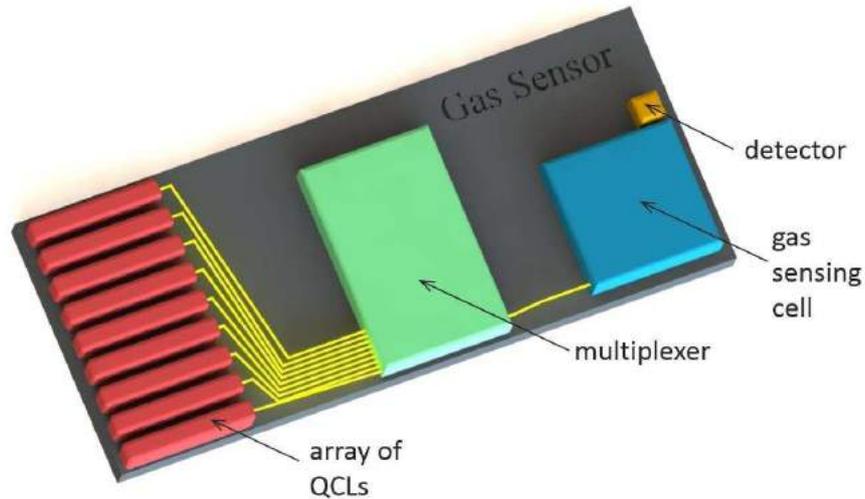

Fig. 1.2 Scheme of optical gas sensor integrated on a chip.

The pursuits to verify the concept of all optical integration were done throughout the last two decades [10] – [17]. One of the pioneers was Dragone et al., who integrated monolithically 7x7 and 11x11 multiplexers based on arrayed waveguide grating (AWG) configuration. In this work, based on $SiO_2$ /Si technology, they have achieved wave guiding at 1.3 µm wavelength with insertion loss below 2.5 dB and crosstalk less than -25 dB [10]. More recent integration also based on InP waveguides at 1.54 µm wavelength was carried out by Tolstikhin et al. [11], where they have integrated the 44-channel multiplexer based on planar concave grating (PCG) configuration together with photodetectors, achieving a footprint size of 20 mm x 7.3 mm. A compact spectroscopic sensor using AWG multiplexer operating in the visible spectral range was built by Kodate et al. [12]. Ma et al. [13] demonstrated a 9 mm x 6 mm spectrometer integrated on a chip, where they integrated their source with PCG multiplexer operating around 850 nm wavelength. Another spectrometer operating in broad near-infrared region 1500 to 2300 nm wavelength was realized by Ryckeboer et al. [14]. They integrated four PCG multiplexers with GaInAsSb photodetectors on a silicon-on-insulator chip. Lin et al. introduced their lab-on-chip mid-infrared gas sensor concept with PbTe planar detectors and chalcogenide glass ($As_2Se_3$) waveguides designed to operate at around 3.2 µm [16]. Stanton et al. [17] presented a platform for ultra-broad band source that covers a wide spectral range spanning from ultraviolet to mid-infrared, 350 nm – 1500 nm and 1500 nm – 6500 nm. The fabricated chip combined silicon nitride and silicon on insulator technologies, allowing to integrate four AWGs each operating at certain wavelength range. According to their experimental results, the device showed transmission efficiency of 0.9 for wavelength range 780 nm to 3600. In light of recent developments, the forthcoming of powerful ultra-broad band source covering the whole mid-IR spectral range for integrated optical gas sensor becomes feasible.

The low loss mid-IR waveguides became available allowing the design of waveguide based passive integrated optics devices such as multiplexers. Mainly, the waveguide technologies are based on two platforms, one is Ge-on-Si and the other is InGaAs-on-InP. Although both technologies are suitable for infrared wave guiding with acceptable performances, the latter concedes due to higher propagation and coupling losses [37]. The state of the art in this direction is covered in the following chapters.

## 1.3 SiGe waveguides

Over the past three decades, various technologies are being developed towards the increase of functionality and compactness of integrated optics components in order to reduce the net cost of potentially commercial products. The higher contrast waveguides allow to achieve higher confinement by dramatically decreasing the size of waveguides.

The first micrometer size planar lightwave circuits were developed for telecommunication application. The rapid expansion of network traffic due to increasing multimedia communications required the upgrade of network capacity. The earlier network nodes used optical-electrical conversions, electrical (de)multiplexing, which in turn limited the throughput of the nodes [19]. This shortcoming prompted the active development of optical communication systems such as wavelength-division-multiplexing technologies that allowed the wavelength-selective operation and thus were considerably more efficient. The wavelength-division-multiplexing devices, based on silicon-on-insulator technology operating at 1.55 μm wavelength, play the key role in telecommunication applications, whilst they are of limited use in gas sensing due to strong absorption of silica at wavelengths 2.7-2.8 μm and over 4 μm [18].

The great heritage of engineering solutions established for telecom wavelength in its turn accelerated the development of mid-IR waveguide devices. With the recent availability of high refractive index contrast waveguides based on germanium-silicon compound as well as indium gallium arsenide on indium phosphide waveguides, low loss wave guiding in mid-IR became possible [20] – [26].

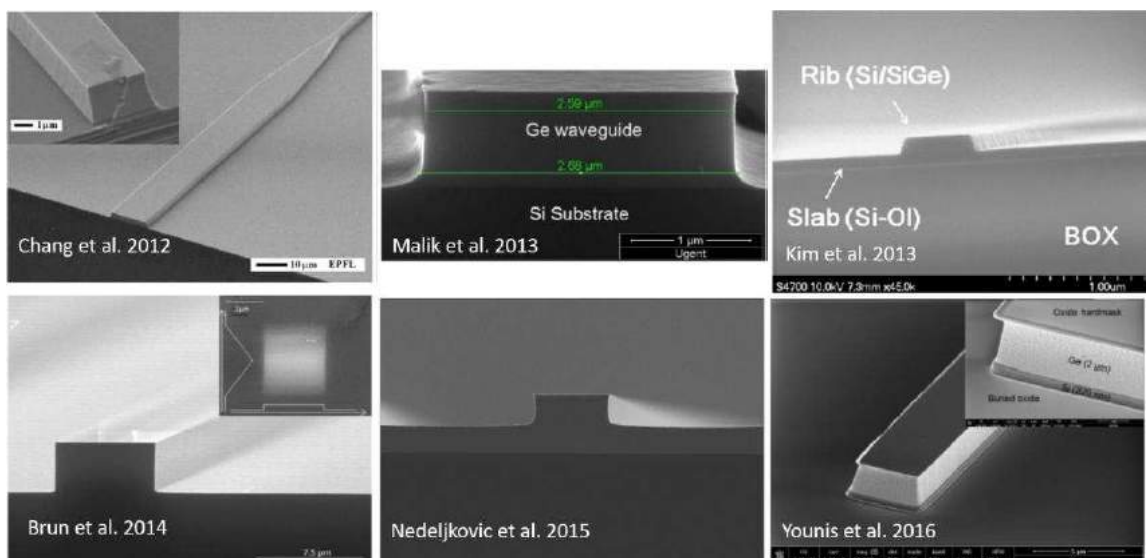

Fig. 1.3. SEM images of mid-IR waveguides cross-sections reported in 2012 - 2016.

The possibility of using SiGe compound as a material for microelectronics was investigated in as early as 1950s [27]. Later with the availability of epitaxial growth on Si substrate in nineties, such advantages as transparency up to 8 µm and higher performance provided by superior non-linear properties were brought forward. The flexibility in choosing germanium concentration in SiGe compound created the opportunity to "tune" the refractive index of mid-IR waveguide. Soref [27] presented theoretical study of single-mode propagation condition for Si/Si$_{0.9}$Ge$_{0.1}$/Si waveguides where germanium concentration of the core was 10 percent.

Within the latest five years, the fabrication and test of waveguides operating at different mid-IR wavelengths were reported, see in Fig. 1.3. Groups of Chang et al. [20] and Nedeljkovic et al. [23] demonstrated open top germanium on silicon waveguides. The strip waveguide designed by Chang et al. guides at 5.8 µm wavelength with 2.5 dB propagation loss. The rib waveguide presented by Nedeljkovic et al. a few years later exhibits 0.6 dB propagation loss at 3.8 µm wavelength.

Kim et al. investigated the possibility to develop a waveguide based on silicon-germanium compound operating at telecom wavelength. They demonstrated Si/Si$_{0.72}$Ge$_{0.28}$/Si photonic wire waveguide with germanium concentration fixed at 28 % for guiding at 1.55 µm with 5.4 dB/cm propagation loss [21]. Malik et al. presented the waveguide made of germanium core sandwiched between silicon cladding layers, operating at 5.2 – 5.4 µm wavelength with propagation loss of 3 dB/cm [22]. The InGaAs/InP waveguides were demonstrated by Gilles et al. at 7.4 µm with propagation loss of 2.9 dB [24]. Luke et al. studied Si$_3$N$_4$/SiO$_2$ waveguides and demonstrated the significant absorption reduction of Si$_3$N$_4$ around 3 µm by means of deposition-anneal cycling [28]. Younis et al. reported germanium on SOI waveguides measured at 3.82 µm with 8 dB/cm propagation loss [25]. Brun et al. from CEA demonstrated first graded index waveguides in mid-IR made of Si/Si$_x$Ge$_{0.4-x}$/Si, operating at 4.5 µm and 7.4 µm wavelengths with propagation losses of 1 dB/cm and 2 dB/cm, respectively [26].

| Waveguide | Reported by / in | Propagation loss | Operational λ | Modal property |
|---|---|---|---|---|
| Air/Ge/Si | Chang et al. 2012 | 2.5 dB/cm (TM) | 5.8 µm | single mode |
| Si/Si$_{0.72}$Ge$_{0.28}$/Si | Kim et al. 2013 | 5.4 dB/cm (TE) | 1.55 µm | single mode |
| Si/Ge/Si | Malik et al. 2013 | 7 dB/cm / 3dB/cm (TE) | 3.7 – 3.8 µm / 5.2 – 5.4 µm | single mode |
| Air/Ge/Si | Nedeljkovic et al. 2015 | 0.6 dB/cm (TE) | 3.8 µm | multi mode |
| InGaAs/InP | Gilles et al. 2015 | 2.9 dB/cm (quasi TM) | 7.4 µm | single mode |
| Ge/SOI | Younis et al. 2016 | 8 dB/cm (TE) | 3.68 µm | single mode |
| Si/Si$_x$Ge$_{0.4-x}$/Si | Brun et al. 2014 | 1 dB/cm / 2 dB/cm (TM) | 4.5 µm / 7.4 µm | single mode |

Tab. 1.2. Mid-infrared waveguides.

The waveguides described above were grown by chemical vapor deposition and reactive ion etching technologies. Tab. 1.2 summarizes the technology, operation wavelengths and propagation losses of the waveguides. It is seen, that guiding at around 3.7 µm with single-mode waveguides was so far demonstrated with high (7–8 dB) losses, compared to 0.6 dB/cm loss of multimode rib waveguide; whereas at wavelengths over 5 µm, the guiding by various waveguides was demonstrates with propagation loss of 2–3 dB/cm.

## 1.4 Multiplexing configurations

A need to integrate on chip increasing number of active and passive devices creates a demand for broader transmission capacities of optical systems. This could be realized by multiplexers. Their architecture might be based on diffraction gratings such as arrayed waveguide grating and planar concave grating or more recent configurations made of ring resonators and angled multimode couplers; however, the optical function remains the same – routing different frequencies into different channels.

The first experimental studies of practically realized diffraction gratings were made by Joseph von Fraunhofer in as early as 1820. Its unsophisticated construction built of fine, parallel wires stretched between two rods allowed to resolve dark lines in lights emitted by sodium source [29]. Franhofer used his gratings to define the refractive indices of glasses and derived the grating equation well familiar nowadays [30]. He also observed the influence of facet shape on the intensity distribution in difference diffraction orders. Thus his work was the first step towards the development of diffraction grating technology.

In the 1870s, the theoretical investigation of Fraunhofer's grating performance done by Rayleigh proved its superior functionality in resolving spectral lines compared to prisms. In those days another milestone in the development of gratings was set by Cornu, who realized the planar grating with variable facet spacing for focusing the diffracted light [30].

The significant highlight grating history was the new configuration with concave gratings instead of plane that was named Rowland after its creator [31]. The design assumes that source and the observer are positioned along a circle with radius twice smaller than the radius of curvature of the grating. The circle touched the grating so that their axes coincided. This configuration allowed to disperse and simultaneously focus the light while keeping the spherical aberration small. The importance of this invention clearly seen from the fact that it is extensively used in present-day devices.

With regard to fabrication technology, the breakthrough technique was suggested by Michelson [30]. The idea was to control the position of grating facets by the interferometer of own design. In the modern fabrication facilities, the principle offered by Michelson is still used. Being upgraded several times with development of technological capabilities, the technique allows us to achieve the nanometer precision.

In the following subchapter, we discuss the progress on two diffraction gratings based on Rowland configuration, as well as other developed multiplexing configurations in integrated optics.

### 1.4.1 Arrayed waveguide grating

The arrayed waveguide grating (AWG) is one of the common multiplexer designs. Its neat configuration allows to direct beams with different wavelengths varying within a certain spectral limit into a single or several channels with a low crosstalk. Fig. 1.4 shows the AWG schema. The multiplexing is realized by means of diffraction of beams that acquire $2\pi$ phase shift while passing through an array of waveguides with constant length difference. Depending on the input wavelength, the beams will pass through the certain output channel.

Proposed by Smit in 1988 [31], the concept of AWG was shortly demonstrated experimentally by Vellekoop and Smit [32]. Takahashi et al. presented a multiple output AWG with a nanometer channel spacing [33]. The next step was taken by Dragon, who extended the concept to NxN wavelength routing device [34]. Adar et al. introduced an asymmetric lay-out allowing broader band multiplexing at lower crosstalk [35]. Amersfoort demonstrated an InP based AWG with flat top response that was achieved using multimode interference (MMI) couplers and integration of AWG with photodetectors showing 3.5 dB transmission loss and 12 dB inter-channel crossing [36]. First AWGs were fabricated based on silica and indium phosphide being mainly oriented for telecom applications [32]–[36]. The devices based on indium phosphide waveguides thanks to higher refractive index contrast enable smaller dimensions that are expected to compensate stronger bending losses in the array waveguides section. However, the main drawback of InP technology is in higher integration complexity compared to that of silicon [37].

Later, the target spectral range shifted towards mid-IR, thus making AWG multiplexers suitable for broad-band source application when combined with miniature and powerful Distributed Feedback Bragg Quantum Cascade Lasers (DFB-ACLs) or Interband Cascade Lasers (DFB-ICLs) for infrared gas sensing [38], [39]. Recently, AWG based on SiGe/Si graded index technology with specific graded index waveguides operating at 4.5 µm wavelength was reported by P.Barritault et al. from CEA-Leti [40]. The 35-ch AWG was developed for optical detection of gases as carbon monoxide, carbon dioxide and nitrous oxide. The normalized transmission spectra measured for outputs of 4.5 µm AWG, the FP-QCL being injected in the central input, showed the insertion loss of 4 dB for TM mode and crosstalk of 20 dB. Interband flatness, which should preferentially be minimized in broad-band source applications, was 5 dB. Recently, IMEC has presented AWGs operating in infrared. The 6-ch AWG operating at 3.8 µm based on silicon-on-insulator technology showed insertion loss of 2 dB for TE polarized light and crosstalk of 25 dB [38]. The 5-ch AWG operating at 5.3 µm based on Ge/Si technology showed insertion loss of 2.5 dB for TE / 3.1 dB for TM polarization of light and crosstalk of 20 dB for TE / 16 dB for TM polarization [39].

Although the "classic" AWG has a banana shape. There were several modifications explored in order to tailor the response towards the desired output. First was Adar [35], who suggested S-shape in order to broaden the spectral coverage of the multiplexer. More recently, Gargallo et al. demonstrated an AWG design combined with Sagnac loop reflectors that allowed to reduce twice the AWG geometry and on the other hand presenting a hybrid configuration utilizing array of waveguides to introduce constant path difference of beams as in AWG and operating rather as a reflecting grating as PCG but with Sagnac loops acting as mirrors [41].

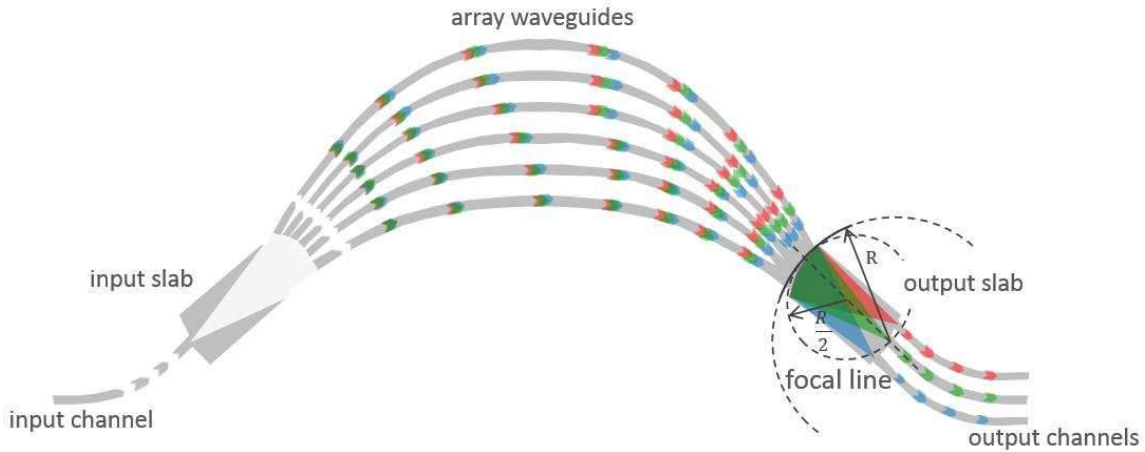

Fig. 1.4. Schema of AWG.

The AWG performance was studied by analyzing phase errors arising in the arrayed waveguide part of the multiplexer. Takada investigated origin of crosstalk in silica-based AWGs by presenting demultiplexers with -30 dB crosstalk [42]. Li and Ma presented analysis of fabrication process parameters that caused phase errors by altering waveguide geometrical parameters as well as refractive index [43]. Quantitative evaluation of both phase and amplitude errors was done by Yamada et al. [44]. Chu et al. presented the analytical investigation of phase errors in AWG [45]. Mask discretization effect on phase errors was also studied [46], [47].

The alternative configurations are further discussed.

### 1.4.2 Planar concave grating

Planar concave grating (PCG) configuration is built of a slab with reflective grating on one side and input/output channels on the other as shown in Fig. 1.5. The phase difference in contrast to AWG, where it appeared in array waveguides, is achieved by positioning the reflective grating facets so that each beam reflected back will have constant path length difference compared to neighbor beams. As a result, the beams constructively interfere in the slab and focus at a certain point along the focal line depending on the input wavelength. The positions of output channels are determined by the target output wavelengths.

First PCGs were made for telecommunication application. They were built of bulk optics and optical fibers [30]. Such optical set ups appeared to be highly challenging due to alignment requirements of the system. In late 1970s one of the first monolithic devices was presented by Tomlinson [48], where the miniature diffraction grating was used together with the prism that provided a tilt. The coupling to input and output fibers was implemented by means of a gradient index lens that served as a collimator. Tangonan et al. were the first to demonstrate the waveguide device based on Rowland configuration [49]. Later, his design was repeated by other research groups [50]–[52].

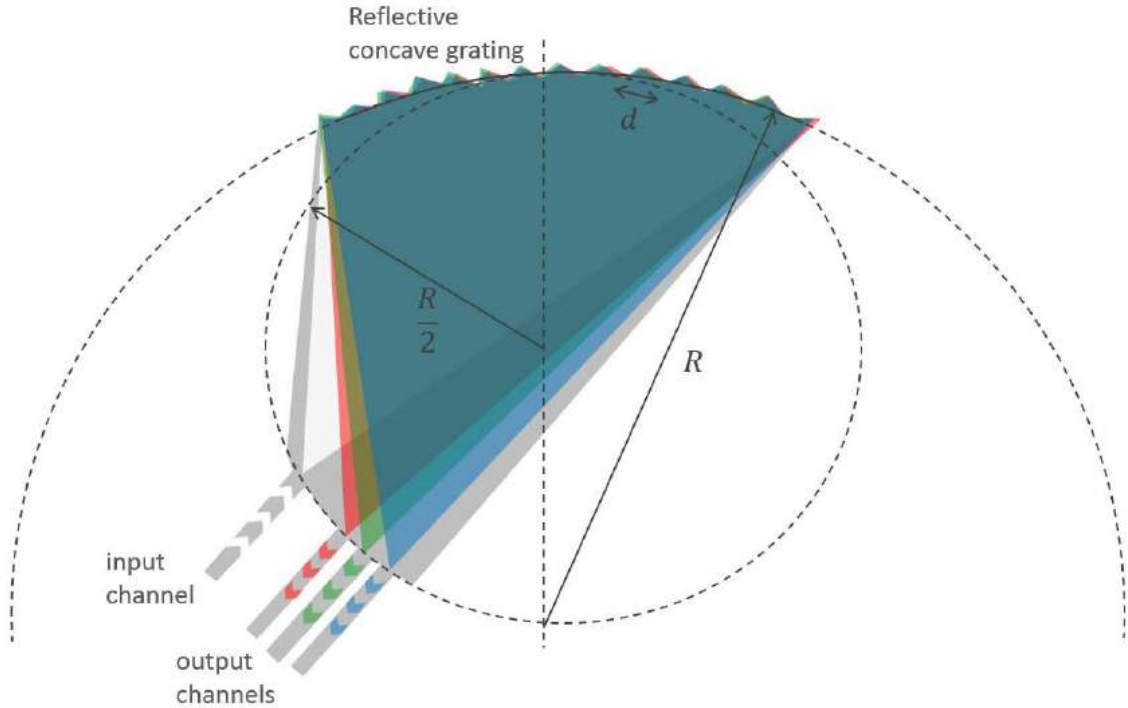

Fig. 1.5. Schema of PCG.

In 1990s, the new technologies, allowing to etch diffraction gratings in a slab waveguide, start to establish. Clemens et al. used silicon dioxide waveguides on silicon substrate [53], Fallahi et al. presented waveguide gratings made of InGaAs/AlGaAs/GaAs and InGa AsP/InP [54], [55]. In the following years, multiple reports were mainly focused on diffraction gratings operating at around 1.55 µm wavelength for telecommunication application.

Recently, with the availability of mid-IR waveguides, the diffraction gratings were designed and characterized for broad band multiplexing in infrared gas sensing application. The PCG operating at 3.8 µm based on silicon-on-insulator technology was presented by Muneb et al. from IMEC. The insertion loss and the crosstalk of the device were 1.6 dB for TE polarized light and 20 dB, respectively [38]. The PCG operating at 5.25 µm based on germanium-on-silicon technology showed insertion loss of 7.6 dB for TE and 6.4 dB for TM polarized light and crosstalk of 27 dB and 21 dB, respectively [56]. The PCG configurations designed at CEA will be discussed in the following chapters of the manuscript.

The design procedure of a multiplexer based on waveguide concave grating was described by Poguntke [57] and McGreer [58]. According to in initial Rowland configuration, the facets were assumed to be positioned such that their projections on the grating tangent were equally spaced. In optimizing the grating, Poguntke refers to the work on high precision calculation of beam path in diffraction gratings presented earlier by Beutler [59]. Horst et al. demonstrated another technique for grating upgrade that is based on grating design with two stigmatic point, which allowed to achieve high compactness while preserving the performance [60]. The influence of phase and amplitude errors on insertion loss and channel crosstalk was studied by Wen et al. using scalar diffraction theory [61]. They conclude that for fabrication of PCG with slab waveguide refractive index of $n_s$ = 1.467 operating at 1.55 μm wavelength, the minimum resolution of photomask of 40 nm is required to obtain the crosstalk below 30 dB. Zhu et al. performed analysis of diffraction efficiency of etched diffraction grating with metallic film coated facets and concluded that the diffraction loss is mainly caused by scattering effect of shaded facets [62]. Another study done by Lin et al. presents the effect of fabrication errors on spectral performance of flat-top planar waveguide demultiplexer [63]. They confirm the conclusion of Wen et al. on photomask requirements by analyzing the performing PCG made of $SiO_2$/SiON/$SiO_2$ waveguides operating at 1.55 μm. Brouckaert studies several factors affecting the PCG spectral response such as facets vertical sidewall tilt, corner rounding, roughness of the grating facet and offset caused by mask pixelation [64]. He shows, that PCGs made of glass have a better tolerance to mask pixelation compared to SOI and InP/InGaAsP waveguides. The analysis of mask grid effect presented in his work were supported by characterization results of PCGs fabricated at 1.55 μm wavelength.

The first two configurations are already been demonstrated for multi-channel multiplexing in infrared. In the following, we briefly consider alternatives.

### 1.4.3 Other configurations

Ring resonator (RR) multiplexer configuration is built of an array of ring-resonators as shown in Fig. 1.6. Each micro ring resonators consist of a ring waveguide as a resonant cavity and bus waveguides, which are coupled to the ring waveguide and provide input and output ports for the device. Each circular length of the ring is chosen in accordance with the resonant wavelength. The main constraint on the given configuration is the choice of waveguide index contrast - the higher index contrast is needed for minimum bending radius supported by the waveguide.

The ring-resonator multiplexers in telecommunication spectral range were presented by CEA [65] and other labs [66]. The demonstration of the concept of Vernier effect array using race track ring resonators in infrared was done by Troia et al. [67] at 3.7-3.8 μm with 1 dB insertion loss and extinction ratio of 30 dB maximum interstitial peak suppression of 10 dB.

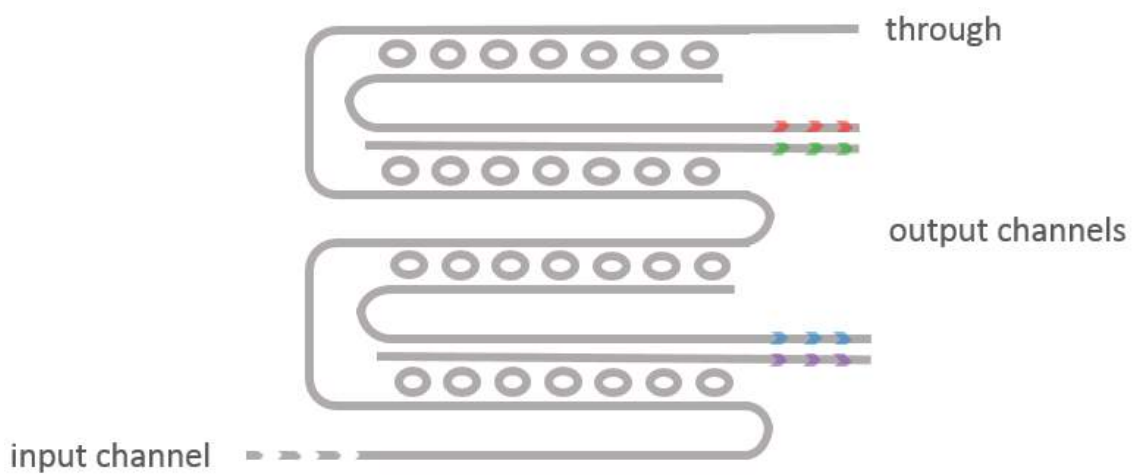

Fig. 1.6. Schema of RR multiplexer.

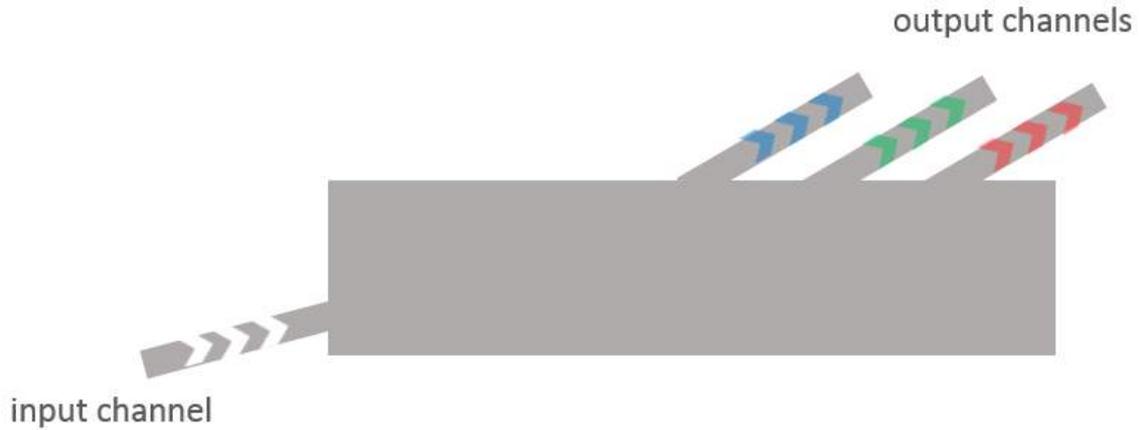

Fig. 1.7. Schema of a-MMI multiplexer.

The mRR configuration can be adapted for multi-channel multiplexing with reasonably compact footprint of the device owing to high index contrast waveguides.

Angled-MMI multiplexer configuration is based on dispersive self-imaging in a multimode waveguide in contrast to other designs where the technology utilizes either gratings or resonant structures [68]. The schema of the device is presented in Fig. 1.7. The angled MMI is built of a multimode waveguide coupled with tilted input/output single-mode waveguides. This approach is found comparatively simple to model and fabricate; however, the quality of self-imaging decreases due to phase errors when the tilt of channels increase.

The multiplexing based on self-imaging in infrared was first demonstrated by Hu et al. They have presented 6-ch angled MMI operating at 3.73 - 3.81 µm [68] with insertion loss of 3 – 4 dB and crosstalk of 15 – 18 dB. As AWG, a-MMI requires only single-step lithography and etching.

The latter configuration is very recent compared to those discussed earlier, and is the subject of further study for its suitability as a multi-channel multiplexer.

| Lab | year | Type | Technology | Δλ | $N_{ch}$ |
|---|---|---|---|---|---|
| IMEC | 2013 | AWG | Si/SiO$_2$ | 3.75 - 3.80 µm | 6 |
| IMEC | 2013 | PCG | Si/SiO$_2$ | 3.77 - 3.80 µm | 8 |
| IMEC | 2013 | AWG | Ge/Si | 5.15 – 5.4 µm | 5 |
| IMEC | 2013 | PCG | Ge/Si | 5.14 – 5.32 µm | 6 |
| ORC | 2014 | MMI | Ge/Si | 3.73 - 3.81 µm | 6 |
| US | 2014 | mRR | Si/SiO$_2$ | 3.72-3.80 µm | 2 |
| CEA | 2015 | AWG | SiGe/Si | 4.37-4.58 µm | 35 |

Tab. 1.3. Multiplexers operating in mid-IR.

Tab. 1.3 summarizes the details of different multiplexer configurations designed and fabricated for operation in mid-IR.

## 1.5 Conclusion

The recent availability of high refractive index mid-IR waveguides triggered the development of multiplexers for mid-IR broad-band source in infrared gas sensing applications. The major technologies used are silicon-germanium in silicon and indium gallium arsenide on indium phosphide. The present achievements in fabrication of low loss mid-IR waveguides together with advancements in optical multiplexing configurations open the way for broad-band multiplexing. Being combined with miniature and powerful sources as DFB-ACLs and DFB-ICLs, such devices potentially constitute the ultra-broad band coverage in order to build a powerful multi-wavelength source for multiple gases detection

The development of a technique combining the ultra-broad band is not trivial task. It could be implemented by assembling several multiplexers, each operating in specific broad-band region. In this connection, the multi-channel devices are preferred. The widely used multiplexing configurations AWG and PCG are promising in the sense of multi-channel operations. Newly developed multiplexer designs are not yet suitable for several tens of output channels.

The goal is to achieve multiplexing in 3 µm to 9 µm spectral range, which covers the majority of absorptions of target gases. The multiplexers reported so far operate in the range between 3.7 µm to 5.4 µm. Therefore, the devices operating in longer wavelength are needed, which seems completely feasible in the light of recent developments.

# 2 Theoretical part

In this chapter, we present the theoretical background of wave guiding that lay the foundations for understanding the results of this work. We start with the principle of wave guiding phenomenon and introduce its electromagnetic description. Then, we discuss basic building blocks of integrated optics followed by detailed study of AWG and PCG multiplexer configurations. For the latter, the basic parameters are derived and spectral characteristics are discussed. The analytical field calculation is presented based on Gaussian approximation and Fourier Optics.

The optimization of spectral response of AWG by means of MMI couplers and tapering is studied. The MMI couplers at the output waveguides allowed to achieve the flattening of the spectra, and the tapers at both ends of array waveguides aimed to increase the coupling intensity. The phase errors analysis is presented for standard deviation of effective index error. The impact of field truncation is studied. The temperature dependence of spectral shift is studied theoretically based on empirical data approximation published by Li earlier.

**Contents**

The PCG geometry based on Rowland configuration is presented in details. Optimization options of PCG response are considered by tailoring the positions of facets.

## 2.1 Waveguides

Waveguides are the structures capable of transporting the energy in the form of illumination. The confinement and guidance of light in integrated optics devices is carried out due to specific structure of the waveguides. It is built of the core and the cladding framing it, both made of dielectric materials (see Fig. 2.1). The core guiding the light has higher refractive index than the cladding. The degree of confinement is defined by geometrical parameters of the waveguide as well as the refractive index contrast of the core and the cladding layers. The index contrast is defined by the choice of materials.

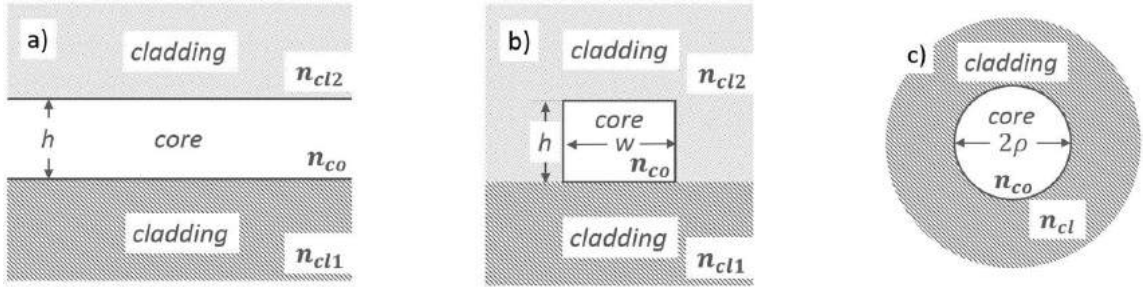

Fig. 2.1. Cross-sections of three waveguide types: a) slab, where h – height of the core; b) rectangular waveguide, where w – width if the waveguide; c) fiber ($\rho$ – radius of the core).

There are several types of waveguides that could be classified into three groups – fibers, rectangular waveguides and the slabs. The simplest guiding structure is the slab, where the light is confined in a single dimension, see Fig. 2.1 (a). It is built of the core plane of height $h$ and refractive index $n_{co}$ sandwiched between two plane layers of cladding of refractive indices $n_{cl2}$ and $n_{cl1}$, where $n_{cl2} \leq n_{cl1} < n_{co}$. In rectangular waveguides, the light is confined in the core of height $h$ and width $w$ and refractive index $n_{co}$ in vertical and horizontal dimensions, with cladding refractive indices $n_{cl1}$ and $n_{cl2}$, where $n_{cl2} \leq n_{cl1} < n_{co}$. Fig. 2.1 (b) shows a simple rectangular waveguide known as strip waveguide. Other modification exists known as rib waveguide [23]. The fiber has a cylindrical shape, where the light is confined in both dimensions as in case of rectangular waveguide, see Fig. 2.1 (c). The core of radius $\rho$ and refractive index $n_{co}$ is concentrically surrounded by the cladding of refractive index $n_{cl}$, where $n_{cl} < n_{co}$.

The refractive index of the core commonly has either uniform or graded profile, as shown by simplified schema in Fig. 2.2. In case of uniform profile of the core index, the waveguide index is addressed as step index profile.

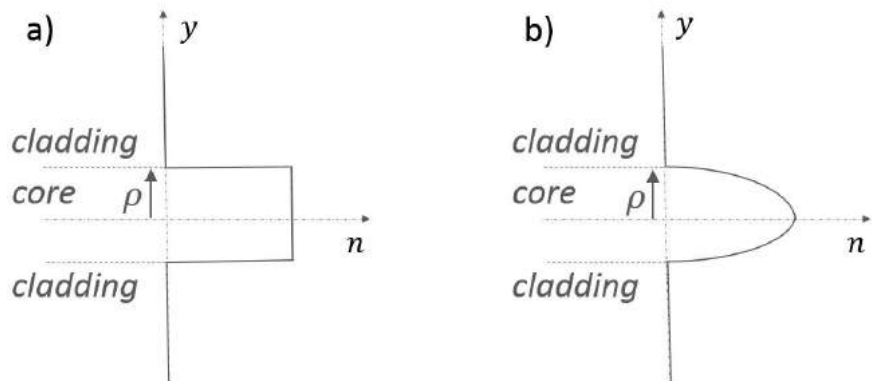

Fig. 2.2. The refractive index profile of the core: a) uniform; b) graded.

### 2.1.1 Basics of optical guiding

The wave guiding inside the core can be viewed as a sequence of multiple internal reflections at the boundaries with claddings, governed by the Snell's law. When input angle $\theta_z$ is greater than $\theta_c$, as shown in Fig. 2.3 (b), the light is partially reflected and partially transmitted to the cladding. In the opposite case, i.e. when the input angle $\theta_z$ lies in the range of values $0 \leq \theta_z < \theta_c$ (Fig. 2.3.a), the rays of light are totally internally reflected, where the critical angle $\theta_c$ is defined as follows:

$$\theta_c = \cos^{-1}\left\{\frac{n_{cl}}{n_{co}}\right\} = \sin^{-1}\left\{1 - \frac{n_{cl}^2}{n_{co}^2}\right\}^{1/2} \tag{2.1}$$

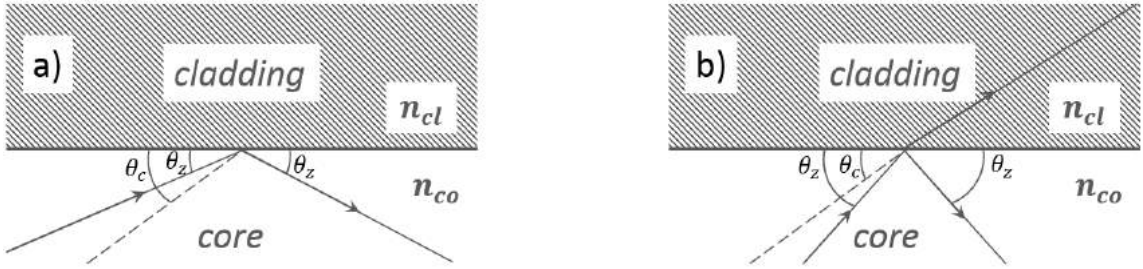

Fig. 2.3. Reflection at planar interface of two dielectric media: a) total internal reflection; b) partial reflection and refraction.

As a consequence, the light rays follow the periodic zigzag path along the waveguide, see Fig. 2.4. The guidance of a wave occurs when the self-consistency condition is satisfied, i.e. when the phases of rays passing via APQB paths and CD path are shifted by a multiple of $2\pi$ [69]:

$$\varphi_{APQB} - \varphi_{CD} = 2\pi n. \tag{2.3}$$

The phase $\varphi_{APQB}$ is the sum of phases of constituent segments:

$$\varphi_{APQB} = \varphi_{AP} + \varphi_{PQ} + \varphi_{QB} + 2\varphi_R \tag{2.4}$$

where $\varphi_{PQ}$ equals to $\varphi_{CD}$ and cancels it out,

$$\varphi_{AP} = \varphi_{QB} = 2k\rho n_{co}\sin\theta_z = 2k\rho\sqrt{n_{co}^2 - \bar{\beta}^2}, \tag{2.5}$$

where $k = \frac{2\pi}{\lambda}$ is the wavenumber, $\lambda$ is the wavelength of light in free space, $\bar{\beta} = n_{co}\cos\theta_z$ is the propagation invariant and $\varphi_R$ is the phase shift that ray obtains after total reflection at the interface of the core and cladding.

The phase shift $\varphi_R$ is defined by Fresnel formulas:

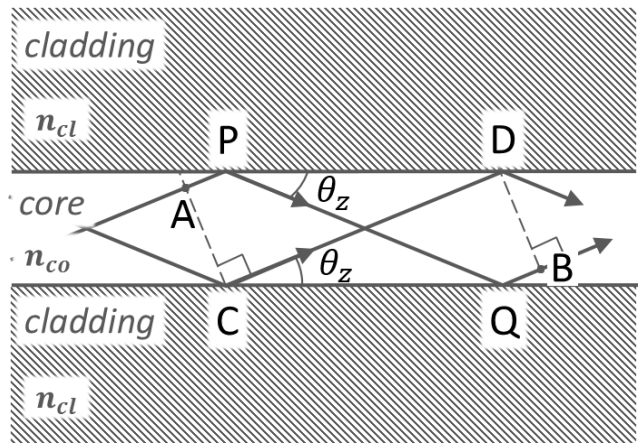

Fig. 2.4. The schematic description of guiding.

$$\varphi_R = -2\text{arctg}\left\{g\frac{\sqrt{\sin^2\theta_c - \sin^2\theta_z}}{\sin\theta_z}\right\} = -2\text{arctg}\left\{g\frac{\sqrt{\bar{\beta}^2 - n_{cl}^2}}{\sqrt{n_{co}^2 - \bar{\beta}^2}}\right\} \quad (2.6)$$

where g is the factor dependent on the polarization of the incident wave that is given by:

$$\begin{aligned} g &= 1; \vec{E} \text{ parallel to the interface (TE mode),} \\ g &= \frac{n_{co}^2}{n_{cl}^2}; \vec{H} \text{ parallel to the interface (TM mode).} \end{aligned} \quad (2.7)$$

Then the final expression for guiding condition is given by:

$$2k\rho\sqrt{n_{co}^2 - \bar{\beta}^2} - 2\text{arctg}\left\{g\frac{\sqrt{\bar{\beta}^2 - n_{cl}^2}}{\sqrt{n_{co}^2 - \bar{\beta}^2}}\right\} = m\pi; \quad (2.8)$$

where $m = 0, 1, 2, \ldots$ is an integer which identifies the mode number. The equation (2.8) is the dispersion equation of waveguide modes. It is clearly seen that the propagation in the waveguide is done only for discrete values of the angle of incidence $\theta_z$ or the propagation constant $\beta$. So the guided mode is a wave with propagation constant that satisfies this dispersion equation.

For guided modes the propagation constant lies in the range $kn_{cl} < \beta < kn_{co}$. The waveguide is single mode when these characteristics are such that it supports only one guided mode. Otherwise, the waveguide is multimode. Fig. 2.5 illustrates the propagation constant profile as a function of $\frac{\rho}{\lambda}$ parameter for slab waveguide with refractive indices $n_{co}$ = 1.48 and $n_{cl}$=1.47.

The cut-off thickness of a waveguide mode can be obtained from the limit value of the guidance condition of the mode:

$$\bar{\beta} = n_{cl}, \quad (2.9)$$

Then the cut-off thickness of the m-th mode $e_m$ is given by:

$$e_m = 2\rho_c = \frac{m\pi}{k\sqrt{n_{co}^2 - n_{cl}^2}}. \quad (2.10)$$

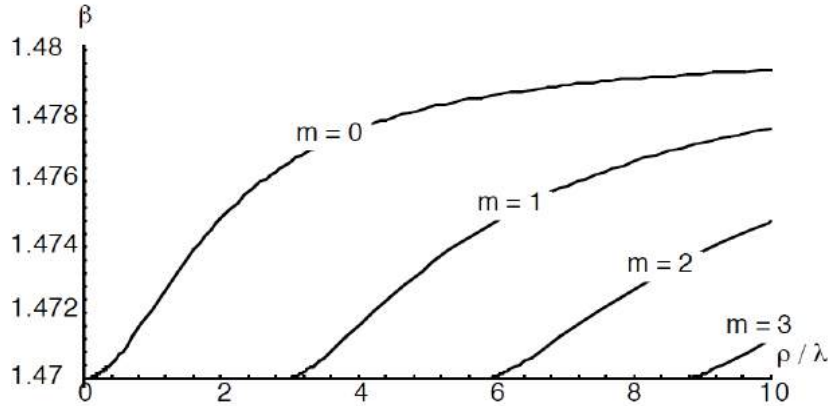

Fig. 2.5. Profile of $\beta$ as a function of $\rho/\lambda$ for $n_{cl}$=1.47, $n_{co}$ = 1.48, adopted from [69].

From (2.10) it is seen, that $e_0 = 0$, i.e. the symmetrical slab waveguide will always support at least one guiding mode. Furthermore, the waveguide is single-mode, when:

$$2k\rho_c\sqrt{n_{co}^2 - n_{cl}^2} < \pi \tag{2.11}$$

It is convenient to use dimensionless waveguide parameter $V$ that combines geometrical dimensions, refractive index contrast and the operational wavelength [70]:

$$V = k\rho\sqrt{n_{co}^2 - n_{cl}^2}, \tag{2.12}$$

Then the single-mode condition for symmetrical slab waveguides can be expressed as $V < \frac{\pi}{2}$.

For propagation constant $\beta$ values below $kn_{cl}$, the part of mode, known as evanescent, decays exponentially along the waveguide. The propagation constant defines the phase rotation per unit propagation distance $\beta = 2\pi n_{eff}/\lambda$, where $n_{eff}$ is the effective index showing the ratio of the wavelength in the waveguide to the wavelength in free space.

The above consideration of guidance was presented in the frame of geometrical optics. It is valid for the cases when the cross-sectional half-width $\rho$ of waveguide is considerably greater than the wavelength, i.e. $V \gg 1$. However, this approach is not sufficient for accurate analysis of guided mode in single-mode waveguides, where the cross-sectional half-width $\rho$ is comparable to wavelength. In that case, we refer to electromagnetic approach based on Maxwell's equations.

### 2.1.2 Electromagnetic description of wave guiding

Maxwell's equations of electromagnetic fields for source free dielectric medium are expressed as follows [71]:

$$\vec{\nabla} \times \vec{E} = -\frac{\partial \vec{B}}{\partial t} \tag{2.13-1}$$

$$\vec{\nabla} \times \vec{H} = \frac{\partial \vec{D}}{\partial t} \tag{2.13-2}$$

$$\vec{\nabla} \cdot \vec{D} = 0 \tag{2.13-3}$$

$$\vec{\nabla} \cdot \vec{B} = 0 \tag{2.13-4}$$

where vector quantities are denoted in bold letters with "→" mark, operator $\vec{\nabla} = \vec{\iota}\frac{\partial}{\partial x} + \vec{j}\frac{\partial}{\partial y} + \vec{k}\frac{\partial}{\partial z}$, $\vec{E}$ and $\vec{H}$ are the electric and magnetic fields, respectively, $\vec{D}$ is the electric displacement, and $\vec{B}$ is the magnetic flux density. The latter two parameters can be represented in terms of electric and magnetic fields.

$$\vec{D} = \epsilon\vec{E} \tag{2.14-1}$$

$$\vec{B} = \mu\vec{H} \tag{2.14-2}$$

where $\epsilon$ and $\mu$ are the electric permittivity and the magnetic permeability of the medium, respectively. Since we consider sources free dielectric waveguides, as effect the electric permittivity is given by $\epsilon = \epsilon_0 n^2$ and the magnetization is ignored $\mu = \mu_0$, where $n$ is the index of refraction, $\epsilon_0$ and $\mu_0$ are the permittivity and the permeability of free space, respectively. For the latter two, we have $\epsilon_0 \mu_0 c^2 = 1$, where $c$ is the speed of light.

We restrict ourselves to harmonic fields in lossless medium. Then the electric and magnetic fields can be expressed in the form:

$$\vec{E}(x,y,z,t) = \vec{E}(x,y,z)\exp(-j\omega t), \tag{2.15-1}$$

$$\vec{H}(x,y,z,t) = \vec{H}(x,y,z)\exp(-j\omega t), \tag{2.15-2}$$

where $\omega$ is the angular frequency of the wave. The wavenumber can be expressed as $k = 2\pi/\lambda = \omega/c$.

Taking this in to account, we can rewrite Maxwell's equations:

$$\vec{\nabla} \times \vec{E} = j\mu_0 \omega \vec{H} = j\sqrt{\frac{\mu_0}{\epsilon_0}} k \vec{H} \tag{2.16-1}$$

$$\vec{\nabla} \times \vec{H} = -j\epsilon_0 n^2 \omega \vec{E} = -j\sqrt{\frac{\epsilon_0}{\mu_0}} k n^2 \vec{E} \tag{2.16-2}$$

$$\nabla \cdot (n^2 \vec{E}) = 0 \tag{2.16-3}$$

$$\nabla \cdot \vec{H} = 0 \tag{2.16-4}$$

The wave equations in terms of electric field can be obtained from (2.16-1) and (2.16-2):

$$\vec{\nabla} \times (\vec{\nabla} \times \vec{E}) = k^2 n^2 \vec{E} \tag{2.17}$$

From equation (2.16-3) it follows:

$$\vec{\nabla} \cdot (n^2 \vec{E}) = \vec{\nabla} n^2 \cdot \vec{E} + n^2 \vec{\nabla} \cdot \vec{E} = 0 \tag{2.18}$$

Using vector identity $\vec{\nabla} \times (\vec{\nabla} \times \vec{A}) = \vec{\nabla}(\vec{\nabla} \cdot \vec{A}) - \nabla^2 \vec{A}$ and equation (2.18), we rewrite equation (2.17):

$$\nabla^2 \vec{E} + k^2 n^2 \vec{E} = -\vec{\nabla}\left(\frac{\vec{\nabla} n^2}{n^2} \cdot \vec{E}\right) \tag{2.19}$$

In the similar manner, the wave equation can be expressed in terms of magnetic field:

$$\vec{\nabla} \times (\vec{\nabla} \times \vec{H}) = -j\sqrt{\frac{\epsilon_0}{\mu_0}} k \vec{\nabla} \times (n^2 \vec{E}) \tag{2.20-1}$$

$$\vec{\nabla}(\vec{\nabla} \cdot \vec{H}) - \nabla^2 \vec{H} = -j\sqrt{\frac{\epsilon_0}{\mu_0}} k (\vec{\nabla} n^2 \times \vec{E} + n^2 \vec{\nabla} \times \vec{E}) \tag{2.20-2}$$

$$\vec{\nabla}^2 \vec{H} + k^2 n^2 \vec{H} = (\vec{\nabla} \times \vec{H}) \times \frac{\vec{\nabla} n^2}{n^2} \qquad (2.20\text{-}3)$$

The equations (2.19) and (2.20-3) are the basis for electromagnetic calculation of any dielectric waveguides.

In the following, we consider the wave propagation in the symmetrical slab waveguide. The height of the core is $2\rho$ and the claddings are spanned to infinity on both sides. Fig. 2.6 illustrates the slab waveguide schema and the choice of coordinate system, where the wave guidance occurs in the z-direction. The refractive index of the waveguide is defined as:

$$n(y) = n_{co} \text{ for } |y| < \rho \qquad (2.21)$$
$$n(y) = n_{cl} \text{ for } |y| > \rho$$

Since the waveguide is plane, we have $\partial/\partial x = 0$. Given this, we can explicate Maxwell's equations for each component of $\vec{E}$ and $\vec{H}$ fields:

$$\frac{\partial E_z}{\partial y} - \frac{\partial E_y}{\partial z} = j \sqrt{\frac{\mu_0}{\epsilon_0}} k H_x \qquad (2.22\text{-}1)$$

$$\frac{\partial E_x}{\partial z} = j \sqrt{\frac{\mu_0}{\epsilon_0}} k H_y \qquad (2.22\text{-}2)$$

$$-\frac{\partial E_x}{\partial y} = j \sqrt{\frac{\mu_0}{\epsilon_0}} k H_z \qquad (2.22\text{-}3)$$

$$\frac{\partial H_z}{\partial y} - \frac{\partial H_y}{\partial z} = -j \sqrt{\frac{\epsilon_0}{\mu_0}} k n^2 E_x \qquad (2.22\text{-}4)$$

$$\frac{\partial H_x}{\partial z} = -j \sqrt{\frac{\epsilon_0}{\mu_0}} k n^2 E_y \qquad (2.22\text{-}5)$$

$$-\frac{\partial H_x}{\partial y} = -j \sqrt{\frac{\epsilon_0}{\mu_0}} k n^2 E_z \qquad (2.22\text{-}6)$$

$$\frac{\partial (n^2 E_y)}{\partial y} + \frac{\partial (n^2 E_z)}{\partial z} = 0 \qquad (2.22\text{-}7)$$

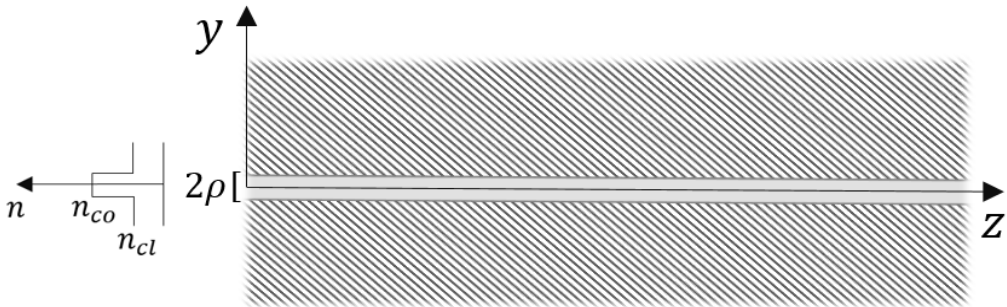

Fig. 2.6. Coordinates for describing a slab waveguide.

$$\frac{\partial H_y}{\partial y} + \frac{\partial H_z}{\partial z} = 0 \tag{2.22-8}$$

One can see that the components of electromagnetic fields appear in two independent sets: $E_x, H_y, H_z$ are interrelated in equations (2.22-2), (2.22-3), (2.22-4), (2.22-8) and $H_x, E_y, E_z$ are interrelated in equations (2.22-1), (2.22-5), (2.22-6), (2.22-7). So we can solve Maxwell's equations separately for these two sets of components and the general solution will be a linear combination of these two types of solutions.

The solutions involving only the components $E_x, H_y, H_z$ are known as transverse electric (TE) mode, and the other solutions are known as transverse magnetic (TM) mode.

We will consider a waveguide slab with constant indices of the core $n_{co}$ and cladding $n_{cl}$, where $n_{co} > n_{cl}$. Then, $\vec{\nabla} n^2 / n^2 = \vec{0}$ for $|y| \neq \rho$.

We first obtain the wave equation for TE modes. We have got $E_y = E_z = 0$, and the solution we look for is of the form $E_x(y, z) = E_x(y)e^{j\beta z}$ with $kn_{cl} < \beta \leq kn_{cl}$. The wave equation will become:

$$\frac{d^2 E_x}{dy^2} + (k^2 n_{co}^2 - \beta^2) E_x = 0 \qquad |y| < \rho \quad (2.23\text{-}1)$$

$$\frac{d^2 E_x}{dy^2} + (k^2 n_{cl}^2 - \beta^2) E_x = 0 \qquad |y| > \rho \quad (2.23\text{-}2)$$

The solutions of these equations are known, and are expressed as follows:

$$E_x(y) = A\cos\left(\sqrt{k^2 n_{co}^2 - \beta^2}\, y\right) + B\sin\left(\sqrt{k^2 n_{co}^2 - \beta^2}\, y\right) \quad |y| < \rho \quad (2.24\text{-}1)$$

$$E_x(y) = C\exp\left(\sqrt{\beta^2 - k^2 n_{cl}^2}\, y\right) + D\exp\left(-\sqrt{\beta^2 - k^2 n_{cl}^2}\, y\right) \quad |y| > \rho \quad (2.24\text{-}2)$$

In order to determine the constants A and B, we first note that the guide is symmetric with respect to the $y$-axis, so the solutions must be symmetric (even) or antisymmetric (odd) with respect to the $y$-axis. On the other hand, the field must vanish at infinity, and, lastly, the field (being tangential) must be continuous at the core-cladding interfaces. This leads to:

$$E_x(y) = \frac{\cos\left(\sqrt{k^2 n_{co}^2 - \beta^2}\, y\right)}{\cos\left(\rho \sqrt{k^2 n_{co}^2 - \beta^2}\right)} \qquad |y| < \rho, \quad (2.25\text{-}1)$$

$$E_x(y) = \frac{\exp\left(-\sqrt{\beta^2 - k^2 n_{cl}^2}\, |y|\right)}{\exp\left(-\rho \sqrt{\beta^2 - k^2 n_{cl}^2}\right)} \qquad |y| > \rho \quad (2.25\text{-}2)$$

for even modes, and:

$$E_x(y) = \frac{\sin\left(\sqrt{k^2 n_{co}^2 - \beta^2}\, y\right)}{\sin\left(\rho \sqrt{k^2 n_{co}^2 - \beta^2}\right)} \qquad |y| < \rho, \quad (2.26\text{-}1)$$

$$E_x(y) = \frac{y \exp\left(-\sqrt{\beta^2-k^2 n_{cl}^2}|y|\right)}{|y| \exp\left(-\rho\sqrt{\beta^2-k^2 n_{cl}^2}\right)} \qquad |y| > \rho \quad (2.26\text{-}2)$$

for odd modes.

Then, we obtain the other components $H_y$ and $H_z$ from the equations (2.22-2), (2.22-3):

$$H_y(y) = -j\sqrt{\frac{\epsilon_0}{\mu_0}}\frac{1}{k}\frac{\partial E_x}{\partial z} = \sqrt{\frac{\epsilon_0}{\mu_0}}\frac{\beta}{k} E_x, \qquad (2.27\text{-}1)$$

$$H_z(y) = j\sqrt{\frac{\epsilon_0}{\mu_0}}\frac{1}{k}\frac{\partial E_x}{\partial y}. \qquad (2.27\text{-}2)$$

For the component $H_y$ we get:

$$H_y(y) = \sqrt{\frac{\epsilon_0}{\mu_0}}\frac{\beta}{k}\frac{\cos\left(\sqrt{k^2 n_{co}^2-\beta^2}\,y\right)}{\cos\left(\rho\sqrt{k^2 n_{co}^2-\beta^2}\right)} \qquad |y| < \rho, \quad (2.28\text{-}1)$$

$$H_y(y) = \sqrt{\frac{\epsilon_0}{\mu_0}}\frac{\beta}{k}\frac{\exp\left(-\sqrt{\beta^2-k^2 n_{cl}^2}|y|\right)}{\exp\left(-\rho\sqrt{\beta^2-k^2 n_{cl}^2}\right)} \qquad |y| > \rho \quad (2.28\text{-}2)$$

for even modes, and:

$$H_y(y) = \sqrt{\frac{\epsilon_0}{\mu_0}}\frac{\beta}{k}\frac{\sin\left(\sqrt{k^2 n_{co}^2-\beta^2}\,y\right)}{\sin\left(\rho\sqrt{k^2 n_{co}^2-\beta^2}\right)} \qquad |y| < \rho, \quad (2.29\text{-}1)$$

$$H_y(y) = \sqrt{\frac{\epsilon_0}{\mu_0}}\frac{\beta}{k}\frac{y \exp\left(-\sqrt{\beta^2-k^2 n_{cl}^2}|y|\right)}{|y| \exp\left(-\rho\sqrt{\beta^2-k^2 n_{cl}^2}\right)} \qquad |y| > \rho \quad (2.29\text{-}2)$$

for odd modes. The component $H_z$:

$$H_z(y) = -j\sqrt{\frac{\epsilon_0}{\mu_0}}\sqrt{n_{co}^2-\beta^2/k^2}\frac{\sin\left(\sqrt{k^2 n_{co}^2-\beta^2}\,y\right)}{\cos\left(\rho\sqrt{k^2 n_{co}^2-\beta^2}\right)} \qquad |y| < \rho, \quad (2.30\text{-}1)$$

$$H_z(y) = -j\sqrt{\frac{\epsilon_0}{\mu_0}}\sqrt{\beta^2/k^2-n_{cl}^2}\,\frac{y}{|y|}\frac{\exp\left(-\sqrt{\beta^2-k^2 n_{cl}^2}|y|\right)}{\exp\left(-\rho\sqrt{\beta^2-k^2 n_{cl}^2}\right)} \qquad |y| > \rho \quad (2.30\text{-}2)$$

for even modes, and:

$$H_z(y) = j\sqrt{\frac{\epsilon_0}{\mu_0}}\sqrt{n_{co}^2-\beta^2/k^2}\frac{\cos\left(\sqrt{k^2 n_{co}^2-\beta^2}\,y\right)}{\sin\left(\rho\sqrt{k^2 n_{co}^2-\beta^2}\right)} \qquad |y| < \rho, \quad (2.31\text{-}1)$$

$$H_z(y) = -j\sqrt{\frac{\epsilon_0}{\mu_0}}\sqrt{\beta^2/k^2 - n_{co}^2}\,\frac{\exp\left(-\sqrt{\beta^2-k^2n_{cl}^2}|y|\right)}{\exp\left(-\rho\sqrt{\beta^2-k^2n_{cl}^2}\right)} \qquad |y|>\rho \quad (2.31\text{-}2)$$

for odd modes.

The shape of the field $E_x$ of the first modes is shown in Fig. 2.7.

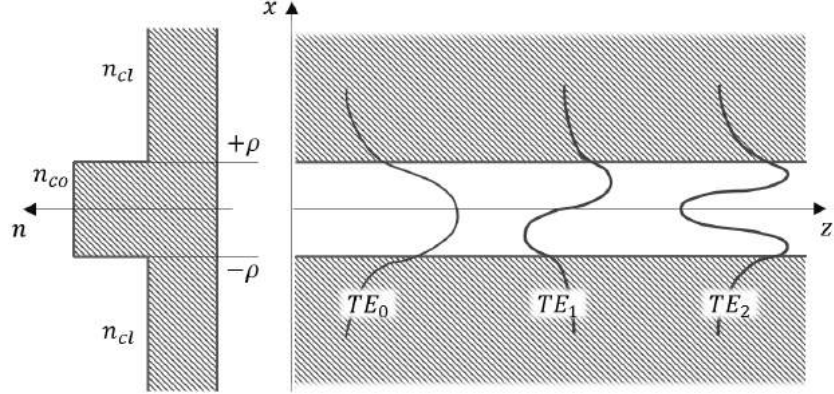

Fig. 2.7. The first TE modes of the symmetrical plane guide.

In order to complete the solution, it remains to determine $\beta$. For it, we note that the tangential component of the magnetic field $H_z$ must be continuous at the core-cladding interfaces. We obtain:

$$\sqrt{\beta^2 - k^2 n_{cl}^2} = \sqrt{k^2 n_{co}^2 - \beta^2}\,tg\left(\rho\sqrt{k^2 n_{co}^2 - \beta^2}\right), \qquad (2.32\text{-}1)$$

for even modes, and

$$\sqrt{\beta^2 - k^2 n_{cl}^2} = -\sqrt{k^2 n_{co}^2 - \beta^2}\,ctg\left(\rho\sqrt{k^2 n_{co}^2 - \beta^2}\right), \qquad (2.32\text{-}2)$$

for odd modes.

The TM guided modes are solved in the same way. In this case, it is preferable to take the wave equation for the magnetic field:

$$\frac{d^2 H_x}{dy^2} + (k^2 n_{co}^2 - \beta^2)H_x = 0 \qquad |y|<\rho \quad (2.33\text{-}1)$$

$$\frac{d^2 H_x}{dy^2} - (\beta^2 - k^2 n_{cl}^2)H_x = 0 \qquad |y|>\rho \quad (2.33\text{-}2)$$

Then, we will get the solutions of these equations in the form:

$$H_x(y) = \sqrt{\frac{\epsilon_0}{\mu_0}}\frac{kn_{co}^2}{\beta}\frac{\cos\left(\sqrt{k^2 n_{co}^2-\beta^2}\,y\right)}{\cos\left(\rho\sqrt{k^2 n_{co}^2-\beta^2}\right)} \qquad |y|<\rho \quad (2.34\text{-}1)$$

$$H_x(y) = \sqrt{\frac{\epsilon_0}{\mu_0}}\frac{kn_{co}^2}{\beta}\frac{\exp\left(-\sqrt{\beta^2-k^2 n_{cl}^2}|y|\right)}{\exp\left(-\rho\sqrt{\beta^2-k^2 n_{cl}^2}\right)} \qquad |y|>\rho \quad (2.34\text{-}2)$$

for even modes, and:

$$H_x(y) = \sqrt{\frac{\epsilon_0}{\mu_0}} \frac{kn_{co}^2}{\beta} \frac{\sin\left(\sqrt{k^2 n_{co}^2 - \beta^2}\, y\right)}{\sin\left(\rho\sqrt{k^2 n_{co}^2 - \beta^2}\right)} \qquad |y| < \rho \quad (2.35\text{-}1)$$

$$H_x(y) = \sqrt{\frac{\epsilon_0}{\mu_0}} \frac{kn_{co}^2}{\beta} \frac{y}{|y|} \frac{\exp\left(-\sqrt{\beta^2 - k^2 n_{cl}^2}\,|y|\right)}{\exp\left(-\rho\sqrt{\beta^2 - k^2 n_{cl}^2}\right)} \qquad |y| > \rho \quad (2.35\text{-}2)$$

for odd modes.

The component $E_y$ is obtained from equation (2.22-5):

$$E_y(y) = j\sqrt{\frac{\mu_0}{\epsilon_0}} \frac{1}{kn^2} \frac{\partial H_x}{\partial z} = -\sqrt{\frac{\mu_0}{\epsilon_0}} \frac{\beta}{kn^2} H_x, \qquad (2.36)$$

which gives:

$$E_y(y) = -\frac{\cos\left(\sqrt{k^2 n_{co}^2 - \beta^2}\, y\right)}{\cos\left(\rho\sqrt{k^2 n_{co}^2 - \beta^2}\right)} \qquad |y| < \rho \quad (2.37\text{-}1)$$

$$E_y(y) = -\frac{n_{co}^2}{n_{cl}^2} \frac{\exp\left(-\sqrt{\beta^2 - k^2 n_{cl}^2}\,|y|\right)}{\exp\left(-\rho\sqrt{\beta^2 - k^2 n_{cl}^2}\right)} \qquad |y| > \rho \quad (2.37\text{-}2)$$

for even modes, and:

$$E_y(y) = -\frac{\sin\left(\sqrt{k^2 n_{co}^2 - \beta^2}\, y\right)}{\sin\left(\rho\sqrt{k^2 n_{co}^2 - \beta^2}\right)} \qquad |y| < \rho \quad (2.38\text{-}1)$$

$$E_y(y) = -\frac{n_{co}^2}{n_{cl}^2} \frac{y}{|y|} \frac{\exp\left(-\sqrt{\beta^2 - k^2 n_{cl}^2}\,|y|\right)}{\exp\left(-\rho\sqrt{\beta^2 - k^2 n_{cl}^2}\right)} \qquad |y| > \rho \quad (2.38\text{-}2)$$

for odd modes.

Finally, the component $E_z$ is obtained from equation (2.22-6):

$$E_z(y) = -j\sqrt{\frac{\mu_0}{\epsilon_0}} \frac{1}{kn^2} \frac{\partial H_x}{\partial y}, \qquad (2.39)$$

We get:

$$E_z(y) = j\sqrt{\frac{k^2}{\beta^2} n_{co}^2 - 1} \frac{\sin\left(\sqrt{k^2 n_{co}^2 - \beta^2}\, y\right)}{\cos\left(\rho\sqrt{k^2 n_{co}^2 - \beta^2}\right)} \qquad |y| < \rho \quad (2.40\text{-}1)$$

$$E_z(y) = j\sqrt{1 - \frac{k^2}{\beta^2} n_{cl}^2} \frac{y}{|y|} \frac{\exp\left(-\sqrt{\beta^2 - k^2 n_{cl}^2}\,|y|\right)}{\exp\left(-\rho\sqrt{\beta^2 - k^2 n_{cl}^2}\right)} \qquad |y| > \rho \quad (2.40\text{-}2)$$

for even modes, and:

$$E_z(y) = -j\sqrt{\frac{k^2}{\beta^2}n_{co}^2 - 1}\frac{\cos\left(\sqrt{k^2n_{co}^2-\beta^2}\,y\right)}{\sin\left(\rho\sqrt{k^2n_{co}^2-\beta^2}\right)} \qquad |y| < \rho \quad (2.41\text{-}1)$$

$$E_z(y) = j\sqrt{1 - \frac{k^2}{\beta^2}n_{cl}^2}\frac{\exp\left(-\sqrt{\beta^2-k^2n_{cl}^2}|y|\right)}{\exp\left(-\rho\sqrt{\beta^2-k^2n_{cl}^2}\right)} \qquad |y| > \rho \quad (2.41\text{-}2)$$

for odd modes.

By taking into account the continuity of the tangential component $E_z$ at the core-cladding interfaces, we get:

$$n_{co}^2\sqrt{\beta^2 - k^2n_{cl}^2} = n_{cl}^2\sqrt{k^2n_{co}^2 - \beta^2}\,tg\left(\rho\sqrt{k^2n_{co}^2 - \beta^2}\right), \qquad (2.42\text{-}1)$$

for even modes, and

$$n_{co}^2\sqrt{\beta^2 - k^2n_{cl}^2} = -n_{cl}^2\sqrt{k^2n_{co}^2 - \beta^2}\,ctg\left(\rho\sqrt{k^2n_{co}^2 - \beta^2}\right), \qquad (2.42\text{-}2)$$

for odd modes. The equations (2.32-1), (2.32-2), (2.42-1), (2.42-2) are transcendental and must be solved numerically.

### 2.1.3 Calculation of effective index of Si/Si$_{0.6+x}$Ge$_{0.4-x}$/Si waveguide in R-soft

In process of designing our waveguide devices, we needed a tool for modeling the effective index of Si/Si$_{0.6+x}$Ge$_{0.4-x}$/Si strip waveguides. The solution gets especially cumbersome given the graded index profile of the waveguide, see Fig. 2.8 (b). The waveguide mode was simulated by building a multilayer structure with the commercial software utility *BeamPROP* of R-soft.

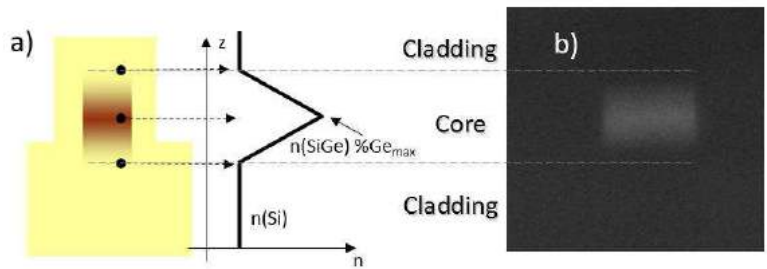

Fig. 2.8. Si$_{0.6+x}$Ge$_{0.4-x}$ graded index waveguide: a) schema; b) SEM image of waveguide realized at CEA-Leti.

#### 2.1.3.1 Refractive index of SiGe

The refractive index of Si$_{0.6+x}$Ge$_{0.4-x}$ waveguide depends on wavelength and germanium concentration given that the dependence is not linear. Based on the refractive index data of several SiGe waveguides with various germanium concentrations, we were able to make the second order polynomial fit [72], as shown in equation (2.43). Fig. 2.9. illustrates the fit.

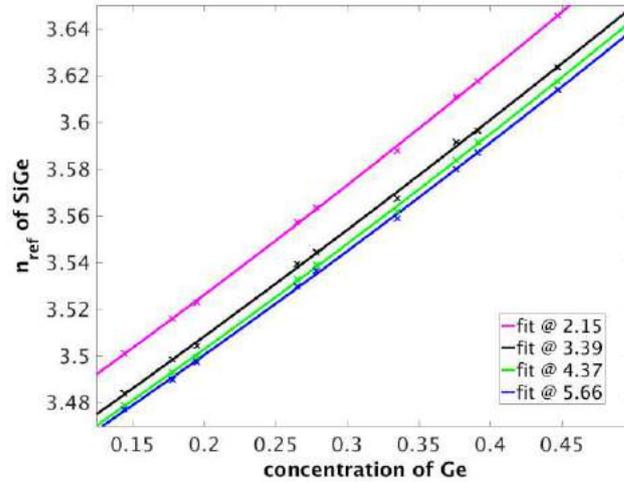

Fig. 2.9. Refractive index of Si$_{0.6+x}$Ge$_{0.4-x}$ as a function of germanium concentration: second order polynomial fit, cross marks correspond to data obtained by M-line measurements.

$$n_{ref} = A_1 C_i^2 + A_2 C_i + A_3, \tag{2.43}$$

where $A_1$, $A_2$, $A_3$ are the coefficients of the fit, $C_i$ is the concentration of germanium.

The fits were available at four different wavelengths. So the coefficients were bound to wavelengths by approximation with second order polynomial as (Fig. 2.10.):

$$A_1 = a_1 \lambda^2 - a_2 \lambda + a_3, \tag{2.44-1}$$

$$A_2 = b_1 \lambda^2 - b_2 \lambda + b_3. \tag{2.44-2}$$

The third coefficient $A_3$ corresponds to pure silicon, without any germanium additive, so it was determined theoretically according to [73]:

$$A_3 = \sqrt{\frac{c_1 \lambda^2}{\lambda^2 - c_4^2} + \frac{c_2 \lambda^2}{\lambda^2 - c_5^2} + \frac{c_3 \lambda^2}{\lambda^2 - c_6^2} + 1}. \tag{2.45}$$

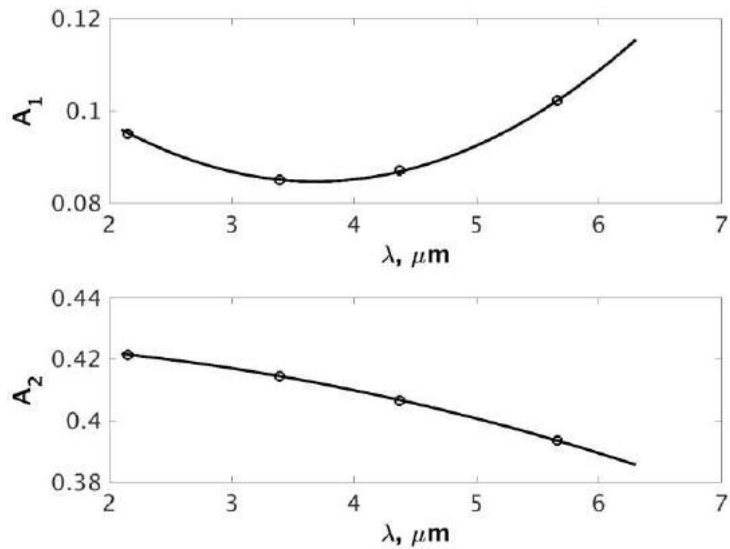

Fig. 2.10. Coefficients $A_1$ and $A_2$ as a function of wavelength.

| coefficient | value |
|---|---|
| $a_1$ | 0.004080 |
| $a_2$ | 0.039522 |
| $a_3$ | 0.403497 |
| $b_1$ | 0.008586 |
| $b_2$ | 0.064886 |
| $b_3$ | 0.392743 |
| $c_1$ | 10.6684293 |
| $c_2$ | 0.00304348 |
| $c_3$ | 1.54133408 |
| $c_4$ | 0.301516485 |
| $c_5$ | 1.13475115 |
| $c_6$ | 1104 |

Tab. 2.1. Coefficients of second order polynomial approximations.

Using the parametrization given above, it is possible to estimate the refractive index of SiGe waveguide for concentration of germanium up to 0.45 in mid-infrared spectral range.

### 2.1.3.2 Graded index profile simulation

In order to have a flexible tool for effective index calculation, we built a multilayer structure Si/Si$_{0.6+x}$Ge$_{0.4-x}$/Si waveguide using parametrized SiGe refractive index in R-soft, see Fig. 2.11 (a).

The principle of discretization is illustrated schematically in Fig. 2.11 (b). The concentration of $i$-th layer is defined as follows:

$$C_i = (i + 1/2)\frac{H}{N}\frac{(C_{max} - C_{min})}{H/2} + C_{min}, \qquad (2.46)$$

where $i$ = 1, …, $\frac{N}{2}$, $N$ is the order of layer, $H$ is the height of the waveguide, $C_{min}$, $C_{max}$ are the minimum and maximum concentrations of germanium, respectively. The refractive indices of layers are defined by the points of intersection of triangular index profile with the bars that are defined by the number of layers.

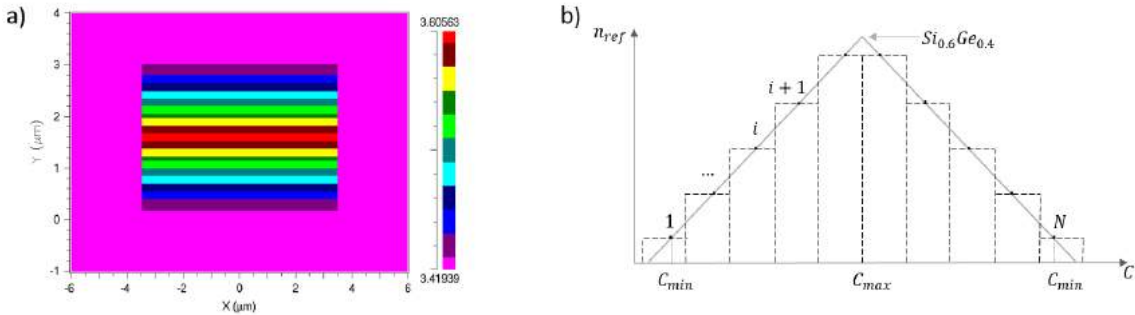

Fig. 2.11. a) Multilayer structure of graded index SiGe waveguide cross section built in R-soft; b) Schema of discretization of graded index SiGe waveguide with linearly varying Ge concentration.

### 2.1.3.3 Effective index of Si/Si$_{0.6+x}$Ge$_{0.4-x}$/Si at 4.5 µm

The effective index calculation tool in R-soft based on multilayer structure with parametrized refractive index was verified by calculating effective index of TM mode of Si/Si$_{0.6+x}$Ge$_{0.4-x}$/Si single-mode waveguide with 3x3.3 µm² cross-section at 4.474 µm wavelength.

In order to define the optimal grid size for calculation, we used R-soft's MOST optimizer utility. Scan of grid size between 0.01 and 0.1 is shown in Fig. 2.12. As it can be seen, the effective index $n_{eff}$ value converges at grid size below 0.02.

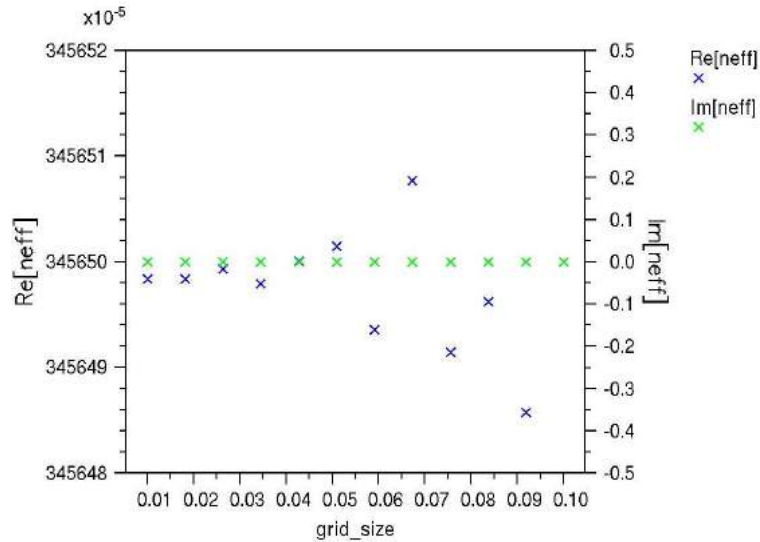

Fig. 2.12. The calculated effective index vs. grid size obtained with R-soft MOST optimizer.

The impact of number of layers was studied by comparing results of 10, 25, 30 and 40 layer discretization. As shown in Tab.2.2, we can expect the stable result for effective index up to 5$^{th}$ significant figure at discretization above 30 layers.

The modeling in R-soft was verified by comparing it to two other effective index calculation methods, Full Vectorial Finite Difference numerical tool developed in Matlab by Dr. Regis Orobtchouk [74] and Field Mode Matching utility in Phoenix Field Designer done by Dr. Pierre Labeye. Tab. 2.3 presents the results of effective index $n_{eff}$ calculation. Full Vectorial Finite Difference mode solver with a 9 point stencil describes accurately discontinuities of tangential component of the electric field. In order to compare with commercial software, an averaging at the interface is also made (noted as "av."). R-soft calculation was done using *Beamprop* utility using full-vectorial (noted as "full") and semi-vectorial (noted as "semi") methods. The difference is that full –vectorial method includes both transverse components of the field, whereas in semi-vectorial method, only one transverse component is considered and the effect of the other is ignored. The latter is used when the single polarization is present. The difference in $n_{eff}$ for full- and semi-vectorial methods is of the order of $10^{-6}$, which is negligible. As it is clearly seen, at step size 0.01, effective indices with precision of up to 4 significant digits after the decimal point, were equal to 3.4565 in both Matlab and R-soft. The calculation using Phoenix showed 3.4565 as well. It should be noted, that calculation times depends on the power of computing machine.

| Number of layers | $n_{eff}$ | |
|---|---|---|
| | TM | TE |
| 10 | 3.455929 | 3.456751 |
| 25 | 3.456666 | 3.457541 |
| 30 | 3.456498 | 3.457356 |
| 40 | 3.456525 | 3.457390 |

Tab. 2.2. Effective indices calculated for TE, TM modes for various $n_{ref}$ discretization for step size 0.01 (R-soft).

| calculation tool | time | step size | $n_{eff}$ | |
|---|---|---|---|---|
| | | | TM | TE |
| Matlab (full vectorial), 30 layers | 2 sec. | 0.1 | 3.460771 | 3.461828 |
| | 1 min 10 s | 0.01 | 3.456542/ av. 3.457008 | 3.457421/ av.3.45788749 |
| | 4 min 17 s, 20GB | 0.005 | 3.456314/ av.3.456334 | 3.457184/ av.3.457202 |
| | > 20 min, > 1 core, > 100GB | 0.001 | - | - |
| R-soft, 30 layers | 1 min /30 s | 0.02 | full 3.456500/ semi 3.456504 | full 3.457354/ semi 3.457360 |
| | 5 min / 2 min 20 s | 0.01 | full 3.456498/ semi 3.456503 | full 3.457356 / semi 3.457363 |
| | 13 min / 6min | 0.005 | full 3.456174 / semi 3.456179 | full 3.457301 / semi 3.457307 |
| | > 13 h. | 0.001 | - | - |

Tab. 2.3. Effective indices calculated for TE, TM modes for various step size (Matlab, R-soft).

### 2.1.4 Bent waveguides

As it was discussed in chapter 2.11, the guiding is associated with total internal reflection phenomenon. When it comes to bent waveguides, the angle of incidence varies along the curvature, and the curved path inevitably introduces losses in the waveguide [70]. In order to evaluate the bending losses quantitatively, one must take into account the coupling between various modes of the guide.

Consider phase velocity of the light wave along the x axis in the bent waveguide shown in Fig. 2.13. At the center of the guide, from the definition of the effective index, we can express the phase velocity as [69]:

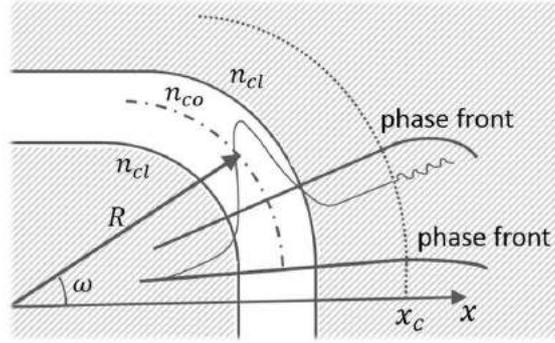

Fig. 2.13. Schema of wave in the bent waveguide.

$$v = \frac{c}{n_{eff}}. \tag{2.47}$$

On the one hand, the wave velocity increases gradually as we move away from the center of the waveguide. At distance $x$ from the center of the waveguide, we get:

$$v(x) = \frac{R+x}{R}\frac{c}{n_{eff}} = \left(1+\frac{x}{R}\right)\frac{c}{n_{eff}}. \tag{2.48}$$

On the other hand, the mode phase velocity $v(x)$ can not exceed the velocity of a wave propagating in the cladding of index $n_{cl}$. Therefore there is a critical distance $x_c$ above which the phase front will undergo distortion that leads to the leakage of a part of the mode power to the cladding. In this critical point, the radial velocity of the wave is equal to the speed of a plane wave in the cladding, which is written as:

$$v(x_c) = \left(1+\frac{x_c}{R}\right)\frac{c}{n_{eff}} = \frac{c}{n_{cl}}, \tag{2.49}$$

which gives:

$$x_c = R\frac{n_{eff} - n_{cl}}{n_{cl}}. \tag{2.50}$$

Qualitatively, we see that the modes with higher effective indices will have critical points further from the center of the waveguide and the power of the modes at these points will be lower. One can therefore predict that bending losses will be higher for lower effective indices and, therefore, the higher order modes will lose the most energy at the bend.

The bent waveguides are solved by method of conformal transformations presented in details in [75]. It consists in obtaining the equivalent straight waveguide structure that permits solution by traditional methods of optical waveguide analysis. We get a propagation of the type:

$$E(x,z) = E_0(x)\exp(j(\beta_r + j\beta_i)z) = E_0(x)\exp(-\beta_i z)\exp(j\beta_r z), \tag{2.51}$$

where $\beta_r$ and $\beta_i$ are real and imaginary components of the propagation constant. The power loss coefficient is then defined as:

$$\alpha = 2\beta_i = \frac{4\pi k_{eff}}{\lambda}, \qquad (2.52)$$

where $k_{eff}$ is the imaginary part of effective index.

## 2.2 Basic building blocks

There are several functional waveguide structures that are widely used in integrated optics. We will discuss a few of them that will be referred to in the following chapters.

### 2.2.1 Tapers

Tapers represent gradient widenings at the ends of single-mode waveguides. Their function is to insure a smooth transition to the slab [76]. By tapering, we can reduce the reflections from the cladding walls between two array waveguides and radiation losses.

Its schema is illustrated in Fig. 2.14. For adiabatic tapering, the local angle of the taper $\alpha$ must be less than a limit value given by [70]:

$$\alpha < \frac{\rho}{2\pi}(\beta - k_0 n_{cl}). \qquad (2.53)$$

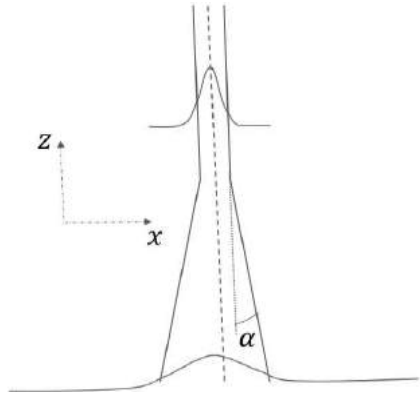

Fig. 2.14. Schema of the taper.

Thus we make sure that only the fundamental mode is excited in the widened part of the waveguide.

### 2.2.2 Y-junctions

*Y-junctions* are used as power dividers or combiners [77]. The schema of Y-junction is shown in Fig. 2.15. Its function can be described as separation of the power in the waveguide in two symmetrically spaced waveguides. Even when losses occur at the separation point, waveguide outputs will have equal power [69].

For the two beam combiner case, it is important that beams in two arms of Y-junction were in phase in order to couple to a single waveguide. If beams are antiphase, the being coupled, they will not be guided, but rather be radiated around the waveguide.

Theoretical part | 2.2 Basic building blocks
35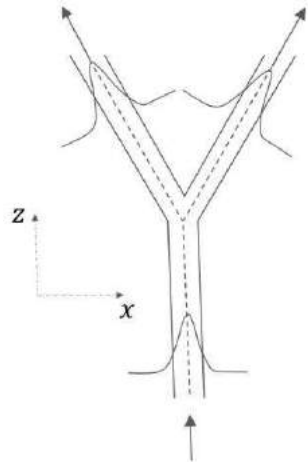

Fig. 2.15. Y-junction schema.

### 2.2.3 Multimode interference (MMI) couplers

Multimode interference (MMI) devices operate based on self-imaging as a result of interference. The self-imaging property of multimode waveguides can be described as the reproduction of initial field profile in single or multiple images at repeating intervals along the direction of propagation.

In our application, we use 1x2 MMI couplers for flattening the multiplexer spectral response, see Fig. 2.16. The length of MMI $L_{mmi}$ is chosen to be equal to beam path where the first two-fold image appears. According to [78], the length is obtained as:

$$L_{mmi} = \frac{3}{8}L_\pi, \qquad (2.54)$$

where $L_\pi \approx \frac{4n_{eff}w_e^2}{3\lambda}$ parameter is the beat length of the two lowest-order modes and $w_e$ is the waist of the beam.

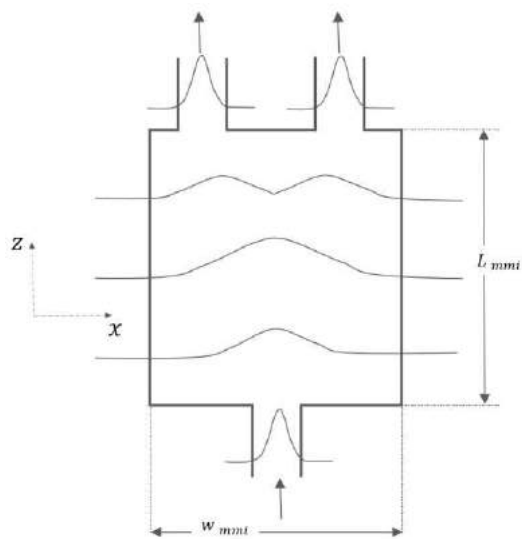

Fig. 2.16. Schema of 1x2 MMI coupler.

## 2.3 AWG operation

Arrayed waveguide grating (AWG) is the integrated optics device used either to multiplex or demultiplex. It is built of input/output channels and slabs as well as an array of waveguides with constant length difference, as illustrated in Fig. 2.17. Owing to symmetrical property of optical systems, the AWG functions as the multiplexer and de-multiplexer depending on the direction of input light. The AWG described in this work is designed as a multiplexer for the broad band source in mid-infrared gas sensing application. It is important to have a powerful mid-infrared multi-wavelength source with a controllable wavelength. For this reason a quantum cascade laser (QCL) array, with each laser having distinct wavelength, is integrated together with the multiplexer that allows scanning a broad spectral region. Since the demultiplexer is considerably more convenient for analytical field calculation rather than multiplexer, the field calculation is performed for the reverse case, i.e. one input channel carrying several wavelengths distributes illumination to the multiple channels.

The AWG operates as follows. The beam from the input channel 1 diverges in the slab 2 due to lack of lateral confinement and couples to waveguides of the array 3. The light path in array waveguides differ by a multiple of central wavelength, so that the beams with central wavelength arrive to the output slab 4 with the $2\pi$ phase difference and focus at the center of the focal line with consequent coupling to the central output channel 5. The distance between array waveguides $d$ is several orders smaller compared to $R$ radius of grating curvature.

The wavelengths other than central one have their own phase distribution. Their phase either outstrips or lags, contributing to a tilt of the phase front at the end of array waveguides section. This condition for constructive interference for AWG is expressed by equation (2.55).

$$d\beta_s sin\theta_{in} + \beta_e \Delta L + d\beta_s sin\theta_{out} = 2\pi m \qquad (2.55)$$

where $d$ is the spacing between array waveguides at the interface with slab regions, $\beta_e = 2\pi n_e/\lambda$, $\beta_s = 2\pi n_s/\lambda$ are the propagation constants of array and slab waveguides, respectively, $n_e$ is the effective index of the single-mode waveguide, $n_s$ is the effective index of the slab, $\lambda$ is the free space wavelength, $\Delta L$ is the length difference between successive array waveguides, $\theta_{in}$ is the angular position of the input, $\theta_{out}$ is the dispersion angle and $m$ order of the phased-array, as shown in Fig. 2.18. In case of single centralized input channel, the $d\beta_s sin\theta_{in}$ on the right side of the equation (2.55) drops.

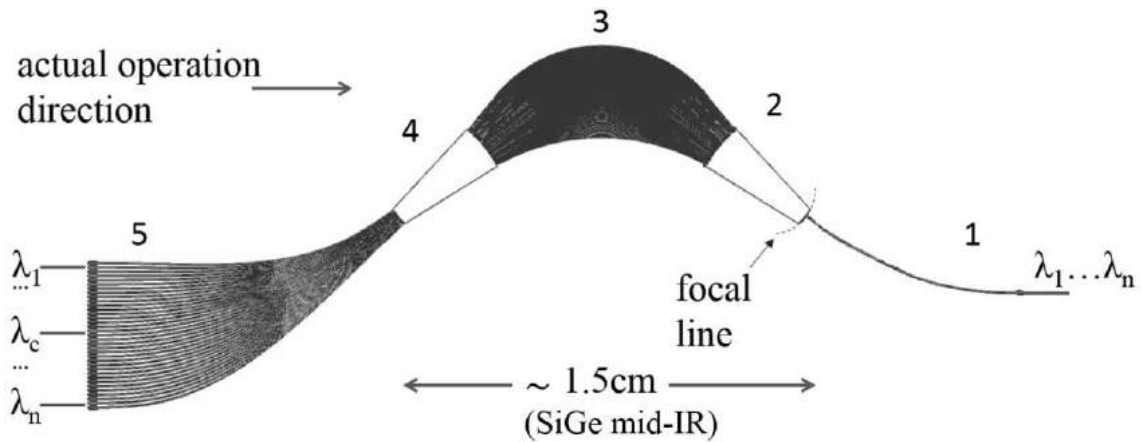

Fig. 2.17. Schema of AWG multiplexer. The analytical calculation is done in reverse order: 1-input waveguide, 2-input slab, 3-arrayed waveguides, 4-output slab, and 5-output waveguides. The inset shows schema of the output slab.

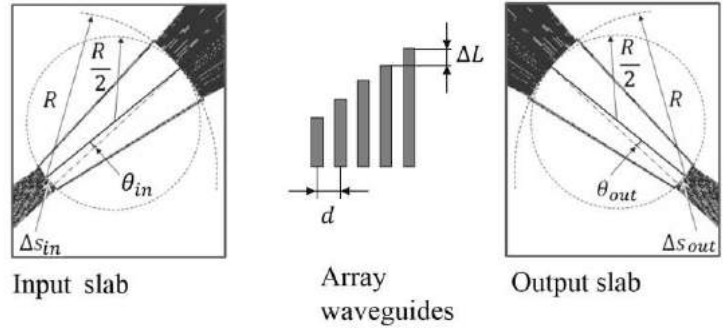

Fig. 2.18. Schemes of AWG input/output slabs and array waveguides.

### 2.3.1 Basic parameters

#### 2.3.1.1 Length difference of array waveguides

Length difference in array waveguides is a constant length different between array waveguides that equals to multiplication of the diffraction order by central wavelength in the waveguide, equation (2.56). The path length difference insures phase difference and allows focusing the beam along the focal line:

$$\Delta L = m \frac{\lambda_c}{n_{ec}}, \qquad (2.56)$$

where the subscript "c" indicates parameters referring to the central wavelength.

#### 2.3.1.2 Dispersion

Dispersion angle describes the wavelength dependent tilt of the beam in the output slab. The dispersion angle is followed from equation (2.55) (the approximation is valid for small angles):

$$\theta_{out} \approx \frac{\left(\frac{\lambda}{\lambda_c} \cdot n_{ec} - n_e\right) \cdot \Delta L}{d \cdot n_s} - \theta_{in}. \qquad (2.57)$$

The dispersion is often represented by two linearly dependent parameters – angular $\frac{d\theta_{out}}{d\lambda}$ and lateral $\frac{dx}{d\lambda} = R \frac{d\theta_{out}}{d\lambda}$ dispersion, where $R$ is the slab length as shown in Fig. 2.18. The angular dispersion as a function of wavelength is obtained by taking the derivative of (2.57) with respect to wavelength.

$$\frac{d\theta_{out}}{d\lambda} = \frac{m n_g}{n_{ec} n_s d} \quad \text{or} \quad \frac{\Delta L n_g}{\lambda_c n_s d} \qquad (2.58)$$

where $n_g = n_{ec} - \lambda_c \frac{dn_e}{d\lambda}$ is the group index.

The lateral dispersion is:

$$\frac{dx_{out}}{d\lambda} = \frac{mRn_g}{n_{ec}n_s d} = \frac{\Delta L R n_g}{\lambda_c n_s d}. \tag{2.59}$$

### 2.3.1.3 Free spectral range

Free spectral region (FSR) is the difference in wavelength (or frequency) between adjacent diffraction orders. It corresponds to the phase shift of $2\pi$. The derivation of FSR follows from its definition:

$$\Delta L\big(\beta_e(\lambda_c + \Delta\lambda_{FSR}) - \beta_e(\lambda_c)\big) = 2\pi \tag{2.60}$$

Taking into account first order Taylor approximation for the given $|\Delta\lambda_{FSR}| \ll \lambda_c$:

$$\beta_e(\lambda_c + \Delta\lambda_{FSR}) - \beta_e(\lambda_c) \approx \frac{d\beta_e}{d\lambda}\Delta\lambda_{FSR}. \tag{2.61}$$

From the definition of FSR, it follows:

$$\Delta L \frac{d\beta_e}{d\lambda}\Delta\lambda_{FSR} = 2\pi, \tag{2.62}$$

where $\frac{d\beta_e}{d\lambda} = -\frac{2\pi}{\lambda^2}n_g$.

We get:

$$\Delta\lambda_{FSR} = -\frac{\lambda_c^2}{n_g \Delta L} \tag{2.63}$$

The angular and lateral distances to the neighbor diffraction order can be directly calculated from (2.58) and (2.59):

$$\Delta\theta_{FSR} = -\frac{\lambda_c}{n_s d} \tag{2.64}$$

$$\Delta x_{FSR} = -\frac{\lambda_c R}{n_s d} \tag{2.65}$$

### 2.3.2 Spectral characteristics
#### 2.3.2.1 Insertion loss

It is important to minimize the insertion loss $L_c$ of the device as in communication application. The main sources of the insertion losses are the couplings between single mode waveguides and slabs, the propagation losses in bent region of array waveguides, the power portion lost in other diffraction orders and scattering losses of the waveguide material. Several techniques are known and widely used in order to take care of this type of losses. The coupling losses at the interfaces, the endings of array waveguides are taken care in process of design. By choosing sufficiently large radius of curvature of array waveguides, it is possible to reduce the losses due to bending to negligible level.

### 2.3.2.2 Non-uniformity

Non-uniformity $L_u$ is the intensity ratio between the outmost and the central channels. In 35 channel AWGs, it typically varies between 1dB and 3dB. In order to minimize the non-uniformity, one should increase the FSR frequency $\Delta f_{FSR}$, which could be implemented by decreasing length difference of array waveguides $\Delta L$ [37]. In process of AWG design, it is convenient to start with the choice of acceptable non-uniformity level that will in turn define the slab length, array waveguides length difference and FSR.

### 2.3.2.3 Flat-top response

In broad-band source application, the flat-top passband is very important as well as low inter-channel crossing. It affects the reduction of inter-channel crossing $X_{ch}$. The common method to achieve it is to introduce MMI couplers at the receiving side of output channels. As the result, the spectral response flattens. It should be noted that there is a tradeoff between flattening the response and the insertion loss. Even though we obtain the reduction of inter-channel crossing, the insertion loss reduces by around 3 dB in simulation.

### 2.3.2.4 Crosstalk

There are several sources that could cause higher crosstalk $L_x$. Possible crosstalk due to geometry, such as coupling between single mode waveguide and slab, truncation of the propagation field at the array interface as well as coupling of the modes between waveguides of the array, could be taken care of by tailoring the design of AWG accordingly. The main sources of phase errors arise due to effective index variation in the array waveguides that could be caused by sidewall variation, variation of germanium concentration in the layers of the waveguide. The phase error analysis is presented in chapter "2.3.6 Phase errors".

### 2.3.3 Analytical field calculation

The calculation of the spectral response is based on Gaussian approximation of the mode and scalar diffraction theory. The field calculation is divided into four parts.

a. Divergence of the wave in the input slab;
b. Coupling of the wave to arrayed waveguides;
c. Diffraction in the output slab;
d. Coupling to the output waveguide.

Fig. 2.17. illustrates the schematic of AWG demultiplexer. It is assumed that at the waveguide-slab and slab-waveguide interfaces, the direction of beam propagation remains along z-axis, and the cross-section of the field is defined in (x, y)-plane. Input (section 1)/ output (section 5) channels and array waveguides (section 3) are single-mode, and slabs (sections 2, 4) are unconfined in horizontal direction. The notations are listed in Tab. 2.4.

Transmission (coupling) coefficient at the interface of two waveguides is calculated as an overlap of the incident field with the mode of the waveguide (both normalized) [70]:

| Notations | |
|---|---|
| $w_{ey}/w_{ex}$ | vertical/horizontal waists of beam in input, output channels and array waveguides (AWs) |
| $w_{sy}/w_{sx}$ | vertical/horizontal waists of beam in slab |
| $w_{my}/w_{mx}$ | vertical/horizontal waists of beam in multimode interference (MMI) coupler |
| $w_{ty}/w_{tx}$ | vertical/horizontal waists of beam in tapered AW |
| $\eta$ | coupling coefficient |
| $R$ | slab length |
| $\lambda$ | wavelength in free space |
| $n_e$ | effective index of input, output channels and AWs |
| $n_s$ | effective index of slab |
| $\beta_e$ | propagation constant of mode in input, output channels and AWs |
| $\beta_s$ | propagation constant of mode in slab |
| $\Psi_1$ | field in input channel |
| $\Psi_2$ | field in input slab |
| $\Psi_3$ | field in AWs |
| $\Psi_4$ | field in output slab |
| $\Psi_5$ | field in output channels |
| $N$ | number of AWs |
| $\chi$ | arc length between neighbor AWs |
| $\Delta L$ | length difference between neighbor AWs |

Tab. 2.4. The list of parameters and their notations used in field calculation.

$$\eta(x,y) = \left| \iint_{-\infty}^{+\infty} dxdy \Psi_1 \Psi_2^* \right|^2 \quad (2.66)$$

The cross-section of the field $\Psi(x,y)$ is represented as a superposition of vertical $\Psi(y)$ and horizontal $\Psi(x)$ field components independent of each other. Vertical dimension of all waveguides is the same, whereas vertical waist of the beam slightly varies due to variations in horizontal dimensions of the waveguides. It allows us to simplify the contribution of the vertical field component to coupling coefficients between sections $\sqrt{\eta_e(y)}\eta_t(y)\sqrt{\eta_m(y)}$ defined by equations (2.67).

$$\eta_e(y) = \frac{2w_{ey}w_{sy}}{(w_{ey}^2+w_{sy}^2)}, \ \eta_t(y) = \frac{2w_{ty}w_{sy}}{(w_{ty}^2+w_{sy}^2)}, \ \eta_m(y) = \frac{2w_{my}w_{sy}}{(w_{my}^2+w_{sy}^2)}. \quad (2.67)$$

The following calculation steps focus on horizontal component of the field.

### 2.3.3.1 Divergence of the wave in the input slab

The wave in the single mode input waveguide $\Psi_1(x)$ is defined by equation (2.68). In the slab, the wave is confined in the vertical y axis, whereas in horizontal x axis it uniformly diverges due to lack of lateral confinement. At distance $r_z$ from the input waveguide, the wave field $\Psi_2(x)$ can be calculated by Fraunhofer diffraction integral [79] as shown in equation (2.69), see Fig. 2.19. The phase-front of far-field has a radial shape and is expressed as exponential multiplication $\exp(-jk\frac{x^2}{2R})$.

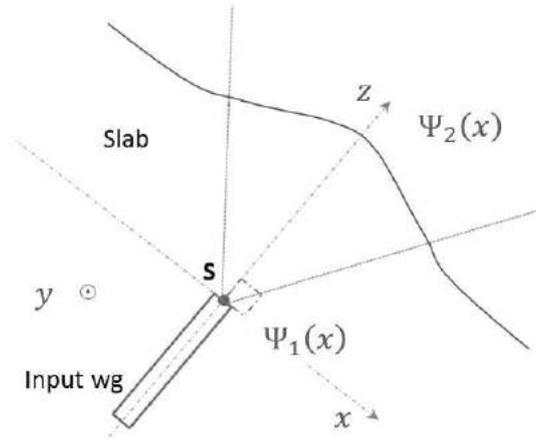

Fig. 2.19. Schema of waveguides orientation at the interface of input channel and slab.

$$\Psi_1(x) = \sqrt[4]{\frac{2}{\pi w_{ex}^2}} \exp(-\frac{x^2}{w_{ex}^2}) \qquad (2.68)$$

$$\Psi_2(x) = \sqrt[4]{\frac{2}{\pi w_{sx}^2}} \exp(-\frac{x^2}{w_{sx}^2}) \qquad (2.69)$$

where $w_{ex}$ is the waist of beam in the input waveguide, $w_{sx}$ is the waist of beam in the slab waveguide, $w_{sx} = w_{ex}\sqrt{1 + \left(\frac{z}{z_0}\right)^2} = w_{ex}\sqrt{1 + \left(\frac{\lambda R}{n_e \pi w_{ex}^2}\right)^2}$, $R \approx z$ is the slab length, and $z_0 = \frac{\pi w_{ex}^2 n_e}{\lambda}$ is called Rayleigh length. The phase front of the wave in the slab has a radial shape given by $\exp(-\frac{j\beta x_2^2}{2z})$, which we omit by assuming that it does not contribute to phase delay.

### 2.3.3.2 Coupling of the wave to arrayed waveguides

The diverged wave couples to all array waveguides (Fig. 2.20). The total field in all array waveguides $\Psi_3(x)$ is expressed as a sum of fields in each array waveguide:

$$\Psi_3(x) = \sum_{p=-M}^{M} \sqrt{\eta(x)} \sqrt[4]{\frac{2}{\pi w_{tx}^2}} \exp(-\frac{(x-p\chi)^2}{w_{tx}^2}) \exp(-jp\beta_e \Delta L) \qquad (2.70)$$

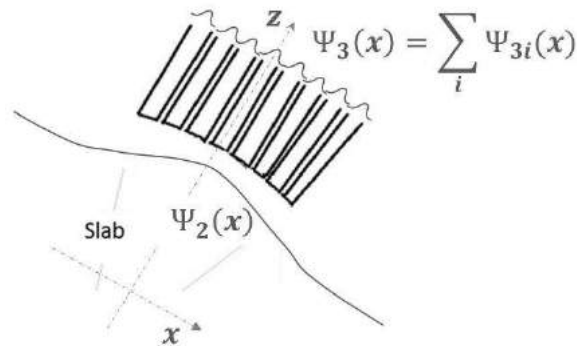

Fig. 2.20. Schema of waveguides orientation at the interface of input channel and slab.

where $M = \frac{N-1}{2}$, $N$ is the total number of array waveguides, $p$ is the order of array waveguide, $\chi \approx d$ is the arc length between neighbor array waveguides, $\beta_e$ is the propagation constant and $\Delta L$ is the difference in length in consecutive array waveguides.

The coupling coefficient $\eta(x)$ is calculated as follows:

$$\eta(x) = \frac{2}{\pi w_{sx} w_{tx}} \left| \int_{-\infty}^{+\infty} \exp\left(-\frac{x^2}{w_{sx}^2}\right) \exp\left(-\frac{(x - p\chi)^2}{w_{tx}^2}\right) dx \right|^2 \quad (2.71\text{-}1)$$

$$\eta(x) = \frac{2 w_{sx} w_{tx}}{a_1^2} \exp\left(-\frac{2(p\chi)^2}{a_1^2}\right) \quad (2.71\text{-}2)$$

where we introduce the notation $a_1 = \sqrt{w_{sx}^2 + w_{tx}^2}$, and use known integral of Gaussian function $\int_{-\infty}^{+\infty} \exp(-ax^2)\exp(-2bx) dx = \sqrt{\frac{\pi}{a}} \exp\left(\frac{b^2}{a}\right)$, with $a > 0$.

### 2.3.3.3 Diffraction in the output slab

The beam that focuses in front of the output waveguide results from diffraction of separate beams coming from array waveguides. The field can be calculated using Fresnel-Kirchhoff diffraction formula (2.72) [58], [80], which is the Fourier transform given as:

$$\Psi_4(x) = a_2 FT[\Psi_3(x)] =$$
$$a_2 \sum_{p=-M}^{M} \sqrt{\eta(x)} FT\left[\sqrt[4]{\frac{2}{\pi w_{tx}^2}} \exp\left(-\frac{(x - p\chi)^2}{w_{tx}^2}\right) \exp(-jp\beta_e \Delta L)\right] \quad (2.72)$$

where $a_2 = \frac{\exp(j2\pi R \frac{n_e}{\lambda})}{j\frac{\lambda}{n_e} R}$ is the multiplication factor. Fourier transform of shifted Gaussian:

$$FT\left[\exp\left(-\frac{(\xi - p\chi)^2}{w_{tx}^2}\right)\right]_{(\xi, u = \frac{x}{\alpha})} = \sqrt{\pi} w_{tx} \exp\left(-\frac{(\pi w_{tx} x)^2}{\alpha^2}\right) \exp\left(-2j\frac{\pi x p\chi}{\alpha}\right) \quad (2.73)$$

where $\alpha = \frac{\lambda}{n_s} R$. Then from equation (2.72) and (2.73), we get a field at the interface of second slab and the output waveguide (Fig. 2.21):

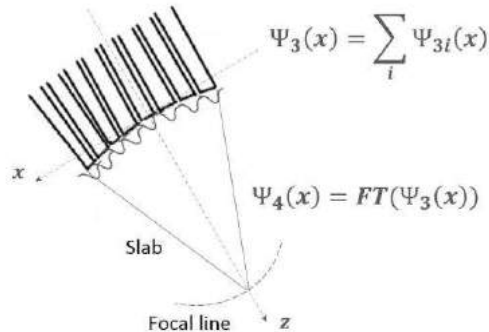

Fig. 2.21. Schema of waveguides orientation at the interface of input channel and slab.

$$\Psi_4(x) = a_2 b \sum_{p=-M}^{M} \sqrt{\eta(x)} \exp(-jp\Delta\varphi) \qquad (2.74)$$

with $b(x, w_{tx}) = \sqrt[4]{2\pi w_{tx}^2} \exp\left(-\frac{(\pi w_{tx} x)^2}{\alpha^2}\right)$ Fourier transform of AW mode Gaussian, and phase shift between neighbor array waveguides $\Delta\varphi = (\beta_e \Delta L + 2\pi\chi\frac{x}{\alpha})$, taken that at the central AW phase shift is zero.

#### 2.3.3.4 Coupling to the output waveguide

Field at the output waveguide:

$$\eta_{out}(x,y) = \left|\int_{-\infty}^{+\infty} \Psi_4(x,y)\Psi_5(x,y)dx\right|^2 \qquad (2.75)$$

where $\Psi_5$ is the Gaussian as $(x, w_{mx})$. The numerical representation of AWG spectral response based on the analytical model discussed above is presented in the chapter "2.3.4 Simulation".

### 2.3.4 Simulation

The design procedures of 33-ch AWG are detailed in Annex I. First, the waveguide width, effective index and fundamental mode waist were defined. Using our multi-layered structure discussed in chapter "2.1.4 Calculation of effective index of Si/Si0.6+xGe0.4-x/Si waveguide in R-soft", we have defined the optimal width of the waveguide that would support single-mode propagation for the whole operational range of our AWG, i.e. 5.65 – 5.88 µm. It was also important to take into account the potential impact of fabrication that could alter the width of waveguide within tenth of micron. Knowing the waveguide width, we can get the effective index of the waveguide for the given wavelengths. The waist of the fundamental mode is defined by Gaussian approximation, see Fig. 2.22.

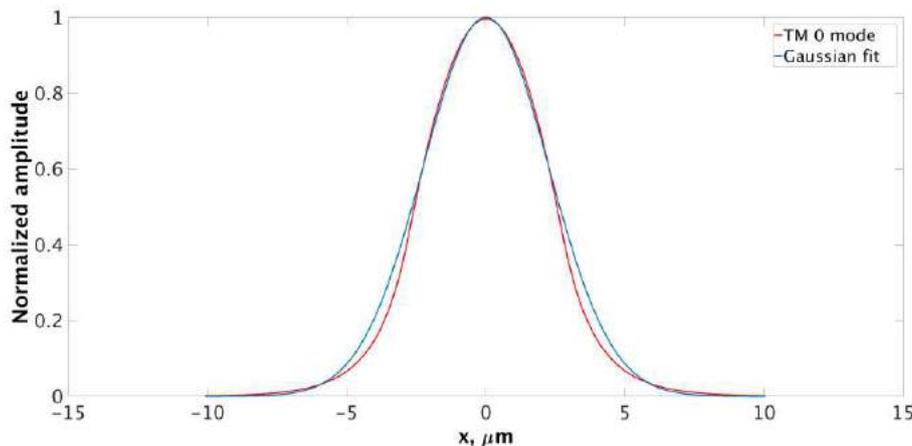

Fig. 2.22. TM fundamental mode of SiGe graded index single-mode waveguide at 5.7 µm and its Gaussian fit.

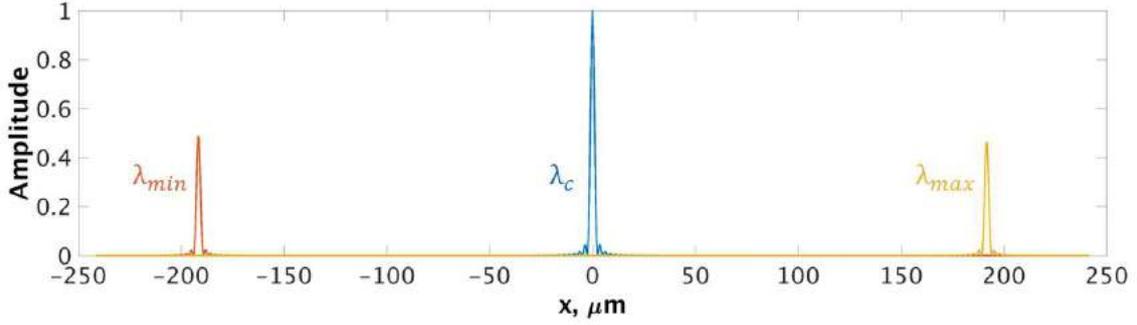

Fig. 2.23. Normalized amplitude and positions of $\lambda_{min}$, $\lambda_c$ and $\lambda_{max}$ on the focal line.

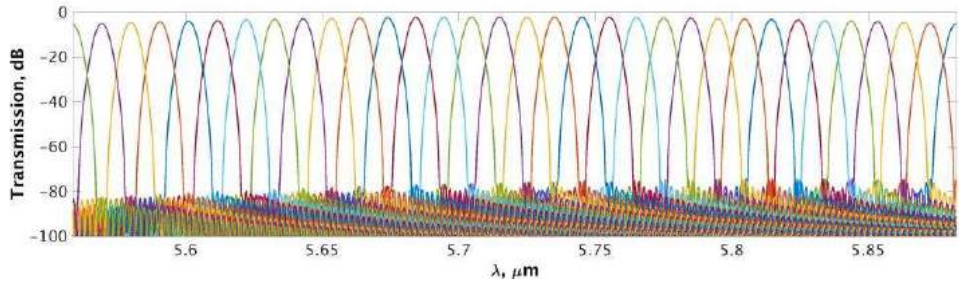

Fig. 2.24. Spectral response of 33-ch AWG at 5.7 μm.

With obtained parameters, we have performed numerical analysis of spectral response using semi-analytical tool developed for Matlab software. Fig. 2.23 shows the positions and normalized amplitudes of $\lambda_{min}$, $\lambda_c$ and $\lambda_{max}$ beams focused along the focal line calculated using equation (2.74). The origin of spatial scale $x$ is fixed at the central output channel. The central wavelength (blue) $\lambda_c$ is 5.72 μm and as expected has the greatest amplitude, i.e. 1. The minimum wavelength (orange) $\lambda_{min}$ is 5.56 μm and has the normalized amplitude of 0.49; and maximum wavelength (yellow) $\lambda_{max}$ is 5.88 μm and has the normalized amplitude of 0.46. The analysis of normalized amplitudes of beams focused on the focal line gives us the information that the non-uniformity is around 3-4dB.

Fig. 2.24 shows the transmission spectra of 33 channel AWGs with central wavelength at 5.72 μm. The simulation showed the following results. The central insertion loss is -2.2 dB, channels crosstalk is below -75 dB, the non-uniformity is 3.6 dB.

### 2.3.5 Optimization

The goal of the multiplexer combined with an array of QCLs in broad-band source application is to achieve a whole coverage of the chosen spectral range uniformly, as depicted with dashed lines in Fig. 2.25. Intersection of neighbor channels $X_{ch}$ should be minimized as well as the non-uniformity $L_u$ and the insertion loss $L_c$. The requirements on the crosstalk ($L_x$) level are more relaxed compared to telecommunication applications, and is acceptable being below -20 dB.

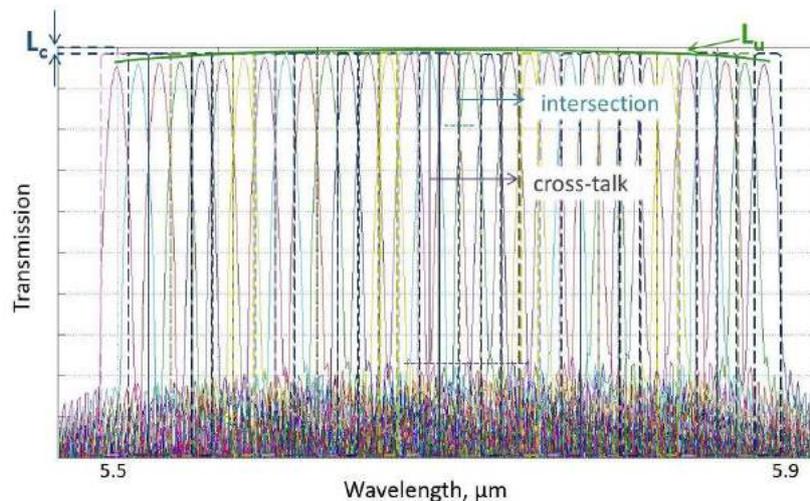

Fig. 2.25. Illustration of main parameters of the multiplexer spectral response: $L_c$ – central insertion loss; $L_u$ – non-uniformity; intersection between neighbor channels, and cross-talk - noise. Spectra in dashed lines shows ideal broad-band source, spectra in solid lines – simulation result for the AWG at 5.7 μm.

In order to optimize the spectral response towards the desired output, we introduced 6 μm tapers at the ends of array waveguides. The geometry of the taper and the AWG spectral response are shown in Fig. 2.26. Compared to simple AWG geometry, the insertion loss has increased by 0.4dB from, i.e. $L_c$ is 2.8 dB. Other improvements to be notes are the decrease of non-uni formity to 2.5 dB as well as the reduction of crosstalk by 2 dB.

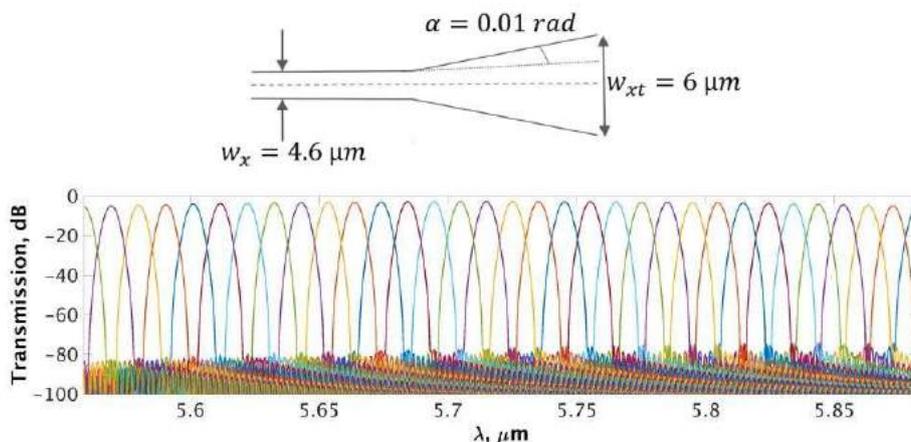

Fig. 2.26. Taper: top – schema; bottom – spectral response of 33-ch AWG at 5.7 μm with 6 μm taper.

We also studied the impact of 1x2 MMI coupler s placed at the entrances to output channels. The length of MMI couplers were defined using home-made tool for Matlab design developed by Dr. Regis Orobtchouk and Beamprop utility in R-soft, see Fig. 2.27. The double self-image position in R-soft is defined as the half of the first single self-image.

Figs. 2.28 and 2.29 present the spectral responses of AWGs with 9 μm and 11 μm MMIs, and their impact on flattening of the top of the response is compared in Fig. 2.30. The insertion losses of AWGs with 9 μm and 11 μm MMIs are -4.8 dB and -6.3 dB, respectively; inter-channel crossing reduced from around 20 dB to 7dB and 8 dB, respectively.

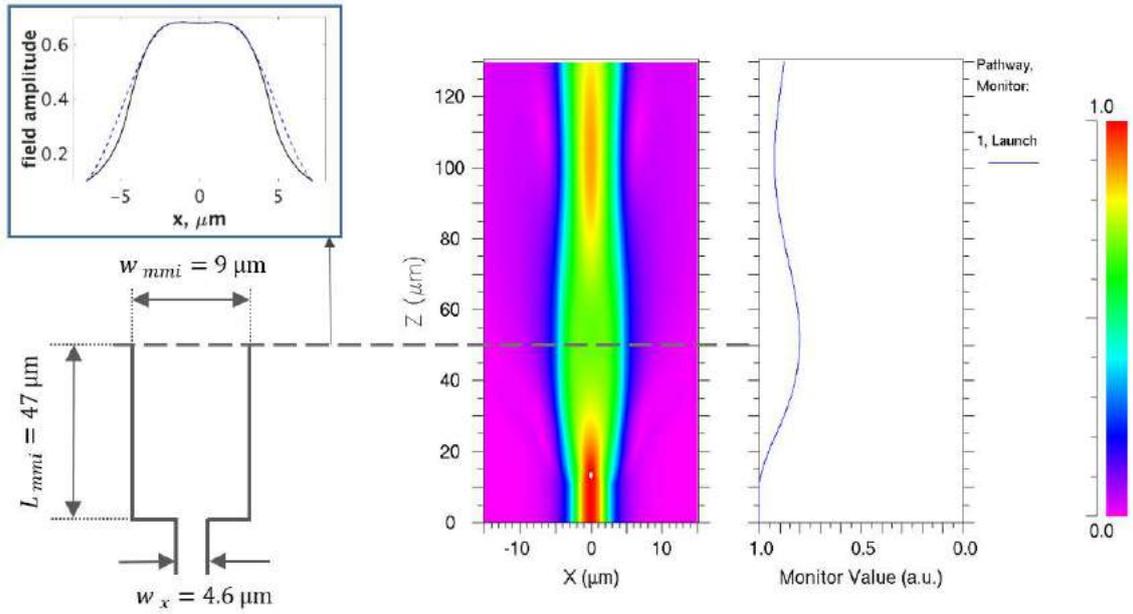

Fig. 2.27. 9 μm MMI coupler and its self-imaging simulation in R-soft; its field amplitude at the cross-section is shown in the inset: the profile obtained by R-soft shown in black solid line, two Gaussian approximation is shown in blue dashed line.

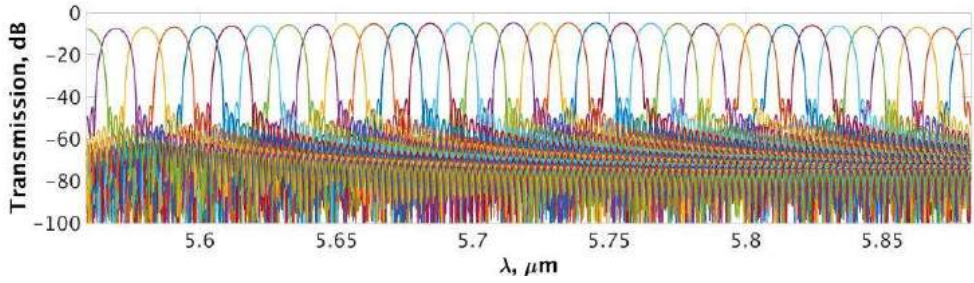

Fig. 2.28. Spectral response of 33-ch AWG at 5.7 μm with 9 μm MMI.

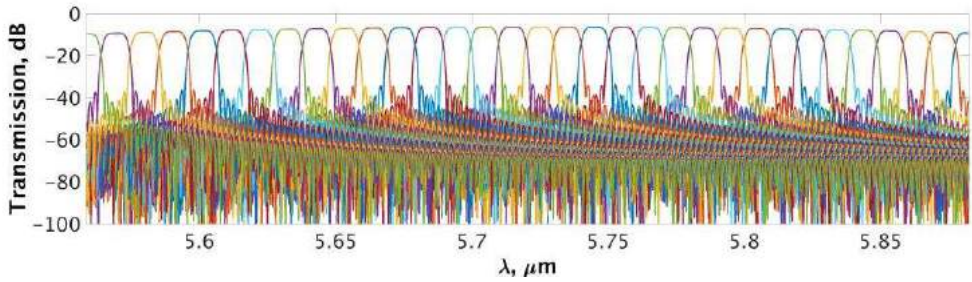

Fig. 2.29. Spectral response of 33-ch AWG at 5.7 μm with 11 μm MMI.

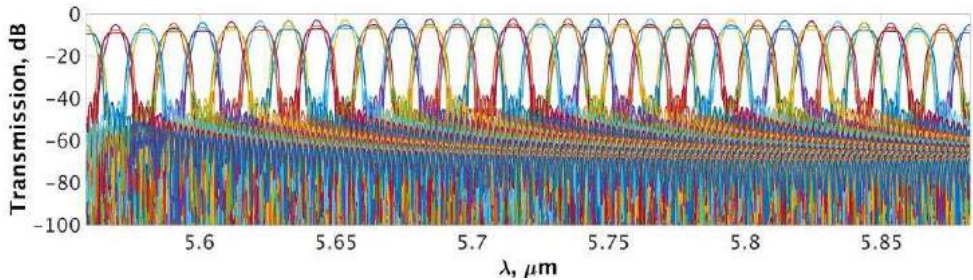

Fig. 2.30. Comparison of impacts of MMIs.

**2.3.6 Phase error**

Quantitative phase error analysis results for 3.4 µm, 4.5 µm, 5.7 µm and 7.6 µm AWGs are presented. The effective index variation and equivalent path length deviation are evaluated for 4.5 µm and 7.6 µm AWGs on the basis of experimental results.

Input illumination with certain wavelength ($\lambda_1,\ldots, \lambda_n$) accumulates its corresponding phase difference $\Delta\varphi(\lambda)$ in the array waveguides that causes a tilt of the phase front and defines a position of diffracted beam on the focal curvature according to:

$$\Delta\varphi(\lambda) = \beta_e(\lambda) \cdot \Delta L = 2\pi m - d \cdot \beta_s(\lambda) \cdot (\sin\theta_{in} + \sin\theta_{out}) \qquad (2.76)$$

where $d$ is the waveguides spacing, $\beta_e$ and $\beta_s$ are the rib and slab propagation constants, respectively, $\theta_{in}$ and $\theta_{out}$ are the input and output angles, $m$ is the diffraction order, and $\lambda$ is the free space wavelength.

Single-mode waveguides are designed to support TM fundamental mode that is accurately approximated by Gaussian Fit in home-made semi-analytical tool for AWG spectral response calculation. Using our simulation tool based on Fourier Optics we studied correlation between level of crosstalk and phase error in order to determine the fabrication tolerance required to obtain desired performance of the AWG.

Sidewall roughness arising from fabrication imperfection alters the effective index of the fundamental mode, which in turn causes phase errors [81]. In order to quantitatively evaluate the fabrication tolerance, we assumed phase errors $\delta$ to be random variables of normal distribution with zero mean value.

$$\varphi = \sum_{i=1}^{N} \exp(-j(\Delta\varphi(\lambda)_i + \delta_i)) \qquad (2.77)$$

where $i$ indicates the order of array waveguide.

For the given standard deviation of phase errors $\sigma(\delta)$, one can evaluate effective index variation from:

$$\sigma(\Delta n_e) = \frac{\lambda \sigma(\delta)}{2\pi L_x}, \qquad (2.78)$$

where $L_x$ is the path length segment under consideration.

In AWGs based on SiGe/Si graded index waveguides, the phase errors mainly arise due to effective index variation in waveguides of the array. Reduction of number of array waveguides could be an attractive option for achieving both reduction of phase error level as well as scaling down overall multiplexer dimensions. However, there is a trade-off between diminishing the number of array waveguides and retaining the sufficient amount of input power coupled from the slab. As expected, the former at some point unavoidably leads to power truncation and increase of crosstalk. Using the semi-analytical simulation tool discussed above, the spectral responses of AWGs were evaluated at certain level of phase error distribution (with standard deviations of 0 rad, 0.0002 rad, 0.0009 rad, 0.0017 rad, 0.0052 rad and 0.0087 rad). Fig. 2.31 presents the field amplitudes of $\lambda_c$(5.72), $\lambda_{min}$(5.56) and $\lambda_{max}$(5.88) at the focal line of the output slab before coupling to output waveguides. Three cases are illustrated, the amplitudes of untruncated field with zero level of phase errors (a), the amplitudes of untruncated field with standard deviation of phase errors $\sigma_\varphi = 0.0009$ rad (b), and the amplitudes of 5% truncated field with zero level of phase errors (c). Fig. 2.32 shows the dependence of crosstalk on the effective index variation and level of power truncation for AWGs at four wavelengths in infrared. The graphs illustrate effective index error vs. crosstalk distribution at certain level of power truncation. Each data point represents an average of five separate results. It is clearly seen, that the impact of effective index variation is more critical for AWGs with smaller operational wavelengths.

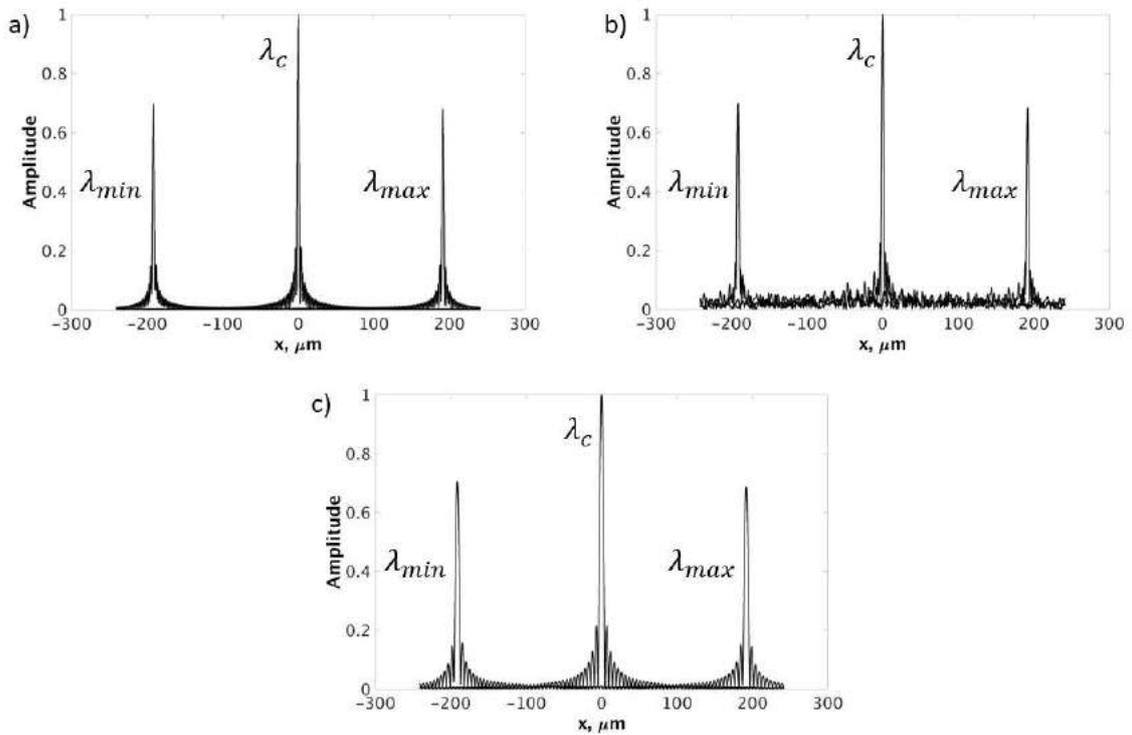

Fig. 2.31. Field amplitudes at the focal line of the output slab before coupling to output waveguides: a) 0% truncation, zero level of phase errors; b) 0% truncation, standard deviation of phase errors $\sigma_\varphi = 0.0009$ rad; and c) 5% truncation, zero level of phase errors.

The power truncation of less than 0.1%, 1% and 5% are presented. The evaluation was done as follows. First the width of diverged beam $w_{sx}$ in the slab waveguide is approximated using Kirchhoff diffraction integral. Then, the power truncation is evaluated from $P = P_0 \cdot (1 - \exp(-2r^2/w_{sx}^2))$ with $P_0$ is the input power, $P$ is the power coupled to array waveguides, and $r$ is the length of the focal line formed by array grating. For the given waveguide width the latter determines the number of array waveguides in the AWG. As shown in Fig. 2.32, the effect of power truncation is noticeably weighty. Given zero phase error level, all four AWGs exhibit minimum 70 dB crosstalk at optimum number of array waveguides, i.e. 185/205/258/259 array waveguides in 3.4 µm/4.5 µm/5.7 µm/7.6 µm AWG, respectively. As the power truncation level reaches 1%, i.e. 111/125/157/157 array waveguides in 3.4 µm/4.5 µm/5.7 µm/7.6 µm AWG, respectively, the crosstalk levels become 34 dB for 3.4 µm and 38 dB for the rest AWGs. At 5% power truncation level, i.e. 90/101/125/127 array waveguides in 3.4 µm/4.5 µm/5.7 µm/7.6 µm AWG, respectively, the crosstalk is around -27 dB. The proximity of number of array waveguides in case of 5.7 µm and 7.6 µm AWGs could be understood by the choice of wider waveguide width for the latter. According to expectations, as the power truncation increases, the crosstalk becomes less sensitive to effective index variation. For the power truncation less than 0.1%, the impact of phase errors is the strongest, and the crosstalk degradation is as high as 30 dB and more for $5 \cdot 10^6$ effective index variation at all four wavelengths. From Fig. 2.32, it is seen that given the crosstalk requirement, there is no need to investigate power truncation over 5%, as the crosstalk as of now approaches the acceptable limit. The spectral responses at various levels of phase error and power truncation are shown in Figs. 2.33, 2.34.

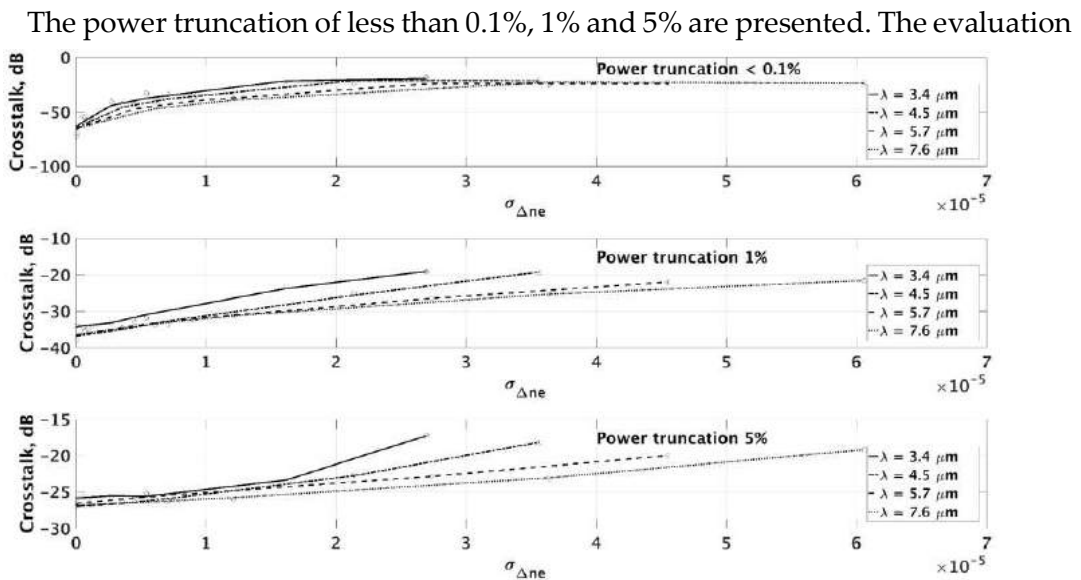

Fig. 2.32. Impact of standard deviation of effective index (10 mm waveguide) and power truncation on crosstalk.

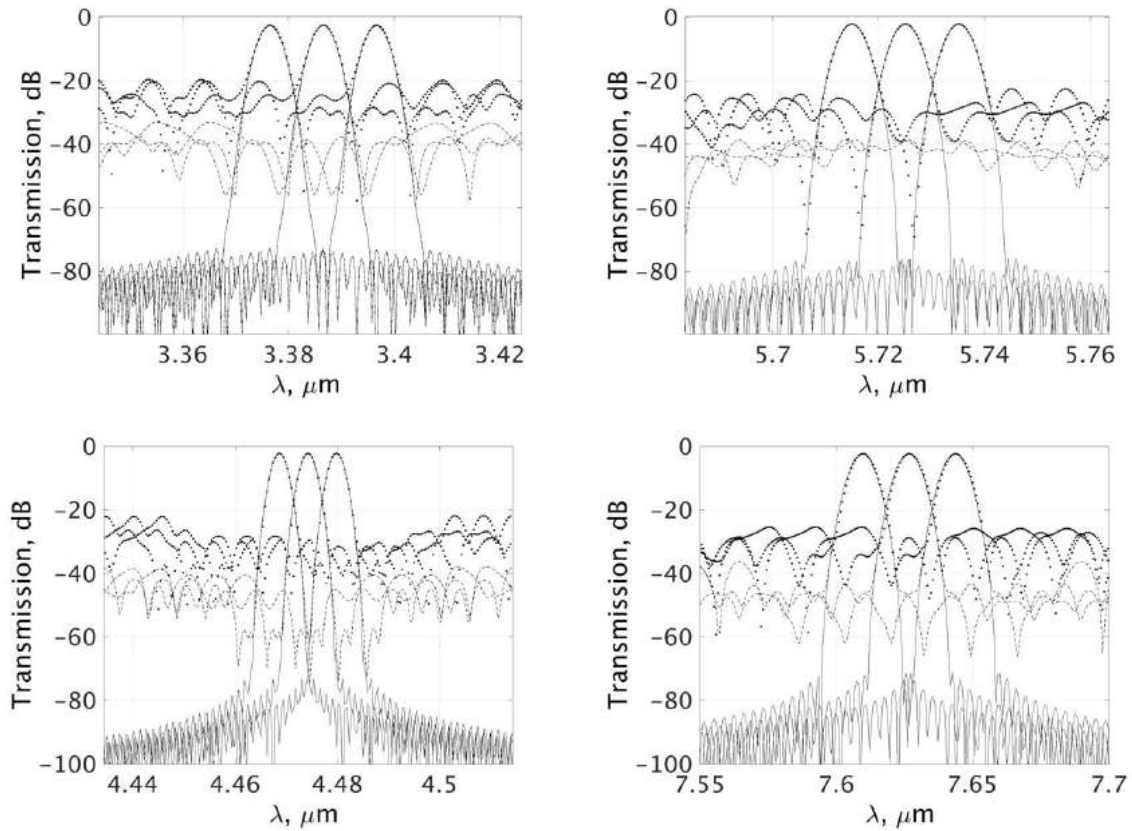

Fig. 2.33. Impact of phase errors in array waveguides (10 mm waveguide): solid line – $\sigma_\varphi = 0$ rad, dashed line – $\sigma_\varphi = 0.0002$ rad, and dotted line – $\sigma_\varphi = 0.0009$ rad.

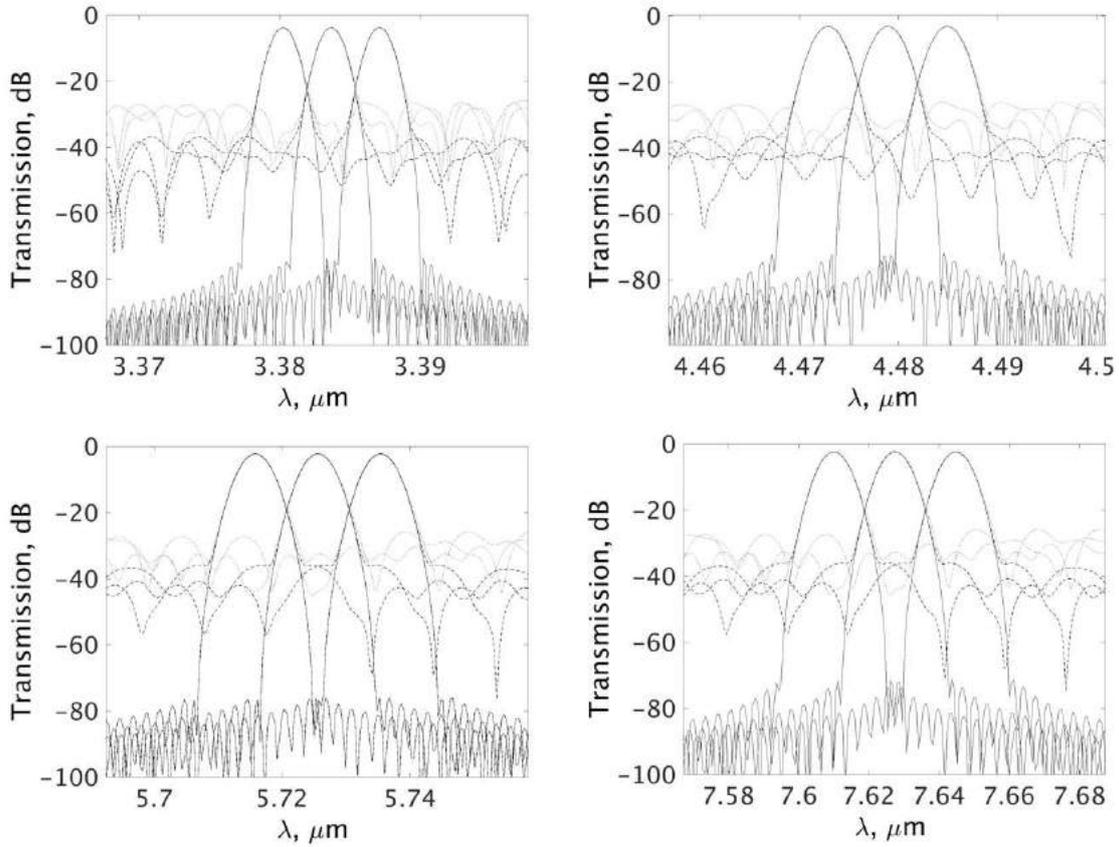

Fig. 2.34. Impact of power truncation: solid line – less than 0.1%, dashed line – 1%, and dotted line 5%.

The AWGs operating at 4.5 μm [82] and 7.6 μm [83] were presented recently. The crosstalk in both cases was around -20 dB. Using our numerical results presented in Fig. 2.30, given 1% of power truncation level, one can estimate the corresponding effective index variation for 10 mm waveguide that is $3.6\times10^{-5}$ in 4.5 μm AWG and $6.1\times10^{-5}$ in 7.6 μm AWG. The phase error deviations could also be represented in equivalent path length deviation that is 0.10 μm in 4.5 μm AWG and 0.20 μm in 7.6 μm AWG length (or ≈λ/40), which is noticeably smaller compared to 10 mm length of the device.

We have also compared the effect of MMI on sensitivities to crosstalk of AWGs.

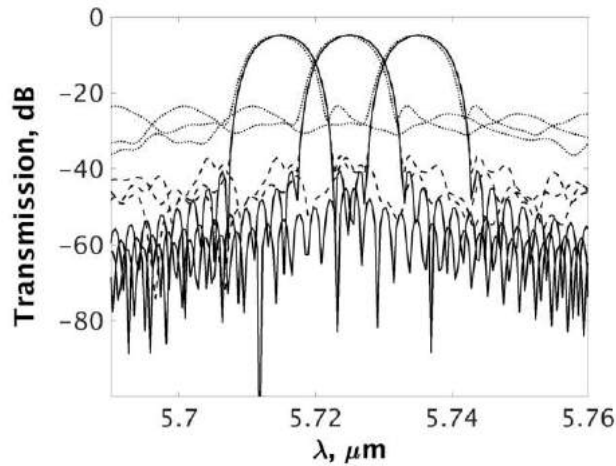

Fig. 2.35. Impact of phase errors on response of AWG with 9 μm MMI: solid line – $\sigma_\varphi = 0$ rad, dashed line – $\sigma_\varphi = 0.0002$ rad, and dotted line – $\sigma_\varphi = 0.0009$ rad.

Fig. 2.35 shows the response of 5.7 μm AWG with 9 μm MMI at various level of phase errors. By comparing with 5.7 μm AWG without MMI from Fig. 2.33, one can see, that despite the initial crosstalk degradation due to introduction of MMI (presented in solid lines in both Figs.), the crosstalk stays around -22 dB and -38 dB when the same level of phase errors, i.e. $\sigma_\varphi = 0.0002$ rad and $\sigma_\varphi = 0.0009$ rad, respectively, is introduced.

### 2.3.7 Temperature dependence

The temperature dependence of spectral response was simulated using the study done by Li [84]. The temperature dependence of silicon and germanium refractive indices are approximated using the following fit:

$$n^2(\lambda, T) = \epsilon(T) + \frac{L(T)}{\lambda^2}(A_0 + A_1 T + A_2 T^2), \tag{2.79}$$

where $\lambda$ is in units of μm, $T$ is in units of K,

$$L(T) = e^{-3\Delta L(T)/L_{293}},$$

$$\epsilon(T) = B_0 + B_1 T + B_2 T^2 + B_3 T^3.$$

The parameters $\Delta L(T)/L_{293}$ varies depending on the material.
For silicon:

$$\frac{\Delta L(T)}{L_{293}} = C_0 + C_1 T + C_2 T^2 + C_3 T^3 \quad \text{at 20-293 K, and}$$

$$\frac{\Delta L(T)}{L_{293}} = D_0 + D_1 T + D_2 T^2 + D_3 T^3 \quad \text{at 293-1600 K.}$$

and germanium:

$$\frac{\Delta L(T)}{L_{293}} = C_0 + C_1(T-100) + C_2(T-100)^2 + C_3(T-100)^3 \quad \text{at 100-293 K, and}$$

$$\frac{\Delta L(T)}{L_{293}} = D_1(T-293) + D_2(T-293)^2 + D_3(T-293)^3 \quad \text{at 293-1200 K,}$$

with coefficients defined in Tab. 2.5.

Using the empirical approximations given above, the 9-layer structure of SiGe graded index waveguide with Ge up to 40% at the core was built in R-soft commercial software according to method described in chapter 2.1.3.

The refractive index of each layer was determined by the linear approximation of germanium concentration in the following manner:

$$n_{Si_{0.6+x}Ge_{0.4-x}} = (1-x)n_{Si} + xn_{Ge}, \tag{2.80}$$

where $x$ is the concentration of Ge varying up to 0.4. Tab. 2.6 specifies the values of refractive indices used in 9-layer SiGe effective index calculation, where the germanium concentration was varied by 0.08 with maximum value at 5th (central) layer.

The simulation results of effective index of TM mode calculated using semi-vectorial Beamprop utility in R-Soft are given in Tab. 2.7.

Using effective indices of SiGe waveguides at various temperature points, we analyzed the corresponding spectral shifts in wavenumbers of AWG response operating at 7.6 µm wavelength. At temperature drop from 20 °C to 12 °C, the effective index variation was $\Delta n_{eff} = 0.001609$ which corresponds to $\Delta \sigma = 0.6$ cm$^{-1}$ spectral shift. At temperature drop from 35 °C to 20 °C, the effective index variation was $\Delta n_{eff} = 0.003157$ which corresponds to $\Delta \sigma = 1.2$ cm$^{-1}$ shift.

| coefficient | Si | Ge |
|---|---|---|
| $A_0$ | 0.8948 | 2.5381 |
| $A_1$ | $4.3977 \times 10^{-4}$ | $1.8260 \times 10^{-3}$ |
| $A_2$ | $7.3835 \times 10^{-8}$ | $2.8888 \times 10^{-6}$ |
| $B_0$ | 11.4445 | 15.2892 |
| $B_1$ | $2.7739 \times 10^{-4}$ | $1.4549 \times 10^{-3}$ |
| $B_2$ | $1.7050 \times 10^{-6}$ | $3.5078 \times 10^{-6}$ |
| $B_3$ | $-8.1347 \times 10^{-10}$ | $-1.2071 \times 10^{-9}$ |
| $C_0$ | $-0.021$ | $-0.089$ |
| $C_1$ | $-4.149 \times 10^{-7}$ | $2.626 \times 10^{-6}$ |
| $C_2$ | $-4.620 \times 10^{-10}$ | $1.463 \times 10^{-8}$ |
| $C_3$ | $1.482 \times 10^{-11}$ | $-2.221 \times 10^{-11}$ |
| $D_0$ | $-0.071$ | 0 |
| $D_1$ | $1.887 \times 10^{-6}$ | $5.790 \times 10^{-6}$ |
| $D_2$ | $1.934 \times 10^{-9}$ | $1.768 \times 10^{-9}$ |
| $D_3$ | $-4.544 \times 10^{-13}$ | $-4.562 \times 10^{-13}$ |

Tab. 2.5. Coefficients of Li's approximation of refractive indices of Si and Ge as a function of temperature and wavelength.

| Layers | 293.15K (20 °C) | 300.65K (27.5 °C) | 305.85K (32.7 °C) | 309.75K (36.6 °C) | 314.55K (41.4 °C) |
|---|---|---|---|---|---|
| n(Si) | 3.4667 | 3.5140 | 3.5614 | 3.6088 | 3.6561 |
| n($Si_{92\%}Ge_{8\%}$) | 3.4667 | 3.4680 | 3.4690 | 3.4697 | 3.4706 |
| n($Si_{84\%}Ge_{16\%}$) | 3.5140 | 3.5155 | 3.5166 | 3.5174 | 3.5184 |
| n($Si_{76\%}Ge_{24\%}$) | 3.5614 | 3.5631 | 3.5642 | 3.5651 | 3.5662 |
| n($Si_{68\%}Ge_{32\%}$) | 3.6088 | 3.6106 | 3.6118 | 3.6128 | 3.6140 |
| n($Si_{60\%}Ge_{40\%}$) | 3.6561 | 3.6581 | 3.6595 | 3.6605 | 3.6618 |

Tab. 2.6. SiGe layers and corresponding refractive indices used to calculate effective index using TM semi-vectorial Beamprop tool in RSoft.

Fig. 2.36 illustrates the dependence of effective index vs. temperature, where the solid line represents 3$^{rd}$ order polynomial approximation in a form $f_{poly3} = a_1 T^3 + a_2 T^2 + a_3 T^1 + a_4$, with the coefficients given in Tab. 2.8. The linear approximation, shown with dashed line, is given in a form $f_{lin} = 3.4968 + 2.24\text{e-}4(T - 293.15)$. It is seen from Fig. 2.36 that the strongest discrepancy is at point T = 300.65 K, leads to effective index difference of 8e-5.

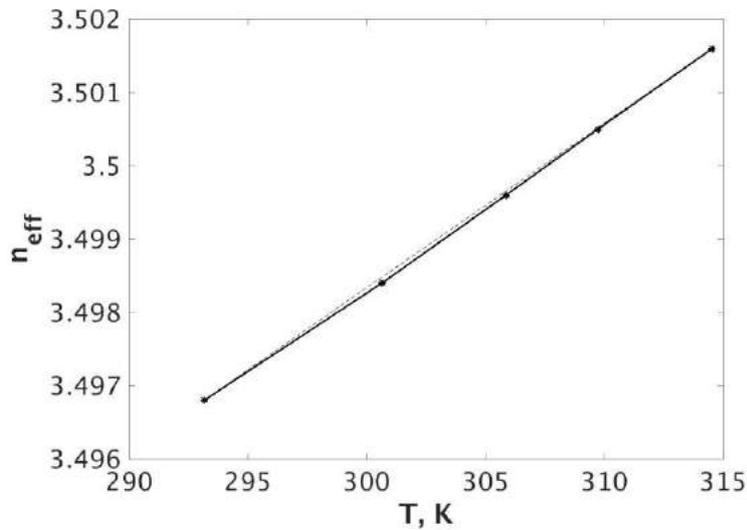

Fig. 2.36. Effective index of SiGe waveguide at 5.7 μm vs. temperature. The solid line shows third order polynomial approximation; dashed line refers to linear approximation.

The spectral shift could be related to difference in rate of thermal expansion of silicon and germanium. This contributes to variation of refractive index and with high probability will cause the certain level of crosstalk degration.

The results of theoretical analysis presented in this chapter will be compared to experimental data in the part "4. Characterization".

|  | $n_{eff}$(SiGe) |
|---|---|
| 293.15K (20 °C) | 3.4968 |
| 300.65K (27.5 °C) | 3.4984 |
| 305.85K (32.7 °C) | 3.4996 |
| 309.75K (36.6 °C) | 3.5005 |
| 314.55K (41.4 °C) | 3.5016 |

Tab. 2.7. The effective index of SiGe waveguide with 40% germanium at the core simulated using TM semi-vectorial Beamprop tool in Rsoft.

| coefficient | value |
|---|---|
| $a_1$ | -5.73e-8 |
| $a_2$ | 5.28e-5 |
| $a_3$ | -1.60e-2 |
| $a_4$ | 5.09 |

Tab. 2.8. The effective index of SiGe waveguide with 40% germanium at the core simulated using TM semi-vectorial Beamprop tool in Rsoft.

## 2.4 PCG operation

As discussed in the previous chapter, the AWG configuration redistributes by wavelength and focuses the input beam by means of array waveguides with constant length difference that introduce the phase delay. As a consequence, the wavelength dependent phase delay depends on the waveguide core thickness and width, and the particular level of sidewall roughness that is unavoidably present in real devices leads to additional source of phase noise. In contrast, in the PCG, the phase delay and the focusing are performed by the slab waveguide and the reflective concave grating. Since the phase delay is sensitive to waveguide core thickness only, the PCG configuration is expected to be more stable to phase noise.

Fig. 2.37 illustrates the schema of PCG configuration. It is built of a reflective grating, input/output waveguides and a slab waveguide in-between.

As in case of AWG, it allows us to direct several input channels having certain variation in wavelength into a single output waveguide. By the symmetry of optical systems, this device can also operate in reverse. In the latter case, it will dispersively split an input beam into several output channels, i.e. demultiplex. Compared to AWG, PCG allows to achieve a smaller footprint for similar wavelength and waveguide geometry.

PCG configuration studied in this work is based on a Rowland mounting illustrated in Fig. 2.37. Its input and output waveguides are placed along the circle, so called Rowland circle, with radius twice smaller than the radius of curvature of the concave grating. The Rowland circle touches the grating curve at its center. Grating facets are positioned so that their projections onto a tangent passing through the grating center are equidistant.

Condition for constructive interference is given by grating equation of the form:

$$d(sin\theta_i + sin\theta_d) = m\frac{\lambda}{n_s}, \qquad (2.81)$$

where $\theta_i$ and $\theta_d$ – angles of incidence and diffraction, respectively, $d$ – grating period, $m$ – diffraction order, $\lambda$ – wavelength of the beam in free space, and $n_s$ – effective index of slab waveguide.

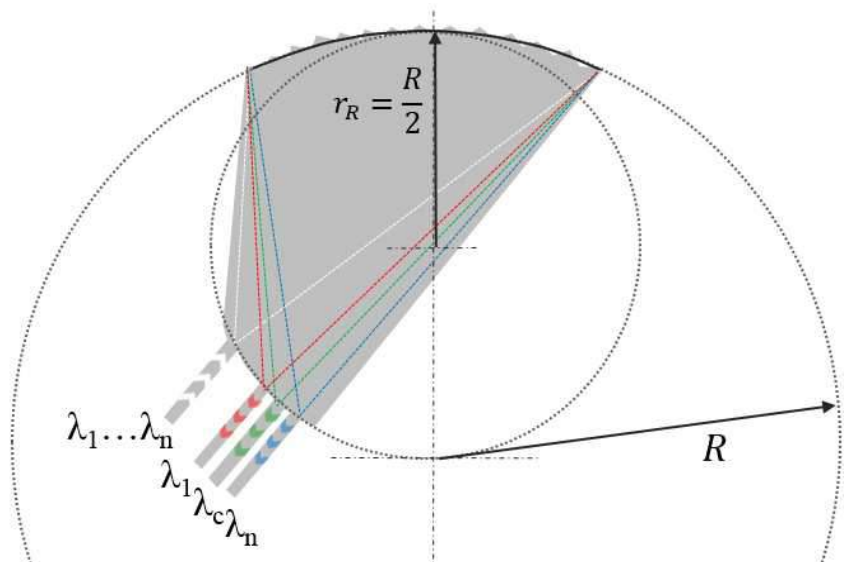

Fig. 2.37. Illustration of PCG operation.

Later, it will be shown, that position of grating facets can be tailored in order to decrease aberrations and improve spectral response of the device by means of 5th order correction of the beam path, and slight modifications of the configuration with one or two stigmatic points.

In case of demultiplexer, PCG functions as follows. The incoming beam confined in the input waveguide freely diverges in the slab due to lack of lateral confinement while propagating towards the concave grating. The grating facets, each acting as a mirror, separate the input beam into multiple components, having constant light path difference in between. The path length ΔL is chosen to be equal to multiple of central wavelength in the slab waveguide $\Delta L = m \frac{\lambda_c}{n_s}$. So each component of the beam at central wavelength acquires $2\pi m$ phase difference from its immediate neighbors that leads to their constructive interference in the slab region and focus at the center of the focal line. While other wavelengths, differing slightly from the central wavelength, will have slightly different phase shift and focus with a shift from the center along the focal line. So the input beam reflected from the grating diffracts and focuses at the corresponding output waveguide.

### 2.4.1 Basic parameters

There is a set of certain parameters that are handy in designing PCG: dispersion, FSR, diffraction order.

The input beam of central wavelength is focused at the central output channel. The other wavelengths will be focused at slightly shifted positions along the focal line. The position of the focused line as a function of wavelength can be calculated by dispersion parameter.

Design of PCG is always made for one diffraction order. In order to have a good operation, it is crucial to have a large FSR. The principle here is to ensure that the neighbor diffraction orders do not overlap with the main spectra. Diffraction order can be directly derived from the FSR. This parameter defines path lengths difference of beam components prior to diffraction.

#### 2.4.1.1 Dispersion

Dispersion is a parameter that describes the angle of deviation of different wavelengths from the central optical axis of the focal line. In order to define the dispersion for PCG configuration, it is convenient to start with the condition of constructive interference given by grating equation (2.82). For the given angle of incidence, angular dispersion is derived from grating equation (2.81) by taking derivative with respect to wavelength.

$$\frac{d\theta_d}{d\lambda} = \frac{mn_g}{n_s^2 d\cos\theta_d}, \quad (2.82)$$

where $n_g = n_{ec} - \lambda_c \frac{dn_e}{d\lambda}$ is the group effective index in the slab waveguide. It can be seen from equation (2.82) that to obtain high dispersion either the grating facet $d$ should be small or the diffraction order should be of higher orders.

In Fig. 2.38 it is shown that linear dispersion on Rowland circle is $r_R \frac{\varphi_{r_R}}{\Delta\lambda_{P_1 P_c}}$, where $\Delta\lambda_{P_1 P_c} = \lambda_{P_c} - \lambda_{P_1}$ is a spectral spacing of channels $P_1$ and $P_c$. On the other hand, $\varphi_{r_R} =$ $2\varphi_R = 2(\theta_{dP_c} - \theta_{dP_1})$, where, $r_R$ – Rowland radius, $R$ – radius of grating curvature. As it can be seen,

$$\varphi_{r_R} r_R = \varphi_R R \qquad (2.83)$$

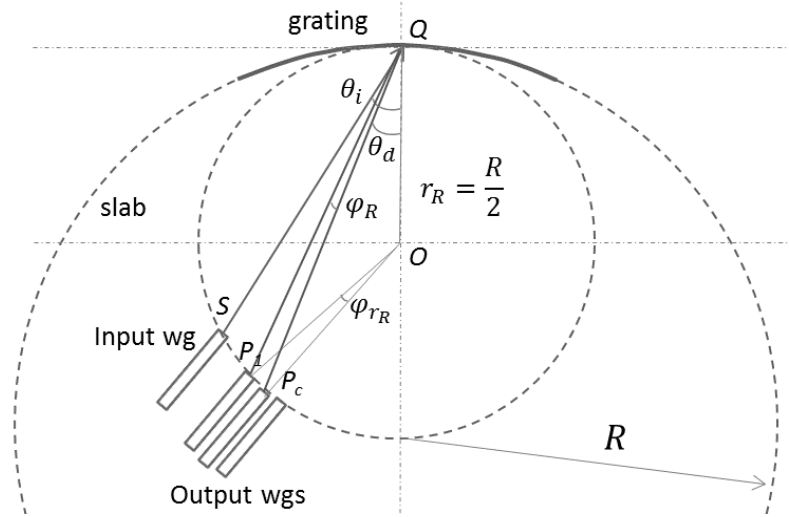

Fig. 2.38. Schema of PCG multiplexer.

Then we get

$$D = R\frac{\varphi_R}{\Delta\lambda_{P_1 P_c}} = \frac{Rmn_g}{n_s^2 d\cos\theta_d} \qquad (2.84)$$

**2.4.1.2 Free spectral range**

Free spectral range (FSR) is a wavelength or frequency shift between adjacent diffraction orders. It could be derived from definition:

$$d(\sin\theta_i + \sin\theta_d) = (m+1)\frac{\lambda}{n_s(\lambda)} \qquad (2.85\text{-}1)$$

$$d(\sin\theta_i + \sin\theta_d) = m\frac{(\lambda + \Delta\lambda_{FSR})}{n_s(\lambda + \Delta\lambda_{FSR})} \qquad (2.85\text{-}2)$$

$$(m+1)\frac{\lambda}{n_s(\lambda)} = m\frac{(\lambda + \Delta\lambda_{FSR})}{n_s(\lambda + \Delta\lambda_{FSR})} \qquad (2.85\text{-}3)$$

Then, using the first order Taylor approximation,

$$n_s(\lambda + \Delta\lambda) = n_s(\lambda) + \frac{dn_s(\lambda)}{d\lambda}\Delta\lambda_{FSR}, \qquad (2.86)$$

we can express the free spectral range from (2.85-3):

$$\Delta\lambda_{FSR} = \frac{\lambda}{m}\left[1 - \frac{(m+1)}{m}\left(1 - \frac{n_g}{n_s}\right)\right]^{-1} \qquad (2.87)$$

where $n_g$ is the group index, defined in 2.4.1.1.

### 2.4.1.3 Diffraction order

In the grating equation (2.81), the sum of sinuses does not exceed 2, so $\left|m\frac{\lambda}{dn_s}\right| < 2$. Explicitly, the allowed diffraction orders $m$ fall into the range $-2d < m\frac{\lambda}{n_s} < 2d$. For $\frac{\lambda}{n_s d} \ll 1$ a large number of diffraction orders will exist.

It can be seen from equation (2.87), that diffraction order is inversely proportional to free spectral range.

$$m = \left(\frac{\lambda_c}{\Delta\lambda_{FSR}} + 1\right)\frac{n_s}{n_g} - 1 \qquad (2.88)$$

From this condition we can deduce optimal diffraction order $m$ for the given design. If $\frac{n_s}{n_g} \approx 1$, the diffraction order simplifies to $\sim \frac{\lambda_c}{\Delta\lambda_{FSR}}$. In order to have smaller device with larger phase shift, higher diffraction order $m$ is preferable. So we choose the smallest FSR given that the spectra of nearest diffraction orders do not overlap.

## 2.4.2 Spectral characteristics

### 2.4.2.1 Insertion loss

Insertion loss $L_c$ in PCG depends on the following factors: the coupling loss between single mode waveguides and the slab, loss in other diffraction orders, reflection loss, grating imperfections such as roughness and corner rounding of the facets, and the diffraction loss that depends on the PCG design [64]. By careful design, we minimize the losses due to coupling from the single-mode to the slab and the diffraction loss. In order to minimize reflection loss, we introduce the metal coating on the facets. We also keep the angle difference $\Delta\theta = (\theta_i - \theta_d)$ small in order to have a maximum diffraction efficiency according to Littrow condition [85].

### 2.4.2.2 Non-uniformity

The non-uniformity loss $L_u$ is bound to the grating period $d$ parameter. The smaller the grating period result in reduction of the $L_u$ loss. This parameter, in its turn, depends on the choice of input $\theta_i$ / output $\theta_d$ angles and the diffraction order. The higher the diffraction order, the stronger is the difference in insertion loss between central and other wavelengths. The diffraction order for specific wavelength and technology is defined by FSR. However, the smaller grating period we have, the more they become sensitive to mask grid resolution. Up to the point, when the grid resolution affects the output, the FSR should be minimized.

### 2.4.2.3 Flat-top response

The reduction of inter-channel crossing $X_{ch}$ is preferred. This is accompanied by the flattening of the spectral response. As in case of AWG, the latter can be achieved by MMI couplers. As expected, the flattening of the response will unavoidable cause the reduction of insertion loss.

### 2.4.2.4 Crosstalk

The crosstalk $L_x$ could be understood as the portion of signal coupled from the neighbor channel. It might arise due to grating facets corner rounding that could cause the dispersion of light into wide range of angles, the portion of which might turn into stray light or undesired modes. The other parameter that affects the crosstalk is the distance between output channels $d_r$. It is optimized by two factors. First, it is chosen smaller for compact geometry of PCG, and large enough to avoid the large cross-talk between output channels. Another source of crosstalk is the effect of mask grid resolution. Since the positions of grating facets are defined only by two points related to the beginning and the end of the facets, the finiteness of the grid size might introduce a slight variation of the facet positions as well as their inclinations. This effect is studied in the chapter "2.4.7. Phase errors".

### 2.4.3 Analytical field calculation

Similarly to AWG, the field calculation of PCG is based on Gaussian approximation of the field and scalar diffraction theory. The field calculation is divided into four parts:
a. Divergence of the TM fundamental mode in the slab;
b. Reflection of the beam from the grating;
c. Diffraction of the field in the output slab;
d. Overlap between the diffracted field and the output waveguide.

Schematics of beam path is illustrated in Fig. 2.38. The SQ line corresponds to the input beam before being reflected from the grating, $SP_1$ and $SP_c$ – after.

The fundamental mode is approximated by power normalized Gaussian. We assume that the input beam propagates in the z direction and its cross-section is defined in (x, y)-plane. Field components are defined in vertical (y) and horizontal (x) directions, respectively.

Transmission coefficient, i.e. portion of transmitted intensity, at the interface of waveguide $A$ and waveguide $B$ is calculated as an overlap of two corresponding normalized fields:

$$\eta(x,y) = \frac{\left|\iint_{-\infty}^{+\infty} dxdy \psi_A \psi_B^*\right|^2}{\iint_{-\infty}^{+\infty} |\psi_A(x)|^2 dxdy \iint_{-\infty}^{+\infty} |\psi_B(x)|^2 dxdy} \qquad (2.89)$$

It is worthwhile noting that vertical dimension of the device does not change. So when the beam diverges in horizontal direction in the slab region, vertical component of fundamental mode changes only a little. It allows us to reduce three-dimensional problem to two-dimensional.

The vertical component of the output field is expressed in equation (2.90). The transmission coefficient $\sqrt{\eta(y)}$, defined by the equation (2.91), is counted twice, since the beam couples from the input waveguide to slab and from slab to output waveguide:

$$\Psi_1(y) = \left(\sqrt{\eta_y}\right)^2 Gauss(y, w_{ey}) \qquad (2.90)$$

$$\eta(y) = \frac{2w_{ey}w_{sy}}{w_{ey}^2 + w_{sy}^2} \qquad (2.91)$$

where $w_{ey}, w_{sy}$ – beam waists of vertical component of the fundamental mode in the input/output waveguide (same) and slab waveguide, respectively.

### 2.4.3.1 Divergence of the beam in the input slab

In horizontal direction, the wave $\Psi_1(x)$ is confined and guided by single mode input waveguide as given by equation (2.92). In the slab, the confinement occurs in y direction, whereas in x direction it uniformly diverges due to lack of lateral confinement. The wave in the slab $\Psi_2(x)$ at distance $z_2$ from the input waveguide can be calculated, similar to AWG, by Fraunhofer diffraction integral [79] and written as shown in equation (2.93), see Fig. 2.39.

$$\Psi_1(x_1) = \sqrt[4]{\frac{2}{\pi w_{ex}^2}} \exp\left(-\frac{x_1^2}{w_{ex}^2}\right), \quad (2.92)$$

$$\Psi_2(x_2) = \sqrt[4]{\frac{2}{\pi w_{sx}^2}} \exp\left(-\frac{x_2^2}{w_{sx}^2}\right), \quad (2.93)$$

where $w_{ex}$ is the beam waist of the horizontal component of the fundamental mode in the input/output waveguide, $w_{sx}$ is the waist of the beam in the slab, that is given by $w_{sx} = \sqrt{\frac{4z_2^2+\beta^2 w_{ex}^4}{\beta^2 w_{ex}^2}} = w_{ex}\sqrt{\left(\frac{2z_2}{\beta w_{ex}^2}\right)^2 + 1}$, $\beta$ is the propagation constant and the component $\frac{\beta w_{ex}^2}{2}$ known as Rayleigh length. The phase front has a radial shape given by $\exp(-\frac{j\beta x_2^2}{2z})$; however, we assume that it does not contribute to phase delay. The amplitude is still Gaussian but with broad beam waist.

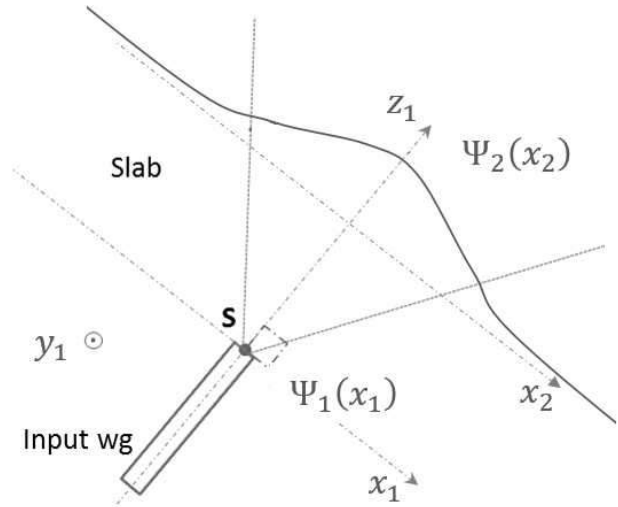

Fig. 2.39. Schema of input channel orientation at the entrance to the slab.

### 2.4.3.2 Reflection of the beam from the grating

The field of the beam reflected from the grating $\Psi_3(x_3)$ can be expressed as a sum of fields reflected from each grating facet. Since the input beam is located on a large distance from the grating compared to grating facet size, the portion of the field encountering the facet is considered to have roughly a plane phase front. However, we have $m\frac{\lambda}{n_s}$ path difference for beams reflected by neighbor facets, so the beams reflected from $M$-th facet will have a phase delay of $M\beta_s \Delta L$. The field $\Psi_3(x_3)$ is given by:

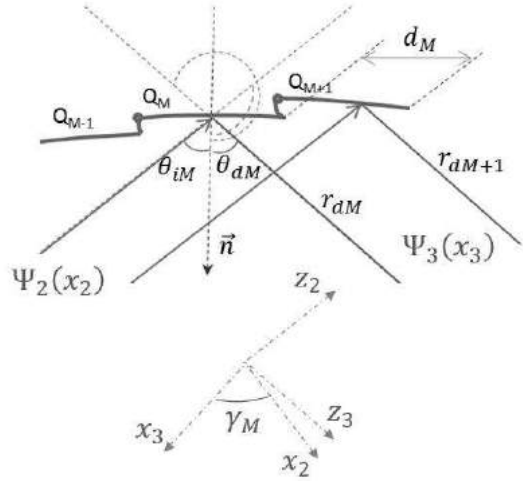

Fig. 2.40. Schema of angles of incidence $\theta_{iM}$ and reflection $\theta_{dM}$, width of the M-th facet $d_M$, and coordinate systems before and after the reflection.

$$\Psi_3(x_3) = \eta_r \sum_{M=-\frac{N-1}{2}}^{\frac{N-1}{2}} \eta(x) rect\left(\frac{x_3 - x_M}{d_M}\right)(\cos(\pi - \theta_{iM}) + \cos(\pi + \theta_{dM}))\exp(-jM\beta_s \Delta L) \quad (2.94)$$

where $\eta_r$ is the reflection coefficient of the grating, $N$ is the number of grating facets, $d_M$ is the width of $M$-th facet, $rect\left(\frac{x_3-x_M}{d_M}\right)$ is the rectangular function describing the beam reflected from $M$-th facet, $x_M$ is the coordinate of the center of $M$-th facet, $\theta_{iM}$ and $\theta_{dM}$ are the angles of input and reflected beams with respect to $M$-th facet normal, see Fig. 2.40.

The coupling coefficient $\eta(x)$ is expressed as follows:

$$\begin{aligned}
\eta(x) &= \sqrt[4]{\frac{2}{\pi w_{sx}^2}} \frac{1}{\sqrt{d_M}} \int_{-\infty}^{+\infty} \exp\left(-\frac{x^2}{w_{sx}^2}\right) rect\left(\frac{x-x_M}{d_M}\right) dx \\
&= \sqrt[4]{\frac{2}{\pi w_{sx}^2}} \frac{1}{\sqrt{d_M}} \int_{x_M - d_M/2}^{x_M + d_M/2} \exp\left(-\frac{x^2}{w_{sx}^2}\right) dx \\
&= \frac{\sqrt[4]{2\pi w_{sx}^2}}{2\sqrt{d_M}}\left[\text{erf}\left(\frac{x_M + \frac{d_M}{2}}{w_{sx}}\right) - \text{erf}\left(\frac{x_M - \frac{d_M}{2}}{w_{sx}}\right)\right]
\end{aligned} \quad (2.95)$$

where $\text{erf}(x) = \frac{2}{\sqrt{\pi}} \int_0^x \exp(-t^2)dt$ is an error function.

The angle between coordinate systems $\{x_2, z_2\}$ and $\{x_3, z_3\}$ is $\gamma_M = \theta_i + \theta_d$. It allows us to relate both sets of coordinates:

$$x_3 = x_2 \cos(\gamma_M) - z_2 \sin(\gamma_M) \quad (2.96)$$

$$z_3 = x_2 \sin(\gamma_M) + z_2 \cos(\gamma_M) \quad (2.97)$$

### 2.4.3.3 Diffraction of the field in the output slab

After being reflected, beam diffracts in the slab region, and the field $\Psi_4(x_4)$ at the entrance to the output waveguide can be calculated by Fourier optics, see Fig. 2.41.

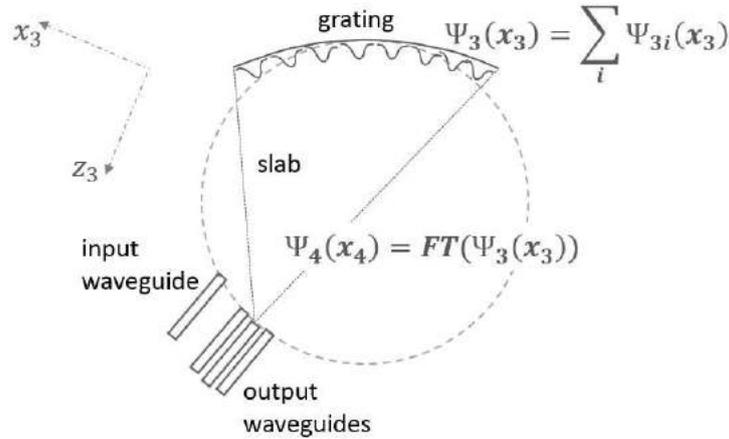

Fig. 2.41. Diffraction in the slab and focusing on one of the output waveguides.

$$\Psi_4(x_4) = FT[\Psi_3(x_3)]\left(f_x = \frac{x_4}{\alpha}\right) \quad (2.98)$$

where $f_x$ is Fourier Transform variable related to $x_4$ through coefficient $\alpha = \frac{\lambda r_{dM}}{n_{eff}}$, and $r_{dM}$ – distance from the $M$-th facet to the central output waveguide. Terms $x_M$, $\theta_{iM}$, $\theta_{dM}$ and $d_M$ are taken as constants with respect to $x_3$ coordinate.

Solving equation (2.98), we get exact expression of the field $\psi_4(x)$

$$\Psi_4(x_4) = \frac{\eta_r \eta(x)}{2} \frac{d_M}{\sqrt{2\pi}} \sum_{M=-\frac{N-1}{2}}^{\frac{N-1}{2}} (\cos\theta_{iM} + \cos\theta_{dM}) \exp\left(-j\left(M\beta_s \Delta L - 2\pi \frac{x_M x_4}{\alpha}\right)\right) sinc\left(\frac{d_M}{2}\frac{x_4}{\alpha}\right) \quad (2.99)$$

The field on the focal line is a sum of sinc-shaped contributions, each representing the far field profile of the individual grating facet.

### 2.4.3.4 Overlap between the diffracted field and the output waveguide

Finally, the output $\eta_{out}(x)$ of PCG is calculated as an overlap of $\Psi_4(x_4)$ and fundamental mode of the output waveguide $\Psi_5(x_5)$.

$$\Psi_5(x_5) = \sqrt[4]{\frac{2}{\pi w_{ex}^2}} \exp(-\frac{x_5^2}{w_{ex}^2}) \quad (2.100)$$

$$\eta_{out}(x) = \frac{\left|\int_{-\infty}^{+\infty} \Psi_4(x_4)\Psi_5(x_5)^* dx\right|^2}{\int_{-\infty}^{+\infty}|\Psi_4(x_4)|^2 dx \int_{-\infty}^{+\infty}|\Psi_5(x_5)|^2 dx} \quad (2.101)$$

The field in the output waveguide $\Psi_5$ in our case is similar to the mode of the input waveguide, as given in equation (2.100). The spectral response of PCG can be obtained by solving equation (2.101).

### 2.4.4 Geometry

In Rowland configuration, the input and output waveguides are placed along the circle with radius $r_R$ twice smaller than the radius of the concave grating $R$ as shown in Fig. 2.42.

#### 2.4.4.1 Rowland radius, positions of input and output waveguides

To start with, we will derive interrelation between geometrical parameters illustrated in Fig. 2.42. The origin of the coordinate system is located at the point $O$.

The positions of input and output waveguides are defined by their centers. The inset of Fig. 2.42 illustrates the cross-section of the input waveguide with intersection of diagonals at point $S$ that defines its position.

As shown in Fig. 2.42, the distance from position of input waveguide $S$ to position of grating center $Q$, is related to Rowland radius by the equation (2.102), that follows from the cosine theorem applied to the isosceles triangle $\Delta SQO$.

$$r_i = SQ = 2r_R cos\theta_i = Rcos\theta_i \qquad (2.102)$$

In similar way, we define the distances to output waveguides. The distance from the grating center $Q$ to the position of central output waveguide $P_c$ is:

$$r_d = QP_c = Rcos\theta_d \qquad (2.103)$$

The arc $SP_c$ connecting input and central output waveguides is given by equation (2.104), where we introduce a new notation for angle difference $\Delta\theta = (\theta_i - \theta_d)$, which should be kept small in order to have maximum diffraction efficiency according to Littrow condition [85].

$$SP_c = 2r_R(\theta_i - \theta_d) = R\Delta\theta \qquad (2.104)$$

Using grating equation (2.81) and equation of linear dispersion (2.84), we can express radius of grating curvature, i.e. twice Rowland circle:

$$R = 2r_R = \frac{Dn_{sc}\lambda_c}{n_g}\left[\frac{sin\theta_i + sin\theta_d}{cos\theta_d}\right]^{-1} \qquad (2.105)$$

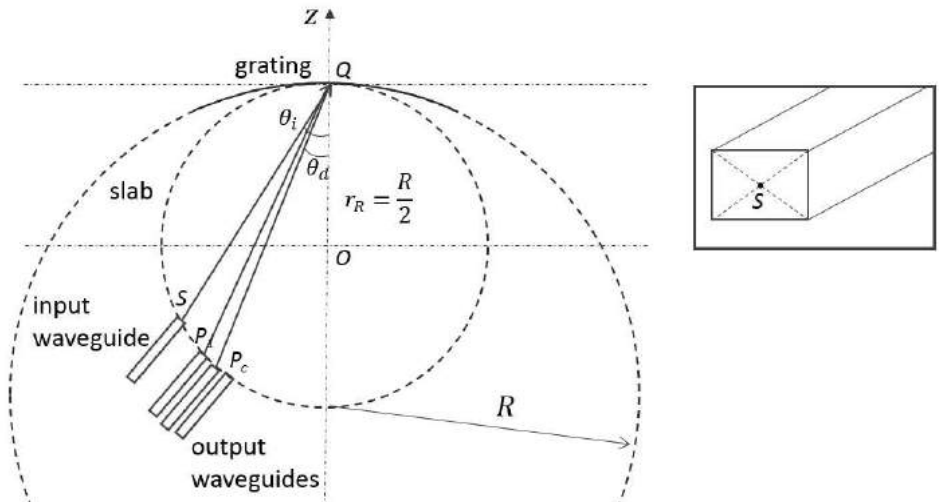

Fig. 2.42. Schema of PCG with Rowland configuration. In the inset: the point $S$ referred to as a position of the input waveguide.

From equations (2.104) and (2.105), we can express incident $\theta_i$ and diffraction $\theta_d$ angles.

$$\theta_d = arctg\left(\frac{(1 - Bsin\Delta\theta)}{B(1 + cos\Delta\theta)}\right) \tag{2.106}$$

$$\theta_i = \theta_d + \Delta\theta \tag{2.107}$$

Where we introduce the notation $B = \frac{n_g SP_c}{D n_{sc} \lambda_c \Delta\theta}$.

Once $SP_c$ and $\Delta\theta$ are chosen, angles $\theta_i$ and $\theta_d$, Rowland radius $r_R$ and grating curve $R$, distances $r_i$ and $r_d$, and grating period $d$ can be calculated.

The coordinates of input waveguide on the Rowland circle $\{x_{inp}, z_{inp}\}$, i.e. of point $S$ in Fig. 2.42, can be defined as a non-zero intersection of input beam direction $z_{in} = r_R + tg\left(\frac{\pi}{2} - \theta_i\right) \cdot x$ and the grating curve $z_{r_R} = \sqrt{r_R^2 - x^2}$.

$$x_{inp} = -2r_R \sin\theta_i \cos\theta_i \tag{2.108}$$

$$z_{inp} = tg\left(\frac{\pi}{2} - \theta_i\right) x_{inp} + r_R \tag{2.109}$$

The coordinates of the central output waveguide on the Rowland circle $\{x_{oc}, z_{oc}\}$, i.e. of point $P_c$ in Fig. 2.42, are defined in the similar manner with $z_d = r_R + tg\left(\frac{\pi}{2} - \theta_d\right) \cdot x$ diffracted beam direction.

$$x_{oc} = -2r_R \sin\theta_d \cos\theta_d \tag{2.110}$$

$$z_{oc} = tg\left(\frac{\pi}{2} - \theta_d\right) x_{dp} + r_R \tag{2.111}$$

After defining the positions of input and output waveguides, we specify the grating.

### 2.4.4.2 Position and length of the grating, number of facets

Length of the grating should be sufficient to reflect as much light as possible. We approximate the propagation of beam in the slab region by Gaussian function with a distance dependent width.

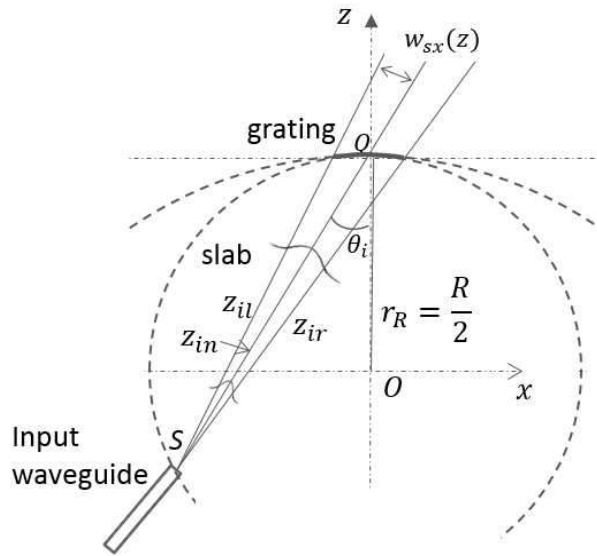

Fig. 2.43. Schema of input beam divergence.

$$w_{sx}(z) = w_{ex}\sqrt{\left(\frac{z}{z_R}\right)^2 + 1} \qquad (2.112)$$

where $z_R = \frac{\beta w_{ex}^2}{2}$ is the distance, known as Rayleigh length, over which the cross-section area of the beam doubles. For large distances ($z \gg z_R$), term $\left(\frac{z}{z_R}\right)^2$ is much greater than 1, and beam waist approaches linear dependences on the distance $z$. This approximation allows us to introduce two lines $z_{il}$ and $z_{ir}$ (Fig. 2.43) corresponding to both sides of the beam waist:

$$z_{il/ir}(x) = z_{in} \pm w_{sx}(z_{in}) \qquad (2.113)$$

$$N \geq \frac{L_g}{d} \qquad (2.114)$$

The intersection of $z_{il}(x)$ and $z_{ir}(x)$ lines with grating curvature allows us to define length of the grating $L_g$. Number of facets is directly determined by the length and period of the grating as given in equation (2.114).

### 2.4.4.3 Facets coordinates and inclination angles

A method of determining the facets inclinations is described in [86]. Inclination of facet faces can be defined by determining two points equidistant from the facet center, for which the beam paths are equal as shown in Fig. 2.44.

$$r_{iM1} + r_{dM1} = r_{iM2} + r_{dM2} \qquad (2.115)$$

where $r_{iM1}$ and $r_{iM2}$ – distances from the input waveguide to the point 1 and 2, respectively, $r_{dM1}$ and $r_{dM2}$ – distances from these points to the output waveguide. After finding the points 1 and 2 of each facet, for which the condition in equation (2.115) is satisfied, we place facet fronts along the lines passing through these points.

Then, angles of facets inclinations $\theta_{fM}$ can be found using coordinates of points 1 ($x_{M1}, z_{M1}$) and 2 ($x_{M2}, z_{M2}$) as follows:

$$\theta_{fM} = arctg\left(\frac{z_{M2} - z_{M1}}{x_{M2} - x_{M1}}\right) \qquad (2.116)$$

The angles of incidence and diffraction with respect to normal of each facet front are defined by equations (2.117) and (2.118).

$$\theta_{iM} = \theta_i - |\theta_{fM}|, \qquad (2.117)$$

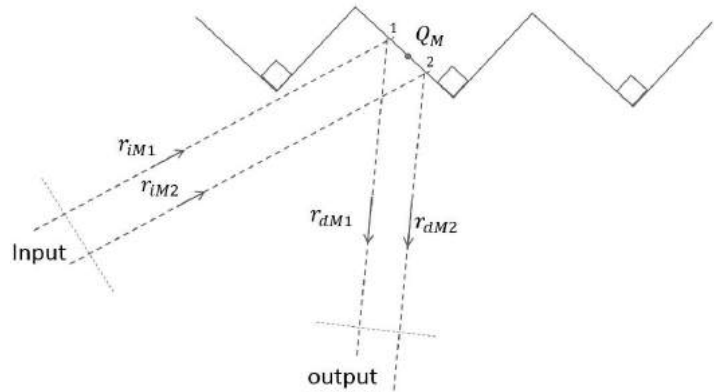

Fig. 2.44. Schema of two points on the facet front determining inclination.

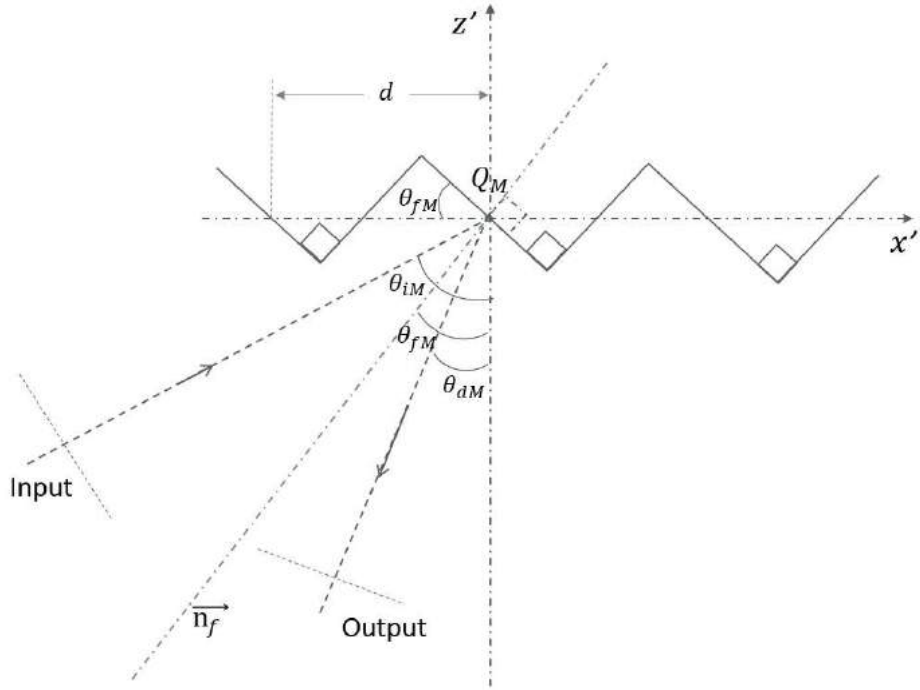

Fig. 2.45. Schema showing the facet inclinations.

$$\theta_{dM} = |\theta_{fM}| - \theta_{dN_{ch}}, \qquad (2.118)$$

where $\theta_{dN_{ch}}$ is the angular position of output channels with respect to main axis.

Points $Q_M$ are located at the centers of the facets of width $d$. Facet edges on its right $\{x_{Mr}, z_{Mr}\}$ and left $\{x_{Ml}, z_{Ml}\}$ sides are defined as follows:

$$x_{Mr} = x_M - \frac{d_M}{2}\cos\theta_{fM}, \qquad (2.119)$$

$$z_{Mr} = z_{M1} + \left(\frac{z_{M2} - z_{M1}}{x_{M2} - x_{M1}}\right)(x_{Mr} - x_{M1}), \qquad (2.120)$$

$$x_{Ml} = x_M + \frac{d_M}{2}\cos\theta_{fM}, \qquad (2.121)$$

$$z_{Ml} = z_{M2} - \left(\frac{z_{M2} - z_{M1}}{x_{M2} - x_{M1}}\right)(x_{Ml} - x_{M2}). \qquad (2.122)$$

Each facet face is chosen to make a right angle with its corresponding side as shown in Fig. 2.45 [86]. The side of each facet starts from its right edge $x_{Mr}$ and ends at the intersection with the neighbor facet face.

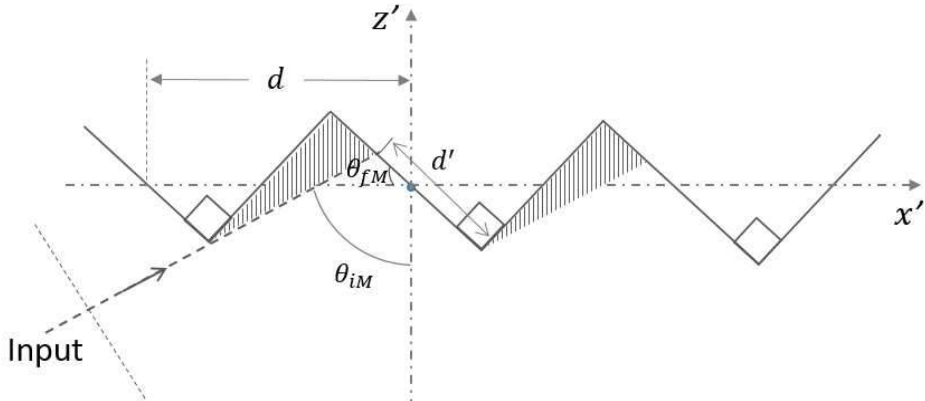

Fig. 2.46. Schema of "shadowing" of the facets effect.

$$d' = d\frac{\sin(\frac{\pi}{2} - \theta_i)}{\sin(\frac{\pi}{2} + \theta_i - \theta_{fM})} = d\frac{\cos\theta_i}{\cos(\theta_i - \theta_{fM})} \quad (2.123)$$

Since facets inclination angles vary from the angle of input beam, part of the facet face will be shadowed by the neighbor facet, see Fig. 2.46. Effective facet width can be found using law of sines, as given by equation (2.123).

**2.4.5 Simulation**

We present the simulation results obtained for PCGs operating at three different wavelengths with parameters detailed in Tab. 2.9. The design procedures of PCG for general case are detailed in Annex II. First, we start with the determination of waveguide width, effective index and waist of the TM fundamental mode. The optimal width of the waveguide that would support single-mode propagation for the whole operational range was chosen similarly to AWG case, using our multilayered structure discussed in chapter "2.1.4 Calculation of effective index of Si/Si$_{0.6+x}$Ge$_{0.4-x}$/Si waveguide in R-soft". It was also important to take into account the potential impact of fabrication that could alter the width of waveguide within tenth of micron. Knowing the waveguide width, we can get the effective index of the waveguide for the given wavelength. The waist of the fundamental mode is defined by Gaussian approximation.

With obtained parameters, we have performed numerical analysis of spectral response using semi-analytical tool developed for Matlab software. Fig. 2.47 shows the positions and normalized amplitudes of $\lambda_{min}$, $\lambda_c$ and $\lambda_{max}$ beams focused along the focal line calculated using equation (2.88) for three PCGs. The origin of spatial scale *x* is fixed at the central output channel. The non-uniformity appeared as follows. For 3.38 μm PCG, the normalized amplitudes of $\lambda_{min}$ and $\lambda_{max}$ wavelengths are 0.55 and 0.66. For 6.85 μm PCG, the normalized amplitudes of $\lambda_{min}$ and $\lambda_{max}$ wavelengths are equal, 0.94. For 8.76 μm PCG, the normalized amplitudes of $\lambda_{min}$ and $\lambda_{max}$ wavelengths are 0.87 and 91.

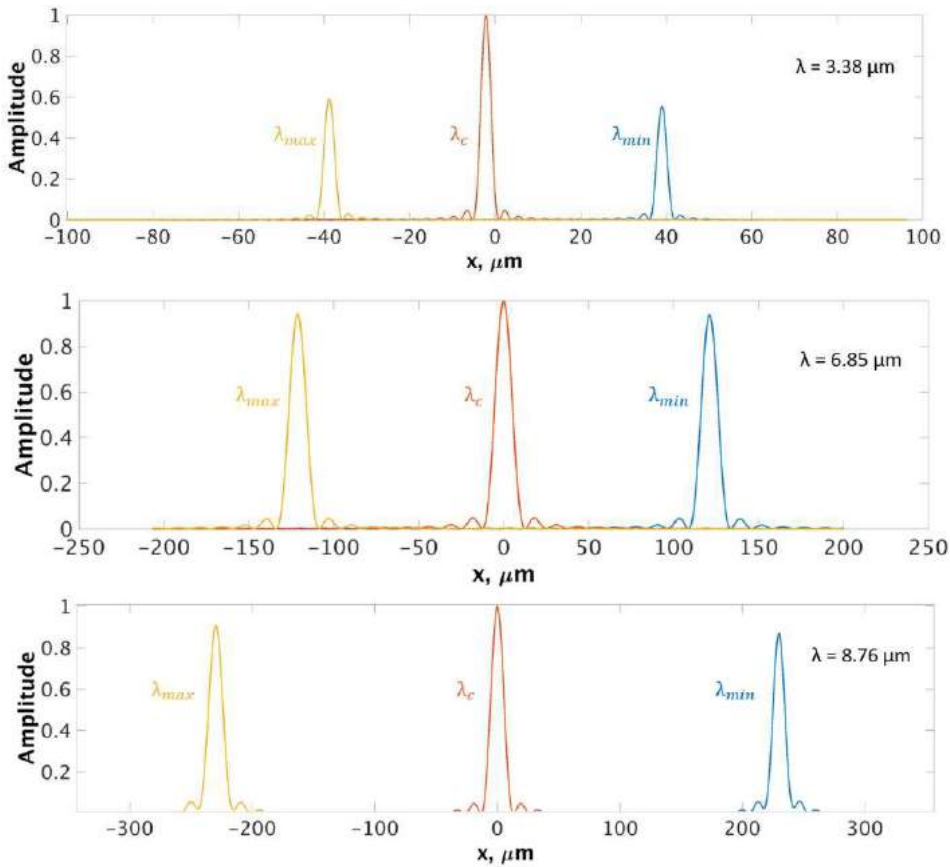

Fig. 2.47. Normalized amplitude and positions of $\lambda_{min}$, $\lambda_c$ and $\lambda_{max}$ on the focal line of PCGs.

Fig. 2.48 shows the transmission spectra of three PCGs. The simulation showed the following results. The PCG at 3.38 μm was designed for specific output wavelengths, so the channel spacing is irregular. The insertion loss is -4 dB, channels crosstalk is -50 dB and the non-uniformity is around 2 dB. In case of PCG at 6.85 μm, the insertion loss is -2.3 dB, channels crosstalk is -39 dB and the non-uniformity is around 0.3 dB. In case of PCG at 6.85 μm, the insertion loss is -2.3 dB, channels crosstalk is -39 dB and the non-uniformity is around 0.3dB.

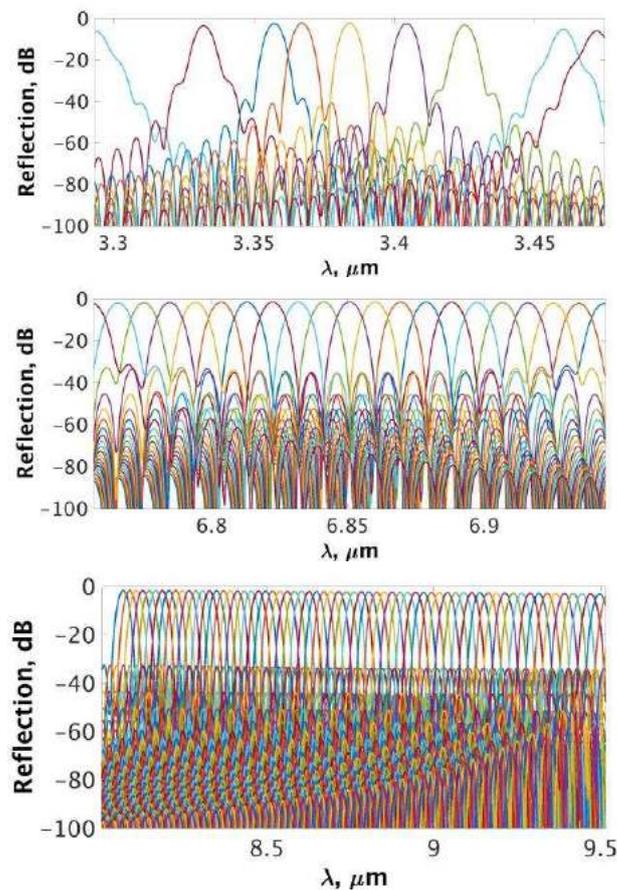

Fig. 2.48. Spectral response of PCGs at 3.38 μm, 6.85 μm and 8.76 μm.

| notation | 3.38 µm | 6.85 µm | 8.76 µm |
|---|---|---|---|
| $N_{ch}$ | 9 (irregular spacing) | 21 | 67 |
| $v_{min}$ - $v_{max}$ | 2878 – 3037 cm$^{-1}$ | 1440 – 1480 cm$^{-1}$ | 1051 – 1249 cm$^{-1}$ |
| technology | Si$_{0.6+x}$Ge$_{0.4-x}$/Si | Si$_{0.6+x}$Ge$_{0.4-x}$/Si | Ge/Si$_{0.6}$Ge$_{0.4}$ |
| $w_x, w_y$ | 2.4 µm | 6.8 µm | 2.8 µm |
| $n_s$ | 3.5146 | 3.4726 | 4.0493 |
| $n_g$ | 3.5947 | 3.4835 | 4.0522 |
| $w_e$ | 1.52 µm | 4.11 µm | 1.88 µm |
| $\Delta\lambda_{FSR}$ | 0.3836 µm | 1.1824 µm | 4.5936 µm |
| m | 8 | 5 | 1 |
| dr | 5.72 µm | 15.45 µm | 7.07 µm |
| D | 566.2184 | 1646.7 | 309.3022 |
| $\Delta\theta$ | 4° | 2° | 3° |
| $SP_c$ | 85.73 µm | 247.25 µm | 353.43 µm |
| $\theta_i / \theta_d$ | 40° / 36° | 39.8° / 37.8° | 12.8° / 9.8° |
| $r_R$ | 613.9722 µm | 3.5416e+03 µm | 3.3751e+03 µm |
| $r_i$ | 939.4697 µm | 5.4398e+03 µm | 6.5819e+03 µm |
| $r_d$ | 992.3393 µm | 5.5948e+03 µm | 6.6513e+03 µm |
| d | 6.2476 µm | 7.8355 µm | 5.4735 µm |
| N | 62 | 141 | 200 |

Tab. 2.9. Parameters of designed PCGs classified by the central wavelength 3.4 µm, 6.9 µm and 8.7 µm.

### 2.4.6 Optimization

#### 2.4.6.1 Optimizing facet positions

Although grating equation (2.81) defines the positions of input and output waveguides, it tells us nothing about the image quality. According to [57], the image quality depends on the precision of light path difference, which should introduce $2\pi$ phase difference. The light path difference for central and $M$-th facets is expressed by equation (2.124):

$$(r_{iM} + r_{dM}) = (r_i + r_d) + \Delta L \qquad (2.124)$$

where $M$ is the order of the grating facet, $r_{iM}$ is a beam path from the input waveguide to the $M$-th facet, $r_{dM}$ is a beam path from the $M$-th facet to the output waveguide, see Fig. 2.49. The points $Q_{Mc}$, $Q_M$ and $Q_{M+1}$ are located at $\{0, r_R\}$, $\{x_M, z_M\}$ and $\{x_{M+1}, z_{M+1}\}$,

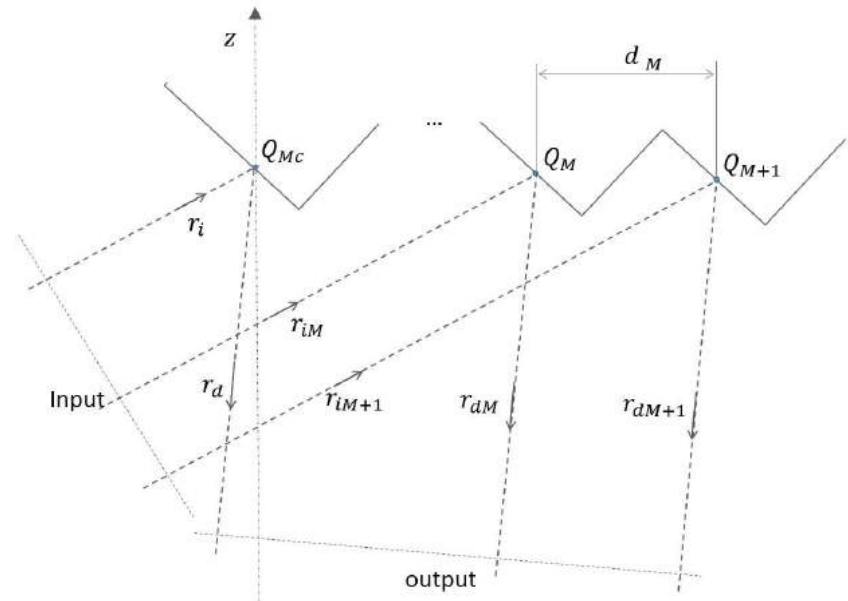

Fig. 2.49. Schema of distances notations.

respectively, and $\Delta L = F_1 + F_2 + F_3 + etc.$, where terms are related to either focal position or the image quality. For Rowland configuration, the expansion up to 5$^{th}$ order is defined as follows:

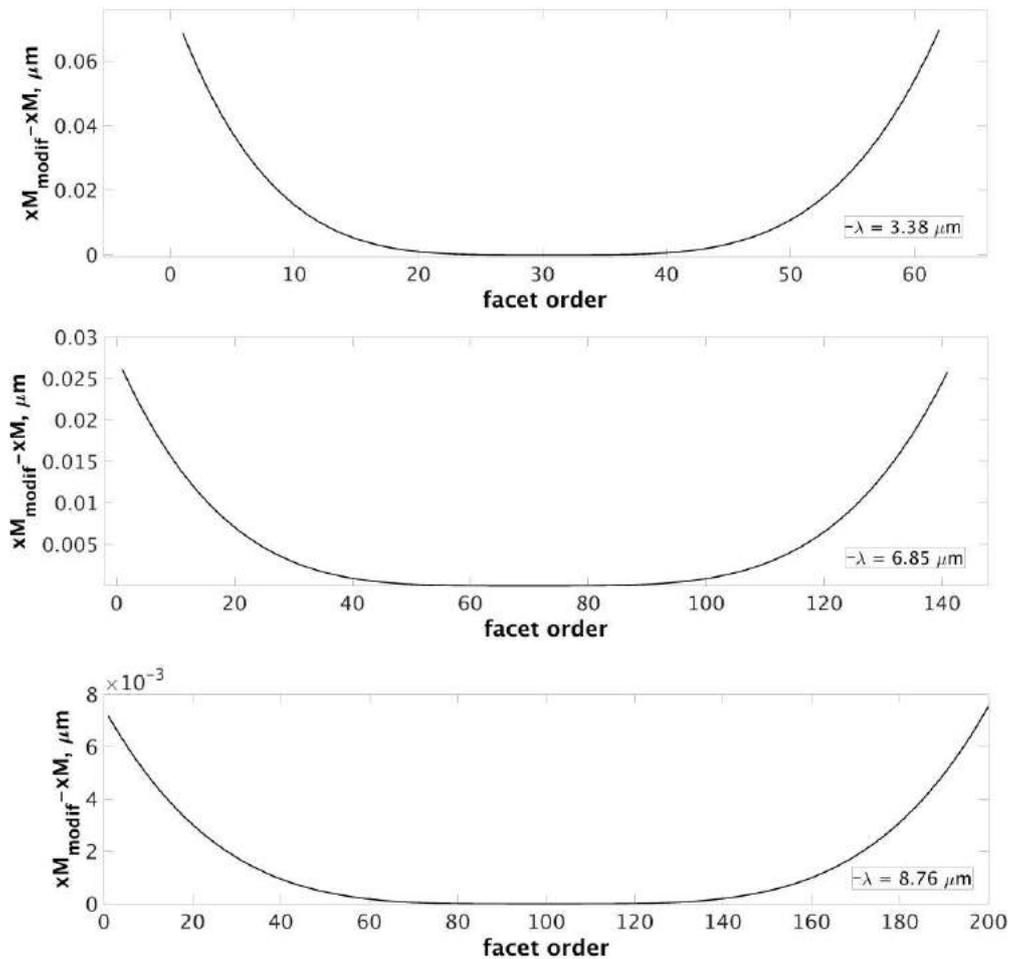

Fig. 2.50. Deviation of facet coordinates $xM$ after optimization for PCGs at 3.38 μm, 6.85 μm and 8.76 μm.

$$\Delta L = -d_M(\sin\theta_i + \sin\theta_d) + \frac{d_M^4}{8R^3}\left[\frac{\sin^2\theta_i}{\cos\theta_i} + \frac{\sin^2\theta_d}{\cos\theta_d}\right]$$
$$+ \frac{d_M^5}{8R^4}\left[\frac{\sin^3\theta_i}{\cos^2\theta_i} + \frac{\sin^3\theta_d}{\cos^2\theta_d}\right]$$
(2.125)

By equating the right side of (2.125) to the integer number of central wavelength in the waveguide, in our case $mM\frac{\lambda_c}{n_{sc}}$, we define the modified grating period $d_M$ that should correct aberrations up to fifth order. Then the facet coordinates change, $x_M$ coordinate is defined by recalculated grating period $d_M$, and the corresponding z-coordinates. The latter is defined by the condition that the facet centers are located along the grating curve with radius R. The positions were defined numerically.

Fig. 2.50 illustrates the change of $x_M$ coordinates for all the facets. It is clearly seen, that the facet positions do not change in the central region of the grating. However, the strongest deviations were 60 nm (3.38 μm PCG with $d$ = 6.25 μm), below 30 nm (6.85 μm PCG with $d$ = 7.84 μm) and below 10 nm (8.76 μm PCG with $d$ = 5.47 μm), too small compared to grating periods. The spectral responses with modified facet positions did not show any visible difference compared to initial results.

### 2.4.6.2 Analysis of facet shape

We have studied the concave facets depth and compared the facet heights $\Delta h$ to that of the straight facets, see Fig. 2.51 (a). The concave facets shown in Fig. 2.51 (b) were defined according to the following condition:

$$|(SQ_{Ml} + PQ_{Ml}) - (SQ_M + PQ_M)|$$
$$= |(SQ_{Mr} + PQ_{Mr}) - (SQ_M + PQ_M)|$$
(2.126)

The differences in height were within several of nanometers, see Tab. 2.10. Given the 5 nm resolution of the fabrication mask, the difference between straight and concave facets is negligible.

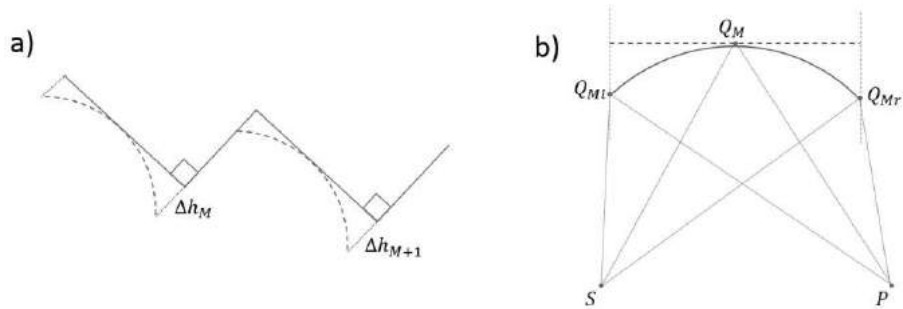

Fig. 2.51. Schema of facet height deviation of concave facets compared to straight.

| PCG λ | leftmost | central | rightmost |
| --- | --- | --- | --- |
| 3.38 μm | 4 nm | 3 nm | 3 nm |
| 6.85 μm | 1 nm | 1 nm | 1 nm |
| 8.76 μm | < 1 nm | < 1 nm | < 1 nm |

Tab. 2.10. The heights of concave facets compared to straight ones given for the outmost right, central and outmost left facets.

## 2.4.7 Phase error analysis

The phase errors deteriorate the spectral response of PCGs. It is especially strongly affects the crosstalk. The major contributions to phase errors arise from mask grid size, tilt and corner rounding of the grating facets and deviations in height of slab waveguide [64]. We have studied the effect of mask resolution with certain grid size. Since the facet positions are defined by its two side point coordinates, the slight alteration of these points leads to shift of the facet position as well as its inclination angle. We studied the effect of 1 nm, 5 nm and 10 nm mask grid size on crosstalk and compared the results of PCGs at three different wavelengths 3.38 μm, 6.85 μm, and 8.76 μm PCGs.

The lateral shift of facet centers contributes to phase errors according to:

$$\sigma_\varphi = 2k\sigma_x \tag{2.127}$$

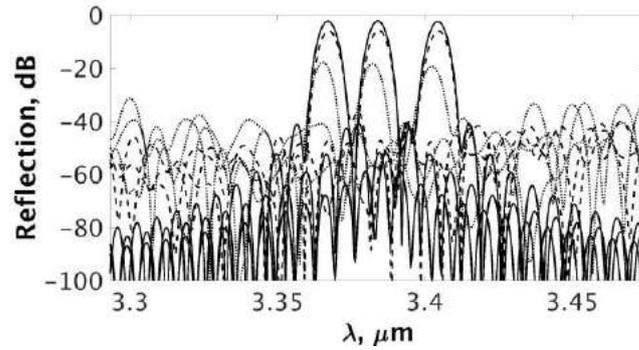

Fig. 2.52. Impact of phase errors in 3.38 μm PCG: solid line $\sigma_\varphi = 0$ rad, dashed line $\sigma_\varphi = 1.7\text{e-}3$ rad, dotted-line $\sigma_\varphi = 3.5\text{e-}3$ rad.

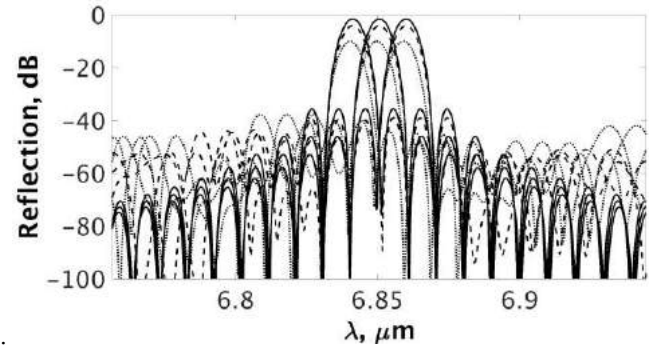

Fig. 2.53. Impact of phase errors in 6.85 μm PCG: solid line $\sigma_\varphi = 0$ rad, dashed line $\sigma_\varphi = 9\text{e-}3$ rad, dottedline $\sigma_\varphi = 17\text{e-}3$ rad.

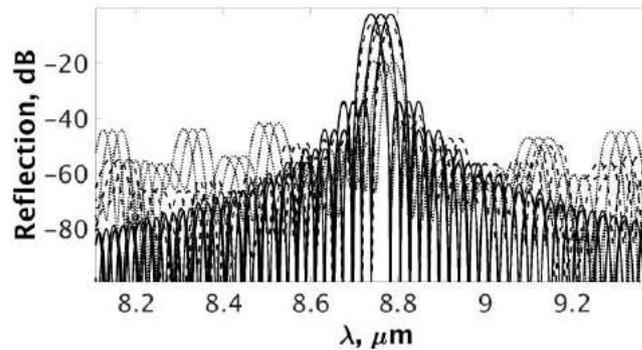

Fig. 2.54. Impact of phase errors in 8.76 μm PCG: solid line $\sigma_\varphi = 0$ rad, dashed line $\sigma_\varphi = 1.7\text{e-}3$ rad, dotted-line $\sigma_\varphi = 3.5\text{e-}3$ rad.

where $\sigma_\varphi$ is the normal distribution of phase errors, $k = \frac{2\pi}{\lambda} n_s$ is the wavenumber in the slab, $\sigma_x$ is the standard deviation of the horizontal coordinate of the facet. Maximum possible phase error caused by 5 nm mask grid is 5.7e-4 $rad$ at 3.38 µm, 2.8e-4 $rad$ at 6.85 µm and 2.5e-4 $rad$ at 8.76 µm. These phase errors are too small to be visible in simulation. Figs. 2.52 – 2.52 illustrate spectral responses of central channels at various levels of phase errors for three PCGs. The configuration at 6.85 µm made of SiGe graded index waveguides is more stable to phase errors compared to PCG at 3.34 µm based on the same waveguide technology, and PCG at 8.76 µm based on Ge/SiGe step-index waveguide technology. The main reason for such difference is in grating period $d$, the larger it is, the more stable is the configuration to phase errors.

## 2.5 Conclusion

In this chapter, we gave the theoretical background of wave guiding phenomenon together with mathematical methods for solving the mode problem.

We presented results of numerical calculation of the transmission of AWG and PCG multiplexers, that are based on analytical field calculation that was done based on Gaussian approximation of the field and Fourier optics. By specifying the desired spectral range and technology, one can follow the design procedures discussed in the chapter to build a multiplexer according to specifications.

We have discussed optimization options of the performance of the multiplexers. We studied the impact of tapering and MMI couplers in AWG as well as grating optimization in PCG. Our semi-analytical tool was used to design an AWG at 5.7 µm [87].

Our semi-analytical field calculation tool allows as to analyze the effect of phase errors on level of crosstalk. By comparing the simulation to experimental data, we can estimate the standard deviation of effective index appearing that could be caused by sidewall roughness.

The fabrication and characterization of multiplexers based on SiGe graded index waveguides will be discussed in the following chapter.

# 3 Fabrication

There are certain fabrication technology requirements for integrated optics components based on high-index contrast waveguides. The main advantage of using silicon as a substrate is in availability of fabrication processes established for electronic integrated circuits. This chapter covers the fabrication of SiGe graded index waveguides. The procedures such as surface smoothening with chemical-mechanical polishing, lithography and patterning of AWG and PCG devices are discussed. The reflecting coating application for the PCG is covered.

**Contents**

## 3.1 SiGe/Si graded index waveguide

The waveguides of the device are built with SiGe graded index core encapsulated in a thick Si cladding on standard silicon substrate. Fabrication is performed on a 200 mm microfabrication pilot line at CEA-LETI [26].

The geometries of the devices to be fabricated were defined in "gds" format files using home-made tool developed in C++ by Dr. P. Labeye for AWG, and home-made tool developed in Matlab during this work for PCG.

In order to obtain optimum performance of the device in the given wavelength range, the material and scattering losses have to be minimized. The use of the graded index structure avoids possible dislocations at the interfaces. Moreover, it requires selection of the proper waveguide width at the designing stage, as well as the fabrication of the perfectly straight waveguide walls, in order to avoid the shift of wavelength due to dispersion.

The widths of our single-mode waveguides were defined using numerical tool based on the multilayer structured waveguide model in R-soft, discussed in details in chapter "2.1.4 Calculation of effective index of Si/Si$_{0.6+x}$Ge$_{0.4-x}$/Si waveguide in R-soft". Since the waveguides were the part of the multiplexer that is typically designed to cover around $\Delta\lambda = 320$ nm spectral range, the choice of width followed from the condition that the waveguide should support single-mode propagations of both $\lambda_{min}$ and $\lambda_{max}$ wavelengths of operational range. Tab. 3.1 presents the results of optimal waveguide width calculations for different wavelengths. It should be noted that above 8 μm, the different technology is used. It is based on step index profile waveguide with Ge core and $Si_{0.6}Ge_{0.4}$ cladding.

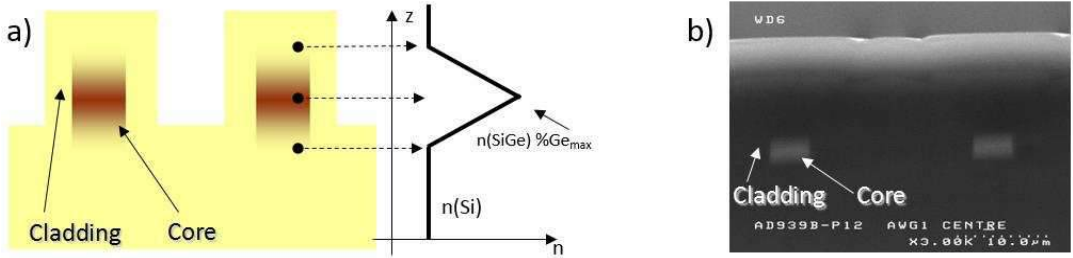

Fig. 3.1. SiGe/Si graded index waveguide: a) sketch of the waveguide stack; b) SEM image of SiGe/Si graded index waveguide designed for 5.7 μm wavelength.

| operational λ-range | technology | waveguide width |
|---|---|---|
| 3.29 – 3.47 μm | $Si_{0.6+x}Ge_{0.4-x}/Si$ | 2.4 μm |
| 5.56 – 5.88 μm | $Si_{0.6+x}Ge_{0.4-x}/Si$ | 4.6 μm |
| 6.76 – 6.94 μm | $Si_{0.6+x}Ge_{0.4-x}/Si$ | 6.8 μm |
| 8.01 – 9.51 μm | $Ge/Si_{0.6}Ge_{0.4}$ | 2.8 μm |

Tab. 3.1. Waveguide widths supporting single-mode propagation in the given range. All the waveguides have 3 μm height.

To reduce further scattering and distribution losses due to potential dislocations, the SiGe core guide composition varies linearly from 0 to 40%, and decrease back symmetrically. This graded index stack thus has a triangular profile of Ge concentration, see Fig. 3.1 (a). The SEM image of the SiGe/Si waveguide is shown in Fig. 3.1 (b). The graded index design thus to obtain the best fit of cladding without any gap or discontinuity. The use of SiGe and Si minimizes interface roughness while SiGe compounds offer low propagation loss and efficient coupling with mid-IR sources.

## 3.2 AWG fabrication process

The AWGs are based on Rowland configuration as discussed in chapter "2.3 AWG operation". Fig. 3.2 shows the chip with AWGs operating at central wavelength of 5.7 μm. The devices vary by three parameters, i.e. number of channels, waveguide width and width of MMI coupler added to output waveguides in order to flatten the response. Tab. 3.2 resumes the characteristic parameters. The idea behind is to test the effect of waveguide width widening by 0.2 μm from the calculated optimum value, and the impact of MMI coupler width.

According to simulation, the dimensions of waveguide width must be controlled to a tenth of a micrometer to achieve the specified performance, so far as the AWG design shows the tolerance to 100 nm waveguide sidewall variations [83].

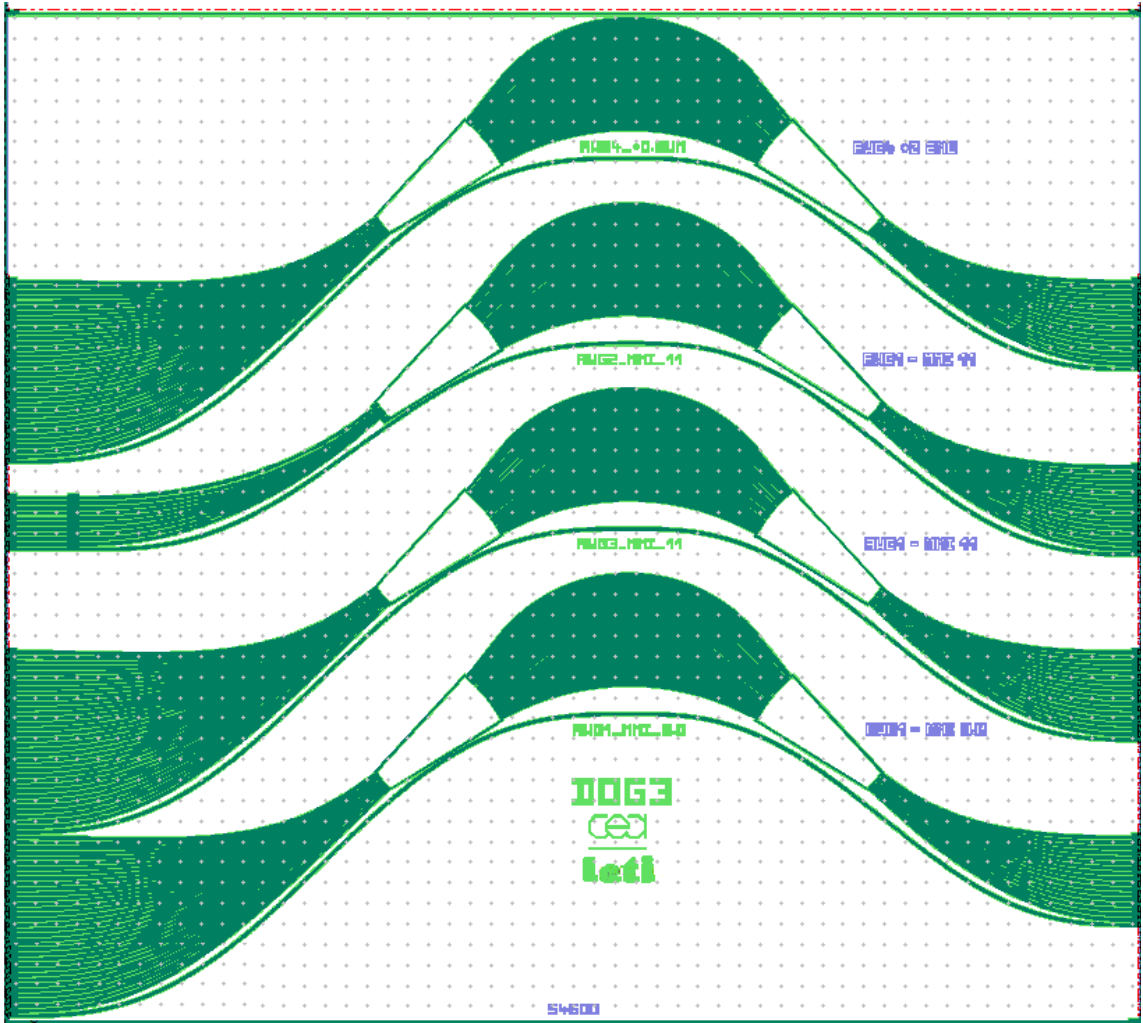

Fig. 3.2. AWGs chip at 5.7 μm on the chip (bottom-up): 17x35 AWG1 with 4.6 μm width of array waveguides, 9 μm MMI width; 17x35 AWG2 with 4.6 μm width of array waveguides, 11 μm MMI width; 10x35 AWG3 with 4.6 μm width of array waveguides, 11 μm MMI width; 17x35 AWG4 with 4.8 μm width of array waveguides, 9 μm MMI width.

| 5.65 μm AWG | N of channels | waveguide width | MMI width |
|---|---|---|---|
| AWG1 | 17x35 | 4.6 μm | 9 μm MMI width |
| AWG2 | 17x35 | 4.6 μm | 11 μm MMI width |
| AWG3 | 10x35 | 4.6 μm | 11 μm MMI width |
| AWG4 | 17x35 | 4.8 μm | 9 μm MMI width |

Tab. 3.2. The list of AWGs on the mask DOG3 of LETI.

The next subchapters detail the fabrication procedures of the device.

### 3.2.1 Epitaxial growth of SiGe graded index waveguide

An epitaxial growth of the SiGe structure in a RP-CVD tool from Applied Materials, at the pressure of 20 Torr, and with high grow rate thanks to a temperature up to 850°C, allows to obtain low-loss fully crystalline structures, Fig. 3.3 (a)-(b). On the top of the 3 µm SiGe layer, additional thick silicon layer is deposited prior to planarization of the surface discussed in the following part.

### 3.2.2 Chemical-mechanical polishing

The chemical-mechanical polishing/planarization (CMP) is a technique of efficient surface smoothening that is achieved by combining the advantages of both rough mechanical and fine chemical treatments. The use of CMP after the epitaxial growth is required in order to get a mirror surface without any non-uniformities.

### 3.2.3 Lithography of waveguides

Waveguides and AWGs are defined using lithography. The waveguide shaping according to the desired size and thickness requires the use of conventional deep UV stepper mask aligner. The minimum resolution on the mask is 1µm and therefore Iline 365nm or DUV 248nm equipment can be used. Adapted to this resolution, the mask grid is then set at 5 nm, which is enough for a good definition of the curves.

The wafers are cleaned after CMP and then the resist is applied on the surface of the wafers by spin coating, see Fig. 3.3 (c). The rate varies between 1000 and 5000 rpm and the spinning time depends on required thickness of the resist, which is normally up to 1 min. In our case, the height of the resist is 820nm. The pattern of AWGs is defined on the resist level by step and repeat technique with a 248 nm DUV equipment. The resist is further developed to reveal the patterns and as shown in Fig. 3.3 (d).

### 3.2.4 Deep Reactive Ion Etching

Due to the range of thickness (around 3µm) of the desired height of the SiGe waveguide, deep reactive ion etching (D-RIE) process is mandatory. The resist thickness is chosen in accordance with the materials etching rate ratio. As the DRIE process has a ratio around 4, (the rate of SiGe etching is around four times faster than that of resist) etching the waveguide of 3 µm required a resist thickness of more than 800nm, see Fig. 3.3 (e).

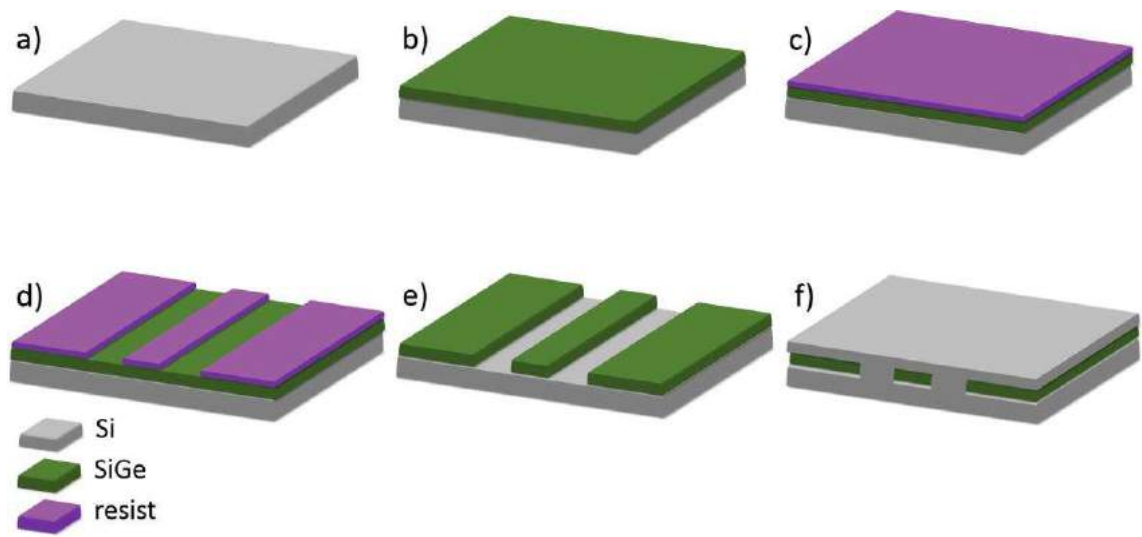

Fig. 3.3. Fabrication process steps.

Finally, a thick silicon layer is epitaxially grown to form the cladding of the waveguide, see Fig. 3.3 (f). By another DRIE etching, input and output facets are defined and covered with an anti-reflective coating of SiN. This step concludes the clean room fabrication. Each wafer is sticked on a dicing frame and diced into separate chips using a diamond saw.

## 3.3 PCG fabrication process

We have designed three PCGs operating at wavelengths 3.4 µm, 6.9 µm and 8.8 µm, as detailed in Tab.3.3. The design is based on Rowland configuration as discussed in chapter "2.4 PCG operation". The devices have been fabricated with the 248 DUV lithography as AWGs.

Figs. 3.4 – 3.6 illustrate the layout of designed PCGs. The PCGs at all three wavelengths were built with two different number of grating facets: one corresponds to the optimum number of facets with respect to design specifications, the other is taken 1.5 times greater for comparison.

| Devices | N of channels | Channels spacing | Technology |
| --- | --- | --- | --- |
| PCG at 3.4 µm | 9 | *irregular (min 8.82 cm$^{-1}$) | Si$_{0.6+x}$Ge$_{0.4-x}$/Si |
| PCG at 6.9 µm | 21 | 2 cm$^{-1}$ | Si$_{0.6+x}$Ge$_{0.4-x}$/Si |
| PCG at 8.7 µm | 67 | 3 cm$^{-1}$ | Ge/Si$_{60}$Ge$_{40}$ |

Tab. 3.3. The list of PCGs fabricated at CEA-Leti.

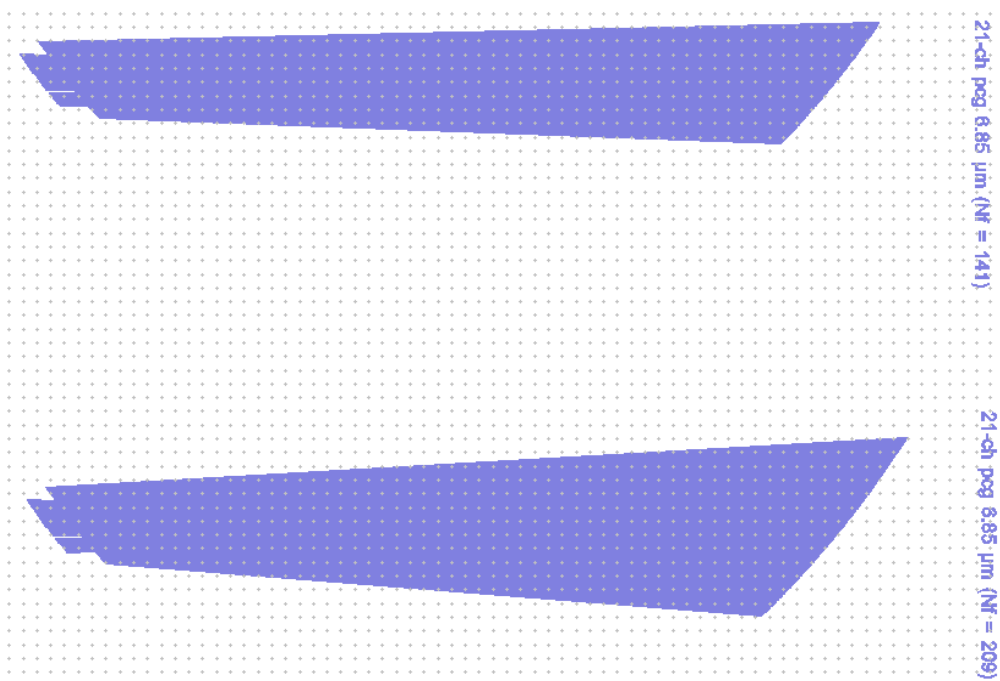

Fig. 3.4. PCGs at 6.85 µm (bottom-up): 11x21 PCG with 209 grating facets, 11x21 PCG with 141 grating facets.

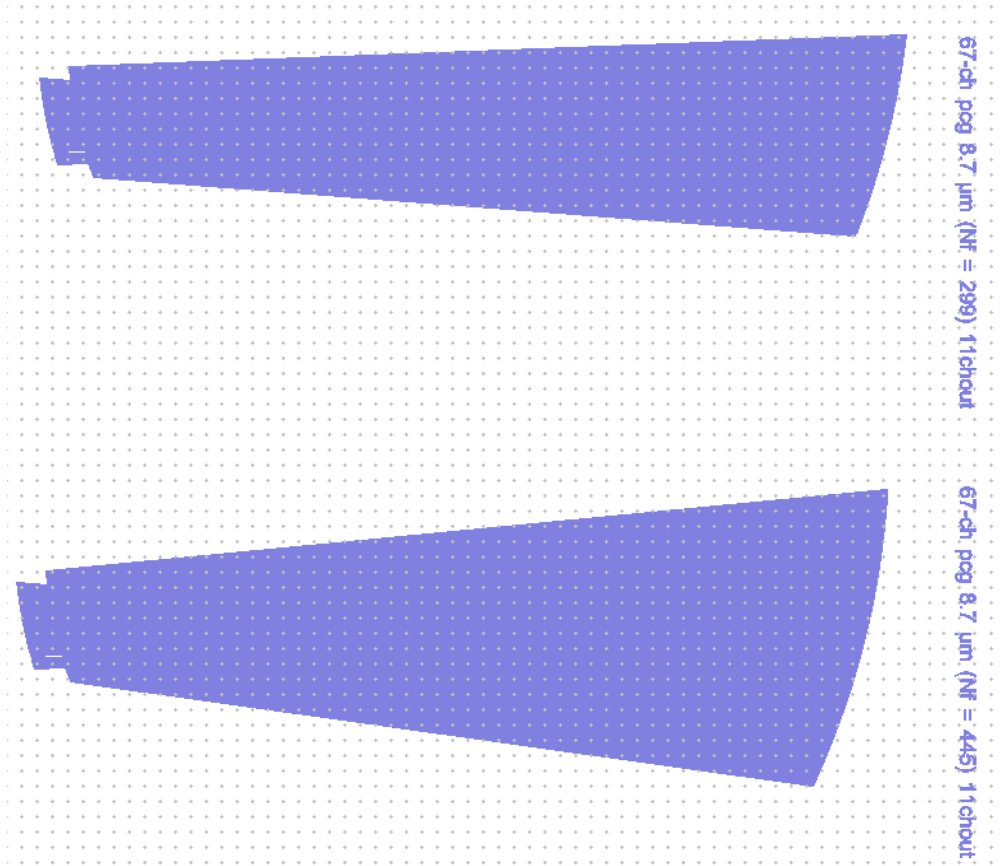

Fig. 3.5. PCGs at 8.76 µm (bottom-up): 11x67 PCG with 445 grating facets, 11x67 PCG with 299 grating facets.

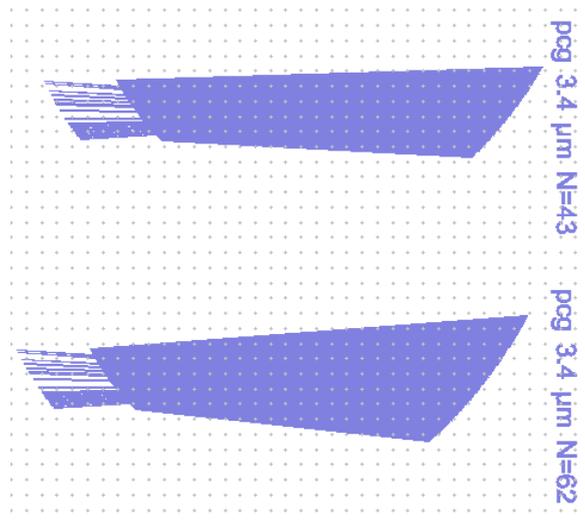

Fig. 3.6. PCGs at 3.34 µm (bottom-up): 9x9 PCG with 62 grating facets, 9x9 PCG with 43 grating facets.

### 3.3.1  Metalization of grating facets

In order to increase the grating reflectivity, the backside of standard flat grating facets are coated with a layer of highly reflecting metal (TiN/W or Ti/Au). The titanium is needed to insure a good adhesion, on the other hand its layer is kept a few nanometers since it has a strong absorbing characteristics which potentially reduces the reflectivity of the facets.

The deposition of metallic layer is carried out by Plasma Vapour Deposition (PVD). The process is undertaken in a vacuum under the low-pressure amount of sputtering gas. Then the voltage is applied to target metal placed parallel in front of the facet surface. The metal molecules are "knocked out" from it bulk piece by ionizing gas molecules and consequently cover uniformly the grating facet by forming a thin film.

## 3.4 Conclusion

In this chapter we have discussed the SiGe graded index technology, the procedures of waveguide device lithography as well as metallization of grating for reflection improvement. The fabrication procedure of AWG and PCG are similar up to the point of grating facet metallization in PCG. Since the AWG requires only one step lithography, it is more fabrication robust. It is clearly seen that the metallization procedure introduces additional fabrication imperfections at the interface of SiGe and metal layers such as non-verticality as well as additional absorption of light by metal.

# 4 Characterization

In this chapter, we present the characterization of AWG operating at 5.7 μm that was fabricated using SiGe graded index waveguides. We start with the description of measurement procedures. The set up and alignment for FTIR spectroscopy measurements are given in details.

The following parts cover analysis of the characterization results. Spectral characteristics such as insertion loss, non-uniformity, crosstalk and inter-channel crossing are measured. The flattening effect of AWG spectral response at varying widths of MMI couplers were compared. Phase errors for the given crosstalk levels were studied. The spectral shifts of AWG responses depending on temperature were measured by analyzing the results at five points between 20 °C and 41 °C, and the results were compared to simulation predictions.

Finally, the impact of non-central input to channel spacing variation is discussed. Experimental and theoretical results are compared.

**Contents**

## 4.1 Measurement methodology

In this chapter, we describe the set up and AWG chips under study, the technique used to measure complete response of AWG multiplexer with single laser is discussed, as well as the procedure of temperature measurements.

### 4.1.1 Experimental set up

The experimental set up was built according to schema shown in Fig. 4.1. The Fabry Perot Quantum Cascade Laser (FP-QCL) was collimated by a lens and directed with the mirror into a focusing lens in order to couple the light into Device Under Test (DUT) using a butt coupling method. Advantage of such coupling technic is that all the wavelengths of sources are effectively injected compared to a grating coupler where response is chromatic. The source was fabricated by the III-V Lab [4]. It can deliver an average power of 50 mW when driven by a pulse generator HP 8114 from Hewlett Packard with a pulse width of 1 μs at a repetition rate of 100 kHz. The specifications of the source are given in Tab. 4.1. The pulse is naturally chirped, which creates a continuous spectrum centered at $\sigma_0$ = 1790 cm$^{-1}$ with a Full Width at Half Maximum (FWHM) of 35 cm$^{-1}$. The lenses are IR1 390037 from LightPath. The light coming out of one output of the DUT is collimated and directed towards a Fourier Transform Infra-Red (FTIR) spectrometer from Thermo Scientific iS50 equipped with a Deuterated Tri Glycine Sulfate (DTGS) detector. A spectral resolution of 0.125 cm$^{-1}$ was used.

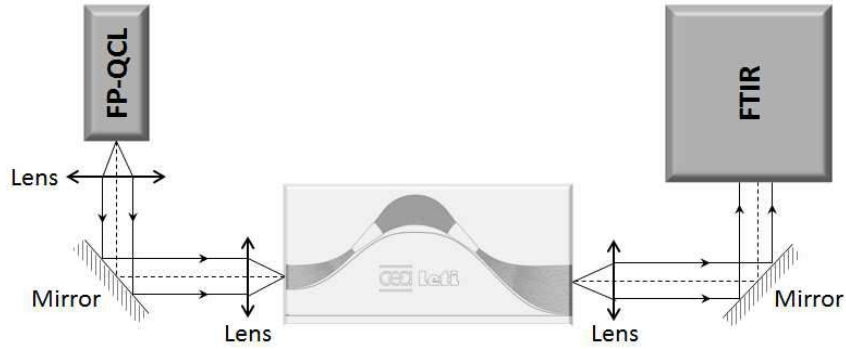

Fig. 4.1. Schema of the experimental set up.

| FP-QCL source | |
|---|---|
| output | 1750 cm$^{-1}$ – 1830 cm$^{-1}$ |
| impulse width | 1 µs |
| repetition frequency | 100 kHz |
| temperature | 20 °C |

Tab. 4.1. Source specifications.

### 4.1.2 Measurement procedure

The AWG described herein is designed to direct 35 QCLs, each having a 2-3 cm$^{-1}$ coverage, to a single detector. Thus, we build a broad band source with spectral range of 100 cm$^{-1}$. At the stage of testing the fabricated device, we measured the transmission in reverse, i.e. output to input direction, with the tunable light source. The bandwidth of our source with a Full Width at Half Maximum (FWHM) of 35 cm$^{-1}$ was not large enough to cover the 35-channel AWG with transmission range of 100 cm$^{-1}$. In order to compensate this shortage, we introduced 16 more output channels spaced by 6 cm$^{-1}$, 12 cm$^{-1}$, 18 cm$^{-1}$ and 24 cm$^{-1}$, 30 cm$^{-1}$, 36 cm$^{-1}$, 42 cm$^{-1}$ and 48 cm$^{-1}$ away on both sides of the central output, hence making it possible to obtain the transmission of all 35 channels. A schema of the AWG with indication of characterization direction is presented in Fig. 4.2.

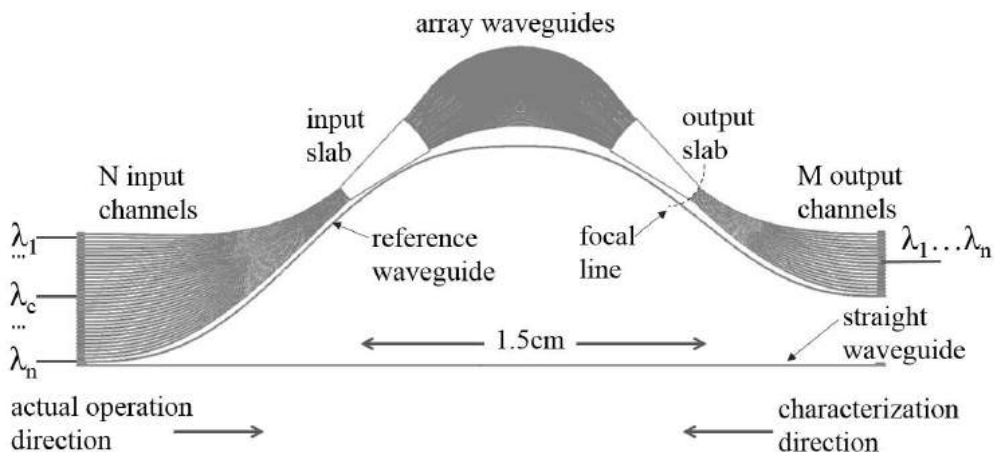

Fig. 4.2. Schema of AWG with indication of the directions of operation and characterization.

The light from a tunable laser is coupled into one of the AWG channels by thorough positioning of the lenses and mirror stages. In the same manner, the illumination transmitted from the multiplexer is directed into FTIR. The measurement of spectrum through the reference waveguide located directly below the AWG allowed us to evaluate absolute transmission of the channels. The straight waveguide was also introduced in order to obtain bend loss of reference waveguide. The parameters of characterized AWGs are given in Tab. 4.2. The AWG2 and AWG3 differ only by the number of channels (same waveguide width and MMI), so the spectral response of AWG2 was chosen to present both multiplexers. The measurements were done for chips from the center (chip 36) and the side (chip 32) of the wafer in order to evaluate the effect of waveguides height fluctuation, Fig. 4.3.

The goal of the multiplexer combined with an array of QCLs in broad-band source application is to achieve a whole coverage of the chosen spectral range uniformly. Inter-channel crossing $X_{ch}$ should be minimized as well as the non-uniformity $L_u$ and the insertion loss $L_c$. The requirements on the crosstalk $L_x$ level are more relaxed compared to telecommunication applications, and is acceptable being below -20dB.

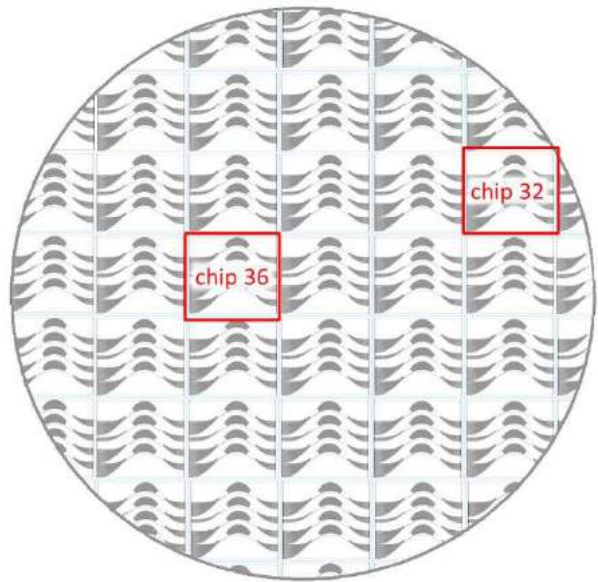

Fig. 4.3. Positions of the central chip 36 and side chip 32 on the wafer.

|  | $w_x$ | $w_{MMI}$ | $\Delta\sigma$ | $N_{ch}$ | $\sigma_{max} - \sigma_{min}$ | chip |
|---|---|---|---|---|---|---|
| AWG 1 | 4.6 µm | 9 µm |  | 17/35 |  |  |
| AWG 2 | 4.6 µm | 11 µm | 3 cm$^{-1}$ | 17/35 | (1700-1799) cm$^{-1}$ | 36 / 32 |
| AWG 3 | 4.6 µm | 11 µm |  | 10/35 |  |  |
| AWG 4 | 4.8 µm | 9 µm |  | 17/35 |  |  |

Tab. 4.2. AWG parameters: widths of waveguides $w_x$ and MMIs $w_{MMI}$, channel spacing $\Delta\sigma$, number of channels $N_{ch}$, spectral range, chip.

### 4.1.3 Temperature measurements

Temperature of the AWG was set by 350B Laser Diode Thermoelectric Temperature Controller. Before proceeding to measurements, preliminary calibration was made by verification of AWG temperature with thermometer Francaise D'Instrumentation FI 307. The temperature stabilized within 3 min, showing variation from the set value within ±1.5ºC.

The spectral responses of AWG2 and AWG4 were measured at five temperature points scattered between 20°C and 42°C. The aim was to compare the temperature dependence of spectral shift for multiplexers with two different waveguide widths (4.6 µm and 4.8 µm). The measurements were done for chips from the center and the side of the wafer to compare the responses.

## 4.2 Experimental measurements

The analysis of spectral responses of AWG1, AWG2 and AWG4 operating at 5.7 µm are presented. The measurements of two chips are compared. Basic parameters at central input were analyzed, the spectral coverage of all channels, FWHM variation depending on widths of waveguides and MMI couplers are discussed.

### 4.2.1 Spectral response of AWG

#### 4.2.1.1 AWG 1

Fig. 4.4 presents the spectral coverage of AWG1, which stretches between 1695 cm$^{-1}$ and 1805 cm$^{-1}$. The vertical axis shows shifts of measured wavenumber values from the target $\Delta\sigma = \sigma_{\text{measured}} - \sigma_{\text{target}}$; and the horizontal axis corresponds to 35 target output wavenumbers $\sigma_{\text{target}}$. The set of outputs that belong to a certain input position is marked with the same symbol. For example, "+" corresponds to the central input. Each data point is defined by peak position of the Gaussian fit, as shown in Fig. 4.5. The data points with $\Delta\sigma$ exceeding 50 cm$^{-1}$, correspond to neighbor diffraction order.

From Fig. 4.4, we note the following. The outputs cover the whole spectral range. The central input data points are shifted by -2.5 cm$^{-1}$ from the target. The measured wavenumbers corresponding to central and near central input channels, are parallel to the vertical axis, which implies the conformity to the designed channel spacing of 3 cm$^{-1}$. Though, as input position is further from the center, the line of output wavenumbers inclines. The impact of non-central input will be evaluated in the following chapter.

In our AWG design, we had extra 8 inputs on each side of the central one that were space by 6 cm$^{-1}$. This allowed to observe the neighbor diffraction order that is spaced by 134 cm$^{-1}$ according to measurement results. The data points that are shifted from target by more than 50 cm$^{-1}$ belong to the neighbor diffraction order.

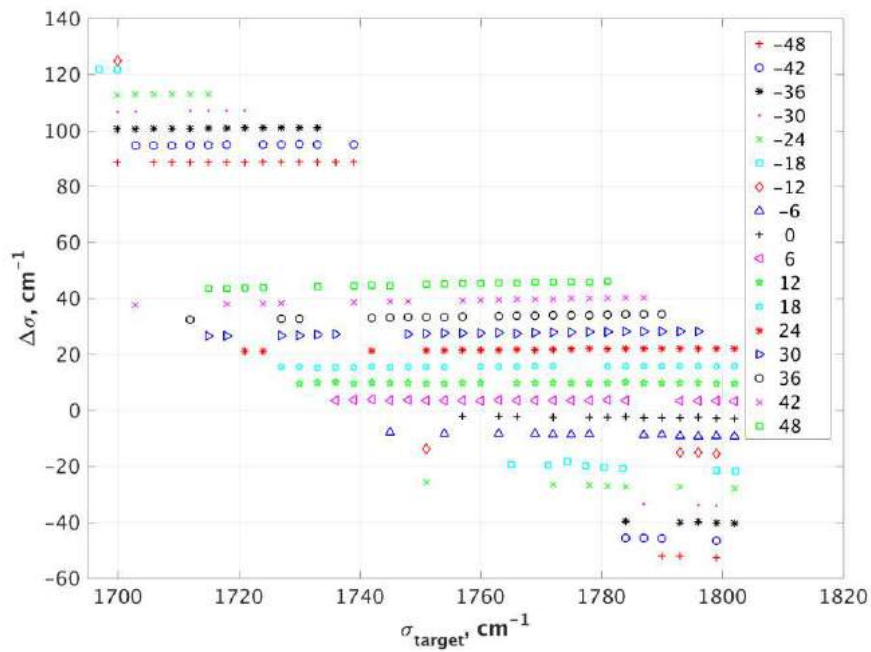

Fig. 4.4. Spectral coverage: target wavenumber vs. difference of measured wavenumber from target ones of AWG1, chip 36. The inset indicates the input positions.

The normalized transmission spectra of AWG1 are shown in Fig. 4.6. The insertion losses of –3.3dB and –4.8dB are measured for chips 36 and 32, respectively, whereas the theoretical expectation is –4.8dB. The crosstalk is below –20dB. Inter-channel crossing, which should preferentially be minimized in broad-band source applications, is around 4.3dB. Increasing crosstalk as well as non-uniformity of insertion loss towards the sides of the spectral response are explained by the rapid decrease of the signal to noise ratio due to narrow band of the QCL source (FWHM = 35 cm$^{-1}$).

Fig. 4.7 presents the transmission (top) and FWHM of Gaussian approximations (bottom) of measured outputs. The central insertion loss is scattered around -5dB with the slight tendency to increase towards the center and sides of spectral response. The average FWHM is 2.2 cm$^{-1}$. Comparison to theoretical expectations is given in "4.2.1.4 Discussion" chapter along with the results of AWG2 and AWG4.

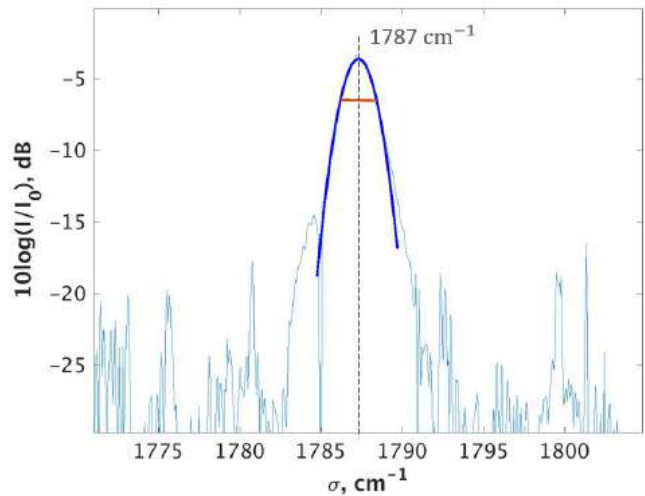

Fig. 4.5. Gaussian fit of output 31 of AWG1 (chip 36).

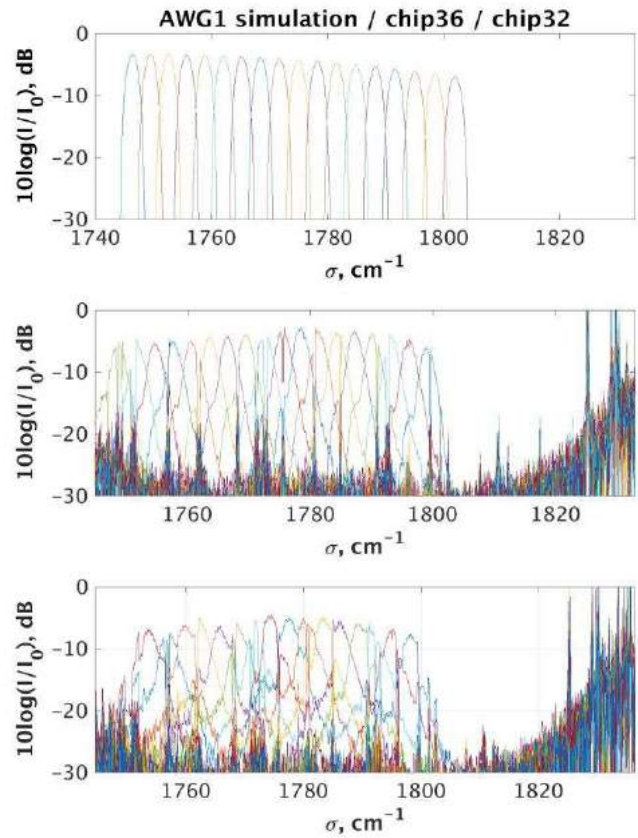

Fig. 4.6. Transmission spectra of outputs 18 to 35 at central input of AWG1 (from top to bottom): simulation, chip 36 and chip 32.

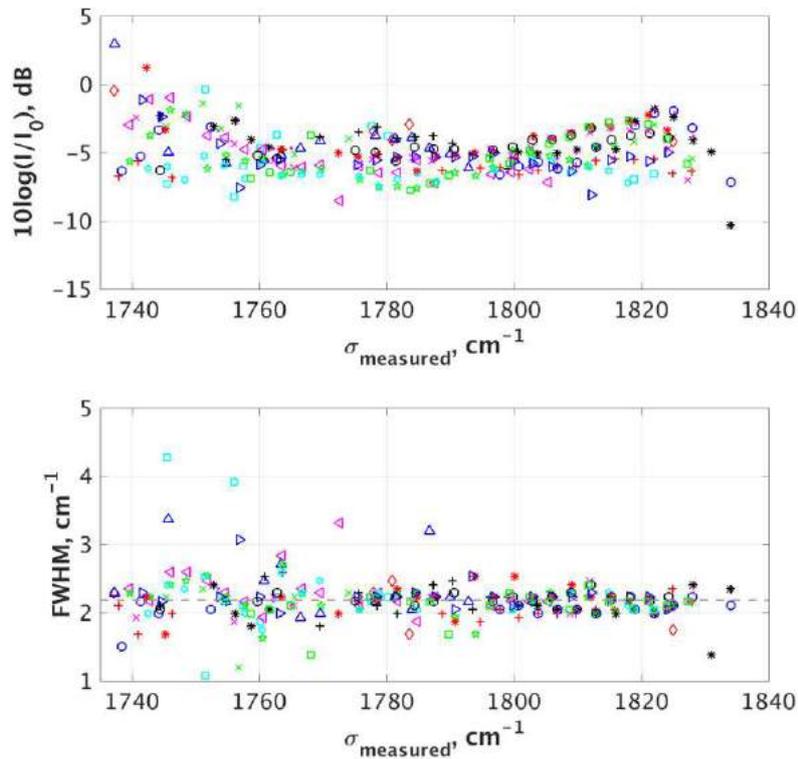

Fig. 4.7. Transmission and FWHM of AWG1 (chip 36).

**4.2.1.2 AWG 2**

Fig. 4.8 presents the spectral coverage of AWG2 in the same manner as it is described for AWG1 in chapter 4.2.1.1. The output wavenumbers cover the whole spectral range as well. The central input data points are shifted by -2.3 cm$^{-1}$ from the target. The measured output wavenumbers are nearly parallel to the vertical axis, which implies the conformity to the designed channel spacing of 3 cm$^{-1}$, the present variations will be analyzed in the following chapter.

The neighbor diffraction order of AWG2 was observed 132 cm$^{-1}$ away from the main order spectra. In Fig. 4.8, its data points are shifted from target by $\Delta\sigma > 50$ cm$^{-1}$.

The normalized transmission spectra of AWG2 shown in Fig. 4.9, evinces the insertion losses of –3.4 dB and –5.7dB for chips 36 and 32, respectively, whereas the theoretical expectation is –6.3dB. The crosstalk is below –20dB. The Inter-channel crossing is 4dB. As in case of AWG1, the increase in crosstalk towards the sides of the spectral response are explained by the rapid decrease of the signal to noise ratio.

Fig. 4.10 shows the transmission (top) and FWHM of Gaussian approximations (bottom) of measured outputs. The central insertion loss has a clear hump at the center of spectral response with average value of -2.3dB, and scatter of values -5dB to -7dB towards the sides. The average FWHM is 2.5 cm$^{-1}$.

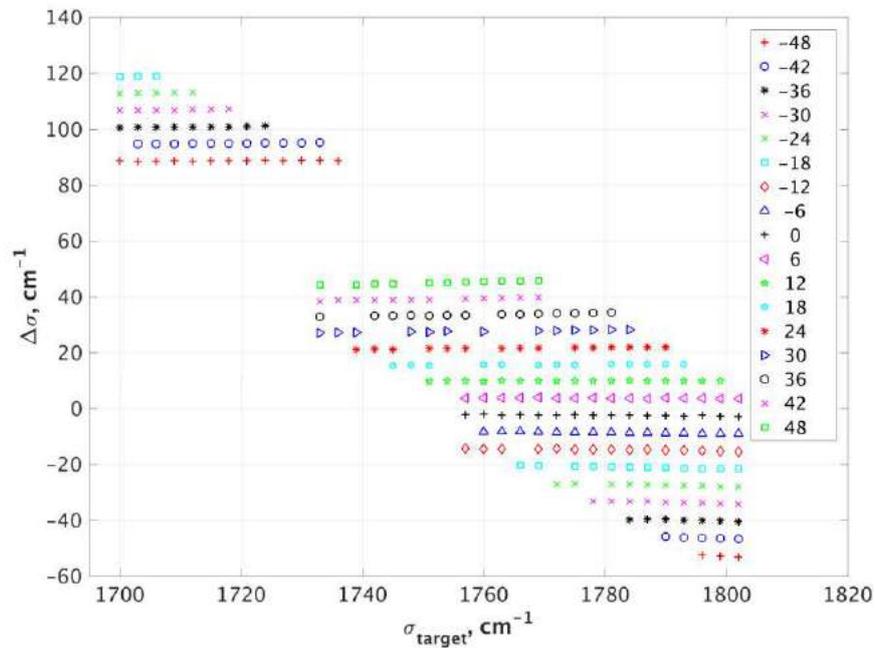

Fig. 4.8. Spectral coverage: target wavenumber vs. difference of measured wavenumber from target ones of AWG2, chip 36.

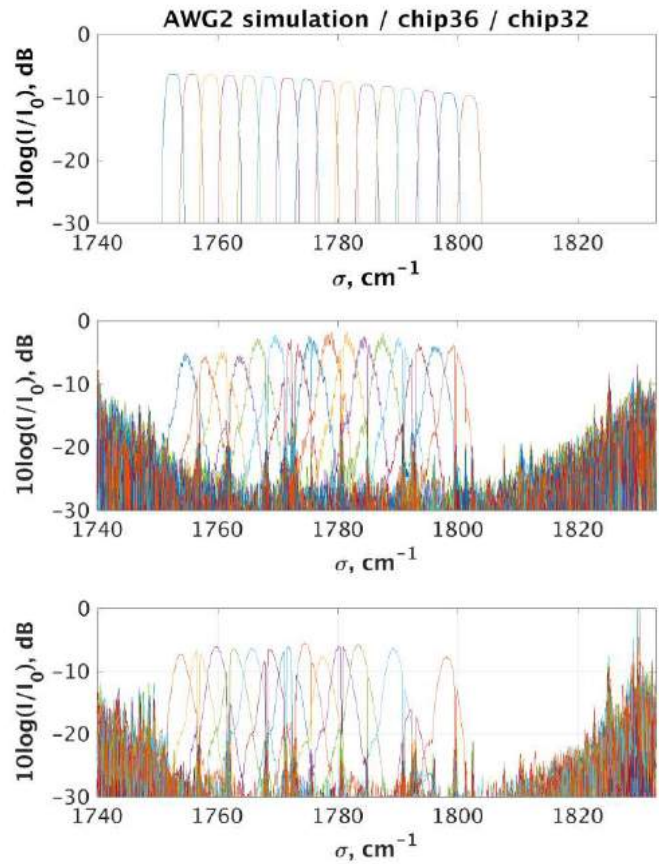

Fig. 4.9. Transmission spectra of outputs 20 to 35 at central input of AWG2 (from top to bottom): simulation, chip 36 and chip 32.

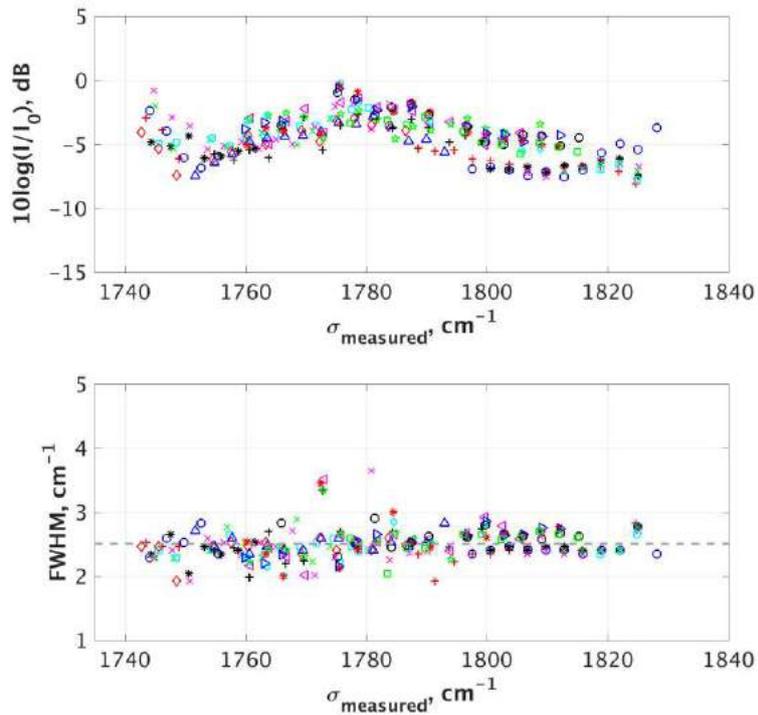

Fig. 4.10. Transmission and FWHM of AWG2, chip 36.

**4.2.1.3 AWG 4**

The spectral coverage of AWG4 is shown in Fig. 4.11. The measured output wavenumbers cover the whole spectral range as well. The central input data points are shifted by -3.4 cm$^{-1}$ from the target. The channel spacing complies with the designed value of 3 cm$^{-1}$ for central and near-central input positions. The variation of channel spacing for input positions further from the center will be analyzed in the following chapter.

As in case of AWG1 and AWG2, the neighbor diffraction order is observed with 132 cm$^{-1}$ shift from the main order spectra, which data points are shifted also by more than 50 cm$^{-1}$ in Fig. 4.11.

Fig. 4.12 presents the normalized transmission spectra of AWG4. The insertion losses are –6.2 dB and –4.2dB for chips 36 and 32, respectively, whereas the theoretical expectation is –5.2dB. The crosstalk is below –20dB. The Inter-channel crossing is 4dB. As in case of AWG1, the increase in crosstalk towards the sides of the spectral response are explained by the rapid decrease of the signal to noise ratio.

The transmission (top) and FWHM of Gaussian approximations (bottom) of measured outputs are shown in Fig. 4.13. The central insertion losses of all outputs were close and scattered around –5dB to -7dB, except an increase towards the sides of spectral response. The average FWHM is 2.3 cm$^{-1}$.

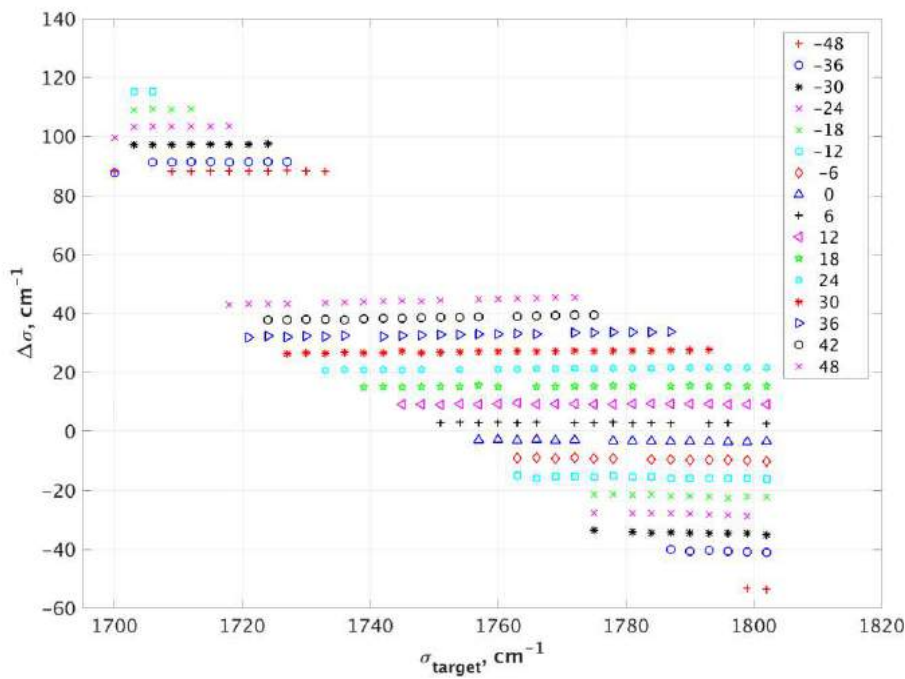

Fig. 4.11. Spectral coverage: target wavenumber vs. difference of measured wavenumber from target ones of AWG4, chip 36.

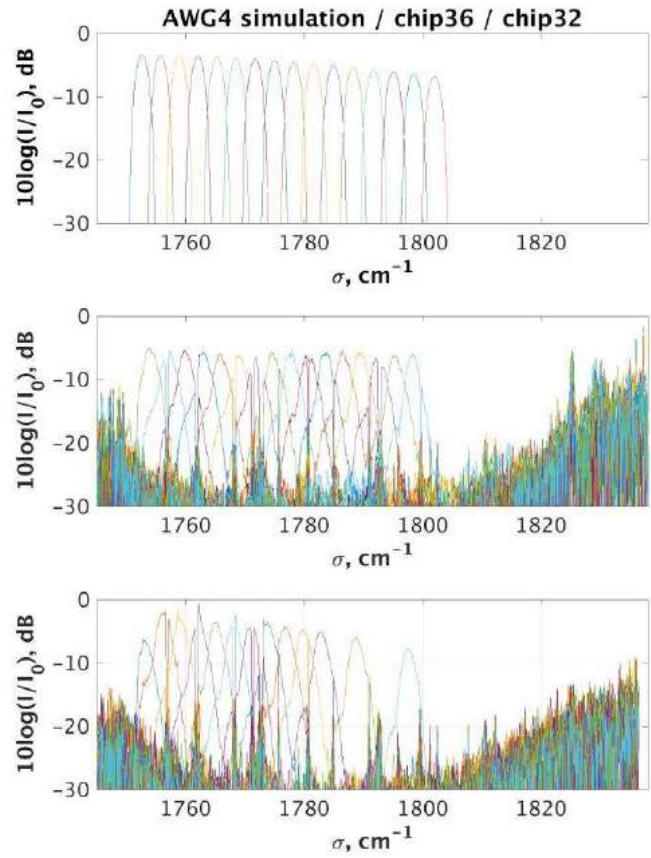

Fig. 4.12. Transmission spectra of channels 20 to 35 at central input of AWG4 (from top to bottom): simulation, chip 36 and chip 32.

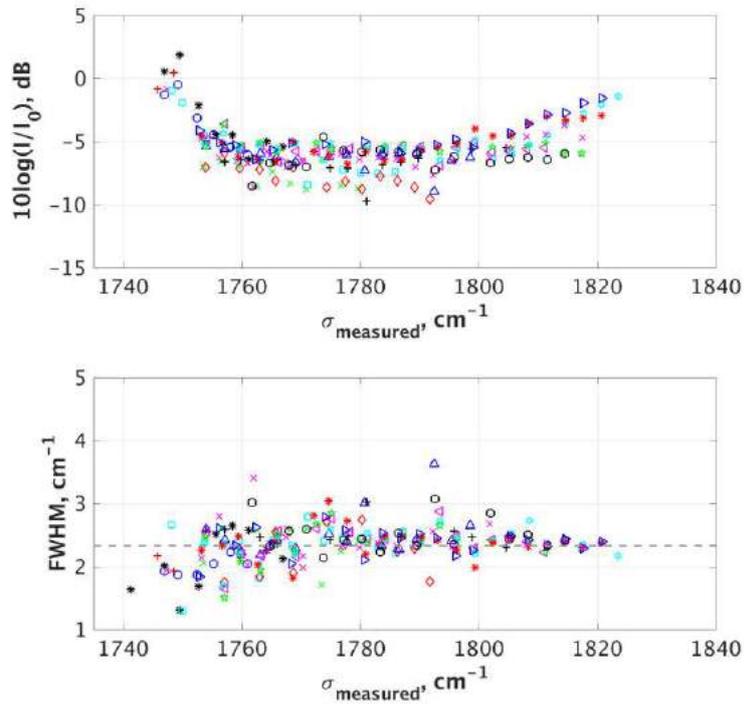

Fig. 4.13. Transmission and FWHM of AWG4, chip 36.

#### 4.2.1.4 Discussion

Tab. 4.3 summarizes the spectral characteristics of DUTs. First, we compare AWG1 and AWG2 that have similar configuration except different widths of MMI couplers at the output waveguides, which are 9 μm and 11 μm, respectively. According to simulation results, the wider MMI needs to be chosen to host the targeted profile of the output spectrum. It is expected that inter-channel crossing $X_{ch}$ improves from 7.8dB to 6.7dB and that will be accompanied by the drop of the insertion loss $L_c$ from -4.8dB to -6.3dB as the MMI width increases. In the experiment, however, we observed the trend on one of the chips only. The inter-channel crossing $X_{ch}$ decreased from 4.6dB to 4dB on chip 36, but was equal to 4dB on chip 32. The deterioration of insertion loss $L_c$ though was by 0.1dB on chip 36, and considerably closer to simulation value, by 0.9dB on chip 32. Both DUTs on chip 36 have insertion losses above the simulation predictions, which could be an effect of fabrication defects in reference waveguide and MMI couplers. As for DUTs on chip 32, the correlation between the MMI width and the insertion loss is evident and close to the expected values.

The transmission spectra shown in Fig. 4.7, Fig. 4.10 and Fig. 4.13 reveal the following. All three DUTs have a common motif towards the outmost wavenumbers in the form of upward bends. This is explained by the operational spectral range of our multiplexer with central wavenumber of 1749.5 cm$^{-1}$. It means that the length difference of array waveguides is chosen so that it leads to $2\pi$ phase difference for beam of central wavenumber, and their constructive interference. Consequently, beam with central wavenumber has the maximum output power, and its neighbor diffraction orders, which are at 1749.5 cm$^{-1}$ ± 157 cm$^{-1}$, have the power transmission maxima slightly lower than that of the main order. The humps around 1777 cm$^{-1}$ are the cause of power distribution of the QCL source used for characterization.

|  | $L_c$, dB | $L_u$, dB | $X_{ch}$, dB | $L_x$, dB | $\Delta f_{FSR}$, cm$^{-1}$ |
|---|---|---|---|---|---|
|  | simulation ||||| 
| AWG1 | -4.8 | 3.5 | 7.8 | 37 | 157 |
| AWG2 | -6.3 | 3.6 | 6.7 | 28 | 157 |
| AWG4 | -5.2 | 3.4 | 7.6 | 38 | 157 |
|  | characterization, chip 36 ||||| 
| AWG1 | -3.3 | 2.9 | 4.6 | 20 | 134 |
| AWG2 | -3.4 | 2.7 | 4 | 18 | 132 |
| AWG4 | -6 | 0.5 | 4.4 | 18 | 132 |
|  | characterization, chip 32 ||||| 
| AWG1 | -4.8 | 3.1 | 4 | 15 | - |
| AWG2 | -5.7 | 2.2 | 4 | 19 | - |
| AWG4 | -4 | 3.8 | 4.7 | 20 | - |

Tab. 4.3. Spectral parameters at central insertion case of AWG 1, AW2, and AWG4 (chip36) experimental and theoretical.

| | $\langle\Delta\nu_c\rangle$, cm$^{-1}$ | $\langle$FWHM$\rangle$, cm$^{-1}$ |
|---|---|---|
| AWG1 (chip36) | -2.5 | 2.2 |
| AWG2 (chip36) | -2.4 | 2.5 |
| AWG4 (chip36) | -3.2 | 2.3 |
| AWG1 (chip32) | -3.5 | 3.0 |
| AWG2 (chip32) | -3.3 | 2.5 |
| AWG4 (chip32) | -4.0 | 2.3 |

Tab. 4.4. Spectral shifts and FWHM of AWG 1, AW2, and AWG4.

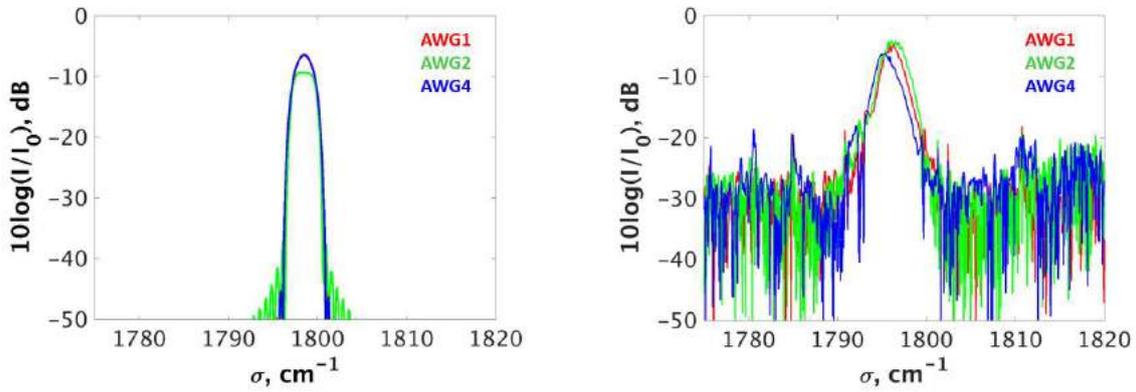

Fig. 4.14. Response of channel 34: simulation (right), characterization (left).

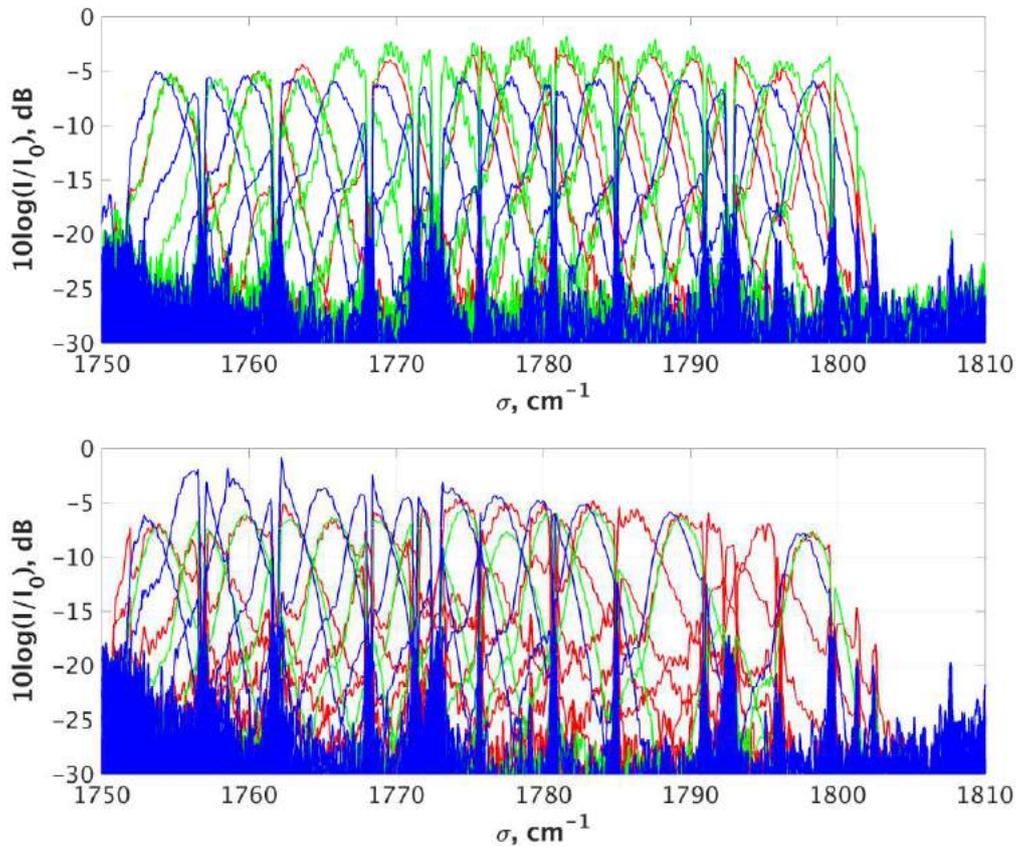

Fig. 4.15. Spectral response of DUTs: chip 36 (top) and chip 32 (bottom).

The non-uniformity loss is affected by the choice of FSR, which is bound to geometrical parameters such as slab length $R$ and difference of length of array waveguides $\Delta L$, and is practically independent of waveguide and MMI widths. So for our three DUTs, the non-uniformity loss is expected to remain around 3.5dB. The measured data are mostly in agreement except for the AWG4 of chip 36. Given the lower (compared to expected) insertion loss of -6dB, the discrepancy could be explained by fabrication defects.

Another important spectral characteristic is the crosstalk. The measurement results did not reveal the correlations predicted by simulation. This indicates that the crosstalk is caused by phase errors. The analysis of phase errors is presented in chapter 4.2.4.

The calculated FSR is 157 cm$^{-1}$. Thanks to having sufficient number of input channels and the wide spectral coverage of QCL source used in characterization, the neighbor-diffraction order was measured for AWGs on chip 36. In average, the FSR appeared to be around 133 cm$^{-1}$.

The widths of spectral responses were evaluated by average FWHM of Gaussian fits. The data are shown in Tab. 4.4. As expected, the widths spectral responses are in a good agreement with the expected values, except AWG1, chip 32. For the majority, the widening of array waveguides as well as MMI couplers is evidently impacts the FWHM. As for AWG1 on chip 32, the Fig. 4.6 (bottom) shows that the response is of noisy, and the peaks are not clear. The AWG with the array waveguide width of 4.6 µm has in average FWHM of 2.2 cm$^{-1}$, increase for wider waveguides, AWG with the array waveguides of 4.8 µm have in average FWHM of 2.3 cm$^{-1}$, with both MMIs of 9 µm. For the AWG with 4.6 µm wide array waveguides with MMI of 11 µm, the average FWHM is 2.5 cm$^{-1}$. The comparison given in Fig. 4.14 (right) exhibits the flattening effect of AWG2 with wider MMI coupler as expected from the theory.

The FSR in simulation was 157 cm$^{-1}$. Thanks to having sufficient number of input channels and the wide spectral coverage of QCL source used in characterization, the neighbor diffraction order was measured for AWGs on chip 36. In average, the FSR appeared to be around 133 cm$^{-1}$. The FSR depends on two parameters, the length difference of array waveguides $\Delta L$ and the group index $n_g$. The discrepancy with the simulation result could be the result of group index variation caused by fabrication defects.

Fig. 4.14 shows the superposed measured transmission spectra of channel 34 for AWG 1, AWG2 and AWG4 multiplexers. It is clearly seen that the spectral response of AWG4 is shifted towards longer wavelengths. The comparison of spectral responses on both chips shown in Fig. 4.15 confirms that the shift is attributed to the effect of wider array waveguides, since it is present on both spectra. This is explained by the change in effective index of the waveguide. The average shift of central wavenumbers from the target values are -2.5 cm$^{-1}$, -2.4 cm$^{-1}$ and -3.2 cm$^{-1}$ for central inputs of AWG1, AWG2 and AWG4, respectively. Thus, the AWG4 is shifted by -0.7 cm$^{-1}$ - 0.8 cm$^{-1}$, which corresponds to 0.0014 – 0.0016 variation in effective index. It is of the same magnitude as the effective index difference 0.001 obtained by numerical calculation using multi-layer R-soft model.

### 4.2.2 Non-central input impact on channel spacing variation

In the array waveguides, the beams of wavenumbers other than central one have the phase distribution either outstripping or lagging, contributing to a tilt of the phase front at the end of array waveguides sections. Consequently, such beams focus at points inclined from the center of the focal line, with the following angle (approximation valid for small angles):

$$\theta_{out}(\sigma) \approx \frac{\left(\frac{\sigma_c}{\sigma} \cdot n_{effc} - n_{eff}(\sigma)\right) \cdot \Delta L}{d \cdot n_s(\sigma)} - \theta_{in} \quad (4.1)$$

where $\sigma$ is the wavenumber, $d$ is the distance between array waveguides at the boundary with slabs, $n_{eff}(\sigma)$ and $n_s(\sigma)$ are the effective indices of array and slab waveguides, respectively, $\theta_{in}$ and $\theta_{out}$ are the angular positions of input and output channels, respectively, $\Delta L$ is the path difference between neighbor array waveguides, the subscript "c" indicates parameters referring to the central wavenumber and the term $\theta_{out}$ is called a dispersion angle.

The position of the diffracted beam is given by:

$$\Delta s = R \cdot \theta_{out} \quad (4.2)$$

where $R$ is the length of the slab and $\Delta s$ is the length of the arc connecting the central point of the focal line to the position of the diffracted beam.

For the given design of AWG, the positions of output channels are determined by equation (4.2), where dispersion angles are calculated for the target wavenumbers using (4.1) with $\theta_{in}$ equal to zero. When the refractive indices of waveguides are independent of input beam frequency, the outmost target wavenumbers $\sigma_{min}$ and $\sigma_{max}$ focus equidistantly on two sides of the focal center. In practice, the dispersion induces slightly different distances from the focal axis for $\sigma_{min}$ and $\sigma_{max}$. The same applies to the whole range of target wavenumbers, and the output channels are placed non-uniformly, as shown in Fig. 4.16.

From (4.1), it can be seen that for non-central input (e.g. $\theta_{in} \neq 0$), the positions of target wavenumbers on the focal line are displaced translationally, since the input angle $\theta_{in}$ is defined solely by the position of input channel and does not depend on the frequency. As illustrated in Fig. 4.16, three central channels designed for $\{\sigma_c - \Delta\sigma_{ch}, \sigma_c, \sigma_c + \Delta\sigma_{ch}\}$ will, for $\theta_{in} \neq 0$, transmit wavenumbers $\{\sigma_c - 2(\Delta\sigma_{ch} + \Delta\sigma), \sigma_c - (\Delta\sigma_{ch} + \Delta\sigma), \sigma_c\}$. The channel spacing changes from the designed value $\Delta\sigma_{ch}$ to $\Delta\sigma_{ch} + \Delta\sigma$.

For the given material, the dispersion follows as:

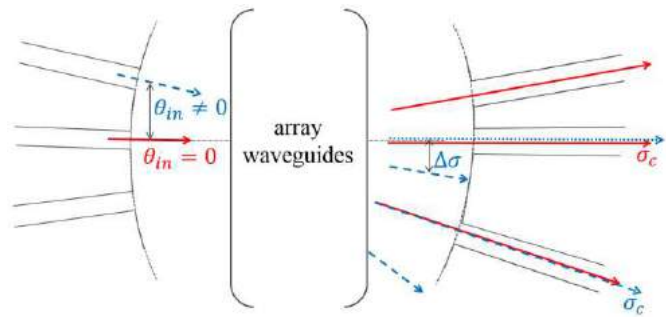

Fig. 4.16. Response of channel 34: simulation (right), characterization (left).

$$D = \frac{ds(\sigma)}{d\sigma} = R \frac{n_g(\sigma) \cdot \Delta L}{\sigma_c n_s d} \qquad (4.3)$$

where $n_g$ is the group index defined as $n_g = n_{effc} - \sigma_c \cdot dn_{eff}(\sigma)/d\sigma$. The dispersion parameter links the shift in frequency (wavenumber) with the corresponding lateral displacement of the diffracted beam on the focal line. From (4.3), it follows that channel spacing can be determined as $\Delta\sigma_{ch} = R \cdot \theta_{out}(\sigma)/D$, where $R$ and $D$ are fixed for the given design and $\theta_{out}(\sigma)$ is defined by the input frequency and effective indices as given in (4.1).

We have evaluated channel spacing variation $\Delta\sigma$ theoretically and compared to experimental measurements for AWGs at three wavelengths.

Fig. 4.17 (top) presents the shifts for AWG at 5.7 μm with designed channel spacing $\Delta\sigma_{ch} = 3$ cm$^{-1}$ which were evaluated using equation (4.3). The points correspond to spectral shifts of outputs and lines indicate input position with central input at channel 9. Points are uniformly distributed along the lines, which show that the channel spacing remains constant within a certain position of injection channel. Negative slopes correspond to channel spacing which are smaller than the designed value $\Delta\sigma_{ch}$, and positive slopes to channel spacing which are larger. The magnitude of the slope is proportional to the shift from channel spacing $\Delta\sigma$. The maximum shift corresponds to channel 17. It is equal to $\Delta\sigma_{Input17} = 0.16$ cm$^{-1}$, which is 5% of the design channel spacing. Overall scatter of channel spacing variation is equal to 4 cm$^{-1}$. As it is seen, the lines that do not intersect at the common center. This is explained by the effect of wavelength dependent refractive index of the waveguide and the positioning of output channels according to spectral symmetry, i.e. spacing between neighbor channels corresponds to 6 cm$^{-1}$. Consequently, the geometrical positions of the channels are not strictly equidistant, but slightly vary.

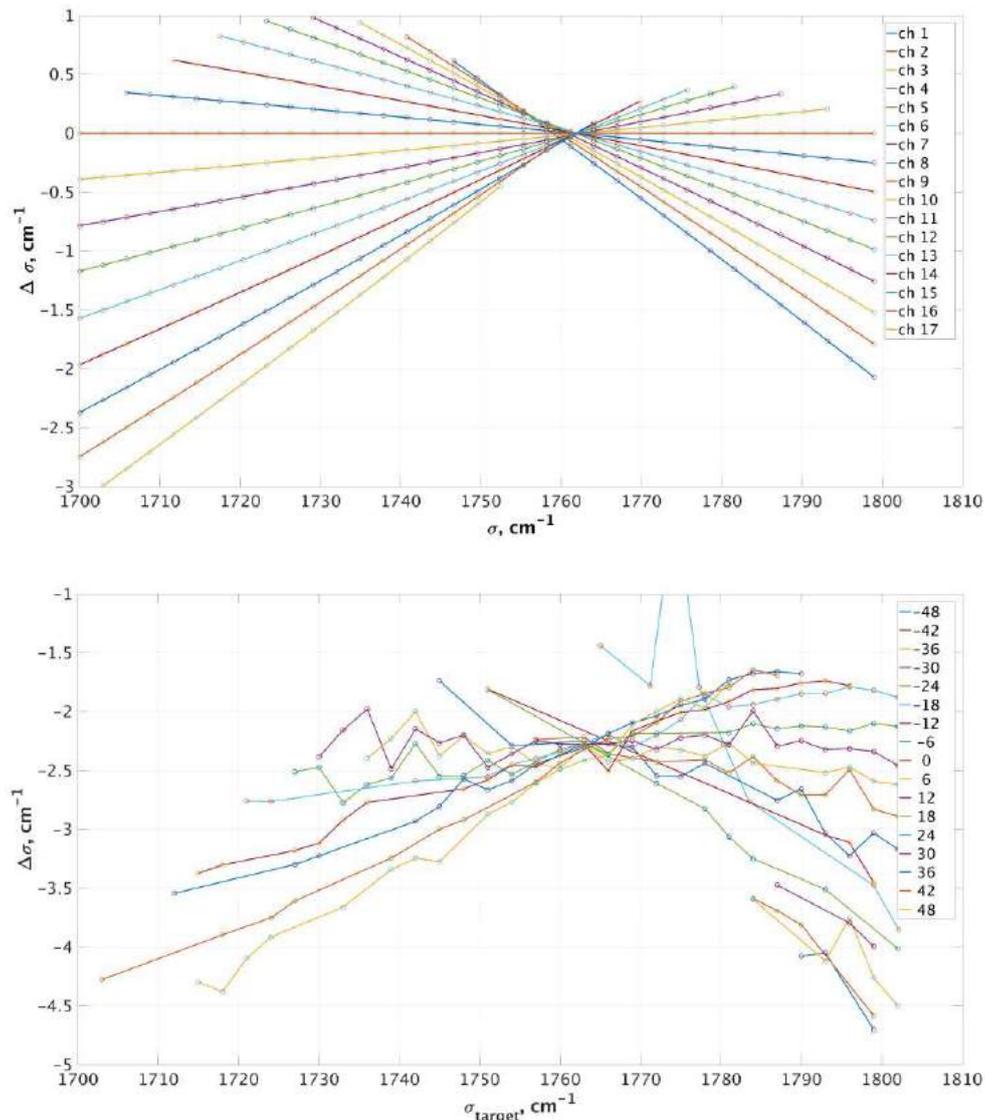

Fig. 4.17. Spectral shift of channel spacing for non-central input of AWG at 5.7 μm: theoretical (top) and experimental (bottom).

Fig. 4.17 (bottom) shows the shift between target and measured output wavenumbers as a function of wavenumber. The lines correspond to input channels with points representing the measured output wavenumbers. Fig. 4.17 (top) and Fig. 4.17 (bottom) clearly present similarities which show a good correlation between experimental measurements and simulations.

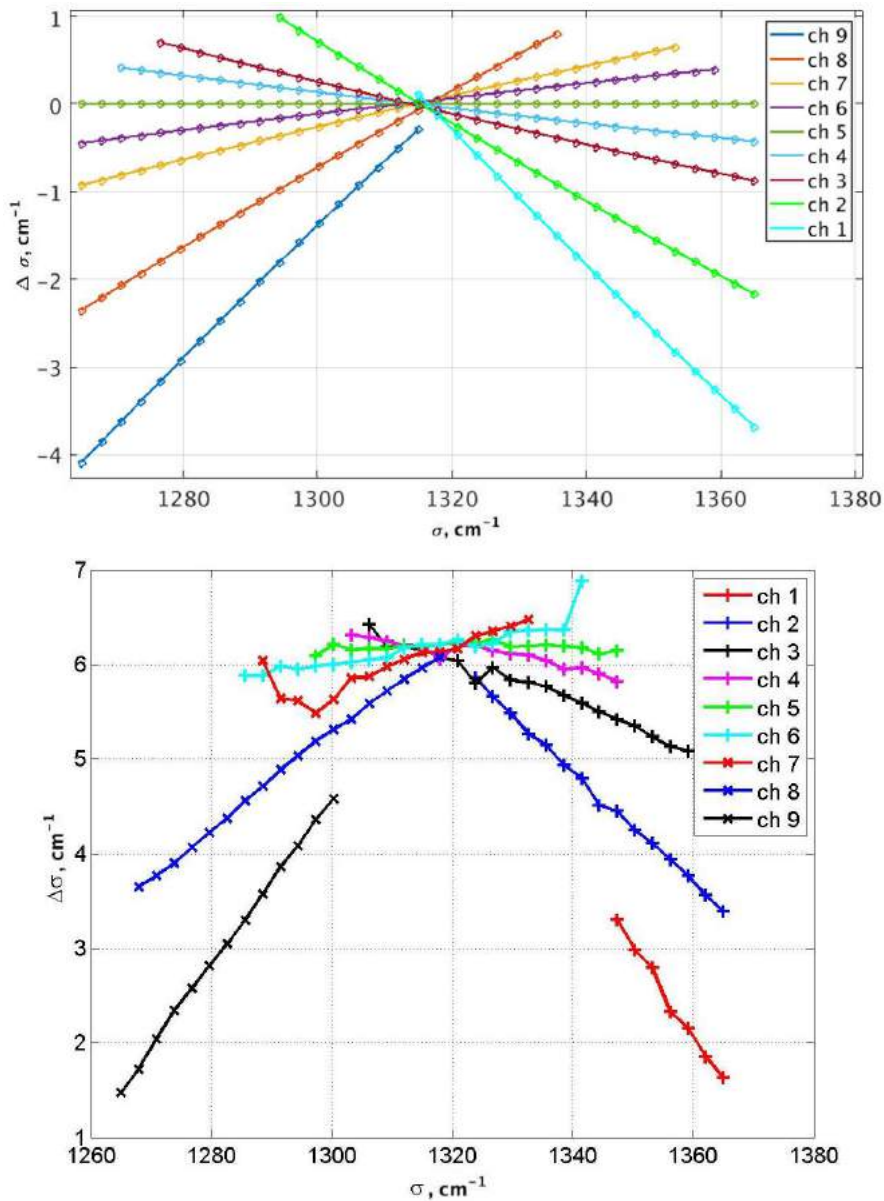

Fig. 4.18. Spectral shift of channel spacing for non-central input of AWG at 7.6 μm: theoretical top) and experimental (bottom).

The analysis of channel spacing variation was studied for two more AWGs. First, it was performed on 9x35 AWG based on SiGe graded index technology with central wavelength at 7.6 µm designed and fabricated at CEA-Leti. The characterized was done by researcher of LCNA, CEA-Leti O.Lartigue [83]. Fig. 4.18 presents the comparison of calculated and measured data, which reveals a good agreement as well. The maximum shift corresponds to channel 9. It is equal to $\Delta\sigma_{Input17}$ = 0.23 cm$^{-1}$, which is 8% of the design channel spacing. Overall scatter of channel spacing variation is equal to 5 cm$^{-1}$.

Another device under study was 17x35 AWG based on SiGe step index technology with central wavelength at 4.5 µm. It was characterized by Ph.D. student of LCNA, CEA-Leti J.Favreau. The results of experimental measurements and their comparison with theoretical expectations are presented in Fig. 4.19. The maximum shift corresponds to channel 1. It is equal to $\Delta\sigma_{Input17}$ = 0.18 cm$^{-1}$, which is 6% of the design channel spacing. Overall scatter of channel spacing variation is around 3 cm$^{-1}$.

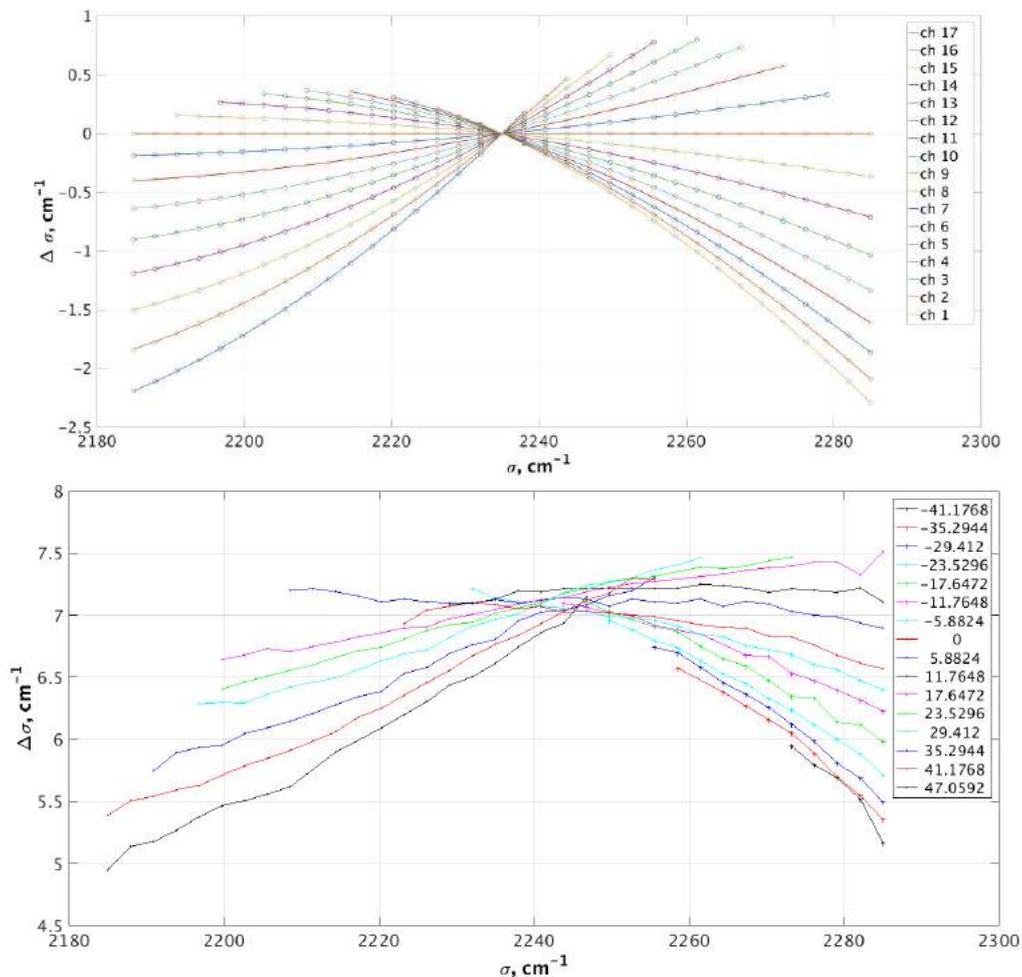

Fig. 4.19. Spectral shift of channel spacing for non-central input of AWG at 4.5 µm: theoretical (top) and experimental (bottom).

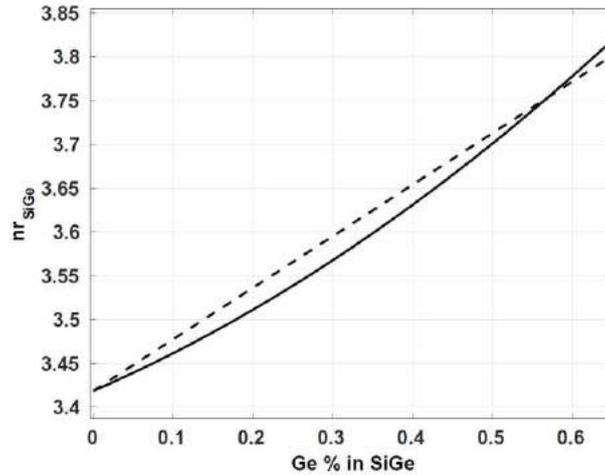

Fig. 4.20. The approximations of refractive index of SiGe at 7.6 µm: dashed line – linear approximation; solid line – parabolic approximation.

As it is seen from Figs. 4.17 – 4.19, the theoretical and measured data are in a good agreement except the overall shifts of experimental spectra which is seen from the intersections displaced from zero. The alteration is caused by effective index $n_{eff}$ errors. For AWG at 5.7 µm, the variation is within -2.3 cm$^{-1}$, which could result in array waveguides widths variations. As it was shown in previous chapter, 200 nm difference in waveguide width causes around -0.7 cm$^{-1}$ spectral shift.

In case of AWG at 7.6 µm, which was designed and fabricated earlier than AWG at 5.7 µm, the displacement of the whole spectra is also affected by the effective index taken for design. Initially, SiGe refractive index was calculated by linear approximation using the germanium and silicon refractive indices and their concentrations. The more recent M-line experimental measurements of SiGe refractive index, revealed a parabolic dependence [72]. Fig. 4.20 illustrates both approximations, linear and parabolic, for SiGe at 7.6 µm. The resulting difference $\Delta n_{SiGe}$ taken into account in simulation yielded 5 cm$^{-1}$ shift of the output spectra, thus showing a good agreement with the measured wavenumber shift. Other potential error study showed insignificant wavenumber changes, specifically, a ±1.2 cm-1 shift for a ±2% alteration of the Ge concentration at the core of the array waveguide as well as ±0.2 cm$^{-1}$ shift for a ±100 nm variation of waveguide width.

The AWG at 4.5 µm is fabricated based on technology with constant SiGe index at the core, which is different from two previous cases. The central wavelength is shifted by 7 cm$^{-1}$. Channel spacing variations theoretical and expreimental are shown in Fig. 4.19. It is noteworthy, that the channel spacing is not constant, which could be seen from the curved profile of the lines.

### 4.2.3 Phase errors

In chapter 2.3.6, we have discussed the phase errors that arise in AWG multiplexer due to sidewall roughness of array waveguides and cause the crosstalk. Fig. 4.21 shows the SEM image of array waveguides with the enlarged cross-section given in the inset. As it can be seen, the walls are not perfectly straight, but rather have a certain level of sidewall roughness. Using our semi-analytical tool in Matlab, we determined the value of standard deviation of phase errors that causes the same level of crosstalk as the measured data. Fig. 4.22 (bottom) shows the measured spectrum of channel 27 of AWG2 (blue solid line) with the simulation with phase error standard deviation of 0.0068 rad (black dashed line) spectrum superposed on the top of it. The red line is Gaussian fit.

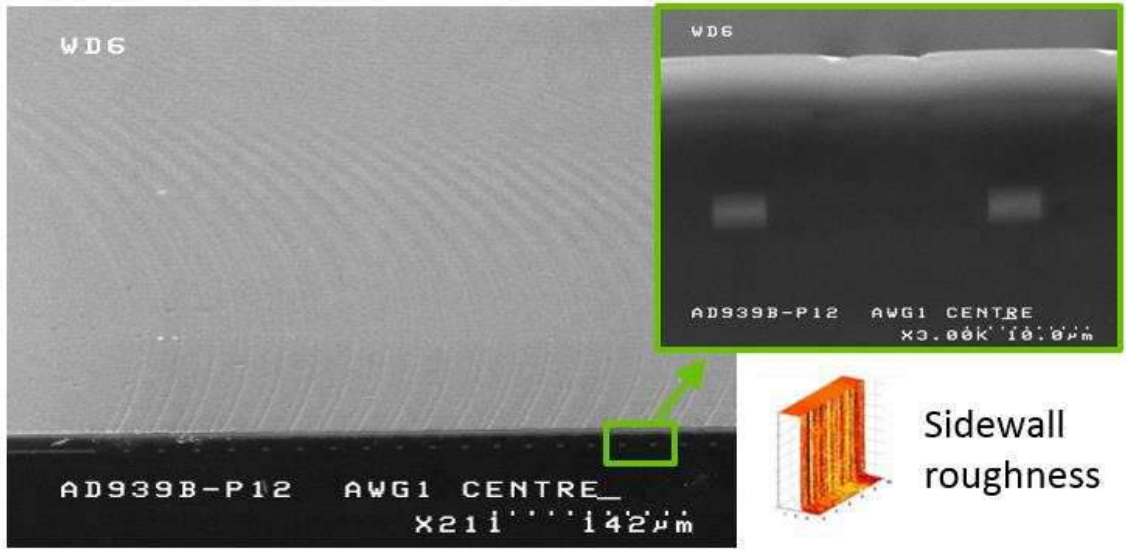

Fig. 4.21. SEM images of array waveguides of AWG at 5.7 µm.

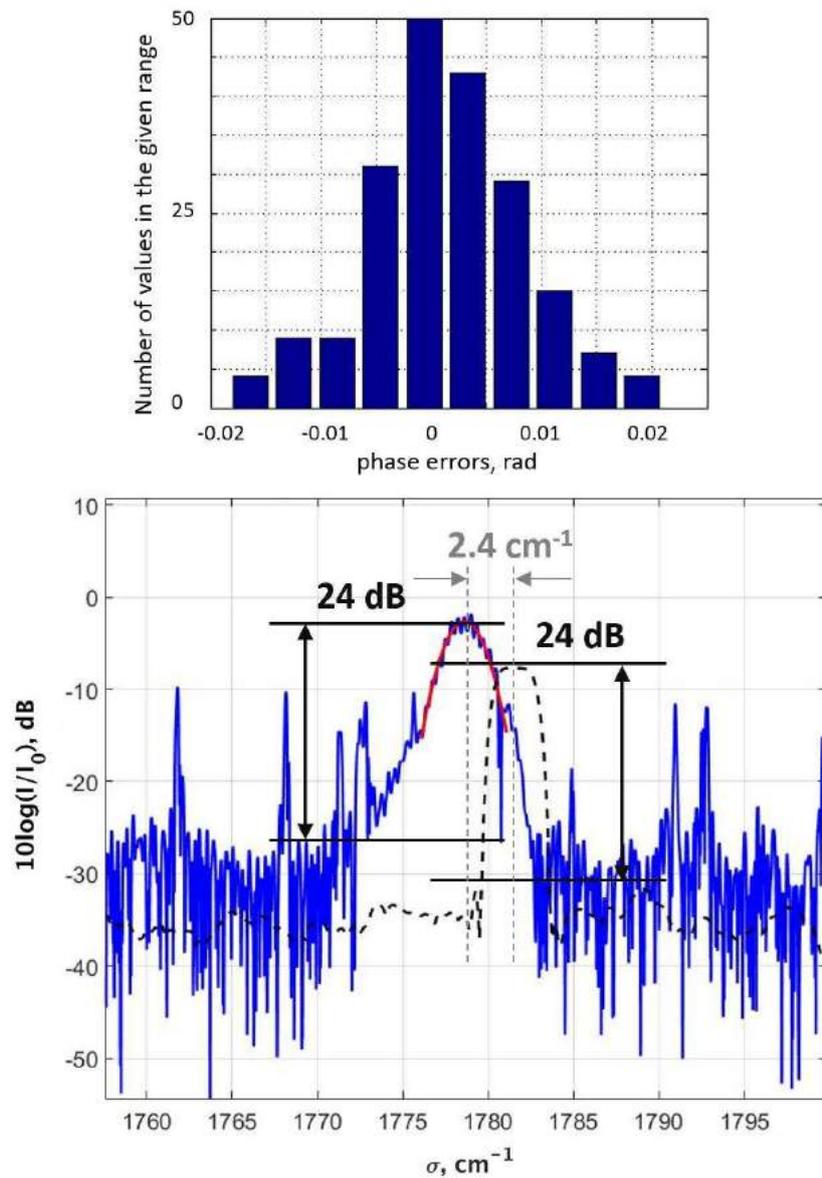

Fig. 4.22. Top: histogram of phase errors distribution; bottom: spectral response of channel 27 of AWG2 (chip 36).

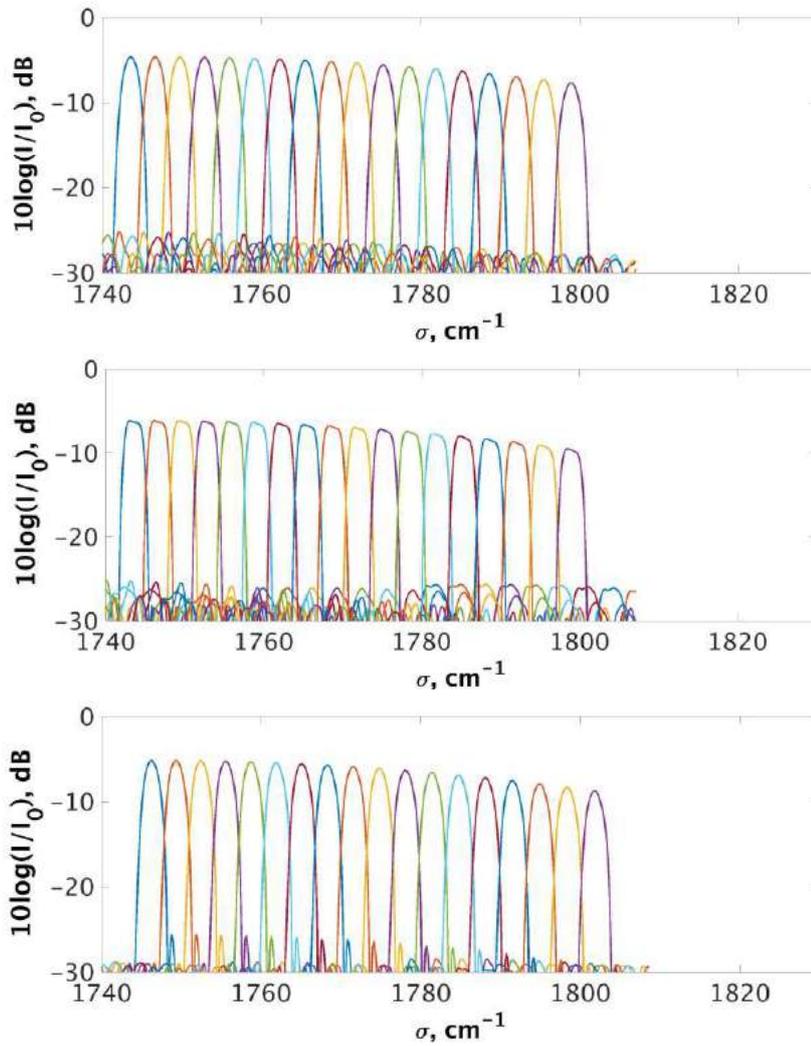

Fig. 4.23. Spectral response top-to-bottom: AWG1, AWG2 and AWG4.

The experimental crosstalk level is 24dB whereas the theoretical expectation in ideal case is 30dB. In order to estimate the impact of fabrication imperfections to crosstalk level, we introduce a normal distribution of phase errors to our simulation as shown in Fig. 4.22 (top). The standard deviation of phase error of 0.0068 rad corresponds to 24 dB crosstalk. The equivalent effective index variation can be estimated from $\Delta n_{eff} = \Delta\varphi\lambda/(2\pi l_x)$, where $l_x$ is the waveguide length. In our case, the estimated standard deviation of effective index due to sidewall roughness of array waveguides is around $3.5 \cdot 10^{-5}$ for 1 cm long waveguide. Fig. 4.23 presents the simulated spectra of AWG1, AWG2 and AWG4 with 0.0068 rad standard deviation of phase errors, which can be compared with the corresponding measured data in Fig. 4.6, Fig. 4.9 and Fig. 4.12.

### 4.2.4 Temperature measurements

The spectral responses of AWG2 and AWG4 were measured at five temperature points 20°, 27.5°, 32.7°, 36.6° and 41.4°. The choice of DUTs is related to the interest to compare the AWGs with two different array waveguide widths. The superposed outputs of channels 29 are presented in Fig. 4.24 for AWG2 (chip 36) and Fig. 4.25 for AWG4

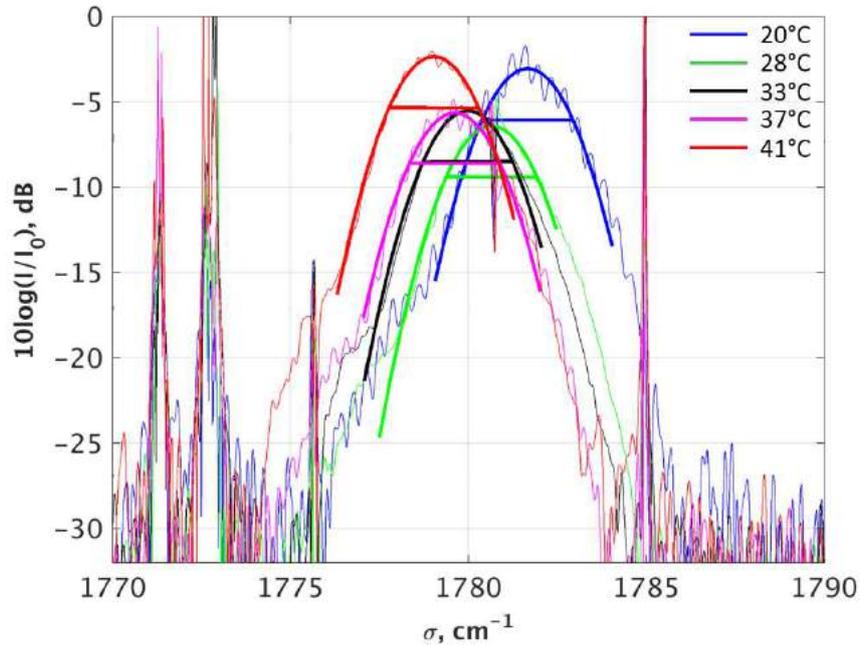

Fig. 4.24. Spectral response of channel 29 of AWG2, chip 36, at temperature points: 20°, 27.5°, 32.7°, 36.6° and 41.4°.

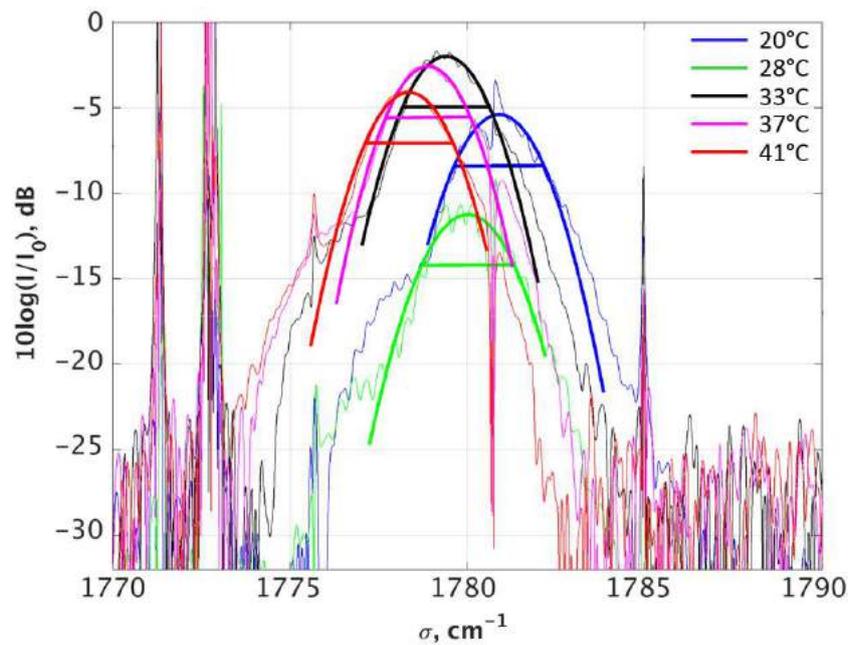

Fig. 4.25. Spectral response of channel 29 of AWG4, chip 36, at temperature points: 20°, 27.5°, 32.7°, 36.6° and 41.4°.

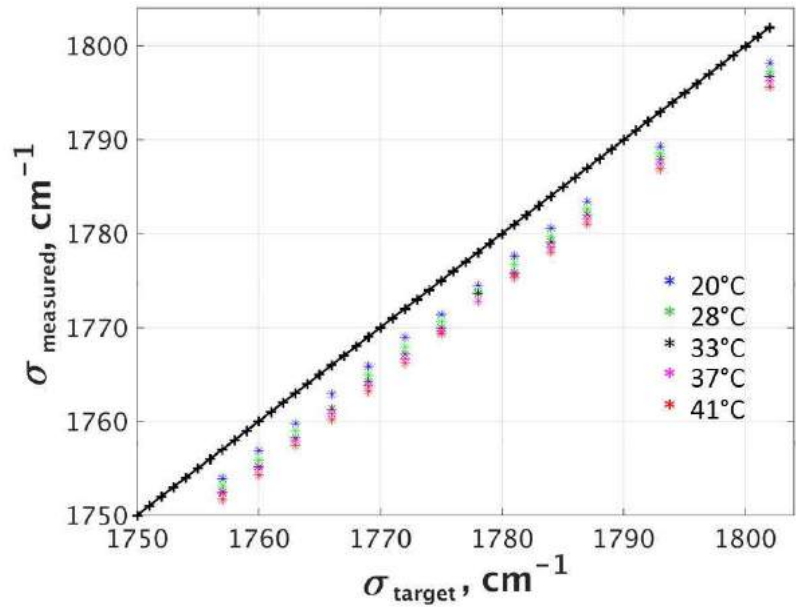

Fig. 4.26. Measured wavenumbers of AWG2, chip 32, at points: 20°, 27.5°, 32.7°, 36.6° and 41.4°.

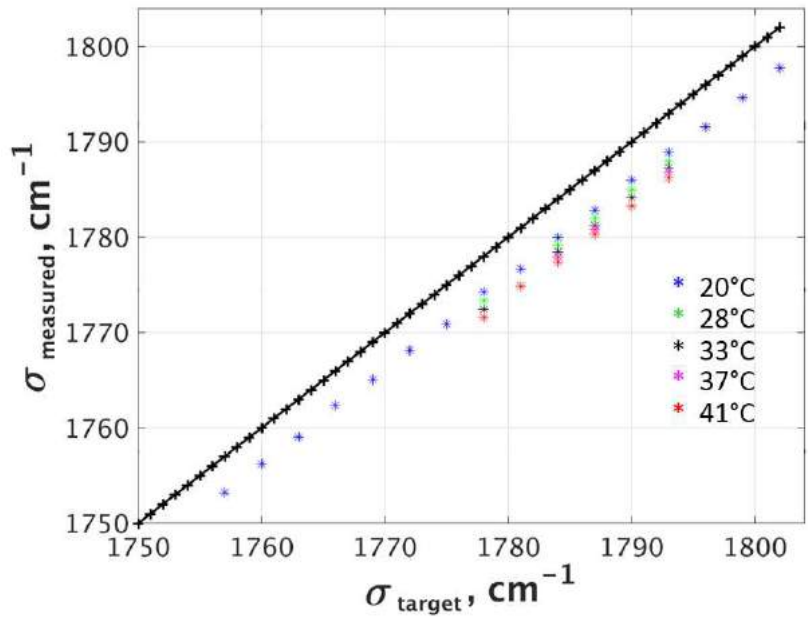

Fig. 4.27. Measured wavenumbers of AWG4, chip 32, at points: 20°, 27.5°, 32.7°, 36.6° and 41.4°.

(chip 32). As it is seen, the spectral shift changes more or less linearly with the difference in temperature. Figs. 4.26 – 4.29 show the measured data points of all outputs for the central input (channel 9). Tab. 4.5 summarizes the results. As it is clearly seen, the measured spectra is in a good agreement with the simulation presented in details in chapter 2.3.7, and exhibit linear dependence of spectral shift with temperature increase in the given range, Fig. 4.30.

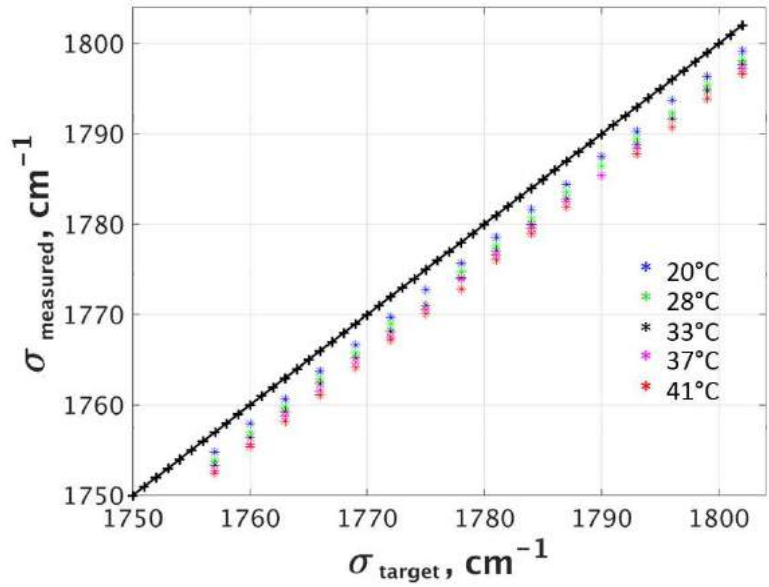

Fig. 4.28. Measured wavenumbers of AWG2, chip 36, at points: 20°, 27.5°, 32.7°, 36.6° and 41.4°.

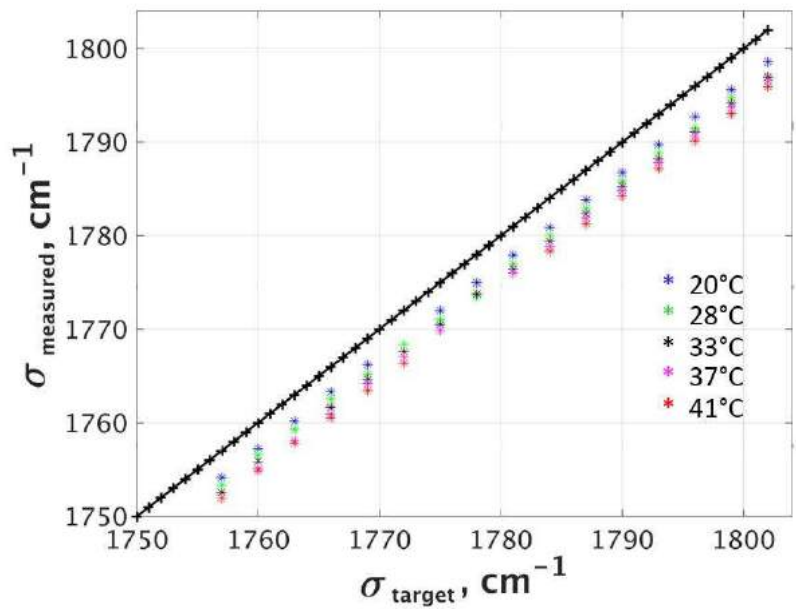

Fig. 4.29. Measured wavenumbers of AWG4, chip 36, at points: 20°, 27.5°, 32.7°, 36.6° and 41.4°.

|      |            | $\langle\Delta\sigma\rangle$, 27.5°C | $\langle\Delta\sigma\rangle$, 32.7°C | $\langle\Delta\sigma\rangle$, 36.6°C | $\langle\Delta\sigma\rangle$, 41.4°C |
|------|------------|-------|-------|-------|-------|
| AWG2 | simulation | 0.834 | 1.405 | 1.856 | 2.408 |
|      | chip 32    | 0.840 | 1.485 | 1.902 | 2.438 |
|      | chip 36    | 0.927 | 1.589 | 1.980 | 2.570 |
| AWG4 | simulation | 0.836 | 1.407 | 1.859 | 2.411 |
|      | chip 32    | 0.897 | 1.723 | 2.037 | 2.485 |
|      | chip 36    | 0.912 | 1.533 | 2.073 | 2.527 |

Tab. 4.5. Average spectral shift due to variation of temperature in cm$^{-1}$.

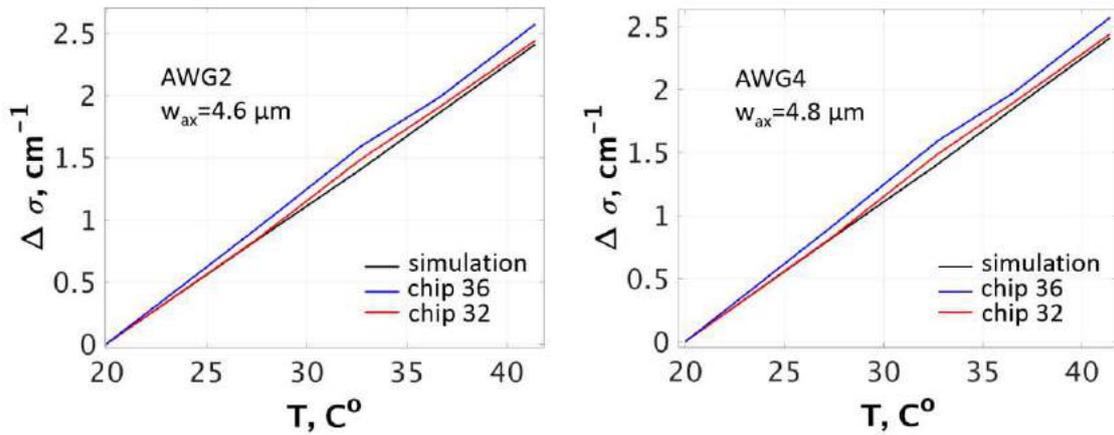

Fig. 4.30. Dependence of the spectral shift as a function of temperature shown for AWG2 with array waveguide width of 4.6 μm and AWG4 with array waveguide width of 4.8 μm.

## 4.3 Conclusion

Compared to the state of the art, straight waveguides fabricated at CEA-Leti demonstrate competitive performance, showing propagation loss of 1 dB/cm at wavelength $\lambda$ = 4.5 μm and 2 dB/cm at $\lambda$ = 7.4 μm. Basic functions transmissions made of SiGe graded index waveguides were studied as well demonstrating nearly theoretical propagation losses [26]. The multiplexers were made of these waveguides.

In this chapter we have presented characterization of AWGs operating at 5.7 μm. It was shown that the width of MMI couplers affects the inter-channel crossing, improving it by around a half of dB.

We have also observed that multiplexers with wider array waveguides (AWG2 vs. AWG4) have total spectral shift of about 0.7 cm$^{-1}$.

The normalized transmission spectra characteristics are close to that of AWG at 7.6 μm presented earlier, i.e. the insertion loss of 3 dB for TM polarized light and crosstalk below -20 dB. The inter-band flatness, which should preferentially be minimized in broad-band source applications, was 3 dB. Spectral coverage of multiplexers reported by IMEC were 0.05 μm for 3.8 μm AWG and 0.25 μm for 5.3 μm AWG. We see that both mid-IR multiplexers are narrower compared to 0.32 μm for 5.7 μm AWG and 0.58 μm for 7.6 μm AWGs presented by CEA.

We studied the temperature dependence of spectral shift of 35-ch AWG operating at 5.7 μm and demonstrated quasi-linear dependence of the spectral response. The simulation based on silicon and germanium refractive index approximations as functions of temperature and wavelength given by [84] are in a good agreement with the measured values.

# 5 Perspectives of potential improvement

In this chapter, we discuss a modified configuration of AWG multiplexer. It is aimed to reduce twice the number of array waveguides and consequently overall footprint of the device by integrating Y-junctions on the input ends.

The geometry is discribed. Bending and Y-junction losses are estimated using multi-layer structure in R-soft. The response of the proposed design is compared to traditional configuration by semi-analytical tool.

**Contents**

## 5.1 U-shape AWG with Y-junctions

One of the current goals in designing an AWG is the reduction of array waveguides number, which potentially improves the crosstalk arising due to impact of sidewall roughness to effective index variation, as well as reduces the device footprint. We propose a U-shape AWG with Y-junctions integrated on the input ends of array waveguides. The schematic comparison with traditional design is presented in Fig. 5.1. The idea is to reduce the crosstalk sensitivity to phase errors without having a sufficient power truncation.

### 5.1.1 Geometry of U-shape AWG with Y-junctions

Fig. 5.2 presents the schema of the design. The major differences of the U-shape design are as follows:

1) input and output slabs are placed so that their axes of symmetry are parallel;

2) the shortest array waveguide represents a half-circle with radius $R_{min}$ chosen to be close to radius of curvature corresponding to 0.1 dB of propagation loss for the given waveguide and wavelength;

3) array waveguides, other than the shortest one, consist of combination of straight and curved waveguides so that the all array waveguides are nearly equidistant;

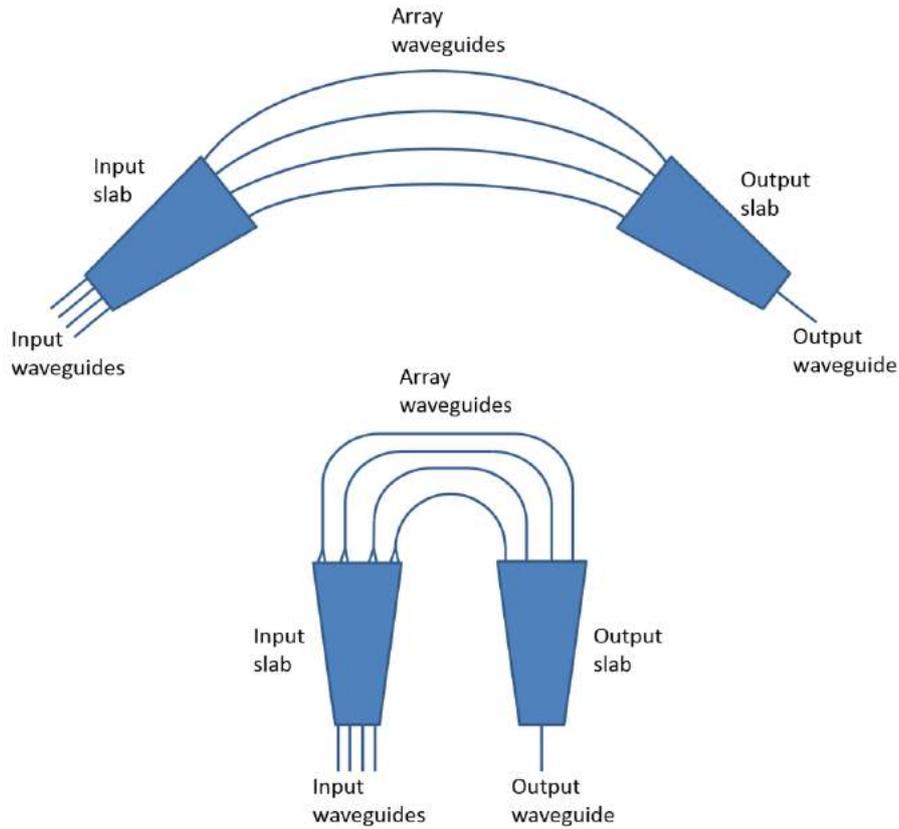

Fig. 5.1. Schematic comparison of two designs: U-shape AWG with Y-junctions (bottom), traditional "banana"-shape AWG (top).

4) in order to capture more light from the input slab and, at the same time, to space array waveguides far enough to minimize the crosstalk, Y-junctions are used at the entrance to array section. It should be noted, that although the array waveguides look parallel on the output side, they get closer towards the output slab so that the distance between waveguides is the same as on the input side and equals to grating period $d$.

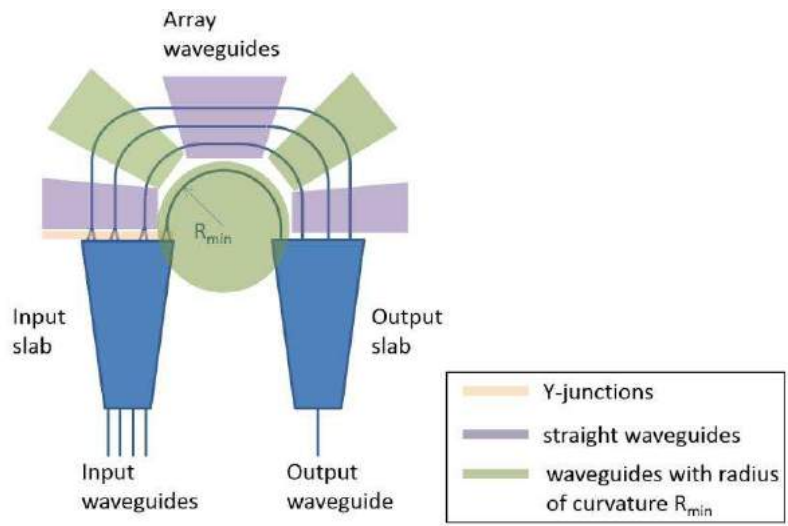

Fig. 5.2. Description of U-shape AWG with Y-junctions.

### 5.1.2 Losses in bends and Y-junctions

The losses were estimated using the multilayer model of graded index single-mode SiGe waveguide built in R-soft. The SiGe graded index waveguide of 4.6 μm width with 906 μm radius of curvature has 0.1 dB loss (TM mode) at 5.7 μm wavelength. We studied the loss in three neighbor array waveguides. As shown in Fig. 5.3, the distance between neighbor waveguides at the interface with slabs is 2 μm, which gives 6.6 μm distance between waveguides central axes. When the number of waveguides is twice reduced, the spacing becomes 8.6 μm.

Fig. 5.4 presents the overview of the studied structure. The outright graph illustrates the amount of power remained in the central waveguide. As it is seen, the coupling between waveguides is insufficient and the transmission is 0.98. Fig. 5.5 and Fig. 5.6 show the power transferred by the outer and inner waveguides, constituting 0.96 and 0.94, respectively.

The losses in Y-junctions were evaluated in the similar manner. Fig. 5.7 shows R-soft simulation of Y-junction. The amount of transmitted power is given by the graph at the left. The 4.6 μm wide SiGe graded index waveguide with 912 μm radius of curvature has 0.4 dB loss for TM mode at 5.7 μm wavelength.

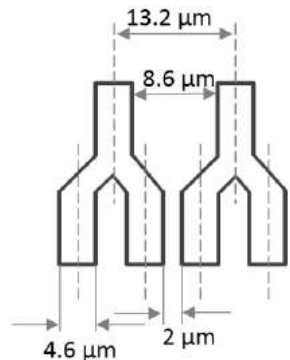

Fig. 5.3. Schema of array waveguide spacing.

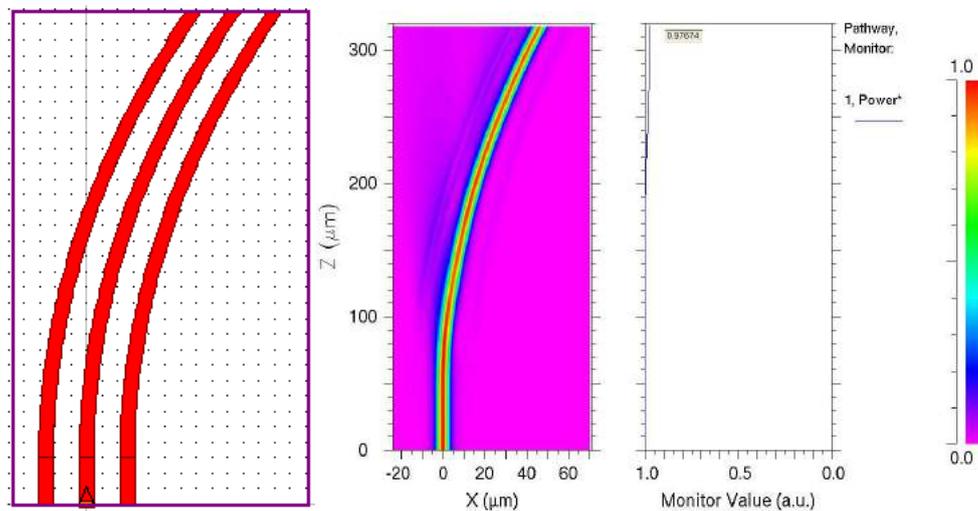

Fig. 5.4. Rsoft simulation of three bent waveguides waveguide of 4.6 μm width with 906 μm radius of curvature. The amount of power remained in the central waveguide is shown on the outright graph.

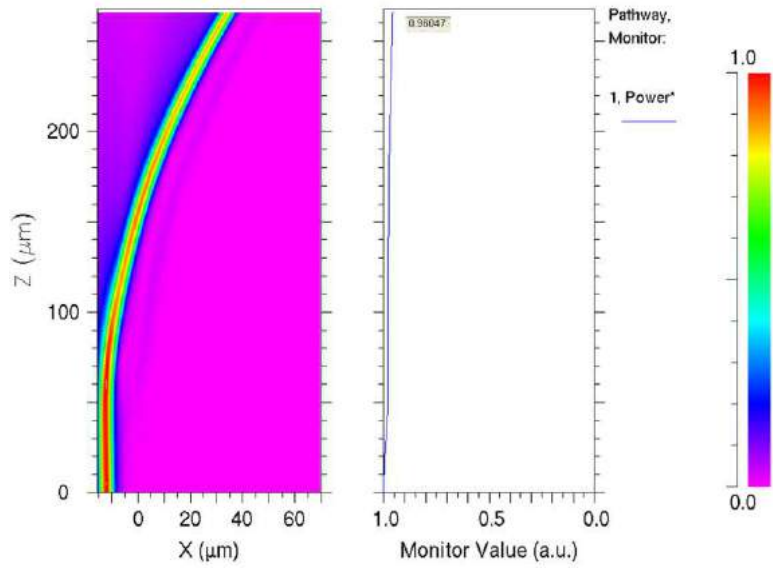

Fig. 5.5. The amount of power remained in the outer waveguide.

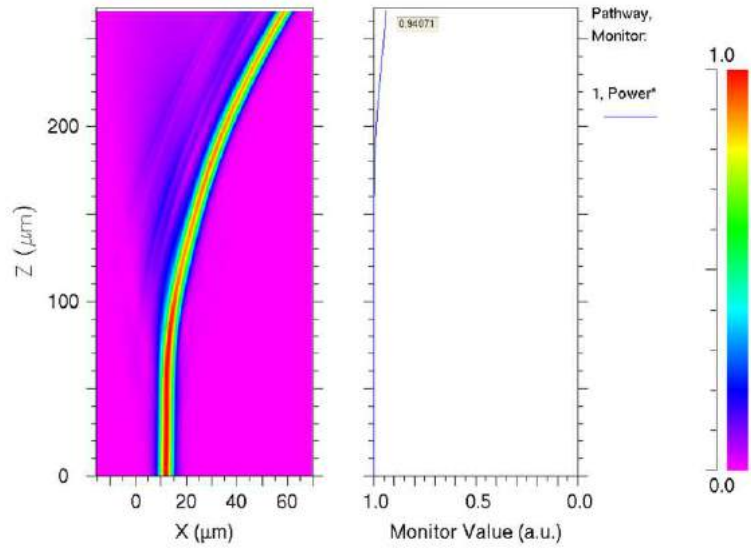

Fig. 5.6. The amount of power remained in the inner waveguide.

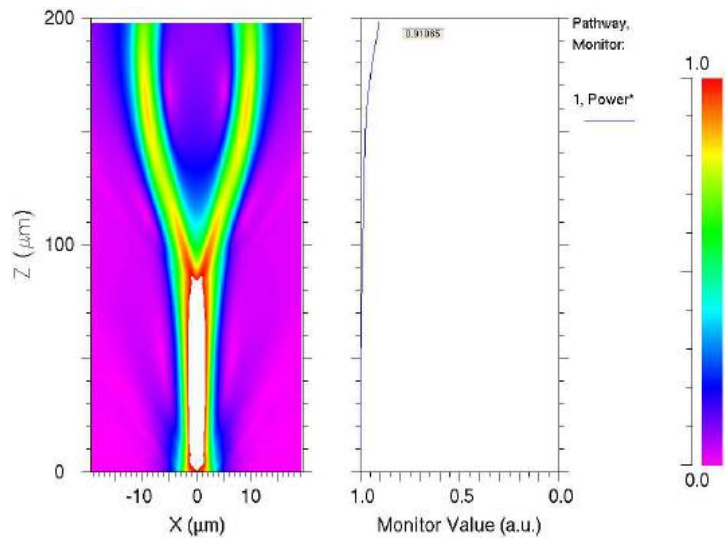

Fig. 5.7. Rsoft simulation of Y-junction. Investigation of power loss.

### 5.1.3 Evaluation of stability to phase errors

The U-shape design was tested for stability to phase errors. We have compared theoretical spectral response of the device operating at 5.7 μm with classic AWG configuration.

For the simulation, we used the same semi-analytical tool in Matlab. Since in our proposed configuration, we consider the distance between array waveguides to be equal to $d$, the dispersion of the device does not change. The major difference is that we have twice lower number of array waveguides at the output.

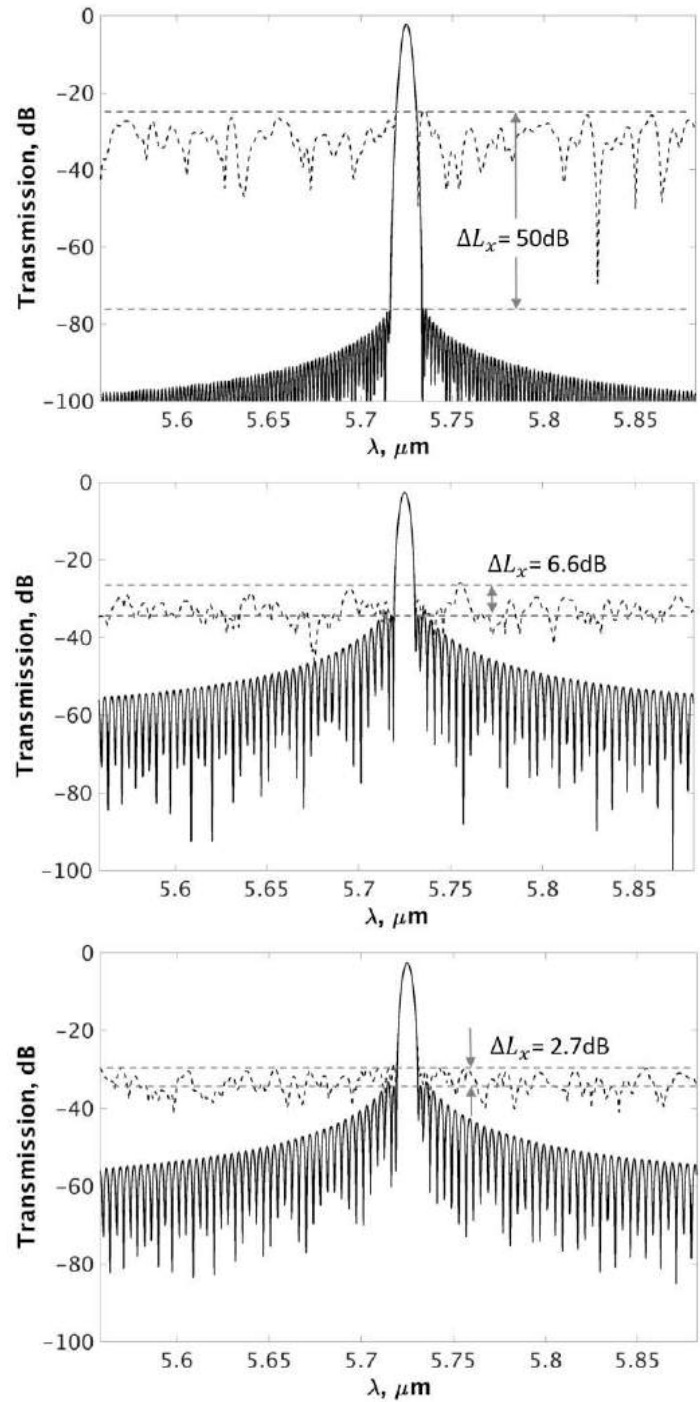

Fig. 5.8. Impact of phase errors with standard deviation of 0.0068 rad on crosstalk: AWG classic, N = 243 (top); AWG classic, N = 130 (center); U shape AWG with Y-junctions, N = 260/130 (center).

|  | AWG classic, N=243 | AWG classic, N=130 | AWG w/Y-junction, N=260/130 |
|---|---|---|---|
| $L_c$ | -2.27 | -2.69 | -2.67 |
| Crosstalk $L_x$ at $<\varphi> = 0$ rad | 73.76 | 29.88 | 30.53 |
| Crosstalk $L_x$ at $<\varphi> = 0.0068$ rad | 23.76 | 23.28 | 27.81 |
| $\Delta L_x$ | 50.00 | 6.60 | 2.72 |

Tab. 5.1. Central insertion loss and crosstalk levels. All data is given in dB.

Fig. 5.8 presents the central channel responses with phase error standard deviation $<\varphi>$ equal to zero (solid line) and 0.0068 rad. The top plot corresponds to classic AWG with optimal number of array waveguides, N = 243. The second plot shows the response of classic AWG with the reduced number of array waveguides, N = 130. Finally, the latter plot is the response of U-shape AWG with Y-junctions, where N = 130, which respectively has 260 Y-junction arms collecting the input illumination from the input slab. The responses at non-zero phase error level are the average of 5 independent numerical calculation runs.

Tab. 5.1 summarizes the insertion losses, crosstalk and its degradation at phase errors level corresponding to our experimental data. It is seen, that U-shape design shows better stability to phase errors.

## 5.2 Conclusion

In this chapter we discussed a potential design of AWG with integrated Y-junctions at the input side of array waveguides. The study of losses at bends and Y-junctions show 0.1dB and 0.4dB, respectively. In total, it might introduce 0.5 dB to typical response of AWG, which is 3dB. The convergence of array waveguides into Y-junctions opens the way to bring the array waveguides closer and thus reduce the overall device size.

The stability to phase errors was analyzed for the standard deviation value that was deduced from experimental crosstalk level in characterization chapter. The comparison of U-shape design to classical AWG both with optimal and reduced number of array waveguides reveals its better stability. Thus the proposed configuration could be a solution to AWGs with lower sensitivity to crosstalk and reduced footprint.

In classic case of AWG multiplexer, as it was discussed in earlier chapters, the design and characterization are done for the reverse case, i.e. demultiplexer. The drawback of U-shape configuration is that, although it can still be designed in the existing way, its characterization requires more routine due to Y-junctions.

The thorough investigation of difference of Y-junction solution with wider taper of MMIs are the subject for further investigation.

# Conclusion

Our work was aimed to develop integrated optics components for infrared sensing on silicon. Based on the powerful QCL sources made by III-V lab and AWG multiplexer at 4.5 µm and 7.6 µm, the goal was to design, implement and characterize a multiplexer for all infrared wavelengths in order to build a broad-band infrared laser source.

The first part of this work was to study the AWG configuration and adapt the existing semi-analytical tool in C++ to MATLAB. The field analysis in AWG was implemented in terms of Gaussians where the diffraction was calculated analytical using Fourier Optics. Our approach allowed to develop a tool in Matlab that unlike the previously available tool, allows to study the effect of certain level of phase errors. Based on calculated parameters, the 35-ch AWG operating at 5.7 µm was developed for optical detection of gases such as chlorine nitrite, formic acid, nitric acid and formaldehyde. The geometry of AWG was defined by the software developed earlier by Dr. P.Labeye.

The AWGs of optimized configurations, i.e. with tapered array waveguides and MMI coupler integrated at the output waveguides, were fabricated and characterized. The fabrication was done by the Silicon Technology Department of CEA-Leti MINATEC. The characterization was performed for two chips, one from the side and one from the center of the wafer. We have studied the impact of MMI coupler width on the flatness of the spectral response. Overall spectral coverage, spectral characteristics such as insertion loss, non-uniformity, inter-channel crossing, crosstalk and FSR were analyzed and proved to be close to theoretical expectations. The temperature dependence of spectral response of AWGs based on graded index SiGe waveguides was studied for the first time. The measurements at five temperature points showed linear dependence, which perfectly fits the theoretical calculations.

Apart from AWG, the alternative configuration PCG of optical multiplexer was investigated. The analytical calculation of the field was performed similar to AWG and based on that, a semi-analytical tool was developed for calculation of spectral response in MATLAB. We have also made a numerical tool that produces a geometry of PCG based on Rowland configuration. By inserting initial parameters, one can obtain the data points of the designed device outline that can be directly read by CleWin software for preparation of fabrication mask. Using our numerical tools, we have prepared PCGs at 3.4 µm, 6.9 µm and 8.7 µm, where the first two were based on SiGe/Si graded index and the latter on Ge/SiGe step index profile technology. The devices are under fabrication and not yet characterized.

With the task at hand to develop a multiplexer for gas sensing application, we have proposed a novel design of AWG with Y-junctions integrated to array waveguides. The aim was to develop a design that would show higher stability to phase errors that are caused by sidewall roughness of real devices. The design was investigated theoretically and showed agreeable results. However, it was not yet tested practically. Preliminary analysis show its feasibility.

Presently, the multiplexers operating at 4.5 µm, 7.6 µm and 5.7 µm are proved their operability and others at 6.9 µm, 3.4 µm and 8.7 µm are either in process of fabrication or to be characterized. After being tested, the devices are to be integrated on chip with miniature narrow band QCL sources with high power output. The next step is to develop an ultra-broad-band combiners that would consist of several multiplexers each covering 100 cm$^{-1}$ spectral range, ultimately allowing multiple gas detecting system.

# Annex I

| Design procedure detained for 33 channel AWG operating at 5.7 μm ||
|---|---|
| Initially specified | Calculated |
| $v_{min}$=1700 cm$^{-1}$, $v_{max}$=1799 cm$^{-1}$, $\Delta v$=3 cm$^{-1}$<br>$\lambda_{min}$ =5.559, $\lambda_{max}$ =5.882,<br>$\lambda_c$=5.716 μm, $\Delta f$=9·10$^{10}$ Hz<br>$f_c$=52.48·10$^{12}$ Hz<br>$\Delta \lambda = 0.01$ | For $\lambda_c$=5.716 μm<br>$n_{eff}$=3.4504 effective index of the rib waveguide (calculated in Matlab mode solver)<br>$n_g$=3.4980 group index of the rib waveguide (calculated in Matlab mode solver)<br>$n_{cl}$=3.4207 effective index of the cladding |
| $w_x$= 4.6 μm waveguide width chosen so that it will support only fundamental mode for the given wavelength and refractive index<br>$w_y$ = 3.0 μm waveguide height | $\Delta L$ length difference between adjacent waveguides is calculated considering the nonuniformity $L_u$ of the order of 3 dB:<br>$\Delta L = \frac{c}{n_g N_{in} \Delta f} = \frac{3 \cdot 10^{14} \mu m/c}{3.4980 \cdot 35 \cdot 9 \cdot 10^{10} \text{ Hz}} = 27.226 \mu m$<br>$m$ order of diffraction: $m = \frac{n_{eff}}{\lambda_c} \Delta L$<br>$m = \frac{27.226 \mu m \cdot 3.4504}{5.716} \mu m \approx 16.4$ |
| $d = 6.6$ μm spacing between array waveguides should be minimal. So it is defined by actual width of the waveguide $w_x$ and technologically acceptable inter distance. | $R$ Rowland radius, i.e. slab length, can be calculated from spatial dispersion per unit length.<br>$D = \frac{dx(f)}{df} = \frac{R \Delta L n_g}{f_c n_s d} \rightarrow R = \frac{f_c n_s dD}{\Delta L n_g}$<br>$R = \frac{11.11 \cdot 10^{-10} \frac{\mu m}{Hz} \cdot 52.48 \cdot 10^{12} \text{ Hz} \cdot 3.4207 \cdot 6.6 \text{ μm}}{26.506 \mu m \cdot 3.4980}$=14197.20 μm<br>R/2 = 7098.60 μm |
| $d_{in}$, $d_{out}$ spacing between input waveguides is 100 μm | $\Delta f_{FSR} = \frac{c}{n_g \Delta L} = \frac{3 \cdot 10^{14} \mu \frac{m}{c}}{3.4980 \cdot 26.506 \text{ μm}} = 3.236 \cdot 10^{12} \text{ } Hz$ |
| $N_{in}$=35 number of input waveguides<br>N out = 17 | $w_{waistx}$ = 3.19 μm, $w_{ey}$ = 2.54 μm waist of the beam (calculated in Phoenix), effective width $w_{ex} = w_{ex}/(2*\sqrt{ln(2)})$ |
| | $D = \frac{dr}{df} = \frac{1.9158 \text{ μm}}{9 \cdot 10^{10} \text{ Hz}} = 21.3 \frac{\mu m}{THz}$ spatial dispersion per unit length defined by spatial distance between input waveguides and wavelength/frequency shift between channels.<br>In Matlab it is calculated as<br>D = 2 * ds$_{max}$ / (c*($\sigma_{max}$ - $\sigma_{min}$)); |

# Annex II

| Procedure of PCG design | | |
|---|---|---|
| parameter | notation | method of calculation |
| number of output channels | $N_{ch}$ | initially specified |
| min/max wavenumber | $\nu_{min}/\nu_{max}$ | initially specified |
| waveguide width, height | $w_x, w_y$ | chosen to insure single-mode propagation |
| effective indices | $n_s, n_{eff}$ | calculated using R-soft multilayer structure |
| group index | $n_g$ | calculated from $n_g = n_{effc} - \lambda_c \frac{dn_{eff}}{d\lambda}$ |
| beam waists | $w_{se}, w_e$ | obtained using Gaussian approximation of the TM fundamental mode |
| free spectral range | $\Delta\lambda_{FSR}$ | chosen minimal before overlap of the main order spectral response with the neighbor |
| diffraction order | m | defined by choosing FSR $$m = \left(\frac{\lambda_c}{\Delta\lambda_{FSR}} + 1\right)\frac{n_s}{n_g} - 1$$ |
| output channel inter-distance | $dr$ | chosen smaller for compact geometry, but large enough not to cause large crosstalk between output channels. Typically $3w_{ex}$ |
| linear dispersion | $D$ | calculated from $D = \frac{dr}{\Delta\lambda_{ch}}$, where $\Delta\lambda_{ch}$ is spectral output channel spacing |
| angular difference $\Delta\theta = \theta_i - \theta_d$ | $\Delta\theta$ | chosen to be small (3° ± 2°), Littrow condition [85] |
| distance between input and central output waveguides | $SP_c$ | chosen smaller for compact geometry. Convenient choice: $SP_c = \frac{(N_{ch}-1)}{2} + 1$ for odd $N_{ch}$, and $SP_c = \frac{N_{ch}}{2} + 1$ for even $N_{ch}$, which means that input waveguide is $dr$ away from the nearest output waveguide |
| angles of incidence and diffraction | $\theta_i, \theta_d$ | calculated using equations: $\theta_d = arctg\left(\frac{(1-B\sin\Delta\theta)}{B(1+\cos\Delta\theta)}\right)$, $B = \frac{n_g SP_c}{Dn_{sc}\lambda_c\Delta\theta}$ and $\theta_i = \theta_d + \Delta\theta$ |
| Rowland radius, grating curvature | $r_R, R$ | calculated from $$R = 2r_R = \frac{Dn_{sc}\lambda_c}{n_g}\left[\frac{\sin\theta_i + \sin\theta_d}{\cos\theta_d}\right]^{-1}$$ |
| distance from input waveguide to the grating center | $r_i$ | calculated from $r_i = R\cos\theta_i$ |
| distance from the grating center to the central output waveguide | $r_d$ | calculated from $r_d = R\cos\theta_d$ |
| grating period | $d$ | calculated from $d(\sin\theta_i + \sin\theta_d) = m\frac{\lambda}{n_s}$ |
| modified grating period, taking into account "shadowing" effect | $d'$ | calculated from $d' = d\frac{\cos\theta_i}{\cos(\theta_i - \theta_{fM})}$ |
| angular dispersion | $\frac{\Delta\theta}{\Delta\lambda}$ | calculated from $\frac{d\theta_d}{d\lambda} = \frac{mn_{sg}}{n_s^2 d\cos\theta_d}$ |
| coordinates on the facet face equidistant from the facet center used to define facet inclination angle | $x_{M1}, z_{M1},$ $x_{M2}, z_{M2}$ | calculated numerically to satisfy condition $r_{iM1} + r_{dM1} = r_{iM2} + r_{dM2}$ |
| angle of facet inclination | $\theta_{fM}$ | calculated from $\theta_{fM} = arctg\left(\frac{z_{M2} - z_{M1}}{x_{M2} - x_{M1}}\right)$ |
| $\theta_i$ with respect to facet normal | $\theta_{iM}$ | calculated from $\theta_{iM} = \theta_i - \left|\theta_{fM}\right|$ |
| $\theta_d$ with respect to facet normal | $\theta_{dM}$ | calculated from $\theta_{dM} = \left|\theta_{fM}\right| - \theta_d$ |
| input waveguide coordinates | $x_{inp}, z_{inp}$ | calculated from $$x_{inp} = -2r_R\sin\theta_i\cos\theta_i;$$ $$z_{inp} = tg\left(\frac{\pi}{2} - \theta_i\right)x_{inp} + r_R$$ |
| central output waveguide coordinates | $x_{oc}, z_{oc}$ | calculated from $$x_{oc} = -2r_R\sin\theta_d\cos\theta_d;$$ $$z_{oc} = tg\left(\frac{\pi}{2} - \theta_d\right)x_{dp} + r_R$$ |
| coordinates of facet centers | $x_M, z_M$ | $x_M$ points are spaced by $d$, $z_M = \sqrt{R^2 - x_M^2}$. |
| right edge coordinates of the facets | $x_{Mr}, z_{Mr}$ | calculated from $$x_{Mr} = x_M - \frac{d_M}{2}\cos\theta_{fM}$$ $$z_{Mr} = z_{M1} + \left(\frac{z_{M2} - z_{M1}}{x_{M2} - x_{M1}}\right)(x_{Mr} - x_{M1})$$ |
| left edge coordinates of the facets | $x_{Ml}, z_{Ml}$ | calculated from $$x_{Ml} = x_M + \frac{d_M}{2}\cos\theta_{fM}$$ $$z_{Ml} = z_{M2} - \left(\frac{z_{M2} - z_{M1}}{x_{M2} - x_{M1}}\right)(x_{Ml} - x_{M2})$$ |

117# List of figures

# List of tables